\documentclass{article}
\usepackage{arxiv}
\usepackage[utf8]{inputenc}
\usepackage[T1]{fontenc}
\usepackage{newtxtext}
\usepackage{amsmath,amssymb,amsfonts}
\usepackage{mathtools}
\usepackage{amsthm}
\usepackage{bm}
\usepackage{graphicx}
\usepackage{array}
\usepackage{booktabs}
\usepackage{algorithm}
\usepackage{algpseudocode}
\usepackage{xcolor}
\usepackage{tikz}
\usepackage{url}
\usepackage{microtype}
\usepackage{natbib}
\usepackage[hidelinks,hypertexnames=false]{hyperref}
\graphicspath{{./ms_figures/}}
\bibpunct[, ]{(}{)}{,}{a}{}{,}

\defcitealias{FederalReserveOCCFDIC2026}{SR~26-2}
\defcitealias{FederalReserveOCC2011}{SR~11-7}
\defcitealias{FederalReserveSR2108}{SR~21-8}
\newcommand{\KL}{\mathrm{KL}}
\newcommand{\kl}[2]{\KL(#1\,\|\,#2)}
\newcommand{\dd}{\mathrm{d}}
\newcommand{\EE}{\mathbb{E}}
\newcommand{\RR}{\mathbb{R}}
\DeclareMathOperator{\Var}{Var}
\DeclareMathOperator{\SR}{SR}
\DeclareMathOperator{\VIX}{VIX}
\newcommand{\cl}{\operatorname{cl}}
\newcommand{\PP}{\mathbb{P}}
\newcommand{\Pz}{\mathbb{P}_0}
\newcommand{\QQ}{\mathbb{Q}}
\newcommand{\QS}{\mathbb{Q}^{S}}
\newcommand{\QE}{\mathbb{Q}_{E}}
\newcommand{\QL}{\mathbb{Q}_{L}}
\newcommand{\QW}{\mathbb{Q}^{W^*}}
\newcommand{\Pc}{\mathbb{P}_{c}}
\newcommand{\Qc}{\mathbb{Q}_{c}}
\newcommand{\Pg}{\mathbb{P}^{(\gamma)}}
\newcommand{\Qg}{\mathbb{Q}^{(\gamma)}}
\newcommand{\Mset}{\mathcal{M}}
\newcommand{\Ball}{\mathcal{B}}
\newcommand{\Bc}{\mathcal{B}_{c}}
\newcommand{\cbar}{\bar{c}}
\newcommand{\defect}{\delta}
\newcommand{\Lent}{L}
\newcommand{\DeltaS}{\Delta S}
\newcommand{\Wstar}{W^{*}}
\newcommand{\hstar}{h^{*}}
\newcommand{\Wg}{W^{(\gamma)}}
\newcommand{\Wc}{W_{c}}

\theoremstyle{plain}
\newtheorem{theorem}{Theorem}
\newtheorem{lemma}{Lemma}
\newtheorem{proposition}{Proposition}
\newtheorem{corollary}{Corollary}
\newtheorem{assumption}{Assumption}
\theoremstyle{definition}
\newtheorem{definition}{Definition}
\newtheorem{remark}{Remark}
\newtheorem{example}{Example}
\newtheoremstyle{restated}{\topsep}{\topsep}{\itshape}{}{\bfseries}{}{ }{\thmnote{#3}}
\theoremstyle{restated}
\newtheorem*{repeattheorem}{Theorem}
\newtheorem*{repeatlemma}{Lemma}
\newtheorem*{repeatproposition}{Proposition}

\newenvironment{proof}[1]{\par\addvspace{6pt plus 3pt}\noindent\textit{#1}\enskip\ignorespaces}{\par\addvspace{6pt plus 3pt}}
\newcommand{\Halmos}{\leavevmode\unskip\penalty9999 \hbox{}\nobreak\hfill\quad\hbox{$\square$}}
\DeclareMathOperator*{\argmin}{arg\,min}
\DeclareMathOperator*{\argmax}{arg\,max}
\newsavebox{\arxivtablebox}
\newcommand{\TABLE}[3]{\caption{#1}\centering\sbox{\arxivtablebox}{#2}%
  \ifdim\wd\arxivtablebox>\linewidth\resizebox{\linewidth}{!}{\usebox{\arxivtablebox}}\else\usebox{\arxivtablebox}\fi\par
  \if\relax\detokenize{#3}\relax\else\vspace{4pt}{\footnotesize\raggedright #3\par}\fi}
\newcommand{\FIGURE}[3]{\caption{#2}\vspace{4pt}\centering #1\par
  \if\relax\detokenize{#3}\relax\else\vspace{4pt}{\footnotesize\raggedright\textit{Note.}\enskip #3\par}\fi}

\title{Model Doubt, Risk Aversion, and the Threshold for Trading: A Geometry of Pricing Measures}

\author{%
Amine Mohamed Aboussalah\\
Department of Finance and Risk Engineering\\
NYU Tandon School of Engineering, New York University\\
Brooklyn, NY, USA\\
\texttt{ama10288@nyu.edu}}

\date{November 2026}
\renewcommand{\shorttitle}{Model Doubt and the Threshold for Trading}
\hypersetup{
pdftitle={Model Doubt, Risk Aversion, and the Threshold for Trading: A Geometry of Pricing Measures},
pdfauthor={Amine Mohamed Aboussalah},
pdfkeywords={model uncertainty, robust pricing, relative entropy, risk premium, incomplete markets}
}

\begin{document}
\maketitle

\begin{abstract}
How much model doubt can an investor bear before they stop trading? Our investor treats every model within a given Kullback--Leibler divergence of their reference model as plausible. In a finite one-period market, they hold no risky asset exactly when this doubt reaches the smallest divergence of a pricing measure from their model, for every smooth, strictly increasing, strictly concave utility; non-traded claims then have intervals of no-trade prices. Below this threshold, for log utility and in a given market, doubt acts exactly like higher constant relative risk aversion; their robust pricing measure moves from the log-utility measure toward the pricing measure at that smallest divergence. With a century of U.S. data, a log investor who doubts the Black--Scholes mean return over its 95\% confidence interval acts like a fully trusting investor with relative risk aversion of about 1.8 instead of 1. Implied doubt turns a quote into a radius; the mean VIX implies doubt beyond the threshold, but whether significantly so depends on the reference model. By a known identity, real-world, pricing, and share (stock-numeraire) measures form a triangle whose Pythagorean defect is a log risk premium; one side is half the squared (ideal) VIX times the horizon. Since 1990, the U.S. defect averages about 10\% per year, about 1 point of it entropy premium.
\end{abstract}


\section{Introduction}\label{sec:intro}

At the heart of investment decision-making lies a fundamental tension: how much uncertainty can an investor tolerate before they stop trading? We model doubt as a radius in the space of probability measures: the investor treats every model within a Kullback--Leibler (KL) divergence $c$ of the reference model $\Pz$ as plausible. Our main result is a threshold $\bar{c}$---the smallest KL divergence of a pricing measure from the reference model---at which the investor stops holding risky assets. The reason is simple: once a pricing measure is plausible, no risky position beats the risk-free bond in the worst case within the radius. The threshold separates two regimes. Below $\bar{c}$ and with log utility, doubt acts like higher constant relative risk aversion; above $\bar{c}$, it gives an interval of no-trade prices for claims that the market does not trade.

The threshold comes from the gap between real-world and pricing measures. In an arbitrage-free market, prices are expected discounted payoffs under a pricing measure $\QQ$, not under the real-world measure $\PP$ \citep{HarrisonKreps1979,HarrisonPliska1981}. The gap has three sources. First, investors demand risk premia, so $\QQ$ puts more weight on bad outcomes than $\PP$. Second, in an incomplete market many pricing measures fit all traded prices, so the value of a non-traded claim depends on which measure is chosen \citep{Davis1997,Frittelli2000,GollRuschendorf2001}. Third, $\PP$ is estimated from limited data and may be wrong, and a wrong $\PP$ leads to wrong portfolios and wrong prices \citep{HansenSargent2001,Cont2006}.

We study all three sources with the KL divergence---also known as relative entropy---defined as $\kl{\PP}{\QQ}:=\EE_{\PP}[\ln(\dd\PP/\dd\QQ)]$, the expected log-likelihood ratio of $\PP$ against $\QQ$ \citep{KullbackLeibler1951}. Each direction of the divergence has an economic meaning. For an investor with log utility who believes in $\PP$ and can trade every claim at the prices of $\QQ$, $\kl{\PP}{\QQ}$ is the best expected log growth in excess of a risk-free bank account. For an investor with exponential utility and unit absolute risk aversion in the same position, $\kl{\QQ}{\PP}$ is the best certainty-equivalent gain, in units of the bank account (Lemma~\ref{lem:duality}). Relative entropy has a long history in asset pricing \citep{Stutzer1996,AlvarezJermann2005,BackusChernovZin2014}; we add its geometry \citep{AmariNagaoka2000,Amari2016}.

In this geometry each probability measure is a point, and two kinds of lines join the points: mixture lines, which average probabilities, and exponential lines, which average log-probabilities and renormalize. The KL divergence is not a distance, but it obeys an exact law of cosines (Lemma~\ref{lem:threepoint}; \citealt{ChenTeboulle1993,Nielsen2021}). Its correction term is the ``Pythagorean defect''---$\defect(\mu,\nu,\rho):=\kl{\mu}{\nu}+\kl{\nu}{\rho}-\kl{\mu}{\rho}$---which measures the inner product at $\nu$ between the mixture line toward $\mu$ and the exponential line toward $\rho$. A zero defect means a right angle. Right angles matter here because they appear at projections onto the set of pricing measures, which is defined by linear equations. For example, the minimal-entropy measure $\QE$ minimizes $\kl{\cdot}{\PP}$ over this set \citep{Frittelli2000}, so for every other pricing measure $\QQ$ the triangle of $\QQ$, $\QE$ and $\PP$ has a right angle at $\QE$ \citep{Csiszar1975}. The log-utility measure $\QL=\PP/\Wstar$ is the other projection: it minimizes $\kl{\PP}{\cdot}$ instead, where $\Wstar$ is the gross return of the growth-optimal portfolio (the portfolio with the largest expected log return); at $\QL$, however, the Pythagorean identity fails in general (Section~\ref{sec:set-proj}). Figure~\ref{fig:triangle} in Section~\ref{sec:setting} shows both projections.

\subsection{Main Results}

We define the set of plausible models as a ball $\Bc := \{\PP : \kl{\PP}{\Pz} \le c\}$, as in \citet{HansenSargent2001}. Under log utility, the investor's best guaranteed growth---the robust value $v(c)$---is the smallest KL divergence between a model in this ball and a pricing measure (Theorem~\ref{thm:minimax}). Geometrically, the threshold is the radius $\bar{c} := \kl{\QE}{\Pz}$ at which the ball first touches the set $\Mset$ of pricing measures, where $\QE$ is the minimal-entropy measure of $\Pz$. From this radius on, the robust investor holds no risky asset. The threshold does not depend on the utility function (Proposition~\ref{prop:utilityfree}). It gives the no-trade mechanism for max--min preferences \citep{DowWerlang1992,EpsteinWang1994} an exact location for the KL ball, $\bar{c}$.

Below $\bar{c}$, log utility gives an explicit description of the solution (Theorem~\ref{thm:robust}). The robust portfolio is the optimal portfolio of an investor who trusts $\Pz$ and has constant relative risk aversion (CRRA) $\gamma_c = 1 + 1/\lambda_c$. Here $\lambda_c = -v'(c)$ is the ``shadow price of doubt''---the loss in guaranteed growth per unit increase of the doubt radius. The map $c \mapsto \gamma_c$ is a continuous, strictly increasing bijection from $[0, \bar{c})$ onto $[1, \infty)$: more doubt means more effective risk aversion. The conversion is exact for every radius below $\bar{c}$, not only for small doubt, and it holds in incomplete one-period markets with finitely many states, so doubt can be reported in a unit that managers already use. The robust pricing measure $\Qc$---this investor's marginal-utility pricing measure \citep{Davis1997}---moves from $\QL(\Pz)$ at $c=0$ toward $\QE$ as $c$ approaches $\bar{c}$.

The threshold $\bar{c}$ is also a boundary for the pricing of non-traded claims $X$ (Theorem~\ref{thm:pricing}). We define a price $p$ as a ``no-trade'' price if offering claim $X$ at that price does not increase the investor's guaranteed growth (Definition~\ref{def:notrade}). Below $\bar{c}$, this no-trade price is unique and coincides with the marginal-utility price $\EE_{\Qc}[X]$ of a CRRA investor with risk aversion $\gamma_c$. So in this regime the robust investor's portfolio and no-trade prices are exactly those of a CRRA investor, with an ``effective'' risk aversion that depends on both the doubt radius $c$ and the market (Remark~\ref{rem:scope}). Above $\bar{c}$, the no-trade prices form an interval $I_c(X)$, again independent of the utility function (Proposition~\ref{prop:utilityfree}). This interval is the ``entropic good-deal interval'' of \citet{Cerny2003}: the prices under pricing measures $\QQ$ satisfying $\kl{\QQ}{\QE} \le c - \bar{c}$.

We define ``implied doubt'' as the smallest radius at which a market quote is a no-trade price (Definition~\ref{def:implieddoubt}). Implied doubt does for model uncertainty what implied volatility does for option prices: it turns a market quote into one number, the radius of doubt that the quote requires for a given reference model. We apply this concept to the VIX---the Cboe volatility index---which gives the only market price in our data for a claim that the stock and bond do not span: the short log contract, which pays minus the stock's log excess return (Section~\ref{sec:numerics-vix}). The mean VIX quote lies above the narrow range of robust prices associated with $c < \bar{c}$, the marginal-utility prices of CRRA investors with relative risk aversion of at least one who trust the empirical return distribution. With the returns of 1990--2026, it lies 60\% above this range (95\% bootstrap interval: 37\% to 83\%); much of this gap comes from the low variance of monthly returns, relative to the sum of daily variances, in these years: with a reference model matched to the daily realized variance, the gap is 17\%, at the edge of significance (Section~\ref{sec:numerics-vix}). Over a century of returns, the gap is 13\%, which is not statistically significant. For a quote above this range, the implied doubt is the threshold $\bar{c}$ plus a quote-specific part. With the 1990--2026 reference model, where the gap is significant, the mean VIX quote is a no-trade price for the robust log investor only at radii where they hold no stock. So the mean VIX carries a premium that KL-ball doubt cannot explain while this investor still holds stock, and the quote-specific part of implied doubt measures how far beyond the threshold the quote lies.

To make the main results concrete, we take the Black--Scholes model with doubt about the drift. There, the results have closed forms: $v(c)=(\sqrt{\bar{c}}-\sqrt c)^2$ and $\gamma_c=1/(1-\sqrt{c/\bar{c}})$ (Proposition~\ref{prop:bsdoubt}). This gives a calibration of the doubt radius to a standard confidence interval. If the doubt radius $c$ corresponds to a confidence interval of $z$ standard errors around a Sharpe ratio estimated from $n$ years of data, then $\gamma_c=\tau/(\tau-z)$, where $\tau$ is the $t$-statistic of the mean excess return. So the robust investor holds stock only if the $t$-statistic exceeds the number of standard errors ($\tau > z$), as in the single-asset rule of \citet{GarlappiUppalWang2007}. We apply this to U.S. market returns from 1926 to 2026. With doubt covering a 95\% confidence interval, these returns give $\gamma_c=1.76$ (or 1.86 when adjusting the standard error for serial correlation; Section~\ref{sec:numerics-doubt}). The example shows the role of data length in this calibration: with the same Sharpe ratio, a robust investor with 18.66 years of data or fewer would hold no stock at all (Table~\ref{tab:emp-doubt}).

We also read the risk premium off a triangle of measures. For the gross return $R := S_T / (S_0 B_T)$ of a stock $S$ relative to the bank account $B$ over the horizon $T$, the share measure $\QS$, with $\dd\QS/\dd\QQ = R$, treats the stock as the unit of account \citep{GemanElKarouiRochet1995}. The Pythagorean defect of the triangle of $\PP$, $\QQ$ and $\QS$ is $\EE_{\PP}[\ln R] - \EE_{\QQ}[\ln R]$ (Theorem~\ref{thm:triangle}), a known identity of \citet{Soklakov2023}. With deterministic interest rates, each side has a financial meaning: $\kl{\PP}{\QQ}$ is the entropy of the pricing kernel, an upper bound on expected log excess returns \citep{BansalLehmann1997,AlvarezJermann2005,BackusChernovZin2014}; $\kl{\PP}{\QS}$ is the exact ``slack'' of this bound for the stock (Proposition~\ref{prop:slack}); and $\kl{\QQ}{\QS} = -\EE_{\QQ}[\ln R]$ is the forward price of the short log contract, half the squared VIX times the horizon \citep[Result~3]{Martin2017} up to the discretization of the published index \citep{Cboe2026}. This defect splits exactly into two parts (Proposition~\ref{prop:decomp}): a log equity premium $\pi := \ln\EE_{\PP}[R]$ and an entropy premium $\varepsilon$, which is the excess of $\ln\EE[R]-\EE[\ln R]$ under $\QQ$ over that under $\PP$.

We measure this decomposition using U.S. data from 1990 onward. The defect averages approximately 10.1\% per year (95\% interval: 5.0\% to 15.0\%), of which the entropy premium is 1.02\% (interval: 0.81\% to 1.27\%; Table~\ref{tab:emp-triangle}). While these numbers restate the log equity premium and a monthly variance premium, the triangle adds one thing: it anchors one of its sides in an observable market price---the VIX quote analyzed in Section~\ref{sec:numerics-vix}.

A further finding, about one-period incomplete markets with finitely many states, concerns the two projections $\QE$ and $\QL$. For any pricing measure $\QQ$, each direction of the discrepancy between $\PP$ and $\QQ$ splits exactly into two parts (Theorem~\ref{thm:twodecomp}): one is fixed by traded prices, and the other is the value of completing the market at the prices of $\QQ$. While the decomposition around the minimal-entropy measure $\QE$ is well known \citep{Csiszar1975,Frittelli2000}, our decomposition around the log-utility measure $\QL$ is, as far as we know, new in this form; it follows from Proposition~\ref{prop:slack} applied to the growth-optimal portfolio. It gives an exact alternative to the natural Pythagorean form at $\QL$, which fails in general.

\subsection{Contributions and Related Literature}

We see three results that, as far as we know, are new; we list them in order of importance. (1) The threshold $\cbar$ is the minimal relative entropy of the market, $\cbar=\kl{\QE}{\Pz}$, so the ratio $c/\cbar$ is a diagnostic that does not depend on utility. (2) Below $\cbar$, with log utility and in a given finite market, doubt acts exactly like risk aversion: we give the map $c\mapsto\gamma_c$ and show that the robust pricing measure $\Qc$ is the CRRA($\gamma_c$) marginal-utility measure. Because the conversion is exact below $\cbar$, a doubt radius can be reported as relative risk aversion, and it selects one member of a known family of tilts. (3) We define implied doubt, the smallest radius at which a market quote is a no-trade price; the implied ambiguity of \citet{HaraHonda2022}, the smallest ambiguity aversion at which a given portfolio is optimal, is a related inverse problem. For a quote outside the range of robust prices, implied doubt splits into the threshold $\cbar$ and a quote-specific part. Applied to the mean VIX, it gives the doubt that the average market price of the short log contract requires, in units of each reference model's own $\cbar$: $4.99\,\cbar$ with the 1990--2026 reference model and $1.30\,\cbar$ with the 1926--2026 one. Four further results sharpen known results or give them a new form. (4) We prove that, for $c\ge\cbar$, the no-trade interval $I_c(X)$ equals the entropic good-deal interval $J_{c-\cbar}(X)$ with budget $c-\cbar$. (5) In the Black--Scholes model, we give the closed forms of $v(c)$ and $c\mapsto\gamma_c$ for the KL ball and their reading in years of data. (6) We prove Theorem~\ref{thm:twodecomp}(b), and we read and measure the known triangle, with the split $\defect=\pi+\varepsilon$. (7) We give the local rate $1-1/\gamma_c$ of alternating projections, attained when there are at least three more states than risky assets; their convergence is due to \citet{CsiszarTusnady1984}.

While our proofs rest on convex duality---the foundation of the work we build upon---geometry adds three things. First, it explains why the Pythagorean identity holds at $\QE$ but fails in general at $\QL$, and identifies what replaces it (Theorem~\ref{thm:twodecomp}(b)). Second, it places the worst-case model on the exponential line from the reference model to the robust pricing measure, at position $1-1/\gamma_c$ (Theorem~\ref{thm:robust}(iii)); this position fixes the effective risk aversion and bounds the local rate of alternating projections (Proposition~\ref{prop:algorithms}). Finally, it reads the threshold $\cbar$ as the radius at which the ball of plausible models first touches the set of pricing measures (Figure~\ref{fig:triangle}). \citet{GaoLimNg2004} also use differential geometry, to approximate the minimal-distance martingale measure of \citet{GollRuschendorf2001}.

These results build on three literatures: robust utility and robust optimization, non-participation under ambiguity, and robust portfolio choice. Theorem~\ref{thm:minimax} is the log-utility, KL-ball case of robust utility duality \citep{Gundel2005,SchiedWu2005,FollmerGundel2006}, where the worst case is given by the exponential tilt of KL-constrained distributionally robust optimization \citep{HuHong2012,BenTalDenHertogDeWaegenaereEtAl2013}. When robustness is modeled with a penalty instead of a ball, it acts like risk aversion \citep{Tallarini2000,Maenhout2004,BarillasHansenSargent2009}, and these two forms are dual to one another \citep{PetersenJamesDupuis2000,HansenSargentTurmuhambetovaEtAl2006}. For a mean--variance investor, max--min doubt measured by relative entropy is observationally equivalent to higher risk aversion \citep{Taboga2006}. Specific results exist for well-known models: in the Black--Scholes model, the robust weight below the threshold is the one under the related $\kappa$-ignorance of \citet{ChenEpstein2002} and equals the single-asset multi-prior rule of \citet{GarlappiUppalWang2007}, while in Merton's model closed forms are known for a bound on the rate of relative entropy \citep{TrojaniVanini2002}. Section~\ref{sec:ec-rob-literature} relates Theorem~\ref{thm:robust} to this work.

Non-participation under ambiguity has a long history. A max--min investor \citep{GilboaSchmeidler1989} holds no risky asset when some plausible model prices every traded asset at its expected payoff. They also do not trade a new claim at a price that such a model supports \citep{DowWerlang1992,EpsteinWang1994,CaoWangZhang2005}; see also \citet{Illeditsch2011} and, for portfolio choice with multiple priors, \citet{BoyleGarlappiUppalWang2012}. We locate this point exactly for the KL ball, at $\cbar$, and describe what happens on both sides of it. When the reference model is an empirical distribution, the objects at the threshold have names in econometrics: $\QE$ is the canonical valuation measure of \citet{Stutzer1996}, and $\cbar$ is the minimized Kullback--Leibler information criterion of exponential tilting \citep{KitamuraStutzer1997}, which \citet{Stutzer1995} uses as an entropy bound to diagnose asset-pricing models. The measure $\QL$ gives the empirical-likelihood weights, and the CRRA measures between $\QL$ and $\QE$ belong to the Cressie--Read family of tilts \citep{CressieRead1984} that \citet{HaleyWalker2010} compare for option pricing; \citet{AlmeidaGarcia2017} derive bounds on pricing kernels from a family of discrepancy functions of this kind. Neural networks give a data-driven route to option prices that keeps the structure of a pricing model: \citet{AboussalahLiChiEtAl2024} embed the pricing equation in the network and read hedge ratios as curvature on the manifold of option prices.

In robust portfolio choice, divergence penalties act like a mean--variance criterion \citep{GotohKimLim2018}, KL ambiguity gives tractable portfolios \citep{Calafiore2007}, and high ambiguity makes the equally weighted portfolio optimal when no riskless asset trades \citep{PflugPichlerWozabal2012}. With a riskless asset, high ambiguity leads to the bond alone (Theorem~\ref{thm:robust}(vi)). Other literature explores specific robustness criteria. For instance, \citet{RujeerapaiboonKuhnWiesemann2016} maximize the worst-case value-at-risk of portfolio growth under moment ambiguity. Similarly, \citet{SunBoyd2018} maximize worst-case expected log growth---the criterion of our investor---over sets of distributions with finitely many outcomes, among which are KL balls; see also \citet{BlanchetChenZhou2022} and \citet{KuhnShafieeWiesemann2025}. Reinforcement learning learns portfolio policies directly from return data \citep{AboussalahLee2020,AboussalahLeeSymmetry2020}, and \citet{AboussalahXuLee2022} train such a policy with a critic that maximizes geometric capital growth, the criterion of \citet{Kelly1956} and of our investor at zero doubt.

\subsection{Why the Results Matter for Managers}

Model risk, the risk of acting on a wrong model, is subject to supervisory guidance for banks \citepalias{FederalReserveOCCFDIC2026}. Our results offer three applications for risk managers, model validators, and trading desks. They apply directly to the models these users already run: historical simulation and scenario analysis value positions over finitely many scenarios with given probabilities, which is our one-period market with reference model $\Pz$; the same finite-scenario models underlie stochastic programming \citep{BirgeLouveaux2011}, and \citet{Birge2000} uses option pricing to bring market risk into capacity planning. First, once a stated rule fixes the radius, model risk becomes a number. The ratio $c/\cbar$ is the first diagnostic for every investor with a smooth, strictly increasing, and strictly concave utility: the investor holds no risky asset exactly when it reaches one. With an empirical reference model and the test radius of Section~\ref{sec:rob-radius}---the radius that matches a standard test of zero risk premia---the ratio is below one exactly when this test rejects. Implied doubt turns a quote into a radius, and a report gives it with the quote's distance from the range of robust prices and its side. Second, with log utility and in a given market, doubt can be reported as risk aversion. Third, a non-traded claim gets one robust price below the threshold and an interval of no-trade prices above it. These diagnostics concern the model risk of one investor; risk that spreads through a network of institutions needs network tools \citep{AboussalahChiLee2023}.

\section{Setting and the Geometry We Use}\label{sec:setting}

This section introduces the geometric tools of the paper, while Section~\ref{sec:ec-geometry} is a short guide to the geometry behind them. Section~\ref{sec:triangle} then reads a known triangle of measures as a risk premium, Section~\ref{sec:incomplete} sets up the incomplete market, Section~\ref{sec:robust} gives the main results on doubt, and Sections~\ref{sec:numerics} and~\ref{sec:implications} measure and apply them. We view each probability measure on a measurable space $(\Omega,\mathcal{F})$ as a point, so that we can draw lines and triangles between measures. Two measures $\mu$ and $\nu$ are equivalent ($\mu\sim\nu$) if they share the same null sets; for such measures, the density $\dd\mu/\dd\nu$ of $\mu$ with respect to $\nu$ exists and is positive almost surely, telling how much more likely each outcome is under $\mu$ than under $\nu$. On a finite state space $\Omega=\{\omega_1,\dots,\omega_M\}$, a measure is represented as a vector of $M$ probabilities, where the vectors with positive entries form the open simplex $\mathcal{P}$ (Figure~\ref{fig:triangle}).

The points in this geometry are models and pricing measures. The real-world measure $\PP$ describes how likely events are, while under a pricing measure $\QQ\sim\PP$, an equivalent martingale measure, a payoff $X$ at horizon $T$ costs $\EE_{\QQ}[X/B_T]$ today, where $B_T$ is the value at time $T$ of one unit invested in the bank account today. We also consider the share measure $\QS$ from Section~\ref{sec:triangle}, which is the pricing measure that uses the traded asset as its unit of account.

\subsection{The Kullback--Leibler Divergence}\label{sec:set-kl}

We measure the gap from $\mu$ to $\nu$ by the Kullback--Leibler (KL) divergence \citep{KullbackLeibler1951}, defined as $\kl{\mu}{\nu}:=\EE_{\mu}[\ln(\dd\mu/\dd\nu)]$ if $\mu\ll\nu$, that is, if every $\nu$-null set is $\mu$-null, and $\kl{\mu}{\nu}:=+\infty$ otherwise. KL is the average log-likelihood ratio of $\mu$ against $\nu$ under $\mu$; by Jensen's inequality, it is nonnegative, and zero only if $\mu=\nu$. It is not a distance: in general $\kl{\mu}{\nu}\neq\kl{\nu}{\mu}$, and the triangle inequality fails. So the order of the arguments matters: we average under the first argument. Each order is also the value of an investment problem (Lemma~\ref{lem:duality}). KL also has an exact law of cosines and exact projections \citep{Csiszar1975,AmariNagaoka2000}, so some bounds become exact decompositions (Proposition~\ref{prop:slack}, Theorem~\ref{thm:twodecomp}).

\subsection{Mixture Lines and Exponential Lines}\label{sec:set-lines}

Let $\mu$, $\nu$, and $\rho$ be mutually equivalent probability measures. The mixture line from $\nu$ toward $\mu$
averages them,
\begin{equation}\label{eq:set-mline}
m_t:=(1-t)\,\nu+t\,\mu,\qquad t\in[0,1],
\end{equation}
so the probability of every event moves linearly from its value under $\nu$ to its value under $\mu$. The exponential
line from $\nu$ toward $\rho$ averages log-densities instead,
\begin{equation}\label{eq:set-eline}
\frac{\dd e_t}{\dd \nu}:=\frac{(\dd \rho/\dd \nu)^{t}}{\EE_{\nu}\bigl[(\dd \rho/\dd \nu)^{t}\bigr]},\qquad t\in[0,1].
\end{equation}
Along it, $\ln(\dd e_t/\dd\nu)$ moves linearly in $t$ from $0$ to $\ln(\dd\rho/\dd\nu)$, up to the normalizing constant
$\EE_{\nu}[(\dd\rho/\dd\nu)^{t}]$. This constant is positive because $\dd\rho/\dd\nu>0$, and at most one by Jensen's
inequality, because $\EE_{\nu}[\dd\rho/\dd\nu]=1$ and $x\mapsto x^{t}$ is concave for $t\in[0,1]$; so each $e_t$ is a
probability measure equivalent to $\nu$. In finance, $e_t$ is an exponential tilt (an Esscher transform) of
$\nu$. We need both lines, because the set $\Mset$ of pricing measures contains the mixture lines between its points but not, in general, the
exponential lines (Section~\ref{sec:set-proj}).

\begin{figure}[t]
\FIGURE{%
\ifdefined\tikzpicture
\begin{tikzpicture}[font=\small, >=stealth, yscale=0.8,
  pt/.style={circle, fill=black, inner sep=0pt, minimum size=4.2pt},
  mline/.style={line width=0.75pt},
  eline/.style={line width=0.75pt, dash pattern=on 4.5pt off 2.5pt},
  vel/.style={->, line width=1.2pt}]
\coordinate (Q) at (0.8,0.45);
\coordinate (P) at (2.0,3.35);
\coordinate (QS) at (6.4,1.3);
\draw[densely dotted, line width=0.6pt] (P) -- (QS)
  node[pos=0.48, above right=-1pt] {$\kl{\PP}{\QS}$};
\draw[mline] (Q) -- (P) node[pos=0.6, left=2pt] {$\kl{\PP}{\QQ}$};
\draw[eline] (Q) .. controls +(24:2.2) and +(170:1.7) .. (QS)
  node[pos=0.55, below=3pt] {$\kl{\QQ}{\QS}$};
\draw[line width=0.5pt] (Q) ++(24:0.62) arc[start angle=24, end angle=67.5, radius=0.62];
\draw[vel] (Q) -- +(67.5:1.2) node[right=1pt] {$\dot m$};
\draw[vel] (Q) -- +(24:1.2) node[below right=-1pt] {$\dot e$};
\node[pt] at (Q) {}; \node[pt] at (P) {}; \node[pt] at (QS) {};
\node[below left=0pt] at (Q) {$\QQ$};
\node[above left=0pt] at (P) {$\PP$};
\node[below right=0pt] at (QS) {$\QS$};
\node at (3.4,-1.1) {(a) The triangle with corner $\QQ$};
\begin{scope}[xshift=8.7cm]
\coordinate (P0) at (1.7,2.1);
\coordinate (QL) at (1.7,0);
\coordinate (QE) at (4.0,0);
\coordinate (Pc) at (2.85,1.05);
\coordinate (Qc) at (2.85,0);
\fill[black!12, rotate around={-24:(2.2671,1.9409)}] (2.2671,1.9409) ellipse [x radius=1.4, y radius=0.75];
\draw[line width=0.5pt, rotate around={-24:(2.2671,1.9409)}] (2.2671,1.9409) ellipse [x radius=1.4, y radius=0.75];
\draw[densely dotted, line width=0.7pt, rotate around={-24:(2.8343,1.7818)}] (2.8343,1.7818)
  ellipse [x radius=2.8, y radius=1.5];
\draw[line width=1.6pt] (-0.3,0) -- (6.2,0) node[right] {$\Mset$};
\draw[mline] (P0) -- (QL);
\draw[eline] (P0) .. controls +(1.4,-0.1) and +(0,1.15) .. (QE);
\draw[eline] (P0) .. controls +(0.7,-0.3) and +(-0.25,0.45) .. (Pc)
  .. controls +(0.25,-0.45) and +(0.22,0.4) .. (Qc);
\draw[line width=1.3pt] (Pc) -- (Qc) node[pos=0.3, left=0pt] {$v(c)$};
\foreach \x in {1.7,4.0} {\draw[line width=0.45pt] (\x,0) ++(0.16,0) -- ++(0,0.16) -- ++(-0.16,0);}
\draw[line width=0.45pt] (2.85,0) ++(-0.16,0) -- ++(0,0.16) -- ++(0.16,0);
\node[pt] at (P0) {}; \node[pt] at (QL) {}; \node[pt] at (QE) {}; \node[pt] at (Pc) {}; \node[pt] at (Qc) {};
\node[above left=-1pt] at (P0) {$\Pz$};
\node[above right=-1pt] at (Pc) {$\Pc$};
\node[below=2pt] at (QL) {$\QL$};
\node[below=2pt] at (Qc) {$\Qc$};
\node[below=2pt] at (QE) {$\QE$};
\node at (2.6,2.4) {$\Bc$};
\node at (5.05,3.05) {$\Ball_{\cbar}$};
\node at (3.0,-1.1) {(b) The set $\Mset$, its two projections, and the ball $\Bc$};
\end{scope}
\end{tikzpicture}%
\else
\fbox{\parbox{0.92\textwidth}{\centering\small Figure~1 is drawn with TikZ: add \texttt{\string\usepackage\{tikz\}}
to the preamble.}}%
\fi}
{Schematic of the Geometry Used in the Paper\label{fig:triangle}}
{Not to scale. The plane is the simplex in mixture coordinates: mixture lines (solid) are straight, exponential
lines (dashed) are curved, and small squares mark right angles in the Fisher inner product. (a) The triangle of the
real-world measure $\PP$, a pricing measure $\QQ$ and the share measure $\QS$; by Lemma~\ref{lem:threepoint}(b), its
defect is the inner product of the velocities $\dot m$ toward $\PP$ and $\dot e$ toward $\QS$ (the dotted segment only
marks the third side). (b) The set $\Mset$ of pricing measures and a reference model $\Pz$. The mixture and exponential
lines from $\Pz$ that meet $\Mset$ at a right angle end at the log-utility measure $\QL$ and at the minimal-entropy
measure $\QE$. The dotted ball $\Ball_{\cbar}$ is the smallest ball around $\Pz$ that touches $\Mset$, at $\QE$ only;
its radius $\cbar$ is the threshold of Section~\ref{sec:robust}. The shaded ball $\Bc$ holds the models within KL
divergence $c<\cbar$ of $\Pz$ (not round in these coordinates); its nearest pair $(\Pc,\Qc)$ with $\Mset$ has
$\kl{\Pc}{\Qc}=v(c)$, the robust value, and $\Pc$ lies on the exponential line from $\Pz$ to $\Qc$
(Theorem~\ref{thm:robust}(iii)).}
\end{figure}

\subsection{The Three-Point Identity}\label{sec:set-threepoint}

Three points $\mu$, $\nu$, and $\rho$ with finite divergences form a triangle with corner $\nu$. Its Pythagorean defect
is
\begin{equation}\label{eq:set-defect}
\defect(\mu,\nu,\rho):=\kl{\mu}{\nu}+\kl{\nu}{\rho}-\kl{\mu}{\rho}.
\end{equation}
It compares the two sides that meet at $\nu$ with the third side. In the plane, the same combination of half squared
lengths is the inner product of the two sides at $\nu$ (the law of cosines), so it vanishes exactly when the angle at
$\nu$ is right. Lemma~\ref{lem:threepoint} shows that KL behaves in the same way. The lemma is known: part (a)
is the three-point property of Bregman divergences in its KL form \citep{ChenTeboulle1993,AmariNagaoka2000}, and
part (b) appears in \citet{Nielsen2021} and \citet{Soklakov2023}; Section~\ref{sec:ec-geometry} gives the proof.

\begin{lemma}[Three-point identity]\label{lem:threepoint}
Let $\mu$, $\nu$, and $\rho$ be mutually equivalent probability measures such that $\kl{\mu}{\nu}$, $\kl{\nu}{\rho}$,
and $\kl{\mu}{\rho}$ are finite, and let $\ell:=\ln(\dd\rho/\dd\nu)$.
\begin{itemize}
\item[(a)] The function $\ell$ is integrable under $\mu$ and under $\nu$, and
\begin{equation}\label{eq:set-threepoint}
\defect(\mu,\nu,\rho)=\EE_{\mu}[\ell]-\EE_{\nu}[\ell].
\end{equation}
Hence $\kl{\mu}{\rho}=\kl{\mu}{\nu}+\kl{\nu}{\rho}$ (exact Pythagoras) if and only if $\EE_{\mu}[\ell]=\EE_{\nu}[\ell]$.
\item[(b)] The defect is the covariance under $\nu$ of $\dd\mu/\dd\nu$ and $\ell$:
$\defect(\mu,\nu,\rho)=\EE_{\nu}\bigl[(\dd\mu/\dd\nu-1)\,(\ell-\EE_{\nu}[\ell])\bigr]$. On a finite state space with
$\mu,\nu,\rho\in\mathcal{P}$, let $\dot m:=\mu-\nu$ and $\dot e:=\nu\,(\ell-\EE_{\nu}[\ell])$ be the velocities at $t=0$
of the mixture line \eqref{eq:set-mline} from $\nu$ toward $\mu$ and of the exponential line \eqref{eq:set-eline} from
$\nu$ toward $\rho$. Then $\defect(\mu,\nu,\rho)=\langle\dot m,\dot e\rangle_{\nu}$, the Fisher inner product at $\nu$,
where $\langle x,y\rangle_{\nu}:=\sum_{\omega\in\Omega}x(\omega)\,y(\omega)/\nu(\omega)$.
\end{itemize}
\end{lemma}

In words, the defect is the change in the average of one function---the log-density $\ell$ of $\rho$ with respect to $\nu$---when we average under $\mu$ instead of under $\nu$. Geometrically, it is the inner product at the corner $\nu$ of the direction toward $\mu$ along a mixture line and the direction toward $\rho$ along an exponential line (Figure~\ref{fig:triangle}(a)). The angle between these two lines is acute, right, or obtuse as the defect is positive, zero, or negative; exact Pythagoras holds if and only if the angle is right.

\subsection{Two Projections onto the Set of Pricing Measures}\label{sec:set-proj}

Lemma~\ref{lem:threepoint} becomes a tool when the corner $\nu$ is a projection. In Section~\ref{sec:incomplete}, we consider a finite state space where $d$ traded assets have excess gains $\DeltaS$, defined as the payoff minus price, in units of the bond. A measure $\QQ \in \mathcal{P}$ prices every traded asset correctly exactly when $\EE_{\QQ}[\DeltaS] = 0$. The pricing measures thus form the set $\Mset := \{\QQ \in \mathcal{P} : \EE_{\QQ}[\DeltaS] = 0\}$, which is nonempty exactly when the market has no arbitrage \citep{HarrisonPliska1981,FollmerSchied2016}. Because linear equations cut $\Mset$ out of the simplex, $\Mset$ is mixture-flat: a linear family \citep{Csiszar1975}.

Because KL is not symmetric, a model $\PP \in \mathcal{P}$ has two closest points in $\Mset$ under no arbitrage ($\Mset\neq\emptyset$), one for each order of the arguments. We name them by the slot of the free measure $\QQ$: an I-projection if $\QQ$ sits in the first slot of KL, and a reverse I-projection if it sits in the second. Lemma~\ref{lem:ec-projections} proves the facts below for any linear family, and Section~\ref{sec:incomplete} gives their explicit forms.

The I-projection $\QE:=\argmin_{\QQ\in\cl\Mset}\kl{\QQ}{\PP}$, taken over the closure $\cl\Mset$ of $\Mset$, is the minimal-entropy martingale measure \citep{Frittelli2000}, which we call the minimal-entropy measure. It is unique, lies in $\Mset$, and is an exponential tilt of the model: $\QE\propto\PP\,e^{-h_E\cdot\DeltaS}$, where $h_E\in\RR^{d}$ is the optimal position of an investor with exponential utility and unit absolute risk aversion who trusts $\PP$. Since $\ln(\dd\PP/\dd\QE)$ is $h_E\cdot\DeltaS$ plus a constant, Lemma~\ref{lem:threepoint}(a) with $(\mu,\nu,\rho)=(\QQ,\QE,\PP)$ gives $\defect(\QQ,\QE,\PP)=h_E\cdot(\EE_{\QQ}[\DeltaS]-\EE_{\QE}[\DeltaS])=0$ for every $\QQ\in\Mset$. This is exact Pythagoras, part (a) of Theorem~\ref{thm:twodecomp}: $\kl{\QQ}{\PP}=\kl{\QQ}{\QE}+\kl{\QE}{\PP}$ for every $\QQ\in\Mset$.

In contrast, the reverse I-projection $\QL:=\argmin_{\QQ\in\cl\Mset}\kl{\PP}{\QQ}$ is the log-utility measure. It is unique and equals $\PP/\Wstar$, where $\Wstar$ is the gross return of the growth-optimal portfolio---the portfolio with the largest expected log return under $\PP$ \citep{Long1990,GollRuschendorf2001}. Here $\defect(\PP,\QL,\QQ)$ need not vanish, so the natural identity $\kl{\PP}{\QQ}=\kl{\PP}{\QL}+\kl{\QL}{\QQ}$ fails in general; Section~\ref{sec:ec-proofs-incomplete} gives an example.

Lemma~\ref{lem:threepoint}(b) explains why the two cases differ. At $\QE$, the exponential line toward $\PP$ is orthogonal to $\Mset$; because the defect pairs it with a mixture direction inside $\Mset$, the defect vanishes. In contrast, at $\QL$, the mixture line toward $\PP$ is orthogonal to $\Mset$, but the defect pairs it with the exponential line toward $\QQ$, which leaves $\Mset$ in general (Section~\ref{sec:ec-geo-projections}). Theorem~\ref{thm:twodecomp}(b) replaces the natural identity with an exact one that uses $\Wstar$ as the unit of account.

Section~\ref{sec:robust} uses the same picture for doubt: a ball of plausible models centered at a reference model $\Pz$ grows toward $\Mset$ and first touches it at the minimal-entropy measure of $\Pz$ (Figure~\ref{fig:triangle}(b)). Before that, Section~\ref{sec:triangle} applies Lemma~\ref{lem:threepoint} to the triangle with a pricing measure at its corner.

\section{The Risk-Premium Triangle}\label{sec:triangle}

This section studies the triangle formed by the real-world measure $\PP$, a pricing measure $\QQ$, and the share measure $\QS$ (Figure~\ref{fig:triangle}(a)). This construction is the simplest use of Lemma~\ref{lem:threepoint}, and each of its sides corresponds to a financial quantity. We show that its defect is a log risk premium, that one side can be read from option prices, and that the defect splits into a log equity premium and an entropy premium. The proofs are in Section~\ref{sec:ec-proofs-triangle}.

\subsection{Market, Share Measure, and Assumptions}\label{sec:triangle-market}

We fix a horizon $T>0$ and a filtered probability space $(\Omega,\mathcal{F},(\mathcal{F}_t)_{t\in[0,T]},\PP)$ with $\mathcal{F}=\mathcal{F}_T$, where $\PP$ is the real-world measure. The market includes a bank account, a positive adapted process $B$ with $B_0=1$, and one traded asset with positive price $S_t$. A pricing measure $\QQ$ is a probability measure on $\mathcal{F}_T$ that is equivalent to $\PP$ and under which $S/B$ is a true martingale on $[0,T]$. The asset enters through its relative gross return $R:=S_T/(S_0B_T)>0$, representing its gross return in units of the bank account. The martingale property gives $\EE_{\QQ}[R]=1$.

Because $R$ is a positive random variable with $\QQ$-mean one, it is a density and defines the share measure $\QS$ by $\dd\QS/\dd\QQ:=R$. The share measure is the pricing measure that uses the asset as the unit of account instead of the bank account \citep{GemanElKarouiRochet1995}, so that a payoff $X$ at time $T$ has price $\EE_{\QQ}[X/B_T]=S_0\,\EE_{\QS}[X/S_T]$. Since $R>0$, all three measures are equivalent. Our object of study is the defect \eqref{eq:set-defect} at the corner $\QQ$: $\defect(\PP,\QQ,\QS)=\kl{\PP}{\QQ}+\kl{\QQ}{\QS}-\kl{\PP}{\QS}$. Three assumptions make its sides finite, make its split in Section~\ref{sec:triangle-vix} finite, and link one side to option prices.

\begin{assumption}[Finite sides]\label{ass:A1}
$\kl{\PP}{\QQ}<\infty$, and $\ln R$ is integrable under $\PP$ and under $\QQ$.
\end{assumption}

\begin{assumption}[Finite expected return]\label{ass:A2}
$\EE_{\PP}[R]<\infty$.
\end{assumption}

\begin{assumption}[Deterministic bank account, no dividends]\label{ass:D}
$B_T$ is constant, and the asset pays no dividends.
\end{assumption}

Assumption~\ref{ass:A1} holds if and only if the three sides are finite (Lemma~\ref{lem:ec-sides}), providing exactly what the defect needs. Assumption~\ref{ass:A2} is used only in Proposition~\ref{prop:decomp}, and it does not follow from Assumption~\ref{ass:A1} (Example~\ref{ex:ec-A2}). Assumption~\ref{ass:D} links the triangle to options: $F_0:=S_0B_T=\EE_{\QQ}[S_T]$ is then the forward price of the asset, $R=S_T/F_0$ is a function of $S_T$, and a European payoff $f(S_T)$ has price $\EE_{\QQ}[f(S_T)]/B_T$. We write $\mathrm{put}(K)$ and $\mathrm{call}(K)$ for the prices of the European put and call with strike $K$ and maturity $T$. With a deterministic dividend yield, all statements below hold for the total-return value of the asset, with $F_0$ the forward price of the asset itself (Remark~\ref{rem:ec-dividends}).

\subsection{The Triangle Identity}\label{sec:triangle-identity}

The first result computes the defect. This is the identity of \citet[eq.~13]{Soklakov2023} applied to the triple $(\PP,\QQ,\QS)$, that is, the three-point identity of Lemma~\ref{lem:threepoint} with $\ln(\dd\QS/\dd\QQ)=\ln R$.

\begin{theorem}[Triangle identity]\label{thm:triangle}
Under Assumption~\ref{ass:A1},
\begin{equation}\label{eq:triangle}
\defect(\PP,\QQ,\QS)=\EE_{\PP}[\ln R]-\EE_{\QQ}[\ln R].
\end{equation}
Let $\mathcal{G}\subseteq\mathcal{F}_T$ be a $\sigma$-field such that $R$ is $\mathcal{G}$-measurable, and compute the three divergences for the restrictions of $\PP$, $\QQ$ and $\QS$ to $\mathcal{G}$. Then Assumption~\ref{ass:A1} holds for the restrictions, $\kl{\QQ}{\QS}$ and $\defect(\PP,\QQ,\QS)$ do not change, and $\kl{\PP}{\QQ}$ and $\kl{\PP}{\QS}$ decrease by the same amount, which is zero if and only if $\dd\PP/\dd\QQ$ is $\mathcal{G}$-measurable. Under Assumption~\ref{ass:D} we may take $\mathcal{G}=\sigma(S_T)$, and then the three divergences are those between the laws of $S_T$ under $\PP$, $\QQ$ and $\QS$.
\end{theorem}

In words, the defect is the expected log excess return of the asset under the real-world measure minus the same expectation under the pricing measure; \citet{Soklakov2023} interprets this spread as the risk, per unit price, of a position with respect to the asset. This identity requires only the laws of $R$ under $\PP$ and $\QQ$: discarding information shrinks the two sides that touch $\PP$ by the same amount and leaves the defect unchanged.

Economically, the defect is a risk premium. Under Assumption~\ref{ass:D}, the claim that pays $\ln R=\ln(S_T/F_0)$ at $T$ is the log contract of \citet{Neuberger1994}, and the defect is its expected payoff $\EE_{\PP}[\ln R]$ minus its forward price $\EE_{\QQ}[\ln R]$. Since $\EE_{\QQ}[\ln R]=-\kl{\QQ}{\QS}$ (Lemma~\ref{lem:ec-sides}), the identity also reads
\begin{equation}\label{eq:triangle2}
\defect(\PP,\QQ,\QS)=\EE_{\PP}[\ln R]+\kl{\QQ}{\QS}.
\end{equation}
We call $\kl{\QQ}{\QS}$ the $\QQ$-side of the triangle. It is the forward price of the short log contract, which pays $-\ln R$, and Section~\ref{sec:rob-pricing} prices this claim under doubt. The $\QQ$-side also equals $\ln\EE_{\QQ}[R]-\EE_{\QQ}[\ln R]$, the volatility drag under $\QQ$, so the defect is the expected log excess return plus the option-implied volatility drag.

In the model of \citet{BlackScholes1973}, with drift $\mu$, volatility $\sigma>0$, interest rate $r$, Sharpe ratio $\SR:=(\mu-r)/\sigma$ and its pricing measure, all four quantities have closed forms (Corollary~\ref{cor:bs} in Section~\ref{sec:ec-tri-bs}):
\[ \kl{\PP}{\QQ}=\frac{\SR^2T}{2},\qquad \kl{\QQ}{\QS}=\frac{\sigma^2T}{2},\qquad \kl{\PP}{\QS}=\frac{(\mu-r-\sigma^2)^2\,T}{2\sigma^2}, \]
and $\defect(\PP,\QQ,\QS)=(\mu-r)\,T$. In this setting, the $\PP$--$\QQ$ side is half the squared Sharpe ratio times the horizon, the $\QQ$-side is half the variance times the horizon, and the defect is the accumulated risk premium. A zero defect forces $\mu=r$, that is, $\PP=\QQ$; so whenever $\mu\neq r$, the Pythagorean relation $\kl{\PP}{\QS}=\kl{\PP}{\QQ}+\kl{\QQ}{\QS}$ fails. The defect also reads as an angle at $\QQ$: its type (acute, right, obtuse) is given by the sign of the defect, while its size depends on the family of distributions in which we draw the lines (Remark~\ref{rem:angles}).

\subsection{Equity Premium, Entropy Premium, and the VIX}\label{sec:triangle-vix}

To split the defect, we measure spread by entropy. For any $X>0$ with $\EE_{\PP}[X]<\infty$, the entropy of $X$ under $\PP$ is $\Lent_{\PP}(X):=\ln\EE_{\PP}[X]-\EE_{\PP}[\ln X]\in[0,\infty]$, and $\Lent_{\QQ}(X)$ is defined in the same way. This entropy is nonnegative by Jensen's inequality, equals zero only for a constant $X$, and is half the variance of $\ln X$ for a lognormal $X$. We call $\pi:=\ln\EE_{\PP}[R]$ the log equity premium and $\varepsilon:=\Lent_{\QQ}(R)-\Lent_{\PP}(R)$ the entropy premium. Part (ii) of the next result also defines $\VIX^2(T)$, the squared VIX of maturity $T$ in its ideal form, in decimal and annual units.

\begin{proposition}[Log equity premium and entropy premium]\label{prop:decomp}
Suppose that Assumption~\ref{ass:A1} holds.
\begin{enumerate}
\item[(i)] If Assumption~\ref{ass:A2} also holds, then $\pi$ and $\varepsilon$ are finite and $\defect(\PP,\QQ,\QS)=\pi+\varepsilon$.
\item[(ii)] $\Lent_{\QQ}(R)=\kl{\QQ}{\QS}=-\EE_{\QQ}[\ln R]$. Under Assumption~\ref{ass:D}, moreover,
\[
\kl{\QQ}{\QS}=-\EE_{\QQ}\Big[\ln\frac{S_T}{F_0}\Big]=B_T\Big(\int_0^{F_0}\frac{\mathrm{put}(K)}{K^2}\,\dd K+\int_{F_0}^{\infty}\frac{\mathrm{call}(K)}{K^2}\,\dd K\Big)\eqqcolon\frac{T}{2}\,\VIX^2(T).
\]
\item[(iii)] If $\ln R$ is normal under $\PP$ and under $\QQ$ with the same variance, then Assumption~\ref{ass:A2} holds, $\varepsilon=0$ and $\defect(\PP,\QQ,\QS)=\pi$.
\end{enumerate}
\end{proposition}

In words, the defect decomposes into two parts: $\pi$, the log of the expected gross excess return, and $\varepsilon$, which compares how spread out $R$ is under $\QQ$ and under $\PP$. Economically, $\pi$ is the equity premium in log units and $\varepsilon$ is a premium for the spread of returns; when the cumulant series of $\ln R$ converges, $\varepsilon$ is half the variance premium of the log return plus premia for its skewness and higher cumulants (Section~\ref{sec:ec-tri-decomp} gives the expansion and relates it to the literature). Part (ii) shows that the $\QQ$-side is both the $\QQ$-entropy of $R$ and the forward price of a static portfolio of out-of-the-money options; so this side can be read from option prices without a model.

Part (ii) collects known facts. Its option formula is the static spanning of the log payoff \citep{BreedenLitzenberger1978,CarrMadan1998}, which uses the portfolio that replicates a variance swap \citep{DemeterfiDermanKamalEtAl1999}. \citet[Result~3]{Martin2017} shows that the squared VIX measures the risk-neutral entropy of the simple return; we read this entropy as the $\QQ$-side. Remark~\ref{rem:svix} identifies Martin's SVIX as a second divergence between $\QQ$ and $\QS$. While part (ii) is exact for the ideal VIX, the published VIX sums over listed S\&P~500 strikes and quotes and interpolates two maturities to 30 days \citep{CboeVIXWhitePaper,Cboe2026}, so $\frac12\VIX_{\mathrm{Cboe}}^2\cdot\frac{30}{365}\approx\kl{\QQ}{\QS}$, with the published index divided by 100, is an approximation (Remark~\ref{rem:ec-cboe}). We use this approximation to measure the U.S. triangle in Section~\ref{sec:numerics-triangle}. The two sides that touch $\PP$ are not observed; the data provide only the lower bound $\EE_{\PP}[\ln R]\le\kl{\PP}{\QQ}$ of Section~\ref{sec:triangle-slack}.

With stochastic volatility, jumps, and a deterministic interest rate, Proposition~\ref{prop:itolevy} in Section~\ref{sec:ec-tri-itolevy} decomposes the defect exactly by source of risk into a drift premium, half the premium on the diffusive variance, and a jump-entropy premium. This decomposition organizes known log-contract formulas \citep{CarrWu2009}. For a jump of log size $y$, the weight in the jump-entropy premium is $k(y):=e^y-1-y\ge0$ and the weight in half the quadratic variation is $y^2/2$. The two weights differ by $y^3/6+y^4/24+\cdots$, that is, by the third and higher powers of the jump size.

\subsection{The Entropy Bound and Its Slack}\label{sec:triangle-slack}

\begin{proposition}[Entropy bound and its slack]\label{prop:slack}
Suppose that Assumption~\ref{ass:A1} holds.
\begin{enumerate}
\item[(i)] $\kl{\PP}{\QQ}=\EE_{\PP}[\ln R]+\kl{\PP}{\QS}$.
\item[(ii)] Hence $\EE_{\PP}[\ln R]\le\kl{\PP}{\QQ}$, with equality if and only if $R=\dd\PP/\dd\QQ$ almost surely, that is, if and only if $\QS=\PP$.
\item[(iii)] Under Assumption~\ref{ass:D}, $\kl{\PP}{\QQ}=\Lent_{\PP}\big(B_T^{-1}\dd\QQ/\dd\PP\big)$, the entropy of the pricing kernel.
\end{enumerate}
\end{proposition}

In plain terms, no asset priced by $\QQ$ has an expected log excess return above $\kl{\PP}{\QQ}$, the entropy of the pricing kernel (part (iii); \citealt{BackusChernovZin2014}). Part (ii) is the entropy bound on expected returns established by \citet{BansalLehmann1997} and \citet{AlvarezJermann2005}, while part (i) identifies its exact slack; for a horse race, part (i) is the doubling-rate identity of \citet[Section~6.1]{CoverThomas2006}. This slack is the $\PP$--$\QS$ side of the triangle and vanishes only for the numeraire portfolio of \citet{Long1990} (Section~\ref{sec:ec-tri-slack}). In the one-period market of Section~\ref{sec:incomplete}, where $\dd\PP/\dd\QQ$ need not be traded, the growth-optimal portfolio has the smallest slack among all traded portfolios. Theorem~\ref{thm:twodecomp}(b) gives this slack, which is zero only at the log-utility measure $\QL$.

This section has taken the pricing measure $\QQ$ as given. In an incomplete market, however, many pricing measures fit the traded prices, and Section~\ref{sec:incomplete} asks which one to use.

\section{Choosing a Pricing Measure in an Incomplete Market}\label{sec:incomplete}

In an incomplete market, many measures price traded assets correctly but disagree on claims that these assets cannot replicate. We consider which of these measures a log investor and an exponential investor would choose. For both investors, the divergence between $\PP$ and a pricing measure $\QQ$ splits exactly into two parts: one is fixed by the traded assets and the other is created by the choice of $\QQ$. These results hold for every strictly positive model $\PP$ (proofs in Section~\ref{sec:ec-proofs-incomplete}).

\subsection{The One-Period Market}

We consider a one-period market with a finite set of states, $\Omega=\{\omega_1,\dots,\omega_M\}$ with $M\ge2$. A model is a probability vector $\PP$ on $\Omega$; the open simplex $\mathcal{P}$ contains the models that give every state positive probability, and the closed simplex $\cl\mathcal{P}$ also allows zeros. Amounts are in units of a bond that costs and pays one, so the interest rate is zero. Risky asset $i=1,\dots,d$ costs $S_0^i$ and pays $S_1^i(\omega)$; its excess gain is $\DeltaS^i:=S_1^i-S_0^i$, and $\DeltaS\in\RR^{d\times M}$ collects the gains. A portfolio $h\in\RR^d$ holds $h_i$ units of asset $i$, financed with the bond, and turns one unit of wealth into $W_h:=1+h\cdot\DeltaS$. It is an arbitrage if $h\cdot\DeltaS\ge0$ in every state and $h\cdot\DeltaS>0$ in some state. Allowing positions that cost nothing or less today gives the same notion of arbitrage (Lemma~\ref{lem:ec-arbitrage}).

\begin{assumption}[Market]\label{ass:market}
(i) Non-redundancy: the vectors $\mathbf{1},\DeltaS^1,\dots,\DeltaS^d$ of $\RR^M$ are linearly independent.
(ii) No arbitrage: no portfolio is an arbitrage.
\end{assumption}

Non-redundancy means that no asset can be built from the bond and other assets; this property makes optimal portfolios unique. The fundamental theorem of asset pricing \citep{HarrisonKreps1979,HarrisonPliska1981,FollmerSchied2016} states that no arbitrage holds if and only if there exist pricing measures---models in $\mathcal{P}$ under which every risky asset is priced at its expected payoff (Theorem~\ref{thm:ec-ftap}). These pricing measures form the set $\Mset$ of Section~\ref{sec:set-proj}. The measures that price every asset correctly but may ignore states, which Section~\ref{sec:robust} needs, form the closure $\cl\Mset=\{\QQ\in\cl\mathcal{P}:\EE_\QQ[\DeltaS]=0\}$.

Each risky asset adds one linear equation to $\sum_\omega\QQ(\omega)=1$, and by non-redundancy these $d+1$ equations are independent. So the set $\Mset$ of pricing measures has dimension $M-1-d$; this is the number of directions in which prices of non-traded claims are free (Proposition~\ref{prop:ec-geometry}). The market is complete if and only if $d+1=M$.

\subsection{Two Investors and Their Pricing Measures}\label{sec:inc-measures}

For an investor with utility $U$ on domain $D\subseteq\RR$ and initial wealth $x$, the value of the market completed at a pricing measure $\QQ$ is given by $u(x;\QQ):=\sup\{\EE_\PP[U(X)]:\ X(\omega)\in D\text{ for all }\omega,\ \EE_\QQ[X]\le x\}$, where every claim $X\in\RR^M$ trades at the price $\EE_\QQ[X]$. We also consider the value $u(x)$, which is the same supremum over traded wealths of the form $X=x+h\cdot\DeltaS$, with $h\in\RR^d$. Because these cost exactly $x$ under every pricing measure, it follows that $u(x)\le u(x;\QQ)$. A pricing measure that minimizes $u(x;\QQ)$ over $\Mset$ is a minimax measure: completing the market at its prices is worth the least to the investor \citep{HePearson1991MF,KaratzasLehoczkyShreveEtAl1991,BelliniFrittelli2002}.

The minimal-entropy measure $\QE$ and the log-utility measure $\QL$ of Section~\ref{sec:set-proj}, the I-projection and
the reverse I-projection of $\PP$ on $\cl\Mset$, are the minimax measures of two standard utilities. Both exist, are
unique, charge every state, and have explicit forms (Propositions~\ref{prop:ec-QE} and~\ref{prop:ec-QL}):
\begin{align*}
\QE&=\frac{\PP\,e^{-h_E\cdot\DeltaS}}{\EE_\PP\bigl[e^{-h_E\cdot\DeltaS}\bigr]}, &
h_E&:=\argmin_{\eta\in\RR^d}\EE_\PP\bigl[e^{-\eta\cdot\DeltaS}\bigr],\\
\QL&=\frac{\PP}{\Wstar},\qquad \Wstar:=1+\hstar\cdot\DeltaS, &
\hstar&:=\argmax_{h:\,W_h>0}\EE_\PP[\ln W_h].
\end{align*}
As in Section~\ref{sec:set-proj}, $h_E$ is the optimal position of the exponential investor, and $\hstar$ is that of the log investor (Lemma~\ref{lem:duality}
below). So each measure shifts weight away from the states in which its optimal portfolio does well. We write $\QE(\PP)$, $h_E(\PP)$, $\QL(\PP)$, $\hstar(\PP)$ and $\Wstar(\PP)$ when the model matters.

\begin{lemma}[Utility duality]\label{lem:duality}
Let Assumption~\ref{ass:market} hold and let $\PP\in\mathcal{P}$. Let $U$ be log utility, $U=\ln$ on $D=(0,\infty)$, or
exponential utility, $U(z)=-e^{-z}$ on $D=\RR$, and let $V(y):=\sup_{z\in D}\{U(z)-zy\}$, $y>0$, be its conjugate. For
every $\QQ\in\Mset$ and every initial wealth $x\in D$,
\begin{equation}\label{eq:duality}
u(x;\QQ)=\min_{y>0}\Bigl\{\EE_\PP\Bigl[V\Bigl(y\,\frac{\dd\QQ}{\dd\PP}\Bigr)\Bigr]+xy\Bigr\}.
\end{equation}
The conjugate, the dual objective and the two values are
\[
\begin{array}{@{}l@{\hspace{1.3em}}c@{\hspace{1.3em}}c@{\hspace{1.3em}}c@{\hspace{1.3em}}c@{}}
 & V(y) & \EE_\PP\bigl[V(y\,\dd\QQ/\dd\PP)\bigr] & u(x;\QQ) & u(x)\\[2pt]
\text{log:} & -\ln y-1 & \kl{\PP}{\QQ}-\ln y-1 & \ln x+\kl{\PP}{\QQ} & \ln x+\kl{\PP}{\QL}\\[2pt]
\text{exponential:} & y\ln y-y & y\ln y-y+y\,\kl{\QQ}{\PP} & -e^{-x-\kl{\QQ}{\PP}} & -e^{-x-\kl{\QE}{\PP}}
\end{array}
\]
In both cases $u(x)=\min_{\QQ\in\Mset}u(x;\QQ)$. The minimax measure, $\QL$ for log and $\QE$ for exponential utility, is
unique and does not depend on $x$. The optimal claim in the market completed at it is the optimal traded wealth:
$x\Wstar$ for log and $x+h_E\cdot\DeltaS$ for exponential utility.
\end{lemma}

Lemma~\ref{lem:duality} expresses the value of a completed market through a divergence: $\kl{\PP}{\QQ}$ for the log investor and $\kl{\QQ}{\PP}$ for the exponential investor. Each minimax measure is the corresponding projection at which completing the market adds nothing \citep{GollRuschendorf2001,BelliniFrittelli2002,Frittelli2000}. The certainty equivalent of a value $u$ is $e^u$ for the log investor and $-\ln(-u)$ for the exponential investor. So $\kl{\PP}{\QL} = \EE_\PP[\ln\Wstar]$ is the certainty-equivalent growth rate of the traded assets, and $\kl{\QE}{\PP}$ is the certainty-equivalent gain from trading, in units of the bond.

\subsection{Two Exact Decompositions}

\begin{theorem}[Two exact decompositions]\label{thm:twodecomp}
Let Assumption~\ref{ass:market} hold and let $\PP\in\mathcal{P}$. For every $\QQ\in\cl\Mset$,
\begin{align*}
\text{(a)}\quad & \kl{\QQ}{\PP}=\kl{\QQ}{\QE}+\kl{\QE}{\PP},\\
\text{(b)}\quad & \kl{\PP}{\QQ}=\kl{\PP}{\QL}+\kl{\PP}{\QW},\qquad
\frac{\dd\QW}{\dd\QQ}:=\Wstar=\frac{\dd\PP}{\dd\QL}.
\end{align*}
Both sides of (b) are $+\infty$ when $\QQ\notin\Mset$, and every claim $X$ satisfies $\EE_\QQ[X]=\EE_{\QW}[X/\Wstar]$.
The term $\kl{\QQ}{\QE}$ in (a) is zero if and only if $\QQ=\QE$, and the term $\kl{\PP}{\QW}$ in (b) is zero if and only if $\QQ=\QL$.
\end{theorem}

Part (a) is known: it is both the Pythagorean identity of the I-projection on a linear family \citep{Csiszar1975} and a characterization of the minimal-entropy martingale measure \citep{Frittelli2000}. Part (b) is Proposition~\ref{prop:slack} with the growth-optimal portfolio as the traded asset. The relative return of this portfolio is $\Wstar$, and its share measure $\QW$---a probability measure because $\EE_\QQ[\Wstar]=1+\hstar\cdot\EE_\QQ[\DeltaS]=1$---is $\QQ$ seen with this portfolio as the unit of account. The expected log return $\EE_\PP[\ln\Wstar]=\kl{\PP}{\QL}$ does not depend on $\QQ$. Identity (a) is the right angle at $\QE$ of Section~\ref{sec:set-proj}, whereas identity (b) does not come from a right angle at $\QL$. The Pythagorean form $\kl{\PP}{\QQ}=\kl{\PP}{\QL}+\kl{\QL}{\QQ}$ fails in general because, for $\QQ\in\Mset$, its defect $\defect(\PP,\QL,\QQ)=\mathrm{Cov}_{\QL}(\Wstar,\ln(\dd\QQ/\dd\QL))$ can take either sign (Example~\ref{ex:ec-counter}, Propositions~\ref{prop:ec-defect} and~\ref{prop:ec-noineq}). Identity (b) avoids this defect by comparing $\PP$ with $\QQ$ measured in units of $\Wstar$, instead of comparing $\QL$ with $\QQ$.

Economically, each identity splits a completion value (Lemma~\ref{lem:duality}). In the case of identity (a), $\kl{\QE}{\PP}$ is the gain the exponential investor achieves from the traded assets, while $\kl{\QQ}{\PP}$ is the investor's gain if all claims were traded at the prices of $\QQ$. So $\kl{\QQ}{\QE}$ is the certainty-equivalent value of completing the market at the prices of $\QQ$. Identity (b) gives the other view: $\kl{\PP}{\QW}$ is the extra growth rate that the same completion gives the log investor. In the terms of Section~\ref{sec:triangle}, identity (b) splits the entropy of the pricing kernel, $\kl{\PP}{\QQ}$, into two components: the growth-optimal bound $\EE_\PP[\ln\Wstar]$ \citep{BansalLehmann1997} and a part created by the choice of $\QQ$.

\subsection{The Entropy-Budget Interval}\label{sec:inc-budget}

Theorem~\ref{thm:twodecomp}(a) also limits the prices of non-traded claims. For a claim $X$ and a budget $b\ge0$, we call
\begin{equation}\label{eq:entropy-budget}
J_b(X):=\bigl\{\EE_\QQ[X]:\ \QQ\in\cl\Mset,\ \kl{\QQ}{\QE}\le b\bigr\}
\end{equation}
the entropy-budget interval of $X$: the prices of $X$ under the pricing measures whose completion term in (a) is at most
$b$. It is a compact interval that contains $\EE_{\QE}[X]$ (Section~\ref{sec:ec-inc-bounds}). For small $b$, its half-width is $\sqrt{2b}\,\mathrm{sd}_{\QE}(\tilde X)+O(b)$, where the
unhedgeable part $\tilde X$ of $X$ is the residual of the $\QE$-weighted least-squares regression of $X$ on $\mathbf{1}$
and $\DeltaS$ (Proposition~\ref{prop:ec-width}). By Lemma~\ref{lem:duality}, the budget caps the exponential investor's
certainty-equivalent value of completing the market at $\QQ$'s prices, so $J_b(X)$ is a good-deal interval
\citep{CochraneSaaRequejo2000} of the entropic kind studied by \citet{Cerny2003}. Section~\ref{sec:robust} uses this
interval at the reference model, $\PP=\Pz$. There identity (a) reads $\kl{\QQ}{\Pz}=\kl{\QQ}{\QE}+\cbar$, where
$\cbar=\kl{\QE}{\Pz}$ is the threshold of Section~\ref{sec:rob-structure}. So a pricing measure lies in the ball
$\Bc$ of plausible models exactly when $\kl{\QQ}{\QE}\le c-\cbar$, and the no-trade interval of
Theorem~\ref{thm:pricing} is $J_{c-\cbar}(X)$ for $c\ge\cbar$.

\section{Robust Pricing Under Model Uncertainty}\label{sec:robust}

While Sections~\ref{sec:triangle} and~\ref{sec:incomplete} take the real-world model as given, in practice it is estimated. This section studies a robust log investor who doubts their reference model $\Pz$, an estimate of the real-world model. They treat every model within a given relative entropy of $\Pz$ as plausible and maximize their worst-case expected log growth (Figure~\ref{fig:triangle}(b); proofs in Section~\ref{sec:ec-proofs-robust}).

\subsection{The Robust Log Investor}\label{sec:rob-setup}

We maintain the market of Section~\ref{sec:incomplete}, where a portfolio $h \in \RR^d$ transforms one unit of wealth into $W_h = 1 + h \cdot \DeltaS$. In this setting, models are now represented as points in the closed simplex $\cl\mathcal P$, because a plausible model may ignore states. The investor treats every model whose relative entropy with respect to the reference model $\Pz$ is at most a radius $c \ge 0$ as plausible; that is, they consider every model in $\Bc = \{\PP \in \cl\mathcal P : \kl{\PP}{\Pz} \le c\}$. The divergence $\kl{\PP}{\Pz}$ is the expected log-likelihood ratio of an alternative model against the reference model, computed under the alternative. This quantity measures how quickly data generated by $\PP$ reveal that $\Pz$ is incorrect, and \citet{HansenSargent2001} use it to measure model misspecification. We use ``doubt'' for the investor's uncertainty about which model is true---a concept referred to as model uncertainty or ambiguity in decision theory---while model risk is its consequence. Finally, we call $c$ the doubt radius and $\Bc$ the ball of plausible models.

The investor uses the portfolios $\mathcal H=\{h\in\RR^d:W_h(\omega)>0\text{ for every }\omega\in\Omega\}$, which keep their wealth positive in every state, also in states that a plausible model ignores, because every state is possible under $\Pz$. The growth rate of $h$ under a model $\PP$ is $G(h,\PP)=\EE_\PP[\ln W_h]$, and the worst-case growth of $h$ is $G_c(h)=\inf_{\PP\in\Bc}G(h,\PP)$. The investor solves
\begin{equation}\label{eq:rob-value}
v(c)=\sup_{h\in\mathcal H}\ \inf_{\PP\in\Bc}\ \EE_\PP\bigl[\ln(1+h\cdot\DeltaS)\bigr],
\end{equation}
and we call $v(c)$ the robust value, the best expected log growth per period that they can guarantee if any model in $\Bc$ may be true. This is max--min expected log utility \citep{Kelly1956} with $\Bc$ as the set of priors \citep{GilboaSchmeidler1989}, the constraint preferences of \citet{HansenSargent2001}.

\begin{assumption}[Robust market]\label{ass:C}
(C1) (Reference model.) $\Pz\in\mathcal P$: every state has a positive reference probability. (C2) (Non-redundancy.) Assumption~\ref{ass:market}(i) holds. (C3) (No arbitrage.) Assumption~\ref{ass:market}(ii) holds, so that the set $\Mset$ of pricing measures is nonempty (Theorem~\ref{thm:ec-ftap}). (C4) (Non-degeneracy.) $\EE_{\Pz}[\DeltaS]\neq0$.
\end{assumption}

Conditions (C1)--(C3) hold throughout our analysis; we invoke condition (C4), which states that $\Pz$ is not a pricing measure, only where we explicitly indicate so. Following Section~\ref{sec:inc-measures}, we use the notation $\QL(\PP)$ for strictly positive $\PP$, as well as $\QE = \QE(\Pz)$ and the exponential investor's position $h_E = h_E(\Pz)$.

\begin{theorem}[Robust value as a KL distance]\label{thm:minimax}
Assume (C1)--(C3) and $c\ge0$. Then
\[
v(c)=\sup_{h\in\mathcal H}\ \min_{\PP\in\Bc}\EE_\PP[\ln W_h]
=\min_{\PP\in\Bc}\ \sup_{h\in\mathcal H}\EE_\PP[\ln W_h]
=\min_{\PP\in\Bc,\ \QQ\in\cl\Mset}\kl{\PP}{\QQ}.
\]
All minima are attained, and the supremum over $h$ is attained in $\mathcal H$. If $h^\dagger$ attains the supremum and $(\PP^\dagger,\QQ^\dagger)$ attains the last minimum, then $(h^\dagger,\PP^\dagger)$ is a saddle point,
\[
G(h,\PP^\dagger)\le G(h^\dagger,\PP^\dagger)=v(c)\le G(h^\dagger,\PP)\qquad\text{for all }h\in\mathcal H\text{ and }\PP\in\Bc,
\]
and $\QQ^\dagger=\QL(\PP^\dagger)$ whenever $\PP^\dagger$ is strictly positive.
\end{theorem}

We define a minimizing pair $(\PP^\dagger,\QQ^\dagger)$ as a ``nearest pair,'' and the minimum as the KL distance between the two sets; like the KL divergence itself, this quantity is not a metric. The investor acts as a log investor who believes the worst-case model $\PP^\dagger$. This theorem is the log-utility, KL-ball case of the duality for robust utility maximization \citep{Gundel2005,SchiedWu2005,FollmerGundel2006}. We prove it on the closed simplex using Sion's minimax theorem \citep{Sion1958}, as the structure results rely on its saddle-point details.

\subsection{The Structure of the Robust Solution}\label{sec:rob-structure}

We now describe the nearest pair as the radius grows. One number organizes the answer, the threshold
\begin{equation}\label{eq:rob-cbar}
\cbar:=\kl{\QE}{\Pz}=\max_{\eta\in\RR^d}\Bigl\{-\ln\EE_{\Pz}\bigl[e^{-\eta\cdot\DeltaS}\bigr]\Bigr\},
\end{equation}
the minimal entropy of the market: the smallest relative entropy of a pricing measure with respect to $\Pz$. The ball $\Bc$ meets $\cl\Mset$ exactly when $c\ge\cbar$ (Lemma~\ref{lem:ec-rob-qe}). In the second form, $-\ln\EE_{\Pz}[e^{-\eta\cdot\DeltaS}]$ is the certainty-equivalent gain of the position $\eta$ under exponential utility (Lemma~\ref{lem:duality}). So $\cbar$ is the largest such gain of an investor with exponential utility and unit absolute risk aversion who trusts $\Pz$ (with absolute risk aversion $\beta$ the largest gain is $\cbar/\beta$). One convex optimization in $d$ variables computes it, at $\eta=h_E$.

We compare the robust investor with investors who trust $\Pz$ and have constant relative risk aversion (CRRA) $\gamma\ge1$, with utility $U_\gamma(x)=x^{1-\gamma}/(1-\gamma)$ for $\gamma>1$ and $U_1(x)=\ln x$. Let $h^{(\gamma)}$ maximize $\EE_{\Pz}[U_\gamma(W_h)]$ over $\mathcal H$, let $\Wg=W_{h^{(\gamma)}}$, and define
\begin{equation}\label{eq:rob-crra}
\frac{\dd\Pg}{\dd\Pz}=\frac{(\Wg)^{1-\gamma}}{\EE_{\Pz}\bigl[(\Wg)^{1-\gamma}\bigr]},\qquad
\frac{\dd\Qg}{\dd\Pz}=\frac{(\Wg)^{-\gamma}}{\EE_{\Pz}\bigl[(\Wg)^{-\gamma}\bigr]},\qquad c(\gamma)=\kl{\Pg}{\Pz}.
\end{equation}
The measure $\Qg$ weights each state by the marginal utility $(\Wg)^{-\gamma}$: it is the investor's marginal-utility pricing measure \citep{Davis1997}. So $\EE_{\Qg}[X]$ is the only price at which a small position in a new claim $X$ leaves their expected utility unchanged to first order. For $\gamma>1$, $\Pg$ tilts $\Pz$ toward the states in which the CRRA portfolio does badly, and $c(\gamma)$ is the relative entropy of this tilt with respect to $\Pz$.

\begin{theorem}[Structure of the robust solution]\label{thm:robust}
Assume (C1)--(C4), so that $\cbar>0$.
\begin{enumerate}
\item[(i)] (Value.) $v$ is convex, nonincreasing and continuous on $[0,\infty)$, with $v(0)=\kl{\Pz}{\QL(\Pz)}=\EE_{\Pz}[\ln\Wstar(\Pz)]$. It is strictly decreasing and strictly convex on $[0,\cbar]$, and $v\equiv0$ on $[\cbar,\infty)$; so $v(c)>0$ iff $c<\cbar$. It is continuously differentiable on $(0,\infty)$, with $v'<0$ on $(0,\cbar)$, $v'=0$ on $[\cbar,\infty)$, and $v'(0^+)=-\infty$.
\item[(ii)] (Nearest pair below the threshold.) For $0\le c<\cbar$ the nearest pair $(\Pc,\Qc)$ is unique. $\Pc$ is strictly positive and $\kl{\Pc}{\Pz}=c$, so the constraint binds. The robust portfolio $h_c$ is unique, $h_c=\hstar(\Pc)\neq0$, and $\Qc=\QL(\Pc)=\Pc/\Wc\in\Mset$, where $\Wc=W_{h_c}$.
\item[(iii)] (Power tilt.) For $0<c<\cbar$ there is a unique $\lambda_c>0$ such that
\[
\frac{\dd\Pc}{\dd\Pz}=\frac{\Wc^{-1/\lambda_c}}{\EE_{\Pz}\bigl[\Wc^{-1/\lambda_c}\bigr]},\qquad
\frac{\dd\Qc}{\dd\Pz}=\frac{\Wc^{-(1+1/\lambda_c)}}{\EE_{\Pz}\bigl[\Wc^{-(1+1/\lambda_c)}\bigr]}.
\]
Equivalently, with $\gamma_c=1+1/\lambda_c$, $\Pc\propto\Pz^{1/\gamma_c}\,\Qc^{1-1/\gamma_c}$: the worst-case model lies on the exponential line from $\Pz$ to $\Qc$. The multiplier is $\lambda_c=-v'(c)$, and $(\Pc,\Qc)$ minimizes $\kl{\PP}{\QQ}+\lambda_c\,\kl{\PP}{\Pz}$ over $\cl\mathcal P\times\cl\Mset$.
\item[(iv)] (Robustness acts as risk aversion in a given market.) For $0\le c<\cbar$, with $\gamma_0=1$, we have $h_c=h^{(\gamma_c)}$, $\Pc=\PP^{(\gamma_c)}$ and $\Qc=\QQ^{(\gamma_c)}$: the robust portfolio is the optimal portfolio under $\Pz$ of a CRRA investor with risk aversion $\gamma_c$, and the robust pricing measure is that investor's marginal-utility pricing measure. Conversely, for every $\gamma\ge1$, $(\Pg,\Qg)$ is the nearest pair for the radius $c(\gamma)$, $h^{(\gamma)}$ is the robust portfolio for that radius, and $\gamma_{c(\gamma)}=\gamma$. The map $\gamma\mapsto c(\gamma)$ is a continuous, strictly increasing bijection from $[1,\infty)$ onto $[0,\cbar)$, with inverse $c\mapsto\gamma_c$. Hence $\lambda_c$ decreases continuously and strictly from $+\infty$ to $0$ as $c$ increases from $0$ to $\cbar$.
\item[(v)] (Endpoints.) At $c=0$ the nearest pair is $(\Pz,\QL(\Pz))$. As $c\uparrow\cbar$: $\Pc\to\QE$, $\Qc\to\QE$, $h_c\to0$ and $h_c/\lambda_c\to h_E$. Equivalently, as $\gamma\to\infty$: $\Pg\to\QE$, $\Qg\to\QE$ and $\gamma h^{(\gamma)}\to h_E$, the optimal position of an investor with exponential utility; all three errors are $O(1/\gamma)$.
\item[(vi)] (No trade at and above the threshold.) For $c\ge\cbar$: $v(c)=0$, the unique robust portfolio is $h_c=0$, and the nearest pairs are the pairs $(\QQ,\QQ)$ with $\QQ\in\cl\Mset\cap\Bc$. This set is nonempty, compact and convex, and it reduces to $\{(\QE,\QE)\}$ at $c=\cbar$.
\end{enumerate}
\end{theorem}

At zero doubt, $v(0)=\kl{\Pz}{\QL(\Pz)}$ is a side of the triangle in Section~\ref{sec:triangle}. For a traded asset with relative return $R>0$ and share measure $\QL^S$ under $\QL(\Pz)$, Proposition~\ref{prop:slack}(i) gives $v(0)=\EE_{\Pz}[\ln R]+\kl{\Pz}{\QL^S}$. So $\EE_{\Pz}[\ln R]$ bounds $v(0)$ from below, and the slack of the entropy bound is the difference (Table~\ref{tab:emp-bridge}).

We call $\lambda_c = -v'(c)$ the shadow price of doubt---the loss of guaranteed growth per unit of extra doubt---and $\gamma_c = 1 + 1/\lambda_c$ the effective risk aversion. Because $v$ is convex with $v'(0^+)=-\infty$, the shadow price of doubt is infinite at zero doubt and falls as doubt grows. As the doubt grows toward $\cbar$, the robust pricing measure travels from $\QL(\Pz)$ to $\QE$. Part (iv) is a short consequence of the KL dual: for fixed $h$, $\min_{\PP\in\Bc}\EE_\PP[\ln W_h]=\sup_{\lambda>0}\{-\lambda\ln\EE_{\Pz}[W_h^{-1/\lambda}]-\lambda c\}$ (Lemma~\ref{lem:ec-rob-gibbs}(c)), where the first term is the log certainty equivalent of $W_h$ for CRRA risk aversion $1+1/\lambda$. For a fixed $\lambda$, the best $h$ is thus the CRRA($1+1/\lambda$) portfolio, and the best $\lambda$ makes the constraint bind such that $c(\gamma)=c$ (Proposition~\ref{prop:algorithms}(b)). Through the Lagrange duality between the constraint and multiplier forms of robustness \citep{PetersenJamesDupuis2000,HansenSargentTurmuhambetovaEtAl2006}, this is the constraint-form counterpart of the known link between the multiplier form and CRRA (Remark~\ref{rem:scope}).

\begin{remark}[Scope of the equivalence]\label{rem:scope}
Theorem~\ref{thm:robust}(iv) is an equivalence of choices and prices in one market. The risk aversion $\gamma_c$ solves $c(\gamma)=c$, and $c(\gamma)$ is computed from the market $(\Pz,\DeltaS)$. So, for a fixed radius, $\gamma_c$ changes with the reference model, the traded assets and the horizon. A new claim $X$ offered at a price other than its robust price $\EE_{\Qc}[X]$ (Section~\ref{sec:rob-pricing}) creates a market with its own map $\gamma\mapsto c(\gamma)$. In that market the robust demand for the claim is in general not that of a CRRA($\gamma_c$) investor; at the robust price nothing changes (Theorem~\ref{thm:pricing}(a), Lemma~\ref{lem:ec-rob-notrade}). So the ball investor has the constraint (max--min) preferences of \citet{HansenSargent2001}, not CRRA preferences. Only the multiplier form, which replaces the ball by a fixed penalty $\theta\,\kl{\PP}{\Pz}$ on the objective, acts like CRRA($1+1/\theta$) in every market (Proposition~\ref{prop:ec-rob-mult}).
\end{remark}

Section~\ref{sec:ec-rob-literature} relates Theorem~\ref{thm:robust} to the literature. Three consequences of the ball form are, to our knowledge, new; the first two require log utility: the exact map $c\mapsto\gamma_c$ in a given market, the identification of $\Qc$ as the CRRA($\gamma_c$) marginal-utility measure on a path from $\QL(\Pz)$ to $\QE$, and the identification $\cbar=\kl{\QE}{\Pz}$ of the threshold. The ball form creates a no-trade region because the guaranteed growth $G_c$ has a kink at $h=0$; no CRRA investor has a no-trade region (Lemma~\ref{lem:ec-rob-crra}(d)).

The structure results also give two ways to compute the robust solution, alternating projections and the KL dual (Proposition~\ref{prop:algorithms} in Section~\ref{sec:ec-rob-algorithms}). The local rate of alternating projections is at most $1-1/\gamma_c$ and equals this bound when $M\ge d+3$; the method then slows down without bound as $c\uparrow\cbar$. Section~\ref{sec:ec-numerics-e4} compares the methods.

\subsection{Robust Prices of New Claims}\label{sec:rob-pricing}

We now add a claim that the market does not trade, with payoff $X \in \RR^M$ (measured in units of the bond) and price $p \in \RR$. Its no-arbitrage interval $\{\EE_\QQ[X]:\QQ\in\Mset\}$ is open when $X$ is not spanned by the bond and the assets, and a point when it is. Access to $X$ at price $p$ adds the excess gain $Y=X-p$, and a strategy is defined as a pair $(h,k) \in \mathcal H^+ = \{(h,k) \in \RR^{d+1} : W_h + kY > 0 \text{ on } \Omega\}$ with $k$ units of $X$. The robust value of the extended market is then $v^+(c;p)=\sup_{(h,k)\in\mathcal H^+}\min_{\PP\in\Bc}\EE_\PP[\ln(W_h+kY)] \ge v(c)$.

\begin{definition}[No-trade price]\label{def:notrade}
We call $p$ a no-trade price of $X$ at radius $c$ if $v^+(c;p)=v(c)$, that is, if access to $X$ at price $p$ does not raise the robust investor's guaranteed growth.
\end{definition}

Theorem~\ref{thm:pricing} relies on three objects. The first is the robust price $p_c(X) := \EE_{\Qc}[X]$ below the threshold $\cbar$. The second is the set $I_c(X) := \{\EE_\QQ[X] : \QQ \in \cl\Mset \cap \Bc\}$ of prices for $X$ under the measures of $\cl\Mset$ in the ball, which may ignore states (Remark~\ref{rem:ec-rob-closure}). The third is the minimal entropy of the market extended by $X$ at price $p$, $\cbar_X(p) := \min\{\kl{\QQ}{\Pz} : \QQ \in \cl\Mset,\ \EE_\QQ[X]=p\}$, with $\min \emptyset = +\infty$. This quantity is the value of the minimum-relative-entropy calibration of $\Pz$ to the traded prices and the quote \citep{Avellaneda1998}; \citet{ChenHansenHansen2020} use the same kind of problem to bound distortions of investors' beliefs. This value is at least $\cbar$, and it equals $\cbar$ only at $p = \EE_{\QE}[X]$.

\begin{theorem}[Robust prices of a new claim]\label{thm:pricing}
Assume (C1)--(C4). Let $X$ be a claim and $p\in\RR$ a price.
\begin{enumerate}
\item[(a)] (Below the threshold.) If $c<\cbar$, then $p$ is a no-trade price of $X$ iff $p=p_c(X)=\EE_{\Qc}[X]$.
\item[(b)] (At and above the threshold.) If $c\ge\cbar$, then $p$ is a no-trade price of $X$ iff $p\in I_c(X)$.
\item[(c)] (The interval.) For $c\ge\cbar$, $I_c(X)$ is a nonempty compact interval, and $I_c(X)=\{p:\cbar_X(p)\le c\}=J_{c-\cbar}(X)$, the entropy-budget interval \eqref{eq:entropy-budget} with budget $c-\cbar$. It equals $\{\EE_{\QE}[X]\}$ at $c=\cbar$, it grows with $c$, and $\bigcap_{c>\cbar}I_c(X)=\{\EE_{\QE}[X]\}$; also $p_c(X)\to\EE_{\QE}[X]$ as $c\uparrow\cbar$.
\item[(d)] (Direction of trade.) Let $c<\cbar$, let $p$ lie in the no-arbitrage interval of $X$, and let $m(k)=\sup_{h:(h,k)\in\mathcal H^+}\min_{\PP\in\Bc}\EE_\PP[\ln(W_h+kY)]$ ($\sup\emptyset=-\infty$), the guaranteed growth with $k$ units of $X$ when the traded position is re-optimized; $m(0)=v(c)$. If $p<p_c(X)$, then $m(k)>v(c)$ for all small $k>0$ and $m(k)<v(c)$ for all $k<0$; if $p>p_c(X)$, the roles of $k>0$ and $k<0$ are exchanged.
\end{enumerate}
\end{theorem}

The switch in Theorem~\ref{thm:pricing}, from one no-trade price below $\cbar$ to an interval of them at and above $\cbar$, has a simple mechanism. Below $\cbar$, the worst case is unique, so the guaranteed growth is differentiable in a small position in $X$. Only one price makes this derivative zero: the marginal-utility price of the CRRA($\gamma_c$) investor \citep{Davis1997,HugonnierKramkovSchachermayer2005}. At and above $\cbar$, the robust portfolio is $h_c=0$. At this portfolio, the guaranteed growth has a kink, as purchases and sales face different worst-case models, so a range of prices leaves both unattractive \citep[as in][]{DowWerlang1992,EpsteinWang1994}. The direction in which doubt moves the robust price below $\cbar$ is not fixed by the theory \citep[for a Wasserstein ball, see][]{OblojWiesel2021}; however, in the data of Section~\ref{sec:numerics-put}, the put price falls slightly.

Above $\cbar$, part (c) identifies the no-trade interval with the entropy-budget interval of Section~\ref{sec:inc-budget}, $I_c(X)=J_{c-\cbar}(X)$, which is the entropic good-deal interval of \citet{Cerny2003} with budget $c-\cbar$ \citep[for good-deal bounds see][]{CochraneSaaRequejo2000,CernyHodges2002}. \citet{BoyarchenkoCerratoCrosbyEtAl2012} and \citet{BechererKentia2017} extend good-deal bounds to model uncertainty. What is new is the form of the switch: within one model, the unique marginal-utility (Davis) price turns into an entropic good-deal interval exactly at the minimal entropy $\cbar$. The entropy budget of Section~\ref{sec:incomplete} then becomes the doubt in excess of the threshold.

Neither the threshold nor the no-trade interval depends on log utility.

\begin{proposition}[Threshold and no-trade interval beyond log utility]\label{prop:utilityfree}
Assume (C1)--(C4). Let $U:(0,\infty)\to\RR$ be strictly increasing, strictly concave, and differentiable at $1$. For $c\ge0$ let $v_U(c)=\sup_{h\in\mathcal H}\min_{\PP\in\Bc}\EE_\PP[U(W_h)]$, and for a claim $X$ and a price $p$ let $v_U^+(c;p)=\sup_{(h,k)\in\mathcal H^+}\min_{\PP\in\Bc}\EE_\PP[U(W_h+k(X-p))]$.
\begin{enumerate}
\item[(a)] If $c\ge\cbar$, then $v_U(c)=U(1)$, and $h=0$ is the unique robust portfolio.
\item[(b)] If $c<\cbar$, then $v_U(c)>U(1)$, so $h=0$ is not a robust portfolio.
\item[(c)] If $c\ge\cbar$, then $v_U^+(c;p)=v_U(c)$ if and only if $p\in I_c(X)$.
\item[(d)] Let $c<\cbar$, and let $U$ be continuously differentiable with $U(x)\to-\infty$ as $x\downarrow0$. Then the robust portfolio $h_c^U$ is unique and nonzero, the worst-case model $\PP_c^U$ against it is unique and charges every state, and $v_U^+(c;p)=v_U(c)$ if and only if $p=\EE_{\PP_c^U}[U'(W_c^U)X]/\EE_{\PP_c^U}[U'(W_c^U)]$, where $W_c^U=W_{h_c^U}$.
\end{enumerate}
\end{proposition}

Parts (a)--(c) apply the non-participation argument for max--min preferences \citep{DowWerlang1992,CaoWangZhang2005} to the KL ball. The ball adds the threshold value $\cbar=\kl{\QE}{\Pz}$ and determines the form of the interval as $I_c(X)=J_{c-\cbar}(X)$. Part (d) identifies a unique no-trade price below $\cbar$, which is the marginal-utility price under the worst-case model; for log utility, this is $p_c(X)=\EE_{\Qc}[X]$. Log utility also gives the closed forms of Theorems~\ref{thm:minimax} and~\ref{thm:robust}. The proof of (a)--(c) relies only on $\Bc$ being compact and convex and meeting $\cl\Mset$ exactly when $c \ge \cbar$. So, for any compact convex set of plausible models, no risky position raises the guaranteed utility above $U(1)$ exactly when the set contains a measure in $\cl\Mset$. The uniqueness of $h=0$ in part (a) also uses the fact that $\cl\Mset \cap \Bc$ contains a strictly positive measure, here $\QE$.

The switch at $\cbar$ also transforms a price quote into a measure of doubt.

\begin{definition}[Implied doubt]\label{def:implieddoubt}
The implied doubt of a quoted price $p$ for the claim $X$ is the smallest radius at which $p$ is a no-trade price of $X$, $c^{\mathrm{imp}}_X(p)=\inf\{c\ge0:\ v^+(c;p)=v(c)\}$ ($\inf\emptyset=+\infty$).
\end{definition}

According to Theorem~\ref{thm:pricing}(a)--(b), the set of radii at which $p$ is a no-trade price of $X$ is $\mathcal N_X(p)\cup[\cbar_X(p),\infty)$, where $\mathcal N_X(p):=\{c\in[0,\cbar):p_c(X)=p\}$ (Corollary~\ref{cor:ec-rob-implied}). So implied doubt has two branches. If $p$ is a robust price $p_c(X)$ for some $c<\cbar$, then the implied doubt is $\min\mathcal N_X(p)$, which rests on log utility. Otherwise, it is $\cbar_X(p)$, the minimal entropy of the extended market; this value does not depend on the choice of utility. Since the first branch lies below $\cbar$ and the second at or above it, implied doubt need not be continuous or monotone in $p$ (Section~\ref{sec:ec-numerics-robust} shows the jump for put quotes).

The reading ``the quote needs at least this much doubt'' holds when $\mathcal N_X(p)$ is empty, which occurs exactly for quotes outside the range of robust prices below $\cbar$. Because this range is narrow in our data (Section~\ref{sec:numerics-put}), we report $\cbar_X(p)$, which is convex in $p$ (Section~\ref{sec:ec-rob-pricing}), as the main reading. For such a quote, we split $\cbar_X(p)$ into two parts. The first part, $\cbar$, is the threshold, which remains the same for every claim and every quote outside the range. The second part, $\cbar_X(p)-\cbar$, is the quote-specific part of implied doubt. When $X$ is not spanned and $p$ lies in its no-arbitrage interval, this difference equals the certainty-equivalent gain that access to $X$ at $p$ gives an exponential investor who trusts $\Pz$ (Lemma~\ref{lem:ec-rob-notrade}). This gain is the entropic good-deal measure of the quote in the sense of \citet{Cerny2003}.

Below $\cbar$, the check of a quote against a radius is a point test. At a radius $c < \cbar$, only $p_c(X)$ is a no-trade price, and every other quote fails, however close it is. So a model-risk report (Section~\ref{sec:impl-radius}) should also give the distance of the quote from $p_c(X)$ and its side.

Implied doubt links robust pricing to the triangle of Section~\ref{sec:triangle}. For a traded asset with relative return $R>0$, consider the short log contract $X=-\ln R$. For every $\QQ\in\cl\Mset$, we have $\EE_\QQ[X]=\kl{\QQ}{\QS}$, which means the price of $X$ is the $\QQ$-side of the triangle. So the implied doubt of an option-implied $\QQ$-side---such as half the squared VIX times the horizon---is the smallest radius at which this quote is a no-trade price for a robust investor with reference model $\Pz$. It is a statement about the quote and the reference model $\Pz$ together (Section~\ref{sec:numerics-vix}).

\subsection{Black--Scholes With Drift Doubt}\label{sec:rob-bs}

Theorems~\ref{thm:minimax} and~\ref{thm:robust} have a continuous-time analogue with closed forms in the Black--Scholes model with doubt about the drift. We fix a horizon $T>0$, a bond $B_t=e^{rt}$, and a stock price $\dd S_t=S_t(\mu\,\dd t+\sigma\,\dd Z_t)$ under $\Pz$, where $Z$ is a standard Brownian motion, $\sigma>0$, and the Sharpe ratio is $\SR=(\mu-r)/\sigma>0$. The investor holds a fraction $w_t$ of their wealth $W_t$ in the stock, such that $\dd W_t=W_t[(r+w_t(\mu-r))\,\dd t+w_t\sigma\,\dd Z_t]$ with initial wealth $W_0=1$. An alternative model $\PP^\phi$ changes the stock drift to $\mu+\sigma\phi_t$, with $\kl{\PP^\phi}{\Pz}=\frac12\EE_{\PP^\phi}[\int_0^T\phi_t^2\,\dd t]$ (Section~\ref{sec:ec-rob-bs} gives the technical conditions). Drift doubt is the set $\mathcal A_c$ of these models with $\kl{\PP^\phi}{\Pz}\le c$, and $v(c)=\sup_w\inf_{\PP^\phi\in\mathcal A_c}\EE_{\PP^\phi}[\ln(W_T/B_T)]$. The market is complete, with a unique pricing measure $\QQ=\PP^{-\SR}$; so $\QE=\QQ$ and $\cbar=\kl{\QQ}{\Pz}=\SR^2T/2$.

\begin{proposition}[Black--Scholes drift doubt]\label{prop:bsdoubt}
Let $a=\sqrt{2c/T}$, the doubt radius in Sharpe-ratio units.
\begin{enumerate}
\item[(a)] If $c<\cbar$, then $v(c)=(\sqrt{\cbar}-\sqrt c)^2$. The robust strategy is unique ($\dd t\otimes\dd\Pz$-a.e.) and constant,
\[
w_c=\frac{\SR-a}{\sigma}=\frac{\mu-r}{\gamma_c\sigma^2},\qquad \gamma_c=\frac{1}{1-\sqrt{c/\cbar}},
\]
which is the Merton fraction of a CRRA investor with risk aversion $\gamma_c$. The worst-case model lowers the drift by the constant $\sigma a$ (that is, $\phi\equiv-a$), and it is the unique worst case against $w_c$.
\item[(b)] If $c\ge\cbar$, then $v(c)=0$, $w\equiv0$ is the unique robust strategy, and $\QQ$ (that is, $\phi\equiv-\SR$) is a worst case.
\item[(c)] (Agreement with Theorem~\ref{thm:robust}.) For $0<c<\cbar$, $\lambda_c=-v'(c)=\sqrt{\cbar/c}-1$, so $\gamma_c=1+1/\lambda_c$, and $c(\gamma)=\cbar(1-1/\gamma)^2$; $v(0)=\cbar=\kl{\Pz}{\QQ}$, and $\QQ$ is the robust pricing measure for every $c$.
\item[(d)] (Calibration to the length of the data.) Suppose the investor estimates $\SR$ from $n$ years of data with $\sigma$ known, so that the estimate has standard error $1/\sqrt n$, and that they doubt every constant drift shift in the confidence interval $|\phi|\le z/\sqrt n$. The matching radius is $c_n=z^2T/(2n)$. With the $t$-statistic $\tau=\SR\sqrt n$ of the mean excess return, $c_n/\cbar=(z/\tau)^2$. If $\tau>z$, then $\gamma_{c_n}=\tau/(\tau-z)$ and $w_{c_n}=w^*(1-z/\tau)$, where $w^*=(\mu-r)/\sigma^2$ is the log-optimal fraction. If $\tau\le z$, that is, if $n\le(z/\SR)^2$, the robust investor holds no stock. The robust strategy is the same whether the doubt set is $\mathcal A_{c_n}$ or only the constant drifts with $|\phi|\le z/\sqrt n$.
\end{enumerate}
\end{proposition}

Parts (a)--(d) are exact; the choice of $c_n$ is a calibration convention used as an illustration. In this model, the threshold is the $\PP$--$\QQ$ side of the Black--Scholes triangle in Section~\ref{sec:triangle-identity}: $\cbar=\SR^2T/2=\kl{\Pz}{\QQ}=\kl{\QQ}{\Pz}$ and $v(0)=\cbar$. Outside this model, however, $v(0)$ and $\cbar$ need not be equal (compare \eqref{eq:rob-cbar} with Theorem~\ref{thm:robust}(i)).

Economically, the worst case lowers the Sharpe ratio by the constant $a$, so the investor acts as if the Sharpe ratio were $\SR-a$. Our additions are the closed forms of $v(c)$ and $c \mapsto \gamma_c$ for the KL ball and their reading in years of data, which matters because drifts are hard to estimate \citep{Merton1980}. Section~\ref{sec:ec-rob-bs} relates the weights to earlier results, and Section~\ref{sec:ec-numerics-robust} measures the accuracy of the closed forms in the empirical market.

\subsection{Choosing the Doubt Radius}\label{sec:rob-radius}

The radius is the only input that the data do not fix. We consider three rules, each based on a first-order large-sample approximation, to address different questions. The full-distribution rule asks for a ball containing the true model. The one-parameter rule asks for a lower confidence bound on a single value: the optimal growth. Finally, under the test rule the investor trades exactly when the data reject zero risk premia. If $\Pz$ is the empirical distribution of $N$ independent observations on $M$ fixed cells, then under the true model $\PP$, $2N\,\kl{\PP}{\Pz}$ is approximately chi-square distributed with $M-1$ degrees of freedom, $\chi^2_{M-1}$ (Proposition~\ref{prop:ec-rob-chi2}; \citealt{BenTalDenHertogDeWaegenaereEtAl2013}). We define $c_{\mathrm{full}}:=\chi^2_{M-1,1-\alpha}/(2N)$, where $\chi^2_{M-1,1-\alpha}$ is the $(1-\alpha)$-quantile of $\chi^2_{M-1}$. This radius gives a ball that contains the true model with probability close to $1-\alpha$ for large $N/M$. This radius grows with $M$. This construction requires independent observations and cells fixed before the data are observed; Section~\ref{sec:ec-numerics-robust} shows its coverage when months depend on each other, while Remark~\ref{rem:ec-rob-cells} provides a first-order heuristic for cells built from sample quantiles.

The one-parameter rule, $c_{\mathrm{one}}:=\chi^2_{1,1-\alpha}/(2N)$, asks less. It matches a confidence interval for one parameter, as in Proposition~\ref{prop:bsdoubt}(d). When $\Pz$ is the empirical distribution of $N$ independent observations, $v(c_{\mathrm{one}})$ is, under regularity conditions, an asymptotic lower confidence bound for the investor's optimal growth under the true model (the value bound), with one-sided level $1-\alpha/2$ \citep{DuchiGlynnNamkoong2021}; at zero premia and with one risky asset this level falls to about $1-\alpha$, and it is lower with more assets (Section~\ref{sec:ec-rob-chi2}). The family of divergence balls of \citet{DuchiGlynnNamkoong2021} includes ours. This rule is the empirical-likelihood calibration of distributionally robust optimization. \citet{Lam2019} studies it with the arguments of KL in the other order and stresses that such balls need not contain the true model. So we can distinguish these two rules by their scope: a ball of radius $c_{\mathrm{full}}$ is likely to contain the true model, which makes every decision and price robust at once, while $c_{\mathrm{one}}$ covers only the value of one decision, the optimal growth. Detection-error probabilities give another calibration (Section~\ref{sec:ec-rob-bs}).

With an empirical reference model, the threshold is a known test statistic. If $\Pz$ is the empirical distribution of $N$ independent observations, $\cbar$ is the minimized Kullback--Leibler information criterion of exponential tilting, and $2N\cbar$ is the exponential-tilting statistic for the hypothesis that all $d$ risk premia are zero, approximately $\chi^2_d$ under this hypothesis \citep{KitamuraStutzer1997}. The test rule uses the radius $\chi^2_{d,1-\alpha}/(2N)$, which we call the test radius: with it, the investor trades exactly when the exponential-tilting test rejects zero premia at asymptotic level $\alpha$. With one risky asset the test radius equals $c_{\mathrm{one}}$, and the rule is the nonparametric form of Proposition~\ref{prop:bsdoubt}(d), where the investor holds stock exactly when $\tau>z$. Section~\ref{sec:ec-rob-chi2} shows why $c_{\mathrm{one}}$ is not a test radius when $d>1$, and Section~\ref{sec:numerics} applies the three rules to U.S. data. The rules only choose $c$; Theorems~\ref{thm:minimax}--\ref{thm:pricing} hold for every $c$.

\section{Measurement and Illustrations}\label{sec:numerics}

This section takes the theory to U.S. data. It measures the risk-premium triangle, calibrates the doubt radius to the length of the data, prices non-traded claims under doubt, and reads the VIX as a market quote. Sections~\ref{sec:numerics-doubt}--\ref{sec:numerics-vix} give the numbers of the model-risk report in Section~\ref{sec:impl-report}. Section~\ref{sec:ec-numerics} adds robustness checks, synthetic examples, and computation.

\subsection{Data}\label{sec:numerics-data}

We use two public data sources. The first is the daily history of the Cboe Volatility Index (VIX) from January 1990 to October 2026 \citep{CboeVIXHistory2026}. Cboe computes the index from out-of-the-money S\&P~500 options with two expirations, interpolated to a constant 30-day horizon \citep{CboeVIXWhitePaper}; as the method has changed over time, our sample spans several methodology vintages (Section~\ref{sec:ec-code}). We write $\VIX_t$ for the close on the last VIX trading day of month $t$, in decimal units. The second source is the Kenneth R.~French Data Library \citep{FrenchDataLibrary2026}, which provides monthly and daily files from July 1926 to September 2026 containing the return of the one-month Treasury bill and the excess return over the bill for the value-weighted U.S. stock market of the Center for Research in Security Prices (CRSP).

We map these data sources to the objects of our theory: the Treasury bill is the bond, and the stock market is the asset $S$ of Section~\ref{sec:triangle}. The relative gross return of month $t+1$ is $R_{t+1}:=(1+r^{m}_{t+1})/(1+r^{f}_{t+1})$, where $r^{m}_{t+1}$ and $r^{f}_{t+1}$ are the simple market and bill returns. The horizon is $T_{t+1}:=(\text{days in month }t+1)/365$, and the realized variance $\mathrm{RV}_{t+1}$ sums the squared daily log market returns of month $t+1$. Our panel pairs each month-end $t$ from January 1990 to August 2026 with the next month, 440 pairs in all. Each $\VIX_t$ is observed before any daily return of month $t+1$.

The two data sources do not refer to the same index: the VIX uses S\&P~500 options, while $R$ is the CRSP market, a total-return series that covers a century. Matched S\&P~500 total returns are not part of our data. If the monthly return variance of the CRSP market exceeds that of the S\&P~500, then the estimate $\hat\Lent_{\PP}$ of the P-entropy in Section~\ref{sec:numerics-triangle} overstates the P-entropy of the S\&P~500, and the estimate $\hat\varepsilon$ of the entropy premium understates its entropy premium. The standard errors below measure sampling error only.

\subsection{The Risk-Premium Triangle of the U.S. Market}\label{sec:numerics-triangle}

Let $\PP_t$ and $\QQ_t$ be the real-world and pricing laws of $R_{t+1}$ given the information at month-end $t$, and $\QS_t$ the share measure, $\dd\QS_t/\dd\QQ_t=R_{t+1}$. Applied month by month, Theorem~\ref{thm:triangle} and Proposition~\ref{prop:decomp} give
\begin{equation}\label{eq:emp-defect}
\defect_t:=\defect(\PP_t,\QQ_t,\QS_t)=\EE_{\PP_t}[\ln R_{t+1}]+\kl{\QQ_t}{\QS_t},
\qquad
\kl{\QQ_t}{\QS_t}\approx\tfrac12\,\VIX_t^2\,T_{t+1}.
\end{equation}
The second term is the $\QQ$-side of the triangle, the forward price of the option portfolio of Proposition~\ref{prop:decomp}(ii), which the squared VIX measures up to discretization \citep{Martin2017}. The first term, the expected log excess return, is not a side of the triangle and is not observed month by month; we call it the P-term. By iterated expectations the average defect is $\EE[\ln R_{t+1}]+\EE[\kl{\QQ_t}{\QS_t}]$, where $\EE$ is the unconditional mean, and we estimate it by $\hat\defect:=\overline{\ln R}+\overline{\tfrac12\VIX^2T}$, where a bar denotes the sample mean. Proposition~\ref{prop:decomp} splits the defect into the log equity premium $\pi=\ln\EE_{\PP}[R]$ and the entropy premium $\varepsilon=\Lent_{\QQ}(R)-\Lent_{\PP}(R)$, with sample analogues $\hat\pi:=\ln\overline{R}$, $\hat\Lent_{\PP}:=\ln\overline{R}-\overline{\ln R}$ and $\hat\varepsilon:=\overline{\tfrac12\VIX^2T}-\hat\Lent_{\PP}$, so that $\hat\defect=\hat\pi+\hat\varepsilon$.

In standard terms, $\hat\defect$ is the mean log excess return plus half the mean option-implied variance, and $\hat\pi$ is the equity premium in logs. The entropy premium $\hat\varepsilon$ is half the difference between the option-implied variance and the variance of monthly log returns, $\frac12(\overline{\VIX^2T}-\widehat{\Var}(\ln R))$, plus a higher-cumulant term of 0.012\% per year (Section~\ref{sec:ec-numerics-vrp}). Annualized, the two variances are 0.0433 and 0.0231, while the mean daily realized variance $\overline{\mathrm{RV}}$ is 0.0327; we report percent per year (monthly values times 12).

\begin{table}[t]
\TABLE{The risk-premium triangle of the U.S. stock market, monthly horizon (\% per year).\label{tab:emp-triangle}}
{\footnotesize\begin{tabular}{@{}lllll@{}}
\toprule
 & Full sample & 1990--2007 & 2008--2026 & Without 2008--2009 \\
\midrule
$\QQ$-side, $\overline{\tfrac12\VIX^2T}$ & 2.16 [1.75, 2.64] & 1.99 [1.48, 2.51] & 2.33 [1.68, 3.13] & 1.96 [1.63, 2.32] \\
P-term, $\overline{\ln R}$ & 7.97 [2.68, 12.93] & 6.51 [0.25, 13.37] & 9.37 [2.73, 17.19] & 9.07 [4.67, 13.43] \\
Defect $\hat\defect$ & 10.14 [4.97, 15.01] & 8.50 [2.44, 15.33] & 11.70 [5.41, 19.43] & 11.04 [6.71, 15.39] \\
Log equity premium $\hat\pi$ & 9.12 [3.96, 13.95] & 7.49 [1.49, 14.22] & 10.67 [4.34, 18.26] & 10.11 [5.80, 14.37] \\
P-entropy $\hat\Lent_{\PP}$ & 1.14 [0.86, 1.46] & 0.98 [0.63, 1.34] & 1.29 [0.85, 1.80] & 1.04 [0.79, 1.30] \\
Entropy premium $\hat\varepsilon$ & 1.02 [0.81, 1.27] & 1.01 [0.77, 1.28] & 1.03 [0.71, 1.48] & 0.93 [0.75, 1.13] \\
Months & 440 & 215 & 225 & 416 \\
\bottomrule
\end{tabular}}
{$\QQ$-side: mean of $\frac12\VIX_t^2T_{t+1}$, which measures $\kl{\QQ_t}{\QS_t}$ from option prices; P-term: mean log return $\overline{\ln R}$ of the CRSP value-weighted market relative to the one-month bill; P-entropy $\hat\Lent_{\PP}=\ln\overline{R}-\overline{\ln R}$. Return months 1990:02--2026:09. Brackets: 95\% moving-block bootstrap percentile intervals \citep[12-month blocks, 10,000 draws;][]{Kunsch1989}; with the block length of \citet{PolitisWhite2004} for the $\QQ$-side, 31 months, its full-sample interval is $[1.63,2.78]$. Full-sample Newey--West standard errors \citep[12 lags, delta method for $\hat\pi$, $\hat\Lent_{\PP}$, $\hat\varepsilon$;][]{NeweyWest1987}, by row: 0.23, 2.61, 2.54, 2.53, 0.15 and 0.12; Table~\ref{tab:ec-emp-robust} gives other bandwidths.}
\end{table}

\looseness=-1 In the full sample of Table~\ref{tab:emp-triangle}, the defect is 10.14\% per year, with a 95\% bootstrap interval of $[4.97,15.01]$. The log equity premium is 9.12\%, representing a share of 0.899, and the entropy premium is 1.02\%, representing a share of 0.101. These two components differ sharply in precision: the entropy premium has a Newey--West standard error of 0.12 percentage points and lies between 0.93\% and 1.03\% in every subsample of Table~\ref{tab:emp-triangle}, while the log equity premium for 1990--2026 is high and imprecise, with a standard error of 2.53 points. So the defect---the premium of the log contract---is mostly drift, and option-implied entropy exceeds monthly-return entropy by about one point per year.

\looseness=-1 The entropy premium has a benchmark in Proposition~\ref{prop:itolevy}: with a deterministic drift, it is half the variance risk premium plus a jump term of third and higher order. The usual measure of half the variance risk premium, $\frac12(\overline{\VIX^2T}-\overline{\mathrm{RV}})$ \citep{CarrWu2009,BollerslevTauchenZhou2009}, is 0.53\% per year (with a $[0.24,0.76]$ interval), which is about half of $\hat\varepsilon$. Almost all of the remaining 0.49 points come from serial dependence. The variance of monthly log returns is only 0.708 times the mean daily realized variance---a variance ratio below one---because daily returns partly reverse within the month, as index returns have tended to do in recent decades \citep{BaltussenVanBekkumDa2019}. The products of daily log returns for the same month average $-0.429\%$ per year, while a deterministic drift implies $0.104\%$. The data reject the deterministic drift (Newey--West $t$-statistic $-2.90$), even when the 1, 5, 10, or 20 highest-variance months are dropped (Section~\ref{sec:ec-numerics-vrp}).

Section~\ref{sec:ec-numerics-vrp} also states three caveats besides the index mismatch. The main one is that $\hat\pi=\ln\overline{R}$ estimates an upper bound on the average conditional log equity premium, so $\hat\varepsilon$ estimates a lower bound on the average conditional entropy premium.

\subsection{How Much Doubt Do the Data Allow?}\label{sec:numerics-doubt}

The length of the data sets a scale for the doubt radius (Proposition~\ref{prop:bsdoubt}(d)). In the Black--Scholes market, an investor who doubts every drift inside a confidence interval of $z$ standard errors, estimated from $n$ years of returns, has the radius $c_n=z^2T/(2n)$. This implies that $c_n/\cbar=(z/\tau)^2$, where $\tau=\SR\sqrt n$ is the $t$-statistic of the mean excess return. If $\tau>z$, the investor's effective risk aversion is $\gamma_{c_n}=\tau/(\tau-z)$ and their stock weight is $w_{c_n}=w^*/\gamma_{c_n}$, where $w^*=\SR/\sigma$ is the log-optimal weight; if $\tau\le z$, they hold no stock.

\begin{table}[t]
\TABLE{Black--Scholes drift doubt calibrated to the length of the data.\label{tab:emp-doubt}}
{\footnotesize\begin{tabular}{@{}lrrrrrrrrrr@{}}
\toprule
Sample & $n$ (years) & $\SR$ & $\sigma$ (\%) & $z$ & $c_n/\cbar$ & $\gamma_{c_n}$ & $\gamma_{c_n}$, HAC & $w^*$ & $w_{c_n}$ & $n^*$ (years) \\
\midrule
1926:07--2026:09 & 100.25 & 0.454 & 18.4 & 1.96 & 0.186 & 1.76 & 1.86 & 2.47 & 1.41 & 18.66 \\
 & & & & 1 & 0.0485 & 1.28 & 1.31 & & 1.93 & 4.86 \\
\addlinespace
1990:01--2026:09 & 36.75 & 0.588 & 15.2 & 1.96 & 0.302 & 2.22 & 2.29 & 3.87 & 1.75 & 11.1 \\
 & & & & 1 & 0.0786 & 1.39 & 1.40 & & 2.79 & 2.89 \\
\bottomrule
\end{tabular}}
{CRSP value-weighted market, simple monthly excess returns over the bill. $\SR$: annualized Sharpe ratio; $\sigma$: annualized standard deviation of simple monthly excess returns; $z$: half-width, in standard errors, of the confidence interval for the Sharpe ratio; $c_n/\cbar$: doubt radius relative to the threshold; $\gamma_{c_n}$: effective risk aversion; $w^*$ and $w_{c_n}$: log-optimal and robust fractions of wealth in the stock; $n^*=(z/\SR)^2$: years of data at or below which the robust investor holds no stock. HAC (heteroskedasticity and autocorrelation consistent): $\tau$ with a Newey--West standard error (12 lags) of the mean excess return; with the Andrews bandwidth, $\gamma_{c_n}=1.81$ and 2.22 at $z=1.96$. With the Black--Scholes-consistent Sharpe ratio from log returns, $\gamma_{c_n}$ is also 1.76 (1926--2026, $z=1.96$). An illustration in the Black--Scholes model with independent and identically distributed (i.i.d.) returns.}
\end{table}

Table~\ref{tab:emp-doubt} applies this rule to the CRSP market. For the 1926--2026 period with $z=1.96$, we find $c_n/\cbar=0.186$ and an effective risk aversion of $\gamma_{c_n}=1.76$. Then the robust investor holds 1.41 times their wealth in stocks, compared to 2.47 for the log investor; with 18.66 years of data or fewer, they would hold no stock. The shorter sample since 1990 allows more doubt and yields $\gamma_{c_n}=2.22$.

This calibration is an illustration in the Black--Scholes model, where returns are independent and identically distributed (i.i.d.). In this setting, the standard error $1/\sqrt n$ of the Sharpe ratio ignores serial correlation and the variance factor $1+\SR_m^2/2$ of \citet{Lo2002} for the monthly Sharpe ratio $\SR_m$; this factor is only about 1.009 here, so serial correlation matters more. With a Newey--West standard error, $\gamma_{c_n}$ rises to 1.86, and over the 95\% interval $\tau\pm1.96$ of the i.i.d.\ $\tau$ itself it ranges from 1.43 to 4.15 (Table~\ref{tab:ec-emp-robust}).

\subsection{Robust Option Prices from the Empirical Distribution}\label{sec:numerics-put}

We now price non-traded claims using Theorem~\ref{thm:pricing}. The reference model $\Pz$ is the empirical distribution of $R$ from July 1926 to September 2026, in which each of the $N=1203$ months is a state with probability $1/N$, and its value is that month's $R$; Section~\ref{sec:ec-numerics-robust} reports binned versions. The traded assets are the bill and the market ($d=1$, $\DeltaS=R-1$). The claims are one-month puts with payoff $(K-R)_+$, $K\in\{0.90, 0.95, 1.00\}$, and a digital claim that pays one if $R<0.90$; prices are in percent of the forward. The log-optimal investor holds $\hstar=2.25$ times their wealth in the market, which is less than the Black--Scholes weight $w^*=2.47$ of Table~\ref{tab:emp-doubt}. This difference arises because $\hstar$ uses the whole empirical distribution, while $w^*=\SR/\sigma$ uses only its mean and variance. The threshold is $\cbar=0.857\%$ per month, with a 95\% block-bootstrap interval of 0.22\% to 1.96\% (Section~\ref{sec:ec-numerics-robust}).

\begin{figure}[t]
\FIGURE{\includegraphics[width=0.76\textwidth]{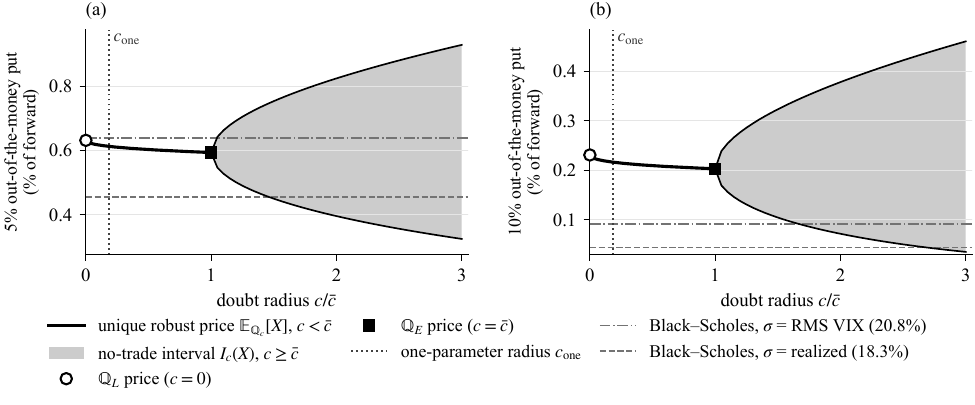}}
{Robust prices of one-month puts struck 5\% and 10\% out of the money, empirical incomplete market.\label{fig:emp-put}}
{$\Pz$: monthly U.S. market returns, 1926:07--2026:09, each of the $N=1203$ months a state with probability $1/N$. (a) $K=0.95$; (b) $K=0.90$; prices in \% of the forward. Below $\cbar$ the robust price is unique; from $\cbar$ on, the no-trade interval. $c_{\mathrm{one}}=\chi^2_{1,0.95}/(2N)=0.186\,\cbar$. Black--Scholes prices at the realized volatility and at the root-mean-square (RMS) VIX are references, not market quotes.}
\end{figure}

Figure~\ref{fig:emp-put} and Table~\ref{tab:emp-put} show that doubt below $\cbar$ moves the unique price of the 5\% out-of-the-money put down, and only a little: from the $\QL$ price of 0.632\% to the $\QE$ price of 0.594\%. This occurs because the log-optimal investor borrows, so their wealth $\Wstar$ is low in crash states, and $\QL=\Pz/\Wstar$ weights these states more than $\QE$ does. The $\QE$ price is the canonical valuation price of \citet{Stutzer1996}, and the $\QL$ price is the empirical-likelihood price; below $\cbar$, the doubt radius selects a tilt between them (Section~\ref{sec:rob-structure}).

This analysis is a partial-equilibrium exercise where $\Pz$ is fixed at the empirical distribution---which already contains the observed crashes---and only the stock and the bond trade. In equilibrium models, aversion to uncertainty about rare events raises out-of-the-money put prices because it changes the pricing measure itself \citep{LiuPanWang2005,Drechsler2013}, and a small chance of a macroeconomic disaster can fit index option skewness and credit spreads together \citep{ChristoffersenDuElkamhi2017}; here, by contrast, doubt below $\cbar$ moves the robust pricing measure from $\QL$ to $\QE$ and lowers the put price, and with CRRA base utilities the put price also ends lower at $\cbar$ than at $c=0$ (Figure~\ref{fig:ec-crra-path}). The prices are an in-sample, one-period illustration that treats months as independent draws (Section~\ref{sec:ec-numerics-robust} checks the radius rules under dependence).

\begin{table}[t]
\TABLE{Robust prices of non-traded claims in the empirical incomplete market (\% of the forward).\label{tab:emp-put}}
{\footnotesize\begin{tabular}{@{}lllll@{}}
\toprule
 & Put, $K=0.90$ & Put, $K=0.95$ & Put, $K=1.00$ & Digital, $R<0.90$ \\
\midrule
$\EE_{\Pz}[X]$ (no risk adjustment) & 0.124 & 0.418 & 1.569 & 2.660 \\
$\EE_{\QL}[X]$, unique price at $c=0$ & 0.231 & 0.632 & 1.976 & 4.102 \\
Unique price at $c=c_{\mathrm{one}}=0.186\,\cbar$ & 0.216 & 0.613 & 1.959 & 3.996 \\
$\EE_{\QE}[X]$, unique price as $c\uparrow\cbar$ & 0.203 & 0.594 & 1.939 & 3.874 \\
\addlinespace
No-trade interval, $c=1.5\,\cbar$ & [0.104, 0.323] & [0.450, 0.754] & [1.779, 2.114] & [2.473, 5.404] \\
No-trade interval, $c=2\,\cbar$ & [0.072, 0.378] & [0.395, 0.825] & [1.717, 2.191] & [1.939, 6.069] \\
No-trade interval, $c=3\,\cbar$ & [0.035, 0.460] & [0.324, 0.930] & [1.632, 2.306] & [1.239, 7.040] \\
No-arbitrage bounds & [0.000, 10.8] & [0.000, 13.6] & [0.007, 16.5] & [0.000, 79.4] \\
\addlinespace
Black--Scholes, $\sigma=18.34\%$ (realized) & 0.044 & 0.456 & 2.111 & 2.476 \\
Black--Scholes, $\sigma=20.81\%$ (RMS VIX) & 0.091 & 0.639 & 2.396 & 4.235 \\
\bottomrule
\end{tabular}}
{$\Pz$: monthly relative returns, 1926:07--2026:09, each of the $N=1203$ months a state with probability $1/N$. For $c\ge\cbar$: $I_c(X)$. $\EE_{\QE}[X]$: canonical valuation \citep{Stutzer1996}; $\EE_{\QL}[X]$: empirical-likelihood weights. Radii (Section~\ref{sec:rob-radius}): $c_{\mathrm{one}}=\chi^2_{1,0.95}/(2N)=0.186\,\cbar$ and $c_{\mathrm{full}}=\chi^2_{N-1,0.95}/(2N)=62.2\,\cbar$. Black--Scholes: one-month prices; realized volatility: annualized standard deviation of monthly $\ln R$; RMS VIX: root mean square of the month-end VIX, 1990--2026.}
\end{table}

Above $\cbar$, the 5\% put has a no-trade interval of $[0.450\%, 0.754\%]$ at $c=1.5\,\cbar$. A trading desk would compare this width with the difference between two Black--Scholes prices: one calculated at the 1926--2026 realized volatility of 18.3\%, that of $\Pz$, and another at the 1990--2026 root-mean-square (RMS) VIX of 20.8\%, which matches the mean $\VIX^2$ behind the $\QQ$-side. These prices are 0.456\% and 0.639\%, respectively. The width is about 1.7 times this difference, a ratio that grows with the radius and varies by strike.

The one-parameter radius $c_{\mathrm{one}} = \chi^2_{1,0.95}/(2N)$ of Section~\ref{sec:rob-radius} is $0.186\,\cbar$. Because this value lies below the threshold, the put option has one robust price of 0.613\% at this radius. There, the effective risk aversion is $\gamma_c=1.79$ and the stock weight is 1.32, not the 2.25 of zero doubt. In absolute terms, $c_{\mathrm{one}}$ equals the radius $c_n$ of Section~\ref{sec:numerics-doubt} with $T$ equal to one month, because $\chi^2_{1,0.95}=1.96^2$. Although the two ratios use different thresholds---$\cbar$ is the minimal entropy of the empirical market here, and in Table~\ref{tab:emp-doubt} it is $\SR^2T/2$---they agree because $2N\cbar=20.6$, the exponential-tilting statistic for a zero premium (Section~\ref{sec:rob-radius}), nearly equals $\tau^2=4.54^2=20.6$ (Table~\ref{tab:emp-doubt}). So the test rejects a zero premium, which is why the investor with radius $c_{\mathrm{one}}$ trades. The full-distribution radius, $62\,\cbar$, does not cover the true model when every month is a state. With 41 or more cells, binning moves the $\QL$ and $\QE$ prices of the tail claims by up to 15\% (Table~\ref{tab:ec-binning}).

\subsection{The VIX as a Market Quote}\label{sec:numerics-vix}

\looseness=-1 The VIX is the only market price in our data of a claim that the stock and the bond do not span. By Proposition~\ref{prop:decomp}(ii), $\frac12\VIX_t^2T_{t+1}$ is the forward price of the short log contract $X=-\ln R$, whose price under a pricing measure $\QQ$ is the $\QQ$-side. Theorem~\ref{thm:robust}(iv) implies that its robust prices below $\cbar$ are the marginal-utility prices of CRRA investors with $\gamma\ge1$ who trust $\Pz$, ranging from $\EE_{\QL}[X]$ to $\EE_{\QE}[X]$. Let $\Pz$ first be the empirical distribution of the 440 panel months, 1990--2026, which match the VIX sample. This range of robust prices is then 0.108\% to 0.113\% per month, and the mean quote $\bar{p}=0.180\%$ lies 60\% above it, with a 95\% block-bootstrap interval of 37\% to 83\%. The mean quote lies above the range in all 2,000 draws (Table~\ref{tab:emp-bridge}). So no CRRA investor with $\gamma\ge1$ who trusts the 1990--2026 distribution prices the short log contract at the average VIX. With the century of Section~\ref{sec:numerics-put}, the range is 0.150\% to 0.160\%, and the gap, 13\% with interval $-17\%$ to 52\%, is not significant.

The implied doubt of the mean quote is $\cbar_X(\bar p)=4.99\,\cbar$ and $1.30\,\cbar$ in the two samples. As in Section~\ref{sec:rob-pricing}, this value has two parts; Table~\ref{tab:emp-bridge} gives intervals for $\cbar_X(\bar p)$ and for both parts. The first part, $\cbar$, is the threshold---the doubt at which a model with no risk premium becomes plausible. It is the same for every quote outside the range, and it alone is 3.4 and 5.4 times the one-parameter radius $c_{\mathrm{one}}$. The second part, $3.99\,\cbar$ and $0.30\,\cbar$, is the quote-specific part. This component is the certainty-equivalent gain per month that $X$ at $\bar p$ gives an investor with exponential utility and unit absolute risk aversion who trusts $\Pz$, the entropic good-deal measure of the quote \citep{Cerny2003}.

With the 1990--2026 reference model, where the gap is significant, the average VIX is thus a no-trade price for the robust log investor only at radii at which they hold no stock (Theorem~\ref{thm:pricing}). So the VIX carries a premium that KL-ball doubt cannot explain while the investor still holds stock, and the quote-specific part of implied doubt, $3.99\,\cbar$, measures how much doubt beyond the threshold the mean quote requires. While equilibrium models relate the variance premium to risk aversion and the higher moments of the real-world distribution \citep{BakshiMadan2006} or to model uncertainty \citep{Drechsler2013}, this static diagnostic adds one number per quote: the doubt it requires in KL units.

In price units, the 1990--2026 gap is 0.81\% per year: the entropy premium $\hat{\varepsilon} = 1.02\%$ of Table~\ref{tab:emp-triangle} less 0.21\%, the entropy premium $\Lent_{\QL}(R)-\Lent_{\Pz}(R)$ of $\QL$, since $\bar p=\Lent_{\Pz}(R)+\hat\varepsilon$ and $\EE_{\QL}[X]=\Lent_{\QL}(R)$ (Proposition~\ref{prop:decomp}(ii)). Much of the gap comes from the within-month reversal of daily returns, which makes $\Pz$ less volatile than daily realized variance suggests (Section~\ref{sec:numerics-triangle}): if the variance of $\Pz$ is set to the mean daily realized variance of its months, the gap falls to 17\%, at the edge of significance, while over the century it rises from 13\% to 23\%, still not significant (Section~\ref{sec:ec-numerics-robust}).

Two caveats apply to these results. First, because the exercise compares average conditional quotes with a static reference model, it cannot separate doubt from misspecification of that model, such as ignored variation in volatility. Second, if the CRSP market is more volatile than the S\&P~500, then the robust prices are too high and the gap understated (Section~\ref{sec:numerics-data}).

The reference model can also follow volatility. Built in real time with a nearly unbiased forecast of next month's variance, it leaves the quote above the range of robust prices in 369 of the 440 months; this count depends on how the forecast is calibrated (Section~\ref{sec:ec-numerics-robust} and Figure~\ref{fig:ec-vix-monthly}).

\begin{table}[t]
\TABLE{One market in one unit: the triangle, the robust objects and the VIX quote.\label{tab:emp-bridge}}
{\footnotesize\begin{tabular}{@{}lrr@{}}
\toprule
$\Pz$: empirical distribution of monthly $R$ & 1990--2026 & 1926--2026 \\
\midrule
P-term $\EE_{\Pz}[\ln R]$ & 0.664 & 0.552 \\
Slack of the entropy bound, $\kl{\Pz}{\QQ^S_L}$ & 0.739 & 0.269 \\
Robust value at zero doubt, $v(0)=\kl{\Pz}{\QL}$ & 1.403 & 0.821 \\
Threshold $\cbar=\kl{\QE}{\Pz}$ (95\% interval) & 1.483 [0.25, 3.91] & 0.857 [0.22, 1.96] \\
$\QQ$-side $\EE_{\QQ}[-\ln R]=\kl{\QQ}{\QS}$: $\QQ=\QL$, $\QE$, market ($\bar p$) & 0.113, 0.108, 0.180 & 0.160, 0.150, 0.180 \\
Defect $\defect(\Pz,\QQ,\QS)$: $\QQ=\QL$, $\QE$, market & 0.777, 0.772, 0.845 & 0.712, 0.702, 0.732 \\
\addlinespace
Radii in units of $\cbar$: $c_{\mathrm{one}}$, $c_n$ (Table~\ref{tab:emp-doubt}), $c_{\mathrm{full}}$ & 0.294, 0.294, 37.4 & 0.186, 0.186, 62.2 \\
Gap of $\bar p$ above the range of robust prices, \% (95\% interval) & 60 [37, 83] & 13 [$-17$, 52] \\
Share of bootstrap draws with $\bar p$ above the range & 1.00 & 0.81 \\
$\cbar_X(\bar p)/\cbar$ (95\% interval), the implied doubt when $\bar p$ lies outside the range & 4.99 [2.25, 27.6] & 1.30 [1.00, 4.70] \\
Quote-specific part $(\cbar_X(\bar p)-\cbar)/\cbar$ (95\% interval) & 3.99 [1.25, 26.6] & 0.30 [0.00, 3.70] \\
\bottomrule
\end{tabular}}
{One-month horizon; KL terms in \% per month. Months: the 440 panel months 1990:02--2026:09, and 1926:07--2026:09 ($N=1203$). $\QQ^S_L$: share measure of $\QL$, so the P-term plus the slack is $v(0)$ (Proposition~\ref{prop:slack}). Market: $\bar p$, the mean of $\frac12\VIX_t^2T_{t+1}$; for 1990--2026 the market defect times 12 is $\hat\defect$ of Table~\ref{tab:emp-triangle}, and in the 1926--2026 column the market entries combine the century P-term with the 1990--2026 mean of $\frac12\VIX_t^2T_{t+1}$. Range of robust prices: from $\EE_{\QE}[X]$ to $\EE_{\QL}[X]$, $X=-\ln R$; gap: $\bar p/\EE_{\QL}[X]-1$. Intervals: 2.5\% and 97.5\% percentiles of a circular block bootstrap with 12-month blocks and 2,000 draws, which redraws reference months and quotes jointly (Section~\ref{sec:ec-numerics-robust}). In 213 of the 2,000 century draws $\bar p$ lies inside the range, where its implied doubt is below $\cbar$. $c_n$: $z=1.96$, in units of this table's $\cbar$.}
\end{table}

\section{Managerial Implications}\label{sec:implications}

Our results give risk managers, model validators, and trading desks the three uses announced in Section~\ref{sec:intro}; Section~\ref{sec:impl-report} puts them together in one report.

\subsection{Choosing and Reporting the Doubt Radius}\label{sec:impl-radius}\label{sec:impl-distance}

A model validator must quantify how wrong a model may be; the ball of plausible models gives one number once a stated rule fixes its radius. KL balls already serve this purpose in stress tests and model-risk measurement \citep{BreuerCsiszar2016,GlassermanXu2014}. The validator should first report $c/\cbar$ because it is utility-free: every investor with a smooth, strictly increasing, and strictly concave utility holds no risky asset exactly when $c/\cbar \ge 1$ (Proposition~\ref{prop:utilityfree}).

The rule should match the question. For the trading decision involving $N$ independent observations of $d$ risky assets, we recommend the test radius $\chi^2_{d,1-\alpha}/(2N)$ described in Section~\ref{sec:rob-radius}. With an empirical reference model, the investor at this radius trades exactly when the exponential-tilting test for zero risk premia rejects at the asymptotic level $\alpha$. In the case of one risky asset, the test radius is $c_{\mathrm{one}}$, and its robust value is also an asymptotic lower confidence bound for the optimal growth. In our century of returns data, $c_{\mathrm{one}}=0.186\,\cbar$ lies far below $\cbar$; however, with the same Sharpe ratio but 18.66 years of data or fewer, the Black--Scholes investor of Table~\ref{tab:emp-doubt} would hold no stock.

For a report that holds for every claim at once, the chosen ball must contain the true model. The full-distribution radius $c_{\mathrm{full}}=\chi^2_{M-1,1-\alpha}/(2N)$ gives such a ball with probability close to $1-\alpha$, but only for independent observations and only on $M$ cells fixed before the data are seen (for example, cells cut at the strikes of the claims). In our data, the choice of $M$ decides whether a claim has one robust price or an interval (Table~\ref{tab:ec-binning}). With blocks of 12 or 36 months, the bootstrap coverage of $c_{\mathrm{full}}$ on 41 cells is 85\% and 75\%, respectively (Section~\ref{sec:ec-numerics-robust}). We therefore recommend $c_{\mathrm{full}}$ only with cells fixed in advance and a block-bootstrap check of its coverage.

A radius can also be derived from a quote $p$ of a non-traded claim $X$, and a report should give three pieces of information (Section~\ref{sec:rob-pricing}). The first is how far the quote lies from the reference model, in price units (the distance from the range of robust prices below $\cbar$) and in entropy units (the quote-specific part of implied doubt, $\cbar_X(p)-\cbar$). The second is whether the quote is above or below that range. The third is the reference model, including its treatment of volatility, because implied doubt is a statement about both the quote and $\Pz$ together. At a radius below $\cbar$, only the robust price is a no-trade price, however close another quote may be. So a desk should also report bootstrap bands for the distance of the quote from the range of robust prices and for $\cbar_X(p)$, as Table~\ref{tab:emp-bridge} does for the VIX.

Revised guidance on model risk management, issued on April 17, 2026 by the Board of Governors of the Federal Reserve System, the Office of the Comptroller of the Currency and the Federal Deposit Insurance Corporation \citepalias{FederalReserveOCCFDIC2026}, supersedes \citetalias{FederalReserveOCC2011} and \citetalias{FederalReserveSR2108}. We do not claim that our measure meets any part of this guidance; it is one input to a validation report.

\subsection{Doubt as Risk Aversion}\label{sec:impl-riskaversion}

A KL radius is not a familiar desk input; risk aversion is. With log utility, Theorem~\ref{thm:robust}(iv) translates one into the other, but only within one market: $\gamma_c$ depends on both the radius and the market (Remark~\ref{rem:scope}). So a desk must recompute $\gamma_c$ when the reference model, assets, or horizon change. CRRA tools miss the no-trade region: a CRRA investor holds risky assets whenever a risk premium is nonzero. While \citet{BarillasHansenSargent2009} reread a large risk aversion as moderate doubt, our map makes this exact for the ball and shows where it ends.

In the calibration of Proposition~\ref{prop:bsdoubt}(d), the robust weight is $w^*(1-z/\tau)$ if $\tau>z$ and zero otherwise, where $w^*$ is the log-optimal weight. This is a fractional Kelly rule \citep{MacLeanZiembaBlazenko1992} with fraction $1/\gamma_c=1-z/\tau$.

\subsection{The Threshold and Price Intervals for Non-Traded Risks}\label{sec:impl-threshold}\label{sec:impl-intervals}

Theorem~\ref{thm:pricing} specifies what to report for a non-traded claim $X$. Below $\cbar$, there is one robust price: $\EE_{\Qc}[X]$ with log utility, and the marginal-utility price under the worst-case model (Proposition~\ref{prop:utilityfree}(d)) for an investor with a continuously differentiable utility that tends to $-\infty$ at zero wealth. How doubt moves this price depends on the market (Section~\ref{sec:numerics-put}).

Above $\cbar$, the model does not single out one price; every price in $I_c(X)$ is a no-trade price. A desk requiring one mark may use a stated rule, such as the minimal-entropy price $\EE_{\QE}[X]$, and report $I_c(X)$ alongside it. The width of this entropic good-deal interval is a range of prices over plausible models, as in \citet{Cont2006}. Section~\ref{sec:numerics-put} compares this width to the difference in Black--Scholes prices at two volatilities. European prudent-valuation rules require a model-risk valuation adjustment for the range of models and calibrations used by market participants \citep[Article~11]{EuropeanCommission2016PV}; $I_c(X)$ can be one input to such an adjustment.

\subsection{A Model-Risk Report in Practice}\label{sec:impl-report}

Table~\ref{tab:impl-report} applies the three uses to one desk. The desk holds the U.S. stock market and the one-month bill, uses the century of monthly returns of Section~\ref{sec:numerics-put} as its reference model, and fixes the radius by the test rule of Section~\ref{sec:rob-radius}, which with one risky asset gives $c_{\mathrm{one}}$. Each line answers a question that a validator or a trader asks, and every number is computed from the data of Section~\ref{sec:numerics}; the effective risk aversion and the stock weight are those of the market of Section~\ref{sec:numerics-put} at $c_{\mathrm{one}}$. The ratio $c/\cbar=0.186<1$ says that the desk should hold stock. The effective risk aversion of 1.79 says what the doubt costs, in a unit the desk already uses. The put lines give one price today and the range that a larger doubt would allow. The last line says how much doubt a market quote implies.

\begin{table}[t]
\TABLE{A model-risk report for a desk holding the U.S. stock market.\label{tab:impl-report}}
{\footnotesize\begin{tabular}{@{}p{0.33\textwidth}p{0.43\textwidth}p{0.18\textwidth}@{}}
\toprule
Question & Answer & Source \\
\midrule
How much doubt does the rule allow? & $c_{\mathrm{one}}=0.186\,\cbar$, with $\cbar=0.857\%$ per month & Section~\ref{sec:rob-radius} \\
Should the desk hold stock? & Yes: $c/\cbar=0.186<1$ & Proposition~\ref{prop:utilityfree} \\
What does the doubt cost? & Effective risk aversion $\gamma_c=1.79$; stock weight 1.32, not 2.25 & Theorem~\ref{thm:robust} \\
Price of the 5\% out-of-the-money put & One robust price, 0.613\% of the forward & Table~\ref{tab:emp-put} \\
Price range if the doubt were $1.5\,\cbar$ & No-trade interval $[0.450\%, 0.754\%]$ & Table~\ref{tab:emp-put} \\
Doubt implied by the mean VIX & $1.30\,\cbar$, gap not significant; with the 1990--2026 reference model, 4.99 times its own threshold & Table~\ref{tab:emp-bridge} \\
\bottomrule
\end{tabular}}
{Reference model $\Pz$: monthly relative returns of the CRSP market over the bill, 1926:07--2026:09 ($N=1203$ months, each a state with probability $1/N$); one-month horizon; log utility; $c_{\mathrm{one}}=\chi^2_{1,0.95}/(2N)$. The effective risk aversion and the stock weight are those of this empirical market at $c_{\mathrm{one}}$; the Black--Scholes calibration of Table~\ref{tab:emp-doubt} gives 1.76 and 1.41.}
\end{table}

\section{Conclusion}\label{sec:conclusion}

Our main result identifies a threshold for trading under doubt. Consider an investor who treats as plausible every model within KL radius $c$ of their reference model, in a one-period market with finitely many states. For every smooth, strictly increasing, and strictly concave utility, they hold no risky asset exactly when $c$ reaches $\cbar=\kl{\QE}{\Pz}$, the minimal entropy of the market. Above $\cbar$, the no-trade prices of a non-traded claim form the entropic good-deal interval with budget $c-\cbar$. With log utility and below $\cbar$, their portfolio and no-trade prices are those of a CRRA($\gamma_c$) investor who trusts $\Pz$, in one market only (Remark~\ref{rem:scope}). In this regime, the robust pricing measure moves from the log-utility to the minimal-entropy measure. Implied doubt turns a quote into a radius: with the 1990--2026 reference model, the mean VIX requires $4.99\,\cbar$, a radius at which the robust investor holds no stock. In a Black--Scholes illustration using U.S. data since 1926, doubt about the mean at 95\% confidence acts like a relative risk aversion of 1.76, or 1.86 with a Newey--West standard error. The defect of the triangle of the real-world, pricing and share measures is a log risk premium; in U.S. data since 1990, it averages about 10\% per year, of which about 1 percentage point is an entropy premium.

These results have limits. The reading of doubt as risk aversion needs log utility, and the analysis is that of one investor who takes prices as given, not an equilibrium. A KL ball cannot add a state that $\Pz$ rules out, while Wasserstein balls can \citep{MohajerinEsfahaniKuhn2018}. The calibration treats months as independent and uses first-order rules. The VIX prices the short log contract only approximately, and on the S\&P~500 rather than the CRSP market. Section~\ref{sec:numerics} is in sample. The VIX reading also compares an average of quotes with a static reference model, and the gap between the mean VIX and the range of robust prices depends on the sample, mainly through the serial dependence of daily returns.

Part of the theory extends beyond one period: robust utility duality holds in semimartingale markets \citep{SchiedWu2005,FollmerGundel2006}; the minimal-entropy measure, and with it $\cbar$, exists in continuous time under integrability conditions \citep{Frittelli2000}; and Proposition~\ref{prop:bsdoubt} gives the threshold and $c\mapsto\gamma_c$ in one complete market. Finiteness enters our proofs through the compactness of the ball and of $\cl\Mset$, the uniform Taylor bounds behind Theorem~\ref{thm:pricing}(a), and the linearization behind the local rate of alternating projections. Dynamic incomplete markets remain open: there the worst case need not be a constant tilt, balls on path space are not rectangular \citep{EpsteinSchneider2003}, and plans must stay consistent over time, as with the robust forward criteria of \citet{KallbladOblojZariphopoulou2018}. Three further extensions are natural. The first replaces KL by an $\alpha$-divergence \citep{Amari2009} and log utility by power utility \citep{GollRuschendorf2001}; the threshold and the interval exist there too (Section~\ref{sec:rob-pricing}), but an exact map onto power utilities is open. The second measures the triangle month by month with option-implied expected returns \citep{Martin2017}. The third uses S\&P~500 returns with option-level data, which would remove the index mismatch, price the short log contract directly, and separate its jump terms \citep{CarrWu2009}.

\bibliographystyle{informs2014}
\bibliography{MS_references}

\clearpage
\appendix
\counterwithin{equation}{section}
\counterwithin{table}{section}
\counterwithin{figure}{section}
\counterwithin{algorithm}{section}
\counterwithin{theorem}{section}
\counterwithin{lemma}{section}
\counterwithin{proposition}{section}
\counterwithin{corollary}{section}
\counterwithin{assumption}{section}
\counterwithin{definition}{section}
\counterwithin{remark}{section}
\counterwithin{example}{section}
\begin{center}{\Large\bfseries Appendix}\end{center}

\section{A Short Guide to Dually Flat Geometry}\label{sec:ec-geometry}

This section collects the geometry behind Section~\ref{sec:setting} and proves Lemma~\ref{lem:threepoint}. Except in
that proof, the state space is finite, $\Omega=\{\omega_1,\dots,\omega_M\}$ with $M\ge2$, as in
Sections~\ref{sec:incomplete} and~\ref{sec:robust}. For $p\in\mathcal{P}$ we write $p_j:=p(\omega_j)$. Standard
references are \citet{AmariNagaoka2000}, \citet{Amari2016}, and, for projections, \citet{Csiszar1975} and
\citet{CsiszarShields2004}. The same geometry is used outside finance; \citet{EddibAboussalah2026}, for example, measure oversmoothing in graph neural networks by the dispersion of probability vectors under the Fisher--Rao metric.

\subsection{Coordinates, the Fisher Metric, and Flat Sets}\label{sec:ec-geo-coords}\label{sec:ec-geo-fisher}\label{sec:ec-geo-lines}

A point $p\in\mathcal{P}$ has two coordinate systems: the mixture coordinates, its probabilities $\eta_j:=p_j$ for
$j<M$, with $p_M=1-\sum_{j<M}\eta_j$, and the exponential coordinates, its log-odds $\theta^j:=\ln(p_j/p_M)$ for $j<M$.
The map $p\mapsto\eta$ is one-to-one onto an open subset of $\RR^{M-1}$, and $p\mapsto\theta$ is one-to-one onto
$\RR^{M-1}$, with inverse $p_j=e^{\theta^j-\psi(\theta)}$ for $j<M$ and $p_M=e^{-\psi(\theta)}$, where
$\psi(\theta):=\ln(1+\sum_{j<M}e^{\theta^j})$. So the simplex is an exponential family with natural parameter $\theta$, and the two systems are Legendre duals. The
normalizer $\psi$ is strictly convex with gradient $\partial\psi/\partial\theta^j=\eta_j$, and its convex conjugate,
the negative entropy $\varphi(\eta):=\sum_{j=1}^{M}p_j\ln p_j$, has gradient $\partial\varphi/\partial\eta_j=\theta^j$.
The two functions are linked at every point by
\begin{equation}\label{eq:ec1-legendre}
\psi(\theta)+\varphi(\eta)=\theta\cdot\eta,
\end{equation}
because $\theta\cdot\eta=\sum_{j}p_j\ln p_j-\ln p_M$ and $\psi(\theta)=-\ln p_M$.

A smooth curve $p(t)$ in $\mathcal{P}$ has a velocity whose entries sum to zero; we call such vectors directions. The
Fisher inner product of two directions $x$ and $y$ at $p$ is $\langle x,y\rangle_{p}:=\sum_{j=1}^{M}x_j\,y_j/p_j$, the
covariance under $p$ of the relative changes $x_j/p_j$ and $y_j/p_j$, which both have mean zero under $p$. Two facts
tie this inner product to the two coordinate systems and to KL. First, a direction $x$ has mixture components
$\dot\eta_j(x)=x_j$ and exponential components $\dot\theta^j(x)=x_j/p_j-x_M/p_M$, and
$\langle x,y\rangle_{p}=\sum_{j<M}\dot\theta^j(x)\,\dot\eta_j(y)$, since
$\sum_{j<M}(x_j/p_j-x_M/p_M)\,y_j=\sum_{j<M}x_jy_j/p_j+x_My_M/p_M$. So the metric pairs the exponential components
of one direction with the mixture components of the other. Second, for a smooth curve with $p(0)=p$ and $\dot p(0)=x$,
a Taylor expansion gives $\kl{p(t)}{p}=\tfrac12\langle x,x\rangle_p\,t^2+o(t^2)$, and the same holds for $\kl{p}{p(t)}$:
the metric is the local quadratic form of KL, in either order of the arguments.

The mixture line \eqref{eq:set-mline} is straight in mixture coordinates, $\eta(m_t)=(1-t)\eta(\nu)+t\eta(\mu)$, and
the exponential line \eqref{eq:set-eline} is straight in exponential coordinates: $e_t\propto\nu^{1-t}\rho^{t}$, so
$\theta(e_t)=(1-t)\theta(\nu)+t\theta(\rho)$. These curves are the geodesics of the mixture and the exponential connection
\citep{AmariNagaoka2000}; we need only their coordinate form. Neither kind is straight in the coordinates of the other,
so the exponential lines of Figure~\ref{fig:triangle}, drawn in mixture coordinates, are curved.

The two coordinate systems also give two kinds of flat sets. A set is mixture-flat if its mixture coordinates fill the
part of an affine subspace that lies in $\eta(\mathcal{P})$; equivalently, it is a linear family
$\{p\in\mathcal{P}:\EE_p[f]=\alpha\}$ for some $f:\Omega\to\RR^k$ and $\alpha\in\RR^k$. A set is exponentially flat if
its exponential coordinates fill an affine subspace; equivalently, it is an exponential family
$\{q\,e^{\lambda\cdot f}/\EE_q[e^{\lambda\cdot f}]:\lambda\in\RR^k\}$ with $q\in\mathcal{P}$. Each kind of set contains
the lines of its own kind between its points. The set $\Mset$ of Section~\ref{sec:set-proj} is mixture-flat, with
$f=\DeltaS$ and $\alpha=0$, and the tilts $\PP\,e^{-\lambda\cdot\DeltaS}/\EE_{\PP}[e^{-\lambda\cdot\DeltaS}]$ of a model
$\PP$ form an exponentially flat set through $\PP$ that contains $\QE(\PP)$ (Lemma~\ref{lem:ec-projections}(a)).

\subsection{KL as the Canonical Divergence}\label{sec:ec-geo-canonical}

The pair $(\psi,\varphi)$ produces KL. For $\mu,\nu\in\mathcal{P}$,
\begin{equation}\label{eq:ec1-canonical}
\kl{\mu}{\nu}=\psi\bigl(\theta(\nu)\bigr)+\varphi\bigl(\eta(\mu)\bigr)-\theta(\nu)\cdot\eta(\mu),
\end{equation}
because $\theta(\nu)\cdot\eta(\mu)=\sum_{j}\mu_j\ln\nu_j-\ln\nu_M$ and $\psi(\theta(\nu))=-\ln\nu_M$. This is the canonical
divergence of the dually flat structure \citep{AmariNagaoka2000}. With the Bregman divergence
$B_F(x,y):=F(x)-F(y)-\nabla F(y)\cdot(x-y)$ of a convex function $F$, \eqref{eq:ec1-legendre} turns
\eqref{eq:ec1-canonical} into $\kl{\mu}{\nu}=B_{\varphi}(\eta(\mu),\eta(\nu))=B_{\psi}(\theta(\nu),\theta(\mu))$. So, in
Bregman form, KL keeps the order of its arguments in mixture coordinates and reverses it in exponential coordinates, and
texts that write the canonical divergence in exponential coordinates reverse the arguments of KL. We avoid this with one
convention: $\kl{\mu}{\nu}=\EE_\mu[\ln(\dd\mu/\dd\nu)]$ throughout, and a projection is named by the slot of the free
measure (Section~\ref{sec:set-proj}).

\subsection{The Generalized Pythagorean Theorem and the Proof of Lemma~\ref{lem:threepoint}}\label{sec:ec-geo-pythagoras}

For $\mu,\nu,\rho\in\mathcal{P}$, insert \eqref{eq:ec1-canonical} into the defect \eqref{eq:set-defect}. The terms in
$\psi$ and $\varphi$ cancel by \eqref{eq:ec1-legendre} at $\nu$, and
\begin{equation}\label{eq:ec1-defect}
\defect(\mu,\nu,\rho)=\bigl(\theta(\rho)-\theta(\nu)\bigr)\cdot\bigl(\eta(\mu)-\eta(\nu)\bigr).
\end{equation}
The factors are the velocities of the exponential line from $\nu$ toward $\rho$ in exponential coordinates and of the
mixture line from $\nu$ toward $\mu$ in mixture coordinates, so, by the pairing of Section~\ref{sec:ec-geo-fisher},
their product is the Fisher inner product of the two velocities at $\nu$. This is the generalized Pythagorean theorem
\citep{AmariNagaoka2000,Amari2016}: $\kl{\mu}{\rho}=\kl{\mu}{\nu}+\kl{\nu}{\rho}$ if and only if the two lines meet at a
right angle at $\nu$. Lemma~\ref{lem:threepoint} states the same fact without coordinates and without a finite state space.

\begin{repeatlemma}[Lemma~\ref{lem:threepoint} (Three-point identity).]
Let $\mu$, $\nu$, and $\rho$ be mutually equivalent probability measures such that $\kl{\mu}{\nu}$, $\kl{\nu}{\rho}$,
and $\kl{\mu}{\rho}$ are finite, and let $\ell:=\ln(\dd\rho/\dd\nu)$.
\begin{itemize}
\item[(a)] The function $\ell$ is integrable under $\mu$ and under $\nu$, and
\begin{equation}\tag{\ref{eq:set-threepoint}}
\defect(\mu,\nu,\rho)=\EE_{\mu}[\ell]-\EE_{\nu}[\ell].
\end{equation}
Hence $\kl{\mu}{\rho}=\kl{\mu}{\nu}+\kl{\nu}{\rho}$ (exact Pythagoras) if and only if $\EE_{\mu}[\ell]=\EE_{\nu}[\ell]$.
\item[(b)] The defect is the covariance under $\nu$ of $\dd\mu/\dd\nu$ and $\ell$:
$\defect(\mu,\nu,\rho)=\EE_{\nu}\bigl[(\dd\mu/\dd\nu-1)\,(\ell-\EE_{\nu}[\ell])\bigr]$. On a finite state space with
$\mu,\nu,\rho\in\mathcal{P}$, let $\dot m:=\mu-\nu$ and $\dot e:=\nu\,(\ell-\EE_{\nu}[\ell])$ be the velocities at $t=0$
of the mixture line \eqref{eq:set-mline} from $\nu$ toward $\mu$ and of the exponential line \eqref{eq:set-eline} from
$\nu$ toward $\rho$. Then $\defect(\mu,\nu,\rho)=\langle\dot m,\dot e\rangle_{\nu}$, the Fisher inner product at $\nu$,
where $\langle x,y\rangle_{\nu}:=\sum_{\omega\in\Omega}x(\omega)\,y(\omega)/\nu(\omega)$.
\end{itemize}
\end{repeatlemma}

\begin{proof}{Proof of Lemma~\ref{lem:threepoint}.}
We use one fact about KL (Lemma~\ref{lem:ec-klfacts}(a)): if one probability measure is absolutely continuous with
respect to another, their KL divergence is well defined in $[0,\infty]$, and it is finite if and only if the log-density
is integrable under the first measure.

(a) The three measures are mutually equivalent, so the chain rule for densities gives
$\ln(\dd\mu/\dd\rho)=\ln(\dd\mu/\dd\nu)-\ell$ almost surely. Since $\kl{\mu}{\nu}$ and $\kl{\mu}{\rho}$ are finite,
$\ln(\dd\mu/\dd\nu)$ and $\ln(\dd\mu/\dd\rho)$ are $\mu$-integrable. Hence $\ell$ is $\mu$-integrable and
$\EE_{\mu}[\ell]=\kl{\mu}{\nu}-\kl{\mu}{\rho}$. Next, $\ln(\dd\nu/\dd\rho)=-\ell$ and $\kl{\nu}{\rho}$ is finite, so
$\ell$ is $\nu$-integrable and $\kl{\nu}{\rho}=-\EE_{\nu}[\ell]$. Adding the two equalities gives
$\defect(\mu,\nu,\rho)=\EE_\mu[\ell]-\EE_\nu[\ell]$; exact Pythagoras is the case $\defect(\mu,\nu,\rho)=0$.

(b) Put $D:=\dd\mu/\dd\nu$. Then $\EE_\nu[D]=1$ and $\EE_\nu[D\,|\ell|]=\EE_\mu[|\ell|]<\infty$ by part (a). So each term of
$(D-1)(\ell-\EE_\nu[\ell])=D\ell-D\,\EE_\nu[\ell]-\ell+\EE_\nu[\ell]$ is $\nu$-integrable, and its $\nu$-expectation is
$\EE_\mu[\ell]-\EE_\nu[\ell]=\defect(\mu,\nu,\rho)$. On a finite state space, $\dot m=\mu-\nu$ because $m_t$ is linear in
$t$. Also $\ln e_t=\ln\nu+t\ell-\ln\EE_\nu[e^{t\ell}]$ has derivative $\ell-\EE_\nu[\ell]$ at $t=0$, so
$\dot e=\nu\,(\ell-\EE_\nu[\ell])$. Then
$\langle\dot m,\dot e\rangle_\nu=\sum_{\omega}(\mu-\nu)(\omega)\,(\ell-\EE_\nu[\ell])(\omega)=\EE_\nu[(D-1)(\ell-\EE_\nu[\ell])]$.
\Halmos
\end{proof}

\subsection{Projections onto a Linear Family and the Slot Rule}\label{sec:ec-geo-projections}

Fix a model $\PP\in\mathcal{P}$, a payoff vector $f:\Omega\to\RR^{k}$, and $\alpha\in\RR^{k}$ such that the linear
family $\mathcal{L}:=\{\QQ\in\mathcal{P}:\EE_{\QQ}[f]=\alpha\}$ is nonempty. Its closure $\cl\mathcal{L}$ adds the
probability vectors with some zero entries that satisfy the same equations, and the directions along $\mathcal{L}$
form the set $\mathcal{T}:=\{x\in\RR^{M}:\sum_jx_j=0,\ \sum_jx_jf(\omega_j)=0\}$. For $\mathcal{L}=\Mset$ take $f=\DeltaS$ and
$\alpha=0$; the two projections below are then $\QE(\PP)$ and $\QL(\PP)$.

\begin{lemma}[Projections onto a linear family]\label{lem:ec-projections}
\begin{itemize}
\item[(a)] (I-projection.) Exactly one $\QQ^{I}\in\cl\mathcal{L}$ minimizes $\kl{\QQ}{\PP}$ over $\cl\mathcal{L}$. It
lies in $\mathcal{L}$, it equals $\PP\,e^{\lambda\cdot f}/\EE_{\PP}[e^{\lambda\cdot f}]$ for some $\lambda\in\RR^{k}$, and
$\kl{\QQ}{\PP}=\kl{\QQ}{\QQ^{I}}+\kl{\QQ^{I}}{\PP}$ for every $\QQ\in\cl\mathcal{L}$.
\item[(b)] (Reverse I-projection.) Exactly one $\QQ^{R}\in\cl\mathcal{L}$ minimizes $\kl{\PP}{\QQ}$ over
$\cl\mathcal{L}$. It lies in $\mathcal{L}$ and equals $\PP/W$, where $W:=1+h\cdot(f-\alpha)>0$ and $h\in\RR^{k}$
maximizes $\EE_{\PP}[\ln(1+h'\cdot(f-\alpha))]$ over all $h'$ with $1+h'\cdot(f-\alpha)>0$.
\item[(c)] (Orthogonality.) At $\QQ^{I}$ the velocity of the exponential line toward $\PP$, and at $\QQ^{R}$ the
velocity $\PP-\QQ^{R}$ of the mixture line toward $\PP$, are Fisher-orthogonal to every direction in $\mathcal{T}$.
\item[(d)] (Defect at the reverse I-projection.) For $\QQ\in\mathcal{L}$, let $e_t\propto(\QQ^{R})^{1-t}\QQ^{t}$,
$t\in\RR$, be the exponential line through $\QQ^{R}$ and $\QQ$. Then
$\defect(\PP,\QQ^{R},\QQ)=-\frac{\dd}{\dd t}\kl{\PP}{e_t}\big|_{t=0}$, and $\defect(\PP,\QQ^{R},\QQ)=0$ if $e_t\in\mathcal{L}$
for all $t$ near $0$.
\end{itemize}
\end{lemma}

\begin{proof}{Proof.}
(a) and (b) are known: (a) is the finite-state case of \citet[Theorem 3.1]{Csiszar1975} with the Pythagorean identity
of \citet[Theorem 3.2]{CsiszarShields2004}, and (b) is the duality between the reverse I-projection and the
growth-optimal portfolio \citep{Long1990,GollRuschendorf2001}. Both also follow from Section~\ref{sec:ec-proofs-incomplete}:
with $\DeltaS:=f-\alpha$ and after dropping every component that is a linear combination of the others (it has no
constant term, since all components have mean zero under a member of $\mathcal{L}$, so $\mathcal{L}$, the tilts and the
wealths do not change), $\mathcal{L}$ is the set $\Mset$ of Section~\ref{sec:incomplete} under
Assumption~\ref{ass:market} (Theorem~\ref{thm:ec-ftap}), and (a) and (b) are Propositions~\ref{prop:ec-QE}
and~\ref{prop:ec-QL}.

(c) Take $x\in\mathcal{T}$. For small $|s|$, both $\QQ^{I}+sx$ and $\QQ^{R}+sx$ lie in $\mathcal{L}$, because the projections are
strictly positive. Since $\sum_jx_j=0$, the first-order conditions at the two minima read
$\sum_jx_j\ln(\QQ^{I}_j/\PP_j)=0$ and $\sum_jx_j\PP_j/\QQ^{R}_j=0$. Let $\sigma:=\ln(\PP/\QQ^{I})$. As in
Lemma~\ref{lem:threepoint}(b), the exponential line from $\QQ^{I}$ toward $\PP$ has velocity
$\dot e=\QQ^{I}(\sigma-\EE_{\QQ^{I}}[\sigma])$, so $\langle x,\dot e\rangle_{\QQ^{I}}=\sum_jx_j\sigma_j=0$. For the mixture
line, $\langle x,\PP-\QQ^{R}\rangle_{\QQ^{R}}=\sum_jx_j\PP_j/\QQ^{R}_j-\sum_jx_j=0$.

(d) Put $\ell:=\ln(\QQ/\QQ^{R})$ and $\chi(t):=\ln\EE_{\QQ^{R}}[e^{t\ell}]$, so that $\ln e_t=\ln\QQ^{R}+t\ell-\chi(t)$ and
$\chi'(0)=\EE_{\QQ^{R}}[\ell]$. Then $\kl{\PP}{e_t}=\kl{\PP}{\QQ^{R}}-t\,\EE_{\PP}[\ell]+\chi(t)$, whose derivative at $t=0$
is $-\EE_{\PP}[\ell]+\EE_{\QQ^{R}}[\ell]=-\defect(\PP,\QQ^{R},\QQ)$ by Lemma~\ref{lem:threepoint}(a). If $e_t\in\mathcal{L}$
near $t=0$, then $t\mapsto\kl{\PP}{e_t}$ has an interior minimum at $t=0$, because $e_0=\QQ^{R}$ minimizes
$\kl{\PP}{\cdot}$ over $\mathcal{L}$; so the derivative is zero. \Halmos
\end{proof}

Parts (c) and (d) make the asymmetry of Section~\ref{sec:set-proj} precise. At $\QQ^{I}$, Lemma~\ref{lem:threepoint}(b)
pairs a direction orthogonal to $\mathcal{T}$ (the exponential direction toward $\PP$) with a direction in $\mathcal{T}$ (the mixture
direction toward $\QQ$), so the defect vanishes. At $\QQ^{R}$, by part (d), the defect is the slope of $\kl{\PP}{\cdot}$
along the exponential line toward $\QQ$. Optimality of $\QQ^{R}$ makes slopes vanish only along curves inside
$\mathcal{L}$, and that line leaves $\mathcal{L}$ in general. On $\Mset$ it leaves as soon as a normalized geometric
average $(\QL)^{1-t}\QQ^{t}$ misprices a traded asset. Section~\ref{sec:ec-proofs-incomplete} gives a three-state market
in which the natural identity $\kl{\PP}{\QQ}=\kl{\PP}{\QL}+\kl{\QL}{\QQ}$ fails and the special markets in which it holds.

These facts give the slot rule its geometric content. A free measure in the first slot, $\min_{\QQ}\kl{\QQ}{\PP}$, gives
an I-projection, an e-projection in the terms of \citet{AmariNagaoka2000}: $\PP$ is joined to it by an orthogonal
exponential line. A free measure in the second slot, $\min_{\QQ}\kl{\PP}{\QQ}$, gives a reverse I-projection, an
m-projection, joined to $\PP$ by an orthogonal mixture line. Exact Pythagoras holds for I-projections onto mixture-flat
sets, and for a reverse I-projection onto a mixture-flat set, such as $\QL$ on $\Mset$, only in special cases. On a
convex set $C$ that is not flat, such as the KL ball of Section~\ref{sec:robust}, the I-projection $\nu^{*}$ of a point
$\nu$ satisfies only the Pythagorean inequality $\kl{\mu}{\nu}\ge\kl{\mu}{\nu^{*}}+\kl{\nu^{*}}{\nu}$ for $\mu\in C$
\citep[Theorem 2.2]{Csiszar1975}; the alternating minimization of Section~\ref{sec:ec-algorithms} builds on it
\citep{CsiszarTusnady1984}.

\section{Proofs for Section~\ref{sec:triangle}}\label{sec:ec-proofs-triangle}

This section proves the results of Section~\ref{sec:triangle} with the three tools of Section~\ref{sec:ec-tri-prelim}, and it states and proves the Black--Scholes closed forms of Section~\ref{sec:triangle-identity} (Corollary~\ref{cor:bs}) and Proposition~\ref{prop:itolevy} (Section~\ref{sec:ec-tri-itolevy}). It also holds the material that Section~\ref{sec:triangle} points to: an example on Assumption~\ref{ass:A2}, the cumulant form of the entropy premium, remarks on dividends and on the published VIX, the numeraire portfolio (Section~\ref{sec:ec-tri-slack}), and the angles of the triangle with the SVIX index (Section~\ref{sec:ec-tri-angles}).

\subsection{Preliminaries}\label{sec:ec-tri-prelim}

The first lemma collects four facts that every proof below uses.

\begin{lemma}[Basic facts]\label{lem:ec-klfacts}
Let $\mu\ll\nu$ be probability measures on a measurable space, and let $D:=\dd\mu/\dd\nu$.
\begin{enumerate}
\item[(a)] $\EE_\mu[(\ln D)^-]\le e^{-1}$. Hence $\kl{\mu}{\nu}=\EE_\mu[\ln D]$ is well defined, and it is finite if and only if $\ln D$ is $\mu$-integrable.
\item[(b)] $\kl{\mu}{\nu}\ge0$, with equality if and only if $\mu=\nu$.
\item[(c)] If $\mu_1\ll\mu_2\ll\mu_3$ are probability measures, then $\dd\mu_1/\dd\mu_3=(\dd\mu_1/\dd\mu_2)(\dd\mu_2/\dd\mu_3)$ almost surely under $\mu_3$; if $\mu_1\sim\mu_2$, then $\dd\mu_2/\dd\mu_1=(\dd\mu_1/\dd\mu_2)^{-1}$.
\item[(d)] Let $\mathcal G$ be a sub-$\sigma$-field, and let $\mu_{\mathcal G}$ and $\nu_{\mathcal G}$ be the restrictions of $\mu$ and $\nu$ to $\mathcal G$. Then $D_{\mathcal G}:=\EE_\nu[D\mid\mathcal G]$ is the density $\dd\mu_{\mathcal G}/\dd\nu_{\mathcal G}$. If $\kl{\mu}{\nu}<\infty$, then $\kl{\mu_{\mathcal G}}{\nu_{\mathcal G}}\le\kl{\mu}{\nu}$, the difference is $\EE_\mu[\ln(D/D_{\mathcal G})]$, and it is zero if and only if $D$ is $\mathcal G$-measurable.
\end{enumerate}
\end{lemma}

\begin{proof}{Proof.}
These facts are standard \citep{KullbackLeibler1951,CsiszarShields2004}. Part (a) follows from $z\ln(1/z)\le e^{-1}$ for $0<z<1$, (b) from Jensen's inequality, (c) is the chain rule for Radon--Nikodym derivatives, and (d) follows from the conditional Jensen inequality for the strictly convex function $u\mapsto u\ln u$, applied to $D/D_{\mathcal G}$, whose conditional $\nu$-mean given $\mathcal G$ is one on $\{D_{\mathcal G}>0\}$. \Halmos
\end{proof}

The second lemma shows that Assumption~\ref{ass:A1} is exactly the finiteness of the three sides.

\begin{lemma}[The sides of the triangle]\label{lem:ec-sides}
Assumption~\ref{ass:A1} holds if and only if $\kl{\PP}{\QQ}$, $\kl{\QQ}{\QS}$ and $\kl{\PP}{\QS}$ are finite. In that case
\begin{equation}\label{eq:ec-sides}
\kl{\QQ}{\QS}=-\EE_{\QQ}[\ln R]\qquad\text{and}\qquad\kl{\PP}{\QS}=\kl{\PP}{\QQ}-\EE_{\PP}[\ln R].
\end{equation}
\end{lemma}

\begin{proof}{Proof.}
By Lemma~\ref{lem:ec-klfacts}(c), $\ln(\dd\QQ/\dd\QS)=-\ln R$ and $\ln(\dd\PP/\dd\QS)=\ln(\dd\PP/\dd\QQ)-\ln R$ almost surely. Suppose that Assumption~\ref{ass:A1} holds. Then $-\ln R$ is $\QQ$-integrable, which gives the first formula in \eqref{eq:ec-sides} with a finite value. The variable $\ln(\dd\PP/\dd\QQ)$ is $\PP$-integrable by Lemma~\ref{lem:ec-klfacts}(a), and so is $\ln R$, which gives the second formula with a finite value. Conversely, suppose that the three sides are finite. Lemma~\ref{lem:ec-klfacts}(a) makes $\ln(\dd\QQ/\dd\QS)=-\ln R$ integrable under $\QQ$, and it makes $\ln(\dd\PP/\dd\QQ)$ and $\ln(\dd\PP/\dd\QS)$ integrable under $\PP$; so their difference, $\ln R$, is integrable under $\PP$. \Halmos
\end{proof}

The third tool writes a smooth payoff as a static portfolio of a bond, a forward and options \citep{BreedenLitzenberger1978,CarrMadan1998}.

\begin{lemma}[Static spanning]\label{lem:ec-spanning}
For $f\in C^2(0,\infty)$ and $x,F>0$,
\[
f(x)=f(F)+f'(F)(x-F)+\int_0^F f''(K)(K-x)^+\,\dd K+\int_F^\infty f''(K)(x-K)^+\,\dd K .
\]
\end{lemma}

\begin{proof}{Proof.}
This is the spanning formula of \citet{CarrMadan1998}: Taylor's formula with integral remainder at $F$, $f(x)=f(F)+f'(F)(x-F)+\int_F^x f''(K)(x-K)\,\dd K$, whose remainder is the second integral of the statement if $x\ge F$ and the first if $x<F$; the other integral is then zero. \Halmos
\end{proof}

\subsection{Proof of Theorem~\ref{thm:triangle}}\label{sec:ec-tri-thm}

\begin{repeattheorem}[Theorem~\ref{thm:triangle} (Triangle identity).]
Under Assumption~\ref{ass:A1},
\begin{equation}\tag{\ref{eq:triangle}}
\defect(\PP,\QQ,\QS)=\EE_{\PP}[\ln R]-\EE_{\QQ}[\ln R].
\end{equation}
Let $\mathcal{G}\subseteq\mathcal{F}_T$ be a $\sigma$-field such that $R$ is $\mathcal{G}$-measurable, and compute the three divergences for the restrictions of $\PP$, $\QQ$ and $\QS$ to $\mathcal{G}$. Then Assumption~\ref{ass:A1} holds for the restrictions, $\kl{\QQ}{\QS}$ and $\defect(\PP,\QQ,\QS)$ do not change, and $\kl{\PP}{\QQ}$ and $\kl{\PP}{\QS}$ decrease by the same amount, which is zero if and only if $\dd\PP/\dd\QQ$ is $\mathcal{G}$-measurable. Under Assumption~\ref{ass:D} we may take $\mathcal{G}=\sigma(S_T)$, and then the three divergences are those between the laws of $S_T$ under $\PP$, $\QQ$ and $\QS$.
\end{repeattheorem}

\begin{proof}{Proof.}
Under Assumption~\ref{ass:A1}, Lemma~\ref{lem:ec-sides} shows that the three sides are finite and gives $\kl{\QQ}{\QS}=-\EE_{\QQ}[\ln R]$ and $\kl{\PP}{\QS}=\kl{\PP}{\QQ}-\EE_{\PP}[\ln R]$. Substituting both into the definition of the defect gives
\[
\defect(\PP,\QQ,\QS)=\kl{\PP}{\QQ}-\EE_{\QQ}[\ln R]-\kl{\PP}{\QQ}+\EE_{\PP}[\ln R]=\EE_{\PP}[\ln R]-\EE_{\QQ}[\ln R].
\]
This is Lemma~\ref{lem:threepoint} with $(\mu,\nu,\rho)=(\PP,\QQ,\QS)$, since $\ln(\dd\QS/\dd\QQ)=\ln R$.

Now let $\mathcal G$ be as in the statement, and let $\PP_{\mathcal G}$, $\QQ_{\mathcal G}$ and $\QS_{\mathcal G}$ be the restrictions. By Lemma~\ref{lem:ec-klfacts}(d), $\dd\QS_{\mathcal G}/\dd\QQ_{\mathcal G}=\EE_{\QQ}[R\mid\mathcal G]=R$, because $R$ is $\mathcal G$-measurable; so $R$ plays the same role on $\mathcal G$ as on $\mathcal F_T$. By Lemma~\ref{lem:ec-klfacts}(d) again, $\kl{\PP_{\mathcal G}}{\QQ_{\mathcal G}}\le\kl{\PP}{\QQ}<\infty$. The $\mathcal G$-measurable variable $\ln R$ has the same integral under a measure and under its restriction to $\mathcal G$. Hence Assumption~\ref{ass:A1} holds on $\mathcal G$, and \eqref{eq:triangle} on $\mathcal G$ gives the same number as on $\mathcal F_T$. Lemma~\ref{lem:ec-sides}, applied on $\mathcal G$ and on $\mathcal F_T$, gives $\kl{\QQ_{\mathcal G}}{\QS_{\mathcal G}}=-\EE_{\QQ}[\ln R]=\kl{\QQ}{\QS}$, and on both $\sigma$-fields it writes the $\PP$--$\QQ$ side as the $\PP$--$\QS$ side plus the same number $\EE_{\PP}[\ln R]$. So $\kl{\PP}{\QQ}-\kl{\PP_{\mathcal G}}{\QQ_{\mathcal G}}=\kl{\PP}{\QS}-\kl{\PP_{\mathcal G}}{\QS_{\mathcal G}}$. By Lemma~\ref{lem:ec-klfacts}(d), this common decrease is nonnegative, and it is zero if and only if $\dd\PP/\dd\QQ$ is $\mathcal G$-measurable.

Under Assumption~\ref{ass:D}, $R=S_T/F_0$ is $\sigma(S_T)$-measurable. A density between two restrictions to $\sigma(S_T)$ is a measurable function $h$ of $S_T$, and $h$ is then the density between the corresponding laws of $S_T$ on $(0,\infty)$; so the divergences between the restrictions equal the divergences between the laws. \Halmos
\end{proof}

\subsection{The Black--Scholes Triangle}\label{sec:ec-tri-bs}

\begin{corollary}[Black--Scholes]\label{cor:bs}
Let $\dd S_t=\mu S_t\,\dd t+\sigma S_t\,\dd Z_t$ and $\dd B_t=rB_t\,\dd t$ with constants $\mu$, $r$ and $\sigma>0$, where $Z$ is a $\PP$-Brownian motion, and let $\dd\QQ/\dd\PP:=\exp(-\SR\,Z_T-\frac12\SR^2T)$. Then Assumptions~\ref{ass:A1}, \ref{ass:A2} and \ref{ass:D} hold, and, both on $\mathcal{F}_T$ and for the laws of $S_T$,
\[
\kl{\PP}{\QQ}=\frac{\SR^2T}{2},\qquad \kl{\QQ}{\QS}=\frac{\sigma^2T}{2},\qquad \kl{\PP}{\QS}=\frac{(\mu-r-\sigma^2)^2\,T}{2\sigma^2},
\]
so that $\defect(\PP,\QQ,\QS)=(\mu-r)\,T$. In particular, $\defect(\PP,\QQ,\QS)>0$ if and only if $\mu>r$, and then the angle of the triangle at $\QQ$ is acute (Remark~\ref{rem:angles}).
\end{corollary}

\begin{proof}{Proof.}
The kernel $\SR$ is constant, so Novikov's condition holds, and Girsanov's theorem makes $Z^{\QQ}_t:=Z_t+\SR\,t$ a $\QQ$-Brownian motion on $[0,T]$. Since $\sigma\SR=\mu-r$,
\[
\frac{S_t}{S_0B_t}=\exp\Big(\big(\mu-r-\tfrac12\sigma^2\big)t+\sigma Z_t\Big)=\exp\Big(\sigma Z^{\QQ}_t-\tfrac12\sigma^2t\Big),
\]
which is a true $\QQ$-martingale, so $\QQ$ is a pricing measure; when the filtration is generated by $Z$, it is the only one \citep{HarrisonPliska1981}. At $t=T$, $\ln R=(\mu-r-\frac12\sigma^2)T+\sigma Z_T=\sigma Z^{\QQ}_T-\frac12\sigma^2T$. So $\ln R$ is normal under both measures, with $\EE_{\PP}[\ln R]=(\mu-r-\frac12\sigma^2)T$ and $\EE_{\QQ}[\ln R]=-\frac12\sigma^2T$. Next, $\ln(\dd\PP/\dd\QQ)=\SR\,Z_T+\frac12\SR^2T$, so $\kl{\PP}{\QQ}=\frac12\SR^2T<\infty$ and Assumption~\ref{ass:A1} holds. Also $\EE_{\PP}[R]=e^{(\mu-r)T}<\infty$ and $B_T=e^{rT}$, so Assumptions~\ref{ass:A2} and \ref{ass:D} hold. Theorem~\ref{thm:triangle} gives $\defect(\PP,\QQ,\QS)=(\mu-r)T$, and Lemma~\ref{lem:ec-sides} gives $\kl{\QQ}{\QS}=\frac12\sigma^2T$ and
\[
\kl{\PP}{\QS}=\frac{(\mu-r)^2T}{2\sigma^2}-\Big(\mu-r-\frac{\sigma^2}{2}\Big)T=\frac{(\mu-r)^2-2(\mu-r)\sigma^2+\sigma^4}{2\sigma^2}\,T=\frac{(\mu-r-\sigma^2)^2\,T}{2\sigma^2}.
\]
The densities $\dd\PP/\dd\QQ$, $\dd\QS/\dd\QQ=R$ and $\dd\PP/\dd\QS$ are functions of $Z_T=\big(\ln(S_T/S_0)-(\mu-\frac12\sigma^2)T\big)/\sigma$, hence of $S_T$. So Theorem~\ref{thm:triangle} with $\mathcal G=\sigma(S_T)$ shows that the laws of $S_T$ give the same four values. Finally, $\defect(\PP,\QQ,\QS)>0$ if and only if $\mu>r$, and the cosine of the angle at $\QQ$ has the sign of the defect (Lemmas~\ref{lem:ec-score} and \ref{lem:ec-expfamily}, Proposition~\ref{prop:ec-bsangles}). \Halmos
\end{proof}

\subsection{Proof of Proposition~\ref{prop:decomp}, and the Published VIX}\label{sec:ec-tri-decomp}

\textit{The entropy premium in cumulants.} Section~\ref{sec:triangle-vix} reads $\varepsilon$ as a premium for the spread of returns; the cumulants of $\ln R$ make this reading precise. Let $\kappa_j^{\PP}$ and $\kappa_j^{\QQ}$ be the $j$-th cumulants of $\ln R$ under $\PP$ and $\QQ$. The function $s\mapsto\ln\EE_{\PP}[R^s]$ is the cumulant generating function of $\ln R$, so its value at $s=1$ is $\ln\EE_{\PP}[R]$, which equals $\sum_{j\ge1}\kappa_j^{\PP}/j!$ when this series converges to it. Since $\kappa_1^{\PP}=\EE_{\PP}[\ln R]$, subtracting gives $\Lent_{\PP}(R)=\sum_{j\ge2}\kappa_j^{\PP}/j!$, the cumulant expansion of entropy \citep{BackusChernovZin2014}, and likewise under $\QQ$. Hence $\varepsilon=\frac12(\kappa_2^{\QQ}-\kappa_2^{\PP})+\frac16(\kappa_3^{\QQ}-\kappa_3^{\PP})+\cdots$: half the variance premium of the log return plus premia for its skewness and higher cumulants. For the variance term, \citet{BakshiMadan2006} relate the spread between risk-neutral and real-world volatility to risk aversion and to the higher moments of the real-world distribution. \citet{ChristoffersenElkamhiFeunouEtAl2010} choose pricing measures under which the risk-neutral return dynamics stay in the family of the real-world dynamics, and show that non-normal innovations are needed to fit the option smirk. The skewness term is related to the skew risk premium of \citet{KozhanNeubergerSchneider2013}, while the divergence risk premia of \citet{SchneiderTrojani2019JFEC} compare option-implied and realized divergences of returns in the same spirit.

Recall from Section~\ref{sec:triangle-vix} the entropies $\Lent_{\PP}(R)=\ln\EE_{\PP}[R]-\EE_{\PP}[\ln R]$ and $\Lent_{\QQ}(R)=-\EE_{\QQ}[\ln R]$, the log equity premium $\pi=\ln\EE_{\PP}[R]$, and the entropy premium $\varepsilon=\Lent_{\QQ}(R)-\Lent_{\PP}(R)$. Part (ii) defines $\VIX^2(T)$.

\begin{repeatproposition}[Proposition~\ref{prop:decomp} (Log equity premium and entropy premium).]
Suppose that Assumption~\ref{ass:A1} holds.
\begin{enumerate}
\item[(i)] If Assumption~\ref{ass:A2} also holds, then $\pi$ and $\varepsilon$ are finite and $\defect(\PP,\QQ,\QS)=\pi+\varepsilon$.
\item[(ii)] $\Lent_{\QQ}(R)=\kl{\QQ}{\QS}=-\EE_{\QQ}[\ln R]$. Under Assumption~\ref{ass:D}, moreover,
\[
\kl{\QQ}{\QS}=-\EE_{\QQ}\Big[\ln\frac{S_T}{F_0}\Big]=B_T\Big(\int_0^{F_0}\frac{\mathrm{put}(K)}{K^2}\,\dd K+\int_{F_0}^{\infty}\frac{\mathrm{call}(K)}{K^2}\,\dd K\Big)\eqqcolon\frac{T}{2}\,\VIX^2(T).
\]
\item[(iii)] If $\ln R$ is normal under $\PP$ and under $\QQ$ with the same variance, then Assumption~\ref{ass:A2} holds, $\varepsilon=0$ and $\defect(\PP,\QQ,\QS)=\pi$.
\end{enumerate}
\end{repeatproposition}

\begin{proof}{Proof.}
(ii) Since $\EE_{\QQ}[R]=1$, $\Lent_{\QQ}(R)=-\EE_{\QQ}[\ln R]$, and this is $\kl{\QQ}{\QS}$ by Lemma~\ref{lem:ec-sides}. Under Assumption~\ref{ass:D}, $R=S_T/F_0$. Apply Lemma~\ref{lem:ec-spanning} to $f(x):=-\ln(x/F_0)$ at $F=F_0$ and $x=S_T$. Here $f(F_0)=0$, $f'(F_0)=-1/F_0$ and $f''(K)=1/K^2$, so
\[
-\ln\frac{S_T}{F_0}=-\frac{S_T-F_0}{F_0}+\int_0^{F_0}\frac{(K-S_T)^+}{K^2}\,\dd K+\int_{F_0}^\infty\frac{(S_T-K)^+}{K^2}\,\dd K .
\]
Take $\QQ$-expectations. The linear term has mean zero because $\EE_{\QQ}[S_T]=F_0$. The two integrands are nonnegative, so Tonelli's theorem lets us exchange $\EE_{\QQ}$ and $\dd K$. Finally, $\EE_{\QQ}[(K-S_T)^+]=B_T\,\mathrm{put}(K)$ and $\EE_{\QQ}[(S_T-K)^+]=B_T\,\mathrm{call}(K)$. This gives the option formula; both sides are finite because the left side is finite under Assumption~\ref{ass:A1}. The last step defines $\VIX^2(T)$.

(i) Under Assumption~\ref{ass:A2}, $\pi=\ln\EE_{\PP}[R]$ is finite, and Assumption~\ref{ass:A1} makes $\EE_{\PP}[\ln R]$ and $\EE_{\QQ}[\ln R]$ finite. Hence $\Lent_{\PP}(R)$ and $\Lent_{\QQ}(R)$ are finite, and so is $\varepsilon$. With $\ln\EE_{\QQ}[R]=0$,
\[
\pi+\varepsilon=\ln\EE_{\PP}[R]-\EE_{\QQ}[\ln R]-\ln\EE_{\PP}[R]+\EE_{\PP}[\ln R]=\EE_{\PP}[\ln R]-\EE_{\QQ}[\ln R],
\]
which is $\defect(\PP,\QQ,\QS)$ by Theorem~\ref{thm:triangle}.

(iii) If $\ln R$ is normal with mean $m$ and variance $s^2$ under some measure, then $R$ has mean $e^{m+s^2/2}$ and entropy $s^2/2$ under that measure. With the same $s^2$ under $\PP$ and $\QQ$, $\EE_{\PP}[R]<\infty$, so Assumption~\ref{ass:A2} holds; moreover $\Lent_{\PP}(R)=\Lent_{\QQ}(R)$, so $\varepsilon=0$, and (i) gives $\defect(\PP,\QQ,\QS)=\pi$. \Halmos
\end{proof}

The next example satisfies Assumption~\ref{ass:A1} but not Assumption~\ref{ass:A2}.

\begin{example}[Assumption~\ref{ass:A2} is needed]\label{ex:ec-A2}
Take one period, so that a measure $\QQ\sim\PP$ is a pricing measure as soon as $\EE_{\QQ}[R]=1$. Let $X:=\ln R$ have, under $\QQ$, the normal density $\varphi_{\QQ}$ with mean $-\frac12s^2$ and variance $s^2>0$, so that $\EE_{\QQ}[R]=1$. Under $\PP$, let $X$ have the density $\varphi_{\PP}(x):=C\varphi_{\QQ}(x)$ for $x\le1$ and $\varphi_{\PP}(x):=Ce^{-x}/x$ for $x>1$, where $C>0$ makes $\varphi_{\PP}$ integrate to one. Both densities are positive, so $\PP\sim\QQ$. The variable $X$ is integrable under $\QQ$, and also under $\PP$ because $\int_1^\infty x\,e^{-x}x^{-1}\,\dd x<\infty$. For $x>1$ the log density ratio is $\ln(\varphi_{\PP}/\varphi_{\QQ})(x)=-x-\ln x+(x+s^2/2)^2/(2s^2)$ plus a constant, so it grows like $x^2/(2s^2)$; since $\int_1^\infty x^2e^{-x}x^{-1}\,\dd x<\infty$, $\kl{\PP}{\QQ}<\infty$. So Assumption~\ref{ass:A1} holds, and Theorem~\ref{thm:triangle} gives a finite defect. But $\EE_{\PP}[R]\ge C\int_1^\infty e^xe^{-x}x^{-1}\,\dd x=\infty$, so Assumption~\ref{ass:A2} fails. Then $\pi$ and $\Lent_{\PP}(R)$ are both infinite, and the split of Proposition~\ref{prop:decomp}(i) would read $\infty-\infty$.
\end{example}

The next two remarks give the details behind two sentences of Section~\ref{sec:triangle}.

\begin{remark}[Dividends]\label{rem:ec-dividends}
Let the asset pay a deterministic dividend yield $q$, and let $\tilde S_t:=S_te^{qt}$ be the value of a position with dividends reinvested, so that $\tilde S/B$ is a true $\QQ$-martingale. With $B_T$ constant, the relative return of this total-return asset is $\tilde S_T/(\tilde S_0B_T)=S_T/F_0$, where $F_0:=S_0e^{-qT}B_T=\EE_{\QQ}[S_T]$ is the forward price of the asset. All results of Section~\ref{sec:triangle} apply to $\tilde S$. Proposition~\ref{prop:decomp}(ii) holds with options on $S$ itself and with this $F_0$, because its proof uses only $R=S_T/F_0$ and $\EE_{\QQ}[S_T]=F_0$. The forward level that the Cboe computes from put--call parity is the empirical counterpart of $F_0$ \citep{Cboe2026}.
\end{remark}

\begin{remark}[The published VIX]\label{rem:ec-cboe}
The VIX is computed from S\&P~500 index options and targets a 30-day horizon \citep{CboeVIXWhitePaper}. For two expiries $T_1<30\text{ days}<T_2$, in years, the Cboe computes
\[
\sigma_i^2=\frac{2}{T_i}\sum_j\frac{\Delta K_j}{K_j^2}\,e^{r_iT_i}\,O_i(K_j)-\frac{1}{T_i}\Big(\frac{F_i}{K_{0,i}}-1\Big)^2,\qquad i=1,2,
\]
where $F_i$ is the forward level from put--call parity, $K_{0,i}$ the first listed strike at or below $F_i$, $O_i(K_j)$ the midpoint quote of the out-of-the-money option with strike $K_j$ (put and call averaged at $K_{0,i}$), $\Delta K_j$ half the distance between the neighbouring strikes, and $r_i$ the risk-free rate to expiry; it then interpolates the total variances $T_i\sigma_i^2$ linearly in time to 30 days, annualizes, and reports $100$ times the square root \citep{Cboe2026}. Compared with Proposition~\ref{prop:decomp}(ii), the index uses a sum over a finite range of listed strikes instead of the integral, quotes instead of prices, and a linear interpolation in time, which is exact only when the $\QQ$-side is affine in the horizon between $T_1$ and $T_2$; it also splits puts and calls at $K_0\le F$ instead of at $F$, and Lemma~\ref{lem:ec-spanning} at $K_0$ gives the exact correction $\frac{2}{T}\big(\ln\frac{F}{K_0}-\frac{F}{K_0}+1\big)$ for this split, whose second-order expansion is the Cboe term $-\frac1T(F/K_0-1)^2$. Hence $\frac12\VIX^2_{\mathrm{Cboe}}\cdot\frac{30}{365}\approx\kl{\QQ}{\QS}$ at a 30-day horizon is an approximation, whose error depends on the strike grid and on the quotes.
\end{remark}

\subsection{The It\^o--L\'evy Split}\label{sec:ec-tri-itolevy}

This subsection states and proves the split by source of risk of Section~\ref{sec:triangle-vix}. Let Assumption~\ref{ass:D} hold with a deterministic short rate $r_t$, so that $B_T=\exp(\int_0^Tr_t\,\dd t)$. Let $Z$ be a $\PP$-Brownian motion; the filtration may carry other risks. Let $N$ be the integer-valued random measure of the jumps of $\ln S$ on $[0,T]\times(\RR\setminus\{0\})$, with compensators $\nu^{\PP}_t(\dd y)\,\dd t$ under $\PP$ and $\nu^{\QQ}_t(\dd y)\,\dd t$ under $\QQ$, and let $\mu_t$ and $\sigma_t$ be predictable processes, with
\begin{align*}
\PP:\quad &\frac{\dd S_t}{S_{t-}}=\mu_t\,\dd t+\sigma_t\,\dd Z_t+\int(e^y-1)\big(N(\dd t,\dd y)-\nu^{\PP}_t(\dd y)\,\dd t\big),\\
\QQ:\quad &\frac{\dd S_t}{S_{t-}}=r_t\,\dd t+\sigma_t\,\dd Z^{\QQ}_t+\int(e^y-1)\big(N(\dd t,\dd y)-\nu^{\QQ}_t(\dd y)\,\dd t\big),
\end{align*}
where $Z^{\QQ}$ is a $\QQ$-Brownian motion; $\sigma_t$ is the same under both measures, because an equivalent change of measure keeps the quadratic variation of the continuous part. Let $k(y):=e^y-1-y\ge0$.

\begin{proposition}[It\^o--L\'evy split]\label{prop:itolevy}
In this setting, assume $\kl{\PP}{\QQ}<\infty$ and
\[
\EE_{\PP}\!\int_0^T\Big(|\mu_t-r_t|+\sigma_t^2+\int\big(y^2+k(y)\big)\,\nu^{\PP}_t(\dd y)\Big)\dd t<\infty,\qquad \EE_{\QQ}\!\int_0^T\Big(\sigma_t^2+\int\big(y^2+k(y)\big)\,\nu^{\QQ}_t(\dd y)\Big)\dd t<\infty .
\]
Then Assumption~\ref{ass:A1} holds and
\begin{align*}
\defect(\PP,\QQ,\QS)={}&\EE_{\PP}\!\int_0^T(\mu_t-r_t)\,\dd t+\frac12\Big(\EE_{\QQ}\!\int_0^T\sigma_t^2\,\dd t-\EE_{\PP}\!\int_0^T\sigma_t^2\,\dd t\Big)\\
&+\EE_{\QQ}\!\int_0^T\!\!\int k(y)\,\nu^{\QQ}_t(\dd y)\,\dd t-\EE_{\PP}\!\int_0^T\!\!\int k(y)\,\nu^{\PP}_t(\dd y)\,\dd t,
\end{align*}
where the first term is the drift premium, the second is half the premium on the diffusive variance, and the difference in the second line is the jump-entropy premium. Moreover, $\kl{\QQ}{\QS}=\frac12\EE_{\QQ}\int_0^T\sigma_t^2\,\dd t+\EE_{\QQ}\int_0^T\!\int k(y)\,\nu^{\QQ}_t(\dd y)\,\dd t=\frac T2\VIX^2(T)$.
\end{proposition}

\begin{proof}{Proof.}
Write $(N-\nu^{\PP})(\dd s,\dd y):=N(\dd s,\dd y)-\nu^{\PP}_s(\dd y)\,\dd s$ for the compensated jump measure under $\PP$. Under $\PP$, $S=S_0\,\mathcal E(Y)$, where $\mathcal E$ is the stochastic exponential and
\[
Y_t:=\int_0^t\mu_s\,\dd s+\int_0^t\sigma_s\,\dd Z_s+\int_0^t\!\!\int(e^y-1)\,(N-\nu^{\PP})(\dd s,\dd y).
\]
A jump of $N$ with mark $y$ makes $Y$ jump by $e^y-1>-1$, so $S>0$. The formula for the stochastic exponential gives
\[
\ln\frac{S_t}{S_0}=Y_t-\frac12\int_0^t\sigma_s^2\,\dd s+\sum_{s\le t}\big(\ln(1+\Delta Y_s)-\Delta Y_s\big),
\]
and at a jump with mark $y$, $\ln(1+\Delta Y_s)-\Delta Y_s=y-(e^y-1)=-k(y)$. The sum converges, because $k\ge0$ and, by the defining property of the compensator for nonnegative predictable integrands, $\EE_{\PP}\int_0^T\!\int k(y)\,N(\dd t,\dd y)=\EE_{\PP}\int_0^T\!\int k(y)\,\nu^{\PP}_t(\dd y)\,\dd t<\infty$. Now split $e^y-1=y+k(y)$ inside the compensated integral. The integral of $y$ against $N-\nu^{\PP}$ is a square-integrable $\PP$-martingale with mean zero, because $\EE_{\PP}\int_0^T\!\int y^2\,\nu^{\PP}_t(\dd y)\,\dd t<\infty$. The integral of $k$ against $N-\nu^{\PP}$ is the difference of the two finite integrals of $k$ against $N$ and against $\nu^{\PP}_t(\dd y)\,\dd t$. With $\ln B_T=\int_0^Tr_t\,\dd t$ this gives
\[
\ln R=\int_0^T\Big(\mu_t-r_t-\frac12\sigma_t^2\Big)\dd t-\int_0^T\!\!\int k(y)\,\nu^{\PP}_t(\dd y)\,\dd t+\int_0^T\sigma_t\,\dd Z_t+\int_0^T\!\!\int y\,(N-\nu^{\PP})(\dd t,\dd y).
\]
The two stochastic integrals are square-integrable $\PP$-martingales with mean zero, since $\EE_{\PP}\int_0^T\sigma_t^2\,\dd t<\infty$, and the two drift terms are $\PP$-integrable by the assumed integrability. Hence $\ln R$ is $\PP$-integrable and
\[
\EE_{\PP}[\ln R]=\EE_{\PP}\!\int_0^T\Big(\mu_t-r_t-\frac12\sigma_t^2\Big)\dd t-\EE_{\PP}\!\int_0^T\!\!\int k(y)\,\nu^{\PP}_t(\dd y)\,\dd t .
\]
The same computation under $\QQ$, with $r_t$, $Z^{\QQ}$, $\nu^{\QQ}$ and the integrability under $\QQ$, gives
\[
\EE_{\QQ}[\ln R]=-\frac12\EE_{\QQ}\!\int_0^T\sigma_t^2\,\dd t-\EE_{\QQ}\!\int_0^T\!\!\int k(y)\,\nu^{\QQ}_t(\dd y)\,\dd t .
\]
Together with $\kl{\PP}{\QQ}<\infty$, Assumption~\ref{ass:A1} holds. Theorem~\ref{thm:triangle} now gives the decomposition, Lemma~\ref{lem:ec-sides} gives $\kl{\QQ}{\QS}=-\EE_{\QQ}[\ln R]$, and Proposition~\ref{prop:decomp}(ii) gives the VIX form. \Halmos
\end{proof}

\subsection{Proof of Proposition~\ref{prop:slack}, and the Numeraire Portfolio}\label{sec:ec-tri-slack}

\begin{repeatproposition}[Proposition~\ref{prop:slack} (Entropy bound and its slack).]
Suppose that Assumption~\ref{ass:A1} holds.
\begin{enumerate}
\item[(i)] $\kl{\PP}{\QQ}=\EE_{\PP}[\ln R]+\kl{\PP}{\QS}$.
\item[(ii)] Hence $\EE_{\PP}[\ln R]\le\kl{\PP}{\QQ}$, with equality if and only if $R=\dd\PP/\dd\QQ$ almost surely, that is, if and only if $\QS=\PP$.
\item[(iii)] Under Assumption~\ref{ass:D}, $\kl{\PP}{\QQ}=\Lent_{\PP}\big(B_T^{-1}\dd\QQ/\dd\PP\big)$, the entropy of the pricing kernel.
\end{enumerate}
\end{repeatproposition}

\begin{proof}{Proof.}
(i) By Lemma~\ref{lem:ec-klfacts}(c), $\ln(\dd\PP/\dd\QQ)=\ln(\dd\PP/\dd\QS)+\ln R$ almost surely. Under Assumption~\ref{ass:A1} all three terms are $\PP$-integrable (Lemma~\ref{lem:ec-sides}), and integration under $\PP$ gives (i).

(ii) By (i) and Lemma~\ref{lem:ec-klfacts}(b), $\kl{\PP}{\QQ}-\EE_{\PP}[\ln R]=\kl{\PP}{\QS}\ge0$, with equality if and only if $\QS=\PP$. Since $\dd\QS/\dd\QQ=R$, $\QS=\PP$ holds if and only if $R=\dd\PP/\dd\QQ$ almost surely.

(iii) Multiplying a positive random variable by a positive constant does not change its entropy. Under Assumption~\ref{ass:D}, $B_T^{-1}$ is a constant, so
$\Lent_{\PP}(B_T^{-1}\dd\QQ/\dd\PP)=\Lent_{\PP}(\dd\QQ/\dd\PP)=\ln\EE_{\PP}[\dd\QQ/\dd\PP]-\EE_{\PP}[\ln(\dd\QQ/\dd\PP)]=0+\kl{\PP}{\QQ}$. \Halmos
\end{proof}

The bound of part (ii) uses entropy where the bound of \citet{HansenJagannathan1991} uses variance, and its slack is zero only for an asset whose relative return is $\dd\PP/\dd\QQ$, the numeraire portfolio of \citet{Long1990}. Its defect is the symmetric (Jeffreys) divergence $\kl{\PP}{\QQ}+\kl{\QQ}{\PP}$ (Lemma~\ref{lem:ec-numeraire}), and in Black--Scholes it keeps the growth-optimal fraction of its wealth in the stock (Lemma~\ref{lem:ec-bsslack}). The equality case of part (ii) has a dynamic reading. An asset $S'$ is priced by $\QQ$ when $S'/B$ is a true $\QQ$-martingale; its relative return is $R':=S'_T/(S'_0B_T)$, and its share measure $\QQ^{S'}$ is defined by $\dd\QQ^{S'}/\dd\QQ:=R'$.

\begin{lemma}[The numeraire portfolio]\label{lem:ec-numeraire}
Suppose that Assumption~\ref{ass:A1} holds and that $R=\dd\PP/\dd\QQ$. Then $S$ is a numeraire portfolio: the price of every claim, measured in units of $S$, is a $\PP$-martingale. Moreover, every other asset priced by $\QQ$ whose relative return $R'$ satisfies Assumption~\ref{ass:A1} has $\EE_{\PP}[\ln R']=\EE_{\PP}[\ln R]-\kl{\PP}{\QQ^{S'}}\le\EE_{\PP}[\ln R]$. Finally, $\defect(\PP,\QQ,\QS)=\kl{\PP}{\QQ}+\kl{\QQ}{\PP}$.
\end{lemma}

\begin{proof}{Proof.}
By Proposition~\ref{prop:slack}(ii), $\QS=\PP$. Let $X$ be a payoff at $T$ with $\EE_{\QQ}[|X|/B_T]<\infty$, and let $V_t:=B_t\,\EE_{\QQ}[X/B_T\mid\mathcal F_t]$ be its price process. Since $S/B$ is a true $\QQ$-martingale, $\EE_{\QQ}[R\mid\mathcal F_t]=S_t/(S_0B_t)$, and Bayes' rule with $R\,X/S_T=X/(S_0B_T)$ gives $\EE_{\QS}[X/S_T\mid\mathcal F_t]=(S_0B_t/S_t)\,\EE_{\QQ}[X/(S_0B_T)\mid\mathcal F_t]=V_t/S_t$, where $\EE_{\QS}[|X|/S_T]=\EE_{\QQ}[|X|/B_T]/S_0<\infty$. Since $\QS=\PP$, this says that $V/S$ is a $\PP$-martingale. Next, let $S'$ be another asset priced by $\QQ$ whose relative return $R'$ satisfies Assumption~\ref{ass:A1}. Proposition~\ref{prop:slack}(i) for $S'$ gives $\kl{\PP}{\QQ}=\EE_{\PP}[\ln R']+\kl{\PP}{\QQ^{S'}}$, and Proposition~\ref{prop:slack}(i) for $S$, with $\kl{\PP}{\QS}=0$, gives $\kl{\PP}{\QQ}=\EE_{\PP}[\ln R]$. Subtracting gives the formula for $\EE_{\PP}[\ln R']$, and the inequality follows from Lemma~\ref{lem:ec-klfacts}(b). Finally, with $\QS=\PP$, $\defect(\PP,\QQ,\QS)=\kl{\PP}{\QQ}+\kl{\QQ}{\PP}-\kl{\PP}{\PP}=\kl{\PP}{\QQ}+\kl{\QQ}{\PP}$, where $\kl{\QQ}{\PP}=\kl{\QQ}{\QS}<\infty$. \Halmos
\end{proof}

The economic reading of Proposition~\ref{prop:slack} in Section~\ref{sec:triangle} uses the following Black--Scholes computation.

\begin{lemma}[Constant-fraction portfolios in Black--Scholes]\label{lem:ec-bsslack}
In the setting of Corollary~\ref{cor:bs}, fix $w\in\RR$, and let $S^{(w)}$ be the value of the self-financing portfolio that keeps the fraction $w$ of its wealth in the stock and the rest in the bank account, with $S^{(w)}_0=1$. Let $R^{(w)}:=S^{(w)}_T/B_T$, let $\QQ^{(w)}$ be its share measure, $\dd\QQ^{(w)}/\dd\QQ:=R^{(w)}$, and let $w^*:=(\mu-r)/\sigma^2$. Then $S^{(w)}/B$ is a true $\QQ$-martingale, $R^{(w)}$ satisfies Assumption~\ref{ass:A1}, and the slack of Proposition~\ref{prop:slack} is
\[
\kl{\PP}{\QQ^{(w)}}=\frac{\sigma^2T}{2}\,(w-w^*)^2 .
\]
For $w=w^*$, $R^{(w^*)}=\dd\PP/\dd\QQ$: this portfolio is the numeraire portfolio, its expected log excess return $\EE_{\PP}[\ln R^{(w^*)}]=\frac12\SR^2T$ is the largest, and its defect is $\SR^2T=\kl{\PP}{\QQ}+\kl{\QQ}{\PP}$. For $w=1$ the slack is the side $\kl{\PP}{\QS}$ of Corollary~\ref{cor:bs}.
\end{lemma}

\begin{proof}{Proof.}
The portfolio satisfies $\dd S^{(w)}_t/S^{(w)}_t=w\,\dd S_t/S_t+(1-w)\,r\,\dd t=\big(r+w(\mu-r)\big)\dd t+w\sigma\,\dd Z_t$, so
\[
R^{(w)}=\exp\Big(\big(w(\mu-r)-\tfrac12w^2\sigma^2\big)T+w\sigma Z_T\Big)=\exp\Big(w\sigma Z^{\QQ}_T-\tfrac12w^2\sigma^2T\Big),
\]
with $Z^{\QQ}$ as in the proof of Corollary~\ref{cor:bs}. The second form shows that $S^{(w)}/B$ is a true $\QQ$-martingale. The variable $\ln R^{(w)}$ is normal under both measures, and $\kl{\PP}{\QQ}<\infty$, so Assumption~\ref{ass:A1} holds, with $\EE_{\PP}[\ln R^{(w)}]=\big(w(\mu-r)-\frac12w^2\sigma^2\big)T$. Proposition~\ref{prop:slack}(i) and $\kl{\PP}{\QQ}=\frac12\SR^2T$ give $\kl{\PP}{\QQ^{(w)}}=\frac{\sigma^2T}{2}\big((\mu-r)^2/\sigma^4-2w(\mu-r)/\sigma^2+w^2\big)=\frac{\sigma^2T}{2}(w-w^*)^2$. For $w=w^*$ we have $w^*\sigma=\SR$ and $w^*(\mu-r)=\SR^2$, so $\ln R^{(w^*)}=\SR\,Z_T+\frac12\SR^2T=\ln(\dd\PP/\dd\QQ)$. Proposition~\ref{prop:slack}(ii) and Lemma~\ref{lem:ec-numeraire} then give the numeraire property, the largest expected log excess return $\kl{\PP}{\QQ}=\frac12\SR^2T$, and the defect $\kl{\PP}{\QQ}+\kl{\QQ}{\PP}$. Here $\kl{\QQ}{\PP}=\frac12\SR^2T$, because $\ln(\dd\QQ/\dd\PP)=-\SR\,Z^{\QQ}_T+\frac12\SR^2T$. For $w=1$, $\QQ^{(1)}=\QS$. \Halmos
\end{proof}

\subsection{Angles and Option-Implied Divergences}\label{sec:ec-tri-angles}\label{sec:triangle-remarks}

For two scores $u$ and $u'$ at $\QQ$, that is, mean-zero elements of $L^2(\QQ)$ that are derivatives of log densities along curves through $\QQ$, the Fisher inner product at $\QQ$ is $\EE_{\QQ}[uu']$ and the Fisher length of $u$ is $(\EE_{\QQ}[u^2])^{1/2}$ (Section~\ref{sec:ec-geometry} gives the background). For probability measures $\mu\ll\nu$, the chi-square divergence is $\chi^2(\mu\,\|\,\nu):=\EE_\nu[(\dd\mu/\dd\nu-1)^2]$.

\begin{lemma}[The defect as an inner product of scores]\label{lem:ec-score}
Suppose that Assumption~\ref{ass:A1} holds, that $\chi^2(\PP\,\|\,\QQ)<\infty$ and $\EE_{\QQ}[(\ln R)^2]<\infty$, and let $D:=\dd\PP/\dd\QQ$. Along the mixture line $m_t:=(1-t)\QQ+t\PP$ and the exponential line $e_t$ with $\dd e_t/\dd\QQ:=R^t/\EE_{\QQ}[R^t]$, $t\in[0,1]$, which join $\QQ$ to $\PP$ and to $\QS$, the scores at $t=0$ are
\[
\frac{\dd}{\dd t}\ln\frac{\dd m_t}{\dd\QQ}\Big|_{t=0}=D-1,\qquad \frac{\dd}{\dd t}\ln\frac{\dd e_t}{\dd\QQ}\Big|_{t=0^+}=\ln R-\EE_{\QQ}[\ln R],
\]
and their Fisher inner product $\EE_{\QQ}\big[(D-1)(\ln R-\EE_{\QQ}[\ln R])\big]=\mathrm{Cov}_{\QQ}(D,\ln R)$ equals $\defect(\PP,\QQ,\QS)$. If $\PP\ne\QQ$ and $R$ is not constant, the angle $\alpha$ at $\QQ$ in the family of all measures equivalent to $\QQ$ (the nonparametric angle) is therefore given by $\cos\alpha=\mathrm{Corr}_{\QQ}(D,\ln R)$, it has the sign of the defect, and $|\defect(\PP,\QQ,\QS)|\le\big(\chi^2(\PP\,\|\,\QQ)\,\Var_{\QQ}(\ln R)\big)^{1/2}$.
\end{lemma}

\begin{proof}{Proof.}
In each state, $\ln(\dd m_t/\dd\QQ)=\ln(1+t(D-1))$ has derivative $D-1$ at $t=0$. For the exponential line, Jensen's inequality gives $\EE_{\QQ}[R^t]\le(\EE_{\QQ}[R])^t=1$ for $t\in[0,1]$, and $\ln(\dd e_t/\dd\QQ)=t\ln R-\ln\EE_{\QQ}[R^t]$. The map $t\mapsto(R^t-1)/t$ is increasing because $t\mapsto R^t$ is convex; so, as $t\downarrow0$, it decreases to $\ln R$, and for $t\le1$ it is bounded above by the $\QQ$-integrable variable $R-1$. Monotone convergence gives $\frac{\dd}{\dd t}\EE_{\QQ}[R^t]\big|_{t=0^+}=\EE_{\QQ}[\ln R]$, which yields the second score. Both scores lie in $L^2(\QQ)$, because $\EE_{\QQ}[(D-1)^2]=\chi^2(\PP\,\|\,\QQ)$ and $\ln R\in L^2(\QQ)$. Since $\EE_{\QQ}[D]=1$, their inner product is $\EE_{\QQ}[D\ln R]-\EE_{\QQ}[\ln R]=\EE_{\PP}[\ln R]-\EE_{\QQ}[\ln R]$, which is $\defect(\PP,\QQ,\QS)$ by Theorem~\ref{thm:triangle}. The two Fisher lengths are $\chi^2(\PP\,\|\,\QQ)^{1/2}$ and $\Var_{\QQ}(\ln R)^{1/2}$, both positive under the stated conditions, so the cosine is the correlation, and the Cauchy--Schwarz inequality gives the bound. \Halmos
\end{proof}

The angle changes when we draw the two lines inside a smaller family of distributions, but the inner product does not. Let $\Phi=(\Phi_1,\dots,\Phi_n)$ be a random vector such that $1,\Phi_1,\dots,\Phi_n$ are linearly independent in $L^2(\QQ)$, and such that the interior $\Theta$ of the set $\{\theta\in\RR^n:\EE_{\QQ}[e^{\theta\cdot\Phi}]<\infty\}$ contains $0$. The exponential family generated by $\Phi$ is $\mathcal K:=\{\QQ_\theta:\theta\in\Theta\}$ with $\dd\QQ_\theta/\dd\QQ:=\exp(\theta\cdot\Phi-\psi(\theta))$ and $\psi(\theta):=\ln\EE_{\QQ}[e^{\theta\cdot\Phi}]$. It contains $\QQ=\QQ_0$. The velocity $\dot\theta$ of a curve in $\mathcal K$ at $\QQ$ has the score $\dot\theta\cdot(\Phi-\EE_{\QQ}[\Phi])$, so the Fisher inner product of two velocities is $\dot\theta^\top\Sigma\,\dot\theta'$ with the positive definite matrix $\Sigma:=\mathrm{Cov}_{\QQ}(\Phi)$. Inside $\mathcal K$, the mixture line from $\QQ$ toward $\PP$ is the curve along which the mean $\EE_{\QQ_\theta}[\Phi]$ moves on the segment from $\EE_{\QQ}[\Phi]$ toward $\EE_{\PP}[\Phi]$; since $\nabla\psi(\theta)=\EE_{\QQ_\theta}[\Phi]$ and $\nabla^2\psi(0)=\Sigma$, its velocity at $\QQ$ is $\Sigma^{-1}(\EE_{\PP}[\Phi]-\EE_{\QQ}[\Phi])$.

\begin{lemma}[The inner product in an exponential family]\label{lem:ec-expfamily}
Suppose that Assumption~\ref{ass:A1} holds, that $\EE_{\PP}[|\Phi|]<\infty$, and that $\QS\in\mathcal K$, that is, $\ln R=\theta_S\cdot\Phi-\psi(\theta_S)$ for some $\theta_S\in\Theta$. Then the exponential line from $\QQ$ to $\QS$ is $t\mapsto\QQ_{t\theta_S}$, which stays in $\mathcal K$, and the Fisher inner product in $\mathcal K$ of the velocities of the mixture line toward $\PP$ and of the exponential line toward $\QS$ equals $\defect(\PP,\QQ,\QS)$.
\end{lemma}

\begin{proof}{Proof.}
Since $\dd\QS/\dd\QQ=R=\exp(\theta_S\cdot\Phi-\psi(\theta_S))$, the exponential line of Lemma~\ref{lem:ec-score} has $\dd e_t/\dd\QQ\propto e^{t\theta_S\cdot\Phi}$, so $e_t=\QQ_{t\theta_S}$. Its velocity is $\theta_S$, with score $\theta_S\cdot(\Phi-\EE_{\QQ}[\Phi])=\ln R-\EE_{\QQ}[\ln R]$, the same as the nonparametric score. The inner product is therefore $\theta_S^\top\Sigma\,\Sigma^{-1}(\EE_{\PP}[\Phi]-\EE_{\QQ}[\Phi])=\EE_{\PP}[\theta_S\cdot\Phi]-\EE_{\QQ}[\theta_S\cdot\Phi]=\EE_{\PP}[\ln R]-\EE_{\QQ}[\ln R]$, which is the defect by Theorem~\ref{thm:triangle}. \Halmos
\end{proof}

\begin{proposition}[Angles in Black--Scholes]\label{prop:ec-bsangles}
In the setting of Corollary~\ref{cor:bs}, let $x:=\SR^2T$ and $\mu\ne r$.
\begin{enumerate}
\item[(a)] In the nonparametric model, $\cos\alpha=\SR\sqrt T/\sqrt{e^x-1}$.
\item[(b)] In the two-parameter normal family of $\ln R$ (equivalently, of $\ln S_T$), $\cos\alpha=\mathrm{sgn}(\mu-r)/\sqrt{1+x/2}$.
\item[(c)] In the one-parameter normal family of $\ln R$ with fixed variance $\sigma^2T$, the three measures lie on one line: $\alpha=0^\circ$ if $\mu>r$ and $\alpha=180^\circ$ if $\mu<r$.
\end{enumerate}
In all three cases the inner product of the two velocities is $\defect(\PP,\QQ,\QS)=(\mu-r)T$, so $\cos\alpha$ has the sign of $\mu-r$, and $|\cos\alpha|$ is smallest in the nonparametric model.
\end{proposition}

\begin{proof}{Proof.}
By the proof of Corollary~\ref{cor:bs}, $\ln R$ is normal with variance $s^2:=\sigma^2T$ under $\PP$, $\QQ$ and $\QS$, with means $m_{\QQ}:=-\frac12\sigma^2T$, $m_{\QQ}+(\mu-r)T$ under $\PP$, and $m_{\QQ}+\sigma^2T$ under $\QS$ (tilting a normal law by $e^{\ln R}$ shifts its mean by $s^2$). Write $\vartheta:=(\mu-r)T$, so that $\vartheta^2/s^2=x$.
(a) $D=\dd\PP/\dd\QQ=\exp(\SR\,Z^{\QQ}_T-\frac12x)$ and $\ln R=\sigma Z^{\QQ}_T-\frac12\sigma^2T$, so $\chi^2(\PP\,\|\,\QQ)=\EE_{\QQ}[D^2]-1=e^x-1$ and $\Var_{\QQ}(\ln R)=s^2$. Lemma~\ref{lem:ec-score} gives the covariance $\vartheta$, so $\cos\alpha=\vartheta/\big(s\sqrt{e^x-1}\big)=\SR\sqrt T/\sqrt{e^x-1}$.
(b) The normal family is the exponential family generated by $\Phi=(\ln R,(\ln R)^2)$, and it contains the three measures. We use the mean and the variance as coordinates; at $\QQ$, where they equal $(m_{\QQ},s^2)$, the Fisher metric is $\mathrm{diag}(1/s^2,1/(2s^4))$. The exponential line from $\QQ$ to $\QS$ keeps the variance and moves the mean, so its velocity is $(s^2,0)$. The mixture line inside the family moves the mean of $\Phi$ linearly: the mean of $\ln R$ is $m_{\QQ}+t\vartheta$ and its second moment is linear in $t$, which gives the variance $s^2+t(1-t)\vartheta^2$ and the velocity $(\vartheta,\vartheta^2)$. The inner product is $\vartheta\,s^2/s^2=\vartheta$, and the lengths are $(\vartheta^2/s^2+\vartheta^4/(2s^4))^{1/2}$ and $s$. Hence $\cos\alpha=\vartheta/\big(\vartheta^2+\vartheta^4/(2s^2)\big)^{1/2}=\mathrm{sgn}(\vartheta)/\sqrt{1+x/2}$.
(c) In the family $\{\mathcal N(m,s^2):m\in\RR\}$ both velocities are multiples of the single coordinate direction, with the signs of $\vartheta$ and of $s^2>0$, so the cosine is $\mathrm{sgn}(\mu-r)$.
The inner product is $\vartheta$ in all three cases. Finally, $e^x-1\ge x+x^2/2$ gives $|\cos\alpha|$ in (a) at most $1/\sqrt{1+x/2}$, the value in (b), which is at most one, the value in (c). \Halmos
\end{proof}

The next remark collects what these results say about the angle of the triangle.

\begin{remark}[Angles]\label{rem:angles}
Under the conditions of Lemmas~\ref{lem:ec-score} and~\ref{lem:ec-expfamily}, the defect is the Fisher inner product at $\QQ$ of the velocities of the mixture line toward $\PP$ and of the exponential line toward $\QS$, both in the family of all measures equivalent to $\QQ$ and in every exponential family that contains $\QQ$ and $\QS$. So, whenever both velocities are nonzero, the type of the angle at $\QQ$ (acute, right, obtuse) is the sign of the defect in each of these families; \citet{Soklakov2023} reads a positive defect as an acute angle. The size of the angle depends on the family: in Black--Scholes its cosine takes three different values in the three families of Proposition~\ref{prop:ec-bsangles}, so we report the defect and treat the angle as a secondary reading.
\end{remark}

Section~\ref{sec:triangle} also mentions a second option-implied divergence between $\QQ$ and $\QS$.

\begin{remark}[VIX and SVIX]\label{rem:svix}
Proposition~\ref{prop:decomp} reads the VIX as the Kullback--Leibler divergence $\kl{\QQ}{\QS}$. The SVIX index of \citet{Martin2017}, $T\cdot\mathrm{SVIX}^2(T):=\Var_{\QQ}(R)$, is a second divergence between the same two measures: because $\dd\QS/\dd\QQ=R$ has $\QQ$-mean one, $T\cdot\mathrm{SVIX}^2(T)=\EE_{\QQ}[(R-1)^2]=\chi^2(\QS\,\|\,\QQ)$. Under Assumption~\ref{ass:D}, Lemma~\ref{lem:ec-spanning} applied to $x\mapsto f(x/F_0)$ at $F=F_0$ shows that every divergence $\EE_{\QQ}[f(R)]$ with convex $f\in C^2(0,\infty)$ and $f(1)=0$ is the forward price of a static portfolio of out-of-the-money options with weight $f''(K/F_0)/F_0^2$ at strike $K$: the VIX ($f(x)=-\ln x$) weights the option with strike $K$ by $1/K^2$, and the SVIX ($f(x)=(x-1)^2$) weights all strikes equally. The member $f(x)=x\ln x$ gives the reverse divergence $\kl{\QS}{\QQ}=\EE_{\QQ}[R\ln R]$, with weight $1/(KF_0)$: up to interest-rate terms, it is the LVIX index of \citet{GaoMartin2021}, who show that it bounds the expected log excess return from below under a modified negative correlation condition. \citet{Martin2017} shows that $\EE_{\PP}[R]-1\ge T\cdot\mathrm{SVIX}^2(T)$ under a negative correlation condition. So the chi-square side bounds the expected simple excess return from below, while \eqref{eq:triangle2} gives the expected log excess return plus the KL side exactly.
\end{remark}

\section{Proofs for Section~\ref{sec:incomplete}}\label{sec:ec-proofs-incomplete}

This section proves the results of Section~\ref{sec:incomplete} and collects the facts about its one-period market that
Section~\ref{sec:robust} also uses. For probability vectors $\mu,\nu$ on $\Omega$,
$\kl{\mu}{\nu}=\sum_{\omega:\,\mu(\omega)>0}\mu(\omega)\ln(\mu(\omega)/\nu(\omega))$, which is $+\infty$ as soon as
$\nu(\omega)=0<\mu(\omega)$ for some state. So $\kl{\QQ}{\PP}<\infty$ for every $\QQ$ when $\PP>0$, while
$\kl{\PP}{\QQ}=+\infty$ when $\PP>0$ and $\QQ$ has a zero. By Gibbs' inequality, $\kl{\mu}{\nu}\ge0$, with equality if
and only if $\mu=\nu$. We write $\mathcal{H}:=\{h\in\RR^d:\ W_h(\omega)>0\text{ for every }\omega\}$ for the portfolios
that keep wealth positive in every state.

\subsection{Arbitrage, the Fundamental Theorem, and the Shape of the Pricing Set}

\begin{lemma}[Arbitrage]\label{lem:ec-arbitrage}
(i) A position with $h^0$ units of the bond and $h\in\RR^d$ units of the risky assets costs $V_0:=h^0+h\cdot S_0$ and pays
$V_1:=h^0+h\cdot S_1=V_0+h\cdot\DeltaS$. A position with $V_0\le0$, with $V_1\ge0$ in every state and with $V_1>0$ in
some state exists if and only if some portfolio is an arbitrage in the sense of Section~\ref{sec:incomplete}. The same
holds with $V_0=0$ in place of $V_0\le0$.
(ii) Under Assumption~\ref{ass:market} there is $\kappa>0$ such that $\min_\omega h\cdot\DeltaS(\omega)\le-\kappa|h|$
for every $h\in\RR^d$. Hence $|h|\le1/\kappa$ whenever $W_h\ge0$ in every state; in particular, $\mathcal{H}$ is
bounded.
\end{lemma}

\begin{proof}{Proof.}
(i) This is the standard one-period equivalence \citep[see][]{FollmerSchied2016}. (ii) Let $m(h):=\min_\omega h\cdot\DeltaS(\omega)$. It is continuous, as a minimum of finitely many linear functions, and
$m(th)=t\,m(h)$ for $t>0$. Take $h\neq0$. By non-redundancy, the rows $\DeltaS^1,\dots,\DeltaS^d$ are linearly
independent, so $h\cdot\DeltaS$ is not the zero vector. By no arbitrage, $h\cdot\DeltaS$ is then negative in some state,
so $m(h)<0$. The unit sphere is compact, so $m$ attains on it a maximum $-\kappa<0$, and
$m(h)=|h|\,m(h/|h|)\le-\kappa|h|$. Finally, if $W_h\ge0$ in every state, then $0\le1+m(h)\le1-\kappa|h|$. \Halmos
\end{proof}

\begin{theorem}[Fundamental theorem, one-period form]\label{thm:ec-ftap}
Let non-redundancy, Assumption~\ref{ass:market}(i), hold. Then the market has no arbitrage if and only if
$\Mset\neq\emptyset$.
\end{theorem}

\begin{proof}{Proof.}
This is the fundamental theorem of asset pricing for a finite one-period market
\citep{HarrisonPliska1981,FollmerSchied2016}. Under no arbitrage, Proposition~\ref{prop:ec-QE} below also builds a measure
$\QE\in\Mset$; its proof uses no arbitrage only through Lemma~\ref{lem:ec-arbitrage}(ii). \Halmos
\end{proof}

The next proposition describes the shape of $\Mset$.  It uses the affine set
\[
\mathcal{A}:=\Bigl\{q\in\RR^M:\ \sum_\omega q(\omega)=1,\ \sum_\omega q(\omega)\DeltaS(\omega)=0\Bigr\}
\]
of signed vectors that sum to one and price the risky assets correctly.

\begin{proposition}[Shape of the set of pricing measures]\label{prop:ec-geometry}
Let Assumption~\ref{ass:market} hold.
\begin{enumerate}
\item[(i)] $\mathcal{A}$ is an affine subspace of dimension $M-1-d$. In particular $d\le M-1$.
\item[(ii)] $\Mset=\mathcal{A}\cap(0,\infty)^M$ is nonempty, convex, and open relative to $\mathcal{A}$. Its affine hull is
$\mathcal{A}$, so $\Mset$ has dimension $M-1-d$.
\item[(iii)] $\cl\Mset=\mathcal{A}\cap[0,\infty)^M=\{\QQ\in\cl\mathcal{P}:\EE_\QQ[\DeltaS]=0\}$, a compact convex
polytope.
\item[(iv)] For $\QQ_1,\QQ_2\in\Mset$, the mixture line $(1-t)\QQ_1+t\QQ_2$ lies in $\mathcal{A}$ for every $t\in\RR$ and in
$\Mset$ for every $t$ in an open interval that contains $[0,1]$. Thus $\Mset$ is the intersection of the open simplex with
an affine subspace, a linear family in the sense of \citet{Csiszar1975}.
\item[(v)] The exponential line between two points $\QQ_1,\QQ_2$ of $\Mset$, which is proportional to
$\QQ_1^{1-t}\QQ_2^{t}$, can leave $\Mset$ (Example~\ref{ex:ec-counter}).
\item[(vi)] The market is complete (every claim $X$ equals $a+h\cdot\DeltaS$ for some $a\in\RR$ and $h\in\RR^d$) if and
only if $d+1=M$, if and only if $\Mset$ is a single point. It is incomplete if and only if $d+1<M$.
\end{enumerate}
\end{proposition}

\begin{proof}{Proof.}
(i) By non-redundancy the linear map $L(q):=(\sum_\omega q(\omega),\,\DeltaS\,q)$ from $\RR^M$ to $\RR^{1+d}$ has rank
$1+d$. So $1+d\le M$, $L$ is onto, and $\mathcal{A}=L^{-1}(1,0)$ is a translate of $\ker L$, of dimension $M-1-d$.
(ii) $\Mset$ is nonempty by Theorem~\ref{thm:ec-ftap}, convex, and open relative to $\mathcal{A}$, so its affine hull is
$\mathcal{A}$. (iii) The set $\mathcal{A}\cap[0,\infty)^M$ is closed, contains $\Mset$, and is the intersection of the
simplex with an affine subspace. Conversely, for $\QQ$ in it and $\QQ^\circ\in\Mset$, the points $(1-t)\QQ+t\QQ^\circ$,
$t\in(0,1]$, lie in $\Mset$ and tend to $\QQ$. (iv) Affine combinations stay in $\mathcal{A}$; strict positivity holds
on $[0,1]$ and is an open condition. (v) See
Example~\ref{ex:ec-counter}. (vi) Under non-redundancy, $\mathbf{1},\DeltaS^1,\dots,\DeltaS^d$ span $\RR^M$ if and only if
$1+d=M$, that is, if and only if $\mathcal{A}$, and hence $\Mset$, is a single point; otherwise $\Mset$ contains a
segment. \Halmos
\end{proof}

In words, every traded asset adds one linear equation, so the pricing measures form a flat slice of the simplex whose
dimension $M-1-d$ counts the free directions of prices for claims that cannot be replicated.

\subsection{Weak Duality with Exact Slack}

The next lemma is the main tool of this section. It holds for every probability vector $\QQ$ with $\EE_\QQ[\DeltaS]=0$; under Assumption~\ref{ass:market} these are the
points of $\cl\Mset$.

\begin{lemma}[Weak duality with exact slack]\label{lem:ec-weakduality}
Let $\QQ\in\cl\mathcal{P}$ satisfy $\EE_\QQ[\DeltaS]=0$.
\begin{enumerate}
\item[(a$'$)] Let $\PP\in\mathcal{P}$. For $\eta\in\RR^d$ let $\varphi(\eta):=\EE_\PP[e^{-\eta\cdot\DeltaS}]$ and
$\PP_\eta(\omega):=\PP(\omega)e^{-\eta\cdot\DeltaS(\omega)}/\varphi(\eta)$. Then
\[
\kl{\QQ}{\PP}=-\ln\varphi(\eta)+\kl{\QQ}{\PP_\eta},\qquad\text{so}\qquad
\kl{\QQ}{\PP}\ge-\ln\EE_\PP\bigl[e^{-\eta\cdot\DeltaS}\bigr].
\]
\item[(b$'$)] Let $\PP\in\cl\mathcal{P}$. For $h\in\mathcal{H}$ let $\QQ^{W_h}(\omega):=W_h(\omega)\QQ(\omega)$. Then
$\QQ^{W_h}\in\cl\mathcal{P}$, it is strictly positive if and only if $\QQ$ is, and
\[
\kl{\PP}{\QQ}=\EE_\PP[\ln W_h]+\kl{\PP}{\QQ^{W_h}}\quad\text{in }[0,\infty],\qquad\text{so}\qquad
\kl{\PP}{\QQ}\ge\EE_\PP[\ln W_h].
\]
\end{enumerate}
\end{lemma}

\begin{proof}{Proof.}
(a$'$) The vector $\PP_\eta$ is strictly positive and sums to one. For each $\omega$ with $\QQ(\omega)>0$,
$\ln(\QQ/\PP)=\ln(\QQ/\PP_\eta)+\ln(\PP_\eta/\PP)=\ln(\QQ/\PP_\eta)-\eta\cdot\DeltaS-\ln\varphi(\eta)$.
Multiply by $\QQ(\omega)$ and sum over the states with $\QQ(\omega)>0$. States with $\QQ(\omega)=0$ contribute nothing to
$\eta\cdot\EE_\QQ[\DeltaS]$, so the middle term sums to $-\eta\cdot\EE_\QQ[\DeltaS]=0$. This gives the identity, and
the inequality follows from $\kl{\QQ}{\PP_\eta}\ge0$.
(b$'$) $\QQ^{W_h}\ge0$, and $\sum_\omega W_h(\omega)\QQ(\omega)=\EE_\QQ[1+h\cdot\DeltaS]=1+h\cdot\EE_\QQ[\DeltaS]=1$. Since
$W_h>0$, $\QQ^{W_h}(\omega)>0$ if and only if $\QQ(\omega)>0$. If $\QQ(\omega)=0<\PP(\omega)$ for some state, then
$\kl{\PP}{\QQ}$ and $\kl{\PP}{\QQ^{W_h}}$ are both $+\infty$ while $\EE_\PP[\ln W_h]$ is finite, so the identity holds.
Otherwise, for each $\omega$ with $\PP(\omega)>0$, $\ln(\PP/\QQ)=\ln W_h+\ln(\PP/(W_h\QQ))$; multiply by $\PP(\omega)$ and
sum over the states with $\PP(\omega)>0$. The inequality follows from $\kl{\PP}{\QQ^{W_h}}\ge0$. \Halmos
\end{proof}

In words, a divergence of a pricing measure is at least the value of every position, with an exact divergence as the
remainder: in (a$'$) the certainty-equivalent gain of an exponential investor who holds $\eta$, and in (b$'$) the
expected log return of the portfolio $h$, as in Proposition~\ref{prop:slack} with $h$ as the share.

\subsection{The Two Projections}

The next two propositions give the two projections of Section~\ref{sec:incomplete} in explicit form, using
Lemma~\ref{lem:ec-weakduality}.

\begin{proposition}[Minimal-entropy measure]\label{prop:ec-QE}
Let Assumption~\ref{ass:market} hold and let $\PP\in\mathcal{P}$.
\begin{enumerate}
\item[(i)] The function $\varphi(\eta)=\EE_\PP[e^{-\eta\cdot\DeltaS}]$ has a unique minimizer $h_E$ on $\RR^d$.
Define the Gibbs measure
\[
\QE(\omega):=\PP(\omega)\,e^{-h_E\cdot\DeltaS(\omega)}/\varphi(h_E).
\]
\item[(ii)] $\QE\in\Mset$; in particular $\QE(\omega)>0$ for every $\omega$.
\item[(iii)] For every $\QQ\in\cl\Mset$,
\[
\kl{\QQ}{\PP}=\kl{\QQ}{\QE}+\kl{\QE}{\PP},\qquad \kl{\QE}{\PP}=-\ln\min_{\eta\in\RR^d}\EE_\PP\bigl[e^{-\eta\cdot\DeltaS}\bigr].
\]
\item[(iv)] $\QE$ is the unique minimizer of $\kl{\cdot}{\PP}$ over $\cl\Mset$, and it lies in $\Mset$.
\item[(v)] For every $x\in\RR$, $h_E$ maximizes $\EE_\PP[-e^{-(x+\eta\cdot\DeltaS)}]$ over $\eta\in\RR^d$: it is
the optimal risky position of an investor with the exponential utility $U$ of Lemma~\ref{lem:duality}. Moreover
$\QE\propto\PP\,U'(x+h_E\cdot\DeltaS)$.
\end{enumerate}
\end{proposition}

\begin{proof}{Proof.}
(i) $\varphi$ is continuous, and it is strictly convex: the exponential is strictly convex, and for $\eta\neq\eta'$
non-redundancy gives a state, of positive probability, in which $\eta\cdot\DeltaS$ and $\eta'\cdot\DeltaS$ differ. By Lemma~\ref{lem:ec-arbitrage}(ii),
every $\eta$ has a state with $\eta\cdot\DeltaS(\omega)\le-\kappa|\eta|$, so $\varphi(\eta)\ge
\min_\omega\PP(\omega)\,e^{\kappa|\eta|}$. So the set $\{\eta:\varphi(\eta)\le\varphi(0)=1\}$ is compact, and $\varphi$
has a unique minimizer $h_E$.
(ii) $\nabla\varphi(\eta)=-\EE_\PP[\DeltaS\,e^{-\eta\cdot\DeltaS}]$ vanishes at $h_E$; dividing by
$\varphi(h_E)>0$ gives $\EE_{\QE}[\DeltaS]=0$. Moreover $\QE>0$ because $\PP>0$.
(iii) Lemma~\ref{lem:ec-weakduality}(a$'$) with $\eta=h_E$ gives $\kl{\QQ}{\PP}=-\ln\varphi(h_E)+\kl{\QQ}{\QE}$.
With $\QQ=\QE$, which is allowed by (ii), this reads $\kl{\QE}{\PP}=-\ln\varphi(h_E)$. Substitute.
(iv) By (iii), $\kl{\QQ}{\PP}-\kl{\QE}{\PP}=\kl{\QQ}{\QE}\ge0$, with equality if and only if $\QQ=\QE$.
(v) $\EE_\PP[-e^{-(x+\eta\cdot\DeltaS)}]=-e^{-x}\varphi(\eta)$ is largest where $\varphi$ is smallest. Since
$U'(y)=e^{-y}$, $\PP\,U'(x+h_E\cdot\DeltaS)=e^{-x}\PP\,e^{-h_E\cdot\DeltaS}\propto\QE$. \Halmos
\end{proof}

\begin{proposition}[Log-utility measure]\label{prop:ec-QL}
Let Assumption~\ref{ass:market} hold, let $\PP\in\mathcal{P}$, and let $G(h):=\EE_\PP[\ln W_h]$ for $h\in\mathcal{H}$.
\begin{enumerate}
\item[(i)] $G$ has a unique maximizer $\hstar$ on $\mathcal{H}$, and $\Wstar:=1+\hstar\cdot\DeltaS>0$.
\item[(ii)] $\EE_\PP[\DeltaS/\Wstar]=0$ and $\EE_\PP[1/\Wstar]=1$. Hence $\QL(\omega):=\PP(\omega)/\Wstar(\omega)$ defines a
measure $\QL\in\Mset$; in particular $\QL(\omega)>0$ for every $\omega$.
\item[(iii)] $\EE_\QQ[\Wstar]=1$ for every $\QQ\in\cl\Mset$. So $\QW:=\Wstar\QQ$ is a probability vector, strictly positive
if and only if $\QQ\in\Mset$.
\item[(iv)] For every $\QQ\in\cl\Mset$, $\kl{\PP}{\QQ}=\kl{\PP}{\QL}+\kl{\PP}{\QW}$, and
$\kl{\PP}{\QL}=\EE_\PP[\ln\Wstar]=\max_{h\in\mathcal{H}}\EE_\PP[\ln W_h]$.
\item[(v)] $\QL$ is the unique minimizer of $\kl{\PP}{\cdot}$ over $\cl\Mset$, and it lies in $\Mset$.
\item[(vi)] (Fibres.) For $\QQ\in\Mset$ and $g\in\mathcal{H}$, the vector $\QQ W_g$ is a model in $\mathcal{P}$ with
$\QL(\QQ W_g)=\QQ$ and $\hstar(\QQ W_g)=g$. Every $\PP\in\mathcal{P}$ equals $\QL(\PP)\,\Wstar(\PP)$. Hence the map
$\PP\mapsto(\QL(\PP),\hstar(\PP))$ is a bijection from $\mathcal{P}$ onto $\Mset\times\mathcal{H}$.
\end{enumerate}
\end{proposition}

\begin{proof}{Proof.}
(i) By Lemma~\ref{lem:ec-arbitrage}(ii), the set $K:=\{h:\ W_h\ge0\text{ in every state}\}$ is compact, and it contains
$\mathcal{H}$. With $\ln0=-\infty$, $G$ extends to an upper semicontinuous function on $K$, which equals $-\infty$ exactly
on $K\setminus\mathcal{H}$ because $\PP>0$. Since $G(0)=0$, its maximum is attained at some $\hstar\in\mathcal{H}$.
Non-redundancy makes $h\mapsto W_h$ one-to-one, so $G$ is strictly concave on the convex set $\mathcal{H}$, and $\hstar$
is unique.
(ii) $\mathcal{H}$ is open and $\nabla G(h)=\EE_\PP[\DeltaS/W_h]$, so $\nabla G(\hstar)=0$. Then
$\EE_\PP[1/\Wstar]=\EE_\PP[(\Wstar-\hstar\cdot\DeltaS)/\Wstar]=1-\hstar\cdot\EE_\PP[\DeltaS/\Wstar]=1$. Thus $\QL$ is strictly
positive, sums to one, and satisfies $\EE_{\QL}[\DeltaS]=\EE_\PP[\DeltaS/\Wstar]=0$.
(iii) $\EE_\QQ[\Wstar]=1+\hstar\cdot\EE_\QQ[\DeltaS]=1$; the rest is Lemma~\ref{lem:ec-weakduality}(b$'$).
(iv) Lemma~\ref{lem:ec-weakduality}(b$'$) with $h=\hstar$ gives $\kl{\PP}{\QQ}=\EE_\PP[\ln\Wstar]+\kl{\PP}{\QW}$. For
$\QQ=\QL$ the measure $\Wstar\QL$ is $\PP$ itself, so $\kl{\PP}{\QL}=\EE_\PP[\ln\Wstar]$. Substitute; the last equality is
(i).
(v) By (iv), $\kl{\PP}{\QQ}-\kl{\PP}{\QL}=\kl{\PP}{\QW}\ge0$, with equality if and only if $\Wstar\QQ=\PP$, that is, if and
only if $\QQ=\QL$.
(vi) $\QQ W_g>0$ and $\sum_\omega\QQ(\omega)W_g(\omega)=\EE_\QQ[W_g]=1$. Under the model $\PP':=\QQ W_g$,
$\EE_{\PP'}[\DeltaS/W_g]=\EE_\QQ[\DeltaS]=0$, so $g$ satisfies the first-order condition of (ii) for $\PP'$. The growth
function of $\PP'$ is strictly concave, so $g$ is its unique maximizer: $\hstar(\PP')=g$ and $\QL(\PP')=\PP'/W_g=\QQ$.
Conversely, $\PP=\QL(\PP)\Wstar(\PP)$ by (ii). \Halmos
\end{proof}

In words, the minimal-entropy measure tilts $\PP$ exponentially against the exponential investor's optimal position,
and the log-utility measure divides $\PP$ by the growth-optimal return; both are interior points of $\Mset$. Part (vi)
says that the models with a given log-utility measure $\QQ$ form the set $\{\QQ W_g:\ g\in\mathcal{H}\}$, one model for
each portfolio.

\subsection{Proof of Lemma~\ref{lem:duality}}

\begin{repeatlemma}[Lemma~\ref{lem:duality} (Utility duality).]
Let Assumption~\ref{ass:market} hold and let $\PP\in\mathcal{P}$. Let $U$ be log utility, $U=\ln$ on $D=(0,\infty)$, or
exponential utility, $U(z)=-e^{-z}$ on $D=\RR$, and let $V(y):=\sup_{z\in D}\{U(z)-zy\}$, $y>0$, be its conjugate. For
every $\QQ\in\Mset$ and every initial wealth $x\in D$,
\begin{equation}\tag{\ref{eq:duality}}
u(x;\QQ)=\min_{y>0}\Bigl\{\EE_\PP\Bigl[V\Bigl(y\,\frac{\dd\QQ}{\dd\PP}\Bigr)\Bigr]+xy\Bigr\}.
\end{equation}
The conjugate, the dual objective and the two values are
\[
\begin{array}{@{}l@{\hspace{1.3em}}c@{\hspace{1.3em}}c@{\hspace{1.3em}}c@{\hspace{1.3em}}c@{}}
 & V(y) & \EE_\PP\bigl[V(y\,\dd\QQ/\dd\PP)\bigr] & u(x;\QQ) & u(x)\\[2pt]
\text{log:} & -\ln y-1 & \kl{\PP}{\QQ}-\ln y-1 & \ln x+\kl{\PP}{\QQ} & \ln x+\kl{\PP}{\QL}\\[2pt]
\text{exponential:} & y\ln y-y & y\ln y-y+y\,\kl{\QQ}{\PP} & -e^{-x-\kl{\QQ}{\PP}} & -e^{-x-\kl{\QE}{\PP}}
\end{array}
\]
In both cases $u(x)=\min_{\QQ\in\Mset}u(x;\QQ)$. The minimax measure, $\QL$ for log and $\QE$ for exponential utility, is
unique and does not depend on $x$. The optimal claim in the market completed at it is the optimal traded wealth:
$x\Wstar$ for log and $x+h_E\cdot\DeltaS$ for exponential utility.
\end{repeatlemma}

\begin{proof}{Proof.}
\emph{Conjugates.} For $y>0$, the strictly concave maps $z\mapsto\ln z-zy$ on $(0,\infty)$ and $z\mapsto-e^{-z}-zy$ on
$\RR$ have maximizers $1/y$ and $-\ln y$ and maxima $-\ln y-1$ and $y\ln y-y$.
\emph{Divergence form.} Write $L:=\dd\QQ/\dd\PP=\QQ/\PP>0$. Then $\EE_\PP[-\ln(yL)-1]=-\ln y-1+\kl{\PP}{\QQ}$, and
$\EE_\PP[yL\ln(yL)-yL]=y\,\EE_\QQ[\ln y+\ln L]-y=y\ln y+y\,\kl{\QQ}{\PP}-y$.
\emph{Weak duality.} Let $X$ be feasible for $u(x;\QQ)$ and $y>0$. Applying the definition of $V$ in each state,
\[
\EE_\PP[U(X)]\le\EE_\PP[U(X)]+y\bigl(x-\EE_\QQ[X]\bigr)=\EE_\PP\bigl[U(X)-X\,yL\bigr]+xy\le\EE_\PP[V(yL)]+xy .
\]
\emph{Dual values.} Log: $y\mapsto\kl{\PP}{\QQ}-\ln y-1+xy$ is strictly convex, with minimizer $y=1/x$ and value
$\ln x+\kl{\PP}{\QQ}$. Exponential: with $K:=\kl{\QQ}{\PP}$, the map $y\mapsto y\ln y-y+yK+xy$ has derivative $\ln y+K+x$,
minimizer $y^*=e^{-x-K}$, and value $y^*(\ln y^*-1+K+x)=-y^*$.
\emph{Attainment.} Log: the claim $X=x\PP/\QQ>0$ has $\EE_\QQ[X]=x$ and $\EE_\PP[\ln X]=\ln x+\kl{\PP}{\QQ}$. Exponential:
the claim $X=x+K+\ln(\PP/\QQ)$ has $\EE_\QQ[X]=x+K-\kl{\QQ}{\PP}=x$ and $\EE_\PP[-e^{-X}]=-e^{-x-K}\EE_\PP[\QQ/\PP]=-e^{-x-K}$.
So the primal values equal the dual values, which proves \eqref{eq:duality} and the two formulas for $u(x;\QQ)$.
\emph{Minimax measures and traded values.} Log: by Proposition~\ref{prop:ec-QL}(v),
$\min_{\QQ\in\Mset}\{\ln x+\kl{\PP}{\QQ}\}=\ln x+\kl{\PP}{\QL}$, attained only at $\QL$, for every $x$. In the traded market
write $h=xg$, so that $x+h\cdot\DeltaS=xW_g$ with $g\in\mathcal{H}$. Then $u(x)=\ln x+\max_{g\in\mathcal{H}}\EE_\PP[\ln W_g]=
\ln x+\EE_\PP[\ln\Wstar]=\ln x+\kl{\PP}{\QL}$ by Proposition~\ref{prop:ec-QL}(iv), and the optimal traded wealth $x\Wstar$
equals the optimal claim $x\PP/\QL$ under $\QL$. Exponential: $k\mapsto-e^{-x-k}$ is increasing, so by
Proposition~\ref{prop:ec-QE}(iv) the minimum over $\Mset$ is $-e^{-x-\kl{\QE}{\PP}}$, attained only at $\QE$, for every $x$.
In the traded market, $u(x)=-e^{-x}\min_\eta\varphi(\eta)=-e^{-x-\kl{\QE}{\PP}}$ by Proposition~\ref{prop:ec-QE}(iii).
Finally, the optimal claim under $\QE$ is
$x+\kl{\QE}{\PP}+\ln(\PP/\QE)=x+\kl{\QE}{\PP}+h_E\cdot\DeltaS+\ln\varphi(h_E)=x+h_E\cdot\DeltaS$, which is the
optimal traded wealth by Proposition~\ref{prop:ec-QE}(v). \Halmos
\end{proof}

\subsection{Proof of Theorem~\ref{thm:twodecomp}}

\begin{repeattheorem}[Theorem~\ref{thm:twodecomp} (Two exact decompositions).]
Let Assumption~\ref{ass:market} hold and let $\PP\in\mathcal{P}$. For every $\QQ\in\cl\Mset$,
\begin{align*}
\text{(a)}\quad & \kl{\QQ}{\PP}=\kl{\QQ}{\QE}+\kl{\QE}{\PP},\\
\text{(b)}\quad & \kl{\PP}{\QQ}=\kl{\PP}{\QL}+\kl{\PP}{\QW},\qquad
\frac{\dd\QW}{\dd\QQ}:=\Wstar=\frac{\dd\PP}{\dd\QL}.
\end{align*}
Both sides of (b) are $+\infty$ when $\QQ\notin\Mset$, and every claim $X$ satisfies $\EE_\QQ[X]=\EE_{\QW}[X/\Wstar]$.
The term $\kl{\QQ}{\QE}$ in (a) is zero if and only if $\QQ=\QE$, and the term $\kl{\PP}{\QW}$ in (b) is zero if and only if $\QQ=\QL$.
\end{repeattheorem}

\begin{proof}{Proof.}
Part (a) is Proposition~\ref{prop:ec-QE}(iii), and part (b) is Proposition~\ref{prop:ec-QL}(iv). By
Proposition~\ref{prop:ec-QL}(iii), $\QW$ is a probability measure, and it is equivalent to $\PP$ if and only if
$\QQ\in\Mset$. If
$\QQ\in\cl\Mset\setminus\Mset$, some state has $\QQ(\omega)=0<\PP(\omega)$ and $\QW(\omega)=0$, so $\kl{\PP}{\QQ}$ and
$\kl{\PP}{\QW}$ are both $+\infty$. For every claim, $\EE_{\QW}[X/\Wstar]=\sum_\omega\Wstar(\omega)\QQ(\omega)X(\omega)/\Wstar(\omega)
=\EE_\QQ[X]$. In particular $\EE_{\QW}[1/\Wstar]=1$ and $\EE_{\QW}[S_1^i/\Wstar]=S_0^i+\EE_\QQ[\DeltaS^i]=S_0^i$: under $\QW$,
prices in units of $\Wstar$ are expectations. The zero cases follow from Gibbs' inequality: $\kl{\QQ}{\QE}=0$ if and only if
$\QQ=\QE$, and $\kl{\PP}{\QW}=0$ if and only if $\Wstar\QQ=\PP$, that is, if and only if $\QQ=\PP/\Wstar=\QL$. \Halmos
\end{proof}

\subsection{The Pythagorean Form Fails at the Log-Utility Measure}

A tempting form of Theorem~\ref{thm:twodecomp}(b) is $\kl{\PP}{\QQ}=\kl{\PP}{\QL}+\kl{\QL}{\QQ}$, a Pythagorean identity
with the right angle at $\QL$. Its error is the defect $\defect(\PP,\QL,\QQ)=\kl{\PP}{\QL}+\kl{\QL}{\QQ}-\kl{\PP}{\QQ}$ of
Section~\ref{sec:setting}. The following example shows that the defect is not zero in general and can have either sign.

\begin{example}[Three states, one stock]\label{ex:ec-counter}
Let $M=3$. The bond has price $1$ and payoff $1$. The stock has price $2$ and payoff $(1,3,4)$, so $\DeltaS=(-1,1,2)$.
The market is non-redundant, has no arbitrage ($\DeltaS$ takes both signs) and is incomplete ($d+1=2<3$), with
$\Mset=\{(\tfrac{1+t}{2},\tfrac{1-3t}{2},t):\ 0<t<\tfrac13\}$. Let $\PP=(0.45,0.25,0.30)$.
\begin{itemize}
\item Growth-optimal portfolio: $\hstar=1/4$, since
$\EE_\PP[\DeltaS/W_{1/4}]=-\frac{0.45}{0.75}+\frac{0.25}{1.25}+\frac{2(0.30)}{1.5}=-0.6+0.2+0.4=0$. Hence
$\Wstar=(\tfrac34,\tfrac54,\tfrac32)$ and $\QL=\PP/\Wstar=(0.6,0.2,0.2)$, the point $t=0.2$.
\item Take $\QQ=(0.55,0.35,0.10)\in\Mset$, the point $t=0.1$. Then
\begin{align*}
\kl{\PP}{\QQ}&=0.45\ln\tfrac{9}{11}+0.25\ln\tfrac57+0.30\ln3=0.155164,\\
\kl{\PP}{\QL}&=0.45\ln\tfrac34+0.25\ln\tfrac54+0.30\ln\tfrac32=0.0479685,\\
\kl{\QL}{\QQ}&=0.60\ln\tfrac{12}{11}+0.20\ln\tfrac47+0.20\ln2=0.0789131.
\end{align*}
The Pythagorean form would give $0.126882\neq0.155164$; the defect is $\defect(\PP,\QL,\QQ)=-0.0282822$.
\item Theorem~\ref{thm:twodecomp}(b) holds. With $\QW=\Wstar\QQ=(0.4125,0.4375,0.15)$,
\[
\kl{\PP}{\QW}=0.45\ln\tfrac{12}{11}+0.25\ln\tfrac47+0.30\ln2=0.107195,
\]
so $\kl{\PP}{\QL}+\kl{\PP}{\QW}=0.155164=\kl{\PP}{\QQ}$.
\item The exponential line from $\QL$ to $\QQ$ leaves $\Mset$: with $Z$ the normalizing constant, its midpoint
$\sqrt{\QL\QQ}/Z=(0.585909,0.269850,0.144241)$ has $\EE[\DeltaS]=-0.0275775\neq0$.
\item Mirror model: $\PP'=2\QL-\PP=(0.75,0.15,0.10)$ has $\hstar(\PP')=-1/4$, $\Wstar(\PP')=(\tfrac54,\tfrac34,\tfrac12)$ and
the same log-utility measure $\QL(\PP')=(0.6,0.2,0.2)$. Now $\kl{\PP'}{\QQ}=0.105522$, while
$\kl{\PP'}{\QL}+\kl{\QL}{\QQ}=0.133804$: the defect is $+0.0282822$, with the opposite sign.
Theorem~\ref{thm:twodecomp}(b) again holds: $0.0548906+0.0506309=0.105522$.
\item Theorem~\ref{thm:twodecomp}(a) in the same market: $e^{-h_E}=0.788781$ is the positive root of
$0.6u^3+0.25u^2-0.45=0$, so $h_E=0.237267$ and $\QE=(0.597791,0.206628,0.195581)$. Then
$\kl{\QE}{\PP}=0.0467264$, $\kl{\QQ}{\QE}=0.0715465$, and their sum is $\kl{\QQ}{\PP}=0.118273$.
\end{itemize}
\end{example}

\begin{proposition}[The defect at the log-utility measure]\label{prop:ec-defect}
Let Assumption~\ref{ass:market} hold and let $\PP\in\mathcal{P}$. For every $\QQ\in\Mset$, with
$\QQ_t\propto\QL^{1-t}\QQ^{t}$ the exponential line from $\QL$ to $\QQ$,
\begin{align*}
\defect(\PP,\QL,\QQ)&=\EE_{\QL}\Bigl[(\Wstar-1)\ln\frac{\dd\QQ}{\dd\QL}\Bigr]
=\hstar\cdot\EE_{\QL}\Bigl[\DeltaS\,\ln\frac{\dd\QQ}{\dd\QL}\Bigr]\\
&=\mathrm{Cov}_{\QL}\Bigl(\Wstar,\ln\frac{\dd\QQ}{\dd\QL}\Bigr)=-\frac{\dd}{\dd t}\kl{\PP}{\QQ_t}\Big|_{t=0}.
\end{align*}
Moreover $\kl{\PP}{\QW}-\kl{\QL}{\QQ}=-\defect(\PP,\QL,\QQ)$. So the Pythagorean form holds for a given $\QQ$ if and only
if the growth-optimal return and the log-likelihood ratio $\ln(\dd\QQ/\dd\QL)$ are uncorrelated under $\QL$.
\end{proposition}

\begin{proof}{Proof.}
The three measures are strictly positive, so Lemma~\ref{lem:threepoint} gives
$\defect(\PP,\QL,\QQ)=\EE_\PP[\ln(\QQ/\QL)]-\EE_{\QL}[\ln(\QQ/\QL)]$. Since $\PP=\Wstar\QL$, $\EE_\PP[f]=\EE_{\QL}[\Wstar f]$
for every $f$, which gives the first form. The second form follows from $\Wstar-1=\hstar\cdot\DeltaS$, and the covariance
form from $\EE_{\QL}[\Wstar-1]=0$. Along the exponential line, $\ln\QQ_t=(1-t)\ln\QL+t\ln\QQ-\psi(t)$ with
$\psi(t):=\ln\sum_\omega\QL(\omega)^{1-t}\QQ(\omega)^{t}$ and $\psi'(0)=\EE_{\QL}[\ln(\QQ/\QL)]$. Hence
$\frac{\dd}{\dd t}\kl{\PP}{\QQ_t}|_{t=0}=-\EE_\PP[\ln(\QQ/\QL)-\psi'(0)]=-\defect(\PP,\QL,\QQ)$. The last identity follows
from Theorem~\ref{thm:twodecomp}(b): $\defect(\PP,\QL,\QQ)=\kl{\QL}{\QQ}-\bigl(\kl{\PP}{\QQ}-\kl{\PP}{\QL}\bigr)=
\kl{\QL}{\QQ}-\kl{\PP}{\QW}$. \Halmos
\end{proof}

\begin{proposition}[No inequality holds in general]\label{prop:ec-noineq}
Let Assumption~\ref{ass:market} hold. Fix $\QQ^\circ,\QQ\in\Mset$, and let
$\xi:=\EE_{\QQ^\circ}[\DeltaS\ln(\dd\QQ/\dd\QQ^\circ)]\in\RR^d$. Every model $\PP_g:=\QQ^\circ W_g$ with $g\in\mathcal{H}$ has
$\QL(\PP_g)=\QQ^\circ$ and $\defect(\PP_g,\QQ^\circ,\QQ)=g\cdot\xi$. Hence, if $\xi\neq0$, the set $\{\PP_g:g\in\mathcal{H}\}$
contains models with a positive defect and models with a negative defect. In particular, neither
$\kl{\PP}{\QQ}\le\kl{\PP}{\QL}+\kl{\QL}{\QQ}$ nor the reverse inequality holds for all models.
\end{proposition}

\begin{proof}{Proof.}
By Proposition~\ref{prop:ec-QL}(vi), $\QL(\PP_g)=\QQ^\circ$ and $\hstar(\PP_g)=g$, so Proposition~\ref{prop:ec-defect}, applied
to the model $\PP_g$, gives $\defect(\PP_g,\QQ^\circ,\QQ)=g\cdot\xi$. The set $\mathcal{H}$ is open and contains $0$, so
$g:=t\xi$ and $-g$ lie in $\mathcal{H}$ for small $t>0$. Then $\defect(\PP_{g},\QQ^\circ,\QQ)=t|\xi|^2>0$ and
$\defect(\PP_{-g},\QQ^\circ,\QQ)=-t|\xi|^2<0$. Example~\ref{ex:ec-counter} has $\xi\neq0$ for $\QQ^\circ=\QL(\PP)$;
there $\PP=\PP_{\hstar}$ and $\PP'=2\QL-\PP=\PP_{-\hstar}$. \Halmos
\end{proof}

The geometry explains the failure. The log-utility measure is an interior minimizer of $\kl{\PP}{\cdot}$ over $\Mset$, so
the slope of this function vanishes along every curve through $\QL$ inside $\Mset$; the Pythagorean form needs the slope
to vanish along the exponential line from $\QL$ to $\QQ$ (Proposition~\ref{prop:ec-defect}, last form), which leaves
$\Mset$ in general. In one special case the defect is zero for every $\QQ$: if every gain $\DeltaS^i$ is nonzero in
exactly two states, each pricing equation fixes the ratio of two state probabilities, so $\Mset$ contains the exponential
lines between its points. Symmetric markets such as $\DeltaS=(-1,0,1)$ are of this type, so they cannot reveal the
failure of the Pythagorean form.

\subsection{The Entropy-Budget Interval}\label{sec:ec-inc-bounds}

The entropy-budget interval $J_b(X)$ of \eqref{eq:entropy-budget} is a compact interval that contains $\EE_{\QE}[X]$.
Indeed, the set $\{\QQ\in\cl\Mset:\ \kl{\QQ}{\QE}\le b\}$ is compact, because $\kl{\cdot}{\QE}$ is continuous on
$\cl\mathcal{P}$ (as $\QE>0$), convex, because KL is convex in its first argument, and it contains $\QE$; and $J_b(X)$ is
its image under the linear map $\QQ\mapsto\EE_\QQ[X]$. The next result gives the end points of $J_b(X)$ to first order
in $b$.

\begin{proposition}[First-order width]\label{prop:ec-width}
Let Assumption~\ref{ass:market} hold, let $\PP\in\mathcal{P}$, and let $X$ be a claim. Let
$\tilde X:=X-\EE_{\QE}[X]-\hat\theta\cdot\DeltaS$ be the residual of the $\QE$-weighted least-squares regression of $X$ on
$\mathbf{1}$ and $\DeltaS$, the unhedgeable part of $X$. For $b\ge0$ let $p^+(b)$ and $p^-(b)$ be the largest and the
smallest value of $\EE_\QQ[X]$ over $\{\QQ\in\cl\Mset:\ \kl{\QQ}{\QE}\le b\}$, the end points of $J_b(X)$. Then, as
$b\downarrow0$,
\[
p^{\pm}(b)=\EE_{\QE}[X]\pm\sqrt{2b}\;\mathrm{sd}_{\QE}(\tilde X)+O(b).
\]
\end{proposition}

\begin{proof}{Proof.}
We treat $p^+$; the case of $p^-$ follows with $-X$ in place of $X$. The normal equations of the regression give
$\EE_{\QE}[\tilde X]=0$ and $\EE_{\QE}[\tilde X\DeltaS]=0$.
\emph{Upper bound.} Let $\QQ$ be feasible and $U:=\dd\QQ/\dd\QE-1$. Then $\EE_{\QE}[U]=0$ and $\EE_{\QE}[U\DeltaS]=0$, so
$\EE_\QQ[X]-\EE_{\QE}[X]=\EE_{\QE}[UX]=\EE_{\QE}[U\tilde X]\le\|U\|_{L^2(\QE)}\,\mathrm{sd}_{\QE}(\tilde X)$. With
$\eta(u):=(1+u)\ln(1+u)-u$ we have $\kl{\QQ}{\QE}=\EE_{\QE}[\eta(U)]$. Let $\bar u:=\max_\omega|U(\omega)|$. Since
$\eta''(v)=1/(1+v)\ge1/(1+\bar u)$ for $v\le\bar u$, we get $\eta(u)\ge u^2/(2(1+\bar u))$ for $-1\le u\le\bar u$, and so
$\|U\|^2_{L^2(\QE)}\le2b(1+\bar u)$. Pinsker's inequality \citep{Csiszar1967} gives
$\sum_\omega|\QQ(\omega)-\QE(\omega)|\le\sqrt{2\kl{\QQ}{\QE}}\le\sqrt{2b}$, so
$\bar u\le\sum_\omega|\QQ(\omega)-\QE(\omega)|/\min_\omega\QE(\omega)\le\sqrt{2b}/\min_\omega\QE(\omega)$. Hence
$p^+(b)-\EE_{\QE}[X]\le\sqrt{2b}\,\mathrm{sd}_{\QE}(\tilde X)+O(b)$.
\emph{Lower bound.} If $\mathrm{sd}_{\QE}(\tilde X)=0$ there is nothing to prove. Otherwise let
$V:=\tilde X/\mathrm{sd}_{\QE}(\tilde X)$, $k:=\max_\omega|V(\omega)|$ and $\QQ^{(s)}:=\QE(1+sV)$ for $0<s<1/k$. Then
$\QQ^{(s)}>0$, it sums to one, and it prices $\DeltaS$ correctly, so $\QQ^{(s)}\in\Mset$. Since $\eta''(v)\le1/(1-sk)$ for
$v\ge-sk$ and $\EE_{\QE}[V^2]=1$, we get $\kl{\QQ^{(s)}}{\QE}\le s^2/(2(1-sk))$. Choosing $s$ with $s^2/(2(1-sk))=b$ makes
$\QQ^{(s)}$ feasible and gives $s=\sqrt{2b(1-sk)}\ge\sqrt{2b}\,(1-sk)=\sqrt{2b}-O(b)$. Finally
$\EE_{\QQ^{(s)}}[X]-\EE_{\QE}[X]=s\,\EE_{\QE}[VX]=s\,\EE_{\QE}[V\tilde X]=s\,\mathrm{sd}_{\QE}(\tilde X)$. \Halmos
\end{proof}

In words, for a small budget the exact price interval is centred at the minimal-entropy price, and its half-width is
$\sqrt{2b}$ times the standard deviation of the unhedgeable part of the claim.

\section{Proofs for Section~\ref{sec:robust}}\label{sec:ec-proofs-robust}

We use the notation of Section~\ref{sec:robust} and the conventions for KL of Section~\ref{sec:ec-proofs-incomplete}; a vector $\xi\in\RR^M$ is also a random variable on $\Omega$, and states with $\PP(\omega)=0$ contribute nothing to expectations under $\PP$. For $\lambda\ge0$ we write $L_\lambda(\PP,\QQ)=\kl{\PP}{\QQ}+\lambda\,\kl{\PP}{\Pz}$.

\subsection{Preliminary Lemmas}\label{sec:ec-rob-prelim}

The first two lemmas collect facts about the sets and about KL that the later proofs use.

\begin{lemma}[Sets]\label{lem:ec-rob-sets}
Assume (C2)--(C3).
(a) $\cl\Mset=\{\QQ\in\cl\mathcal P:\EE_\QQ[\DeltaS]=0\}$ is a nonempty compact convex polytope, and it is the closure of $\Mset$.
(b) $\mathcal H$ is open, convex and bounded, it contains $0$, and its closure is $\{h:W_h\ge0\}$.
(c) The map $h\mapsto W_h$ is injective.
(d) Under (C1), $\Bc$ is compact and convex and contains $\Pz$, and $\kl{\cdot}{\Pz}$ is continuous and strictly convex on $\cl\mathcal P$. If $c\ge\ln(1/\min_\omega\Pz(\omega))$, then $\Bc=\cl\mathcal P$.
\end{lemma}

\begin{proof}{Proof.}
(a) The set is the intersection of the simplex with an affine subspace, so it is a compact convex polytope; it is closed and contains $\Mset$. Conversely, for $\QQ$ in the set and $\QQ^\circ\in\Mset$, $(1-s)\QQ+s\QQ^\circ\in\Mset$ for $s\in(0,1]$, and it tends to $\QQ$ as $s\downarrow0$. (b) $\mathcal H$ is open and convex, as a finite intersection of open half-spaces that contain $0$. If $h_n\in\mathcal H$ and $|h_n|\to\infty$, a limit point $e$ of $h_n/|h_n|$ has $|e|=1$ and $e\cdot\DeltaS\ge0$ in every state; then $\EE_{\QQ^\circ}[e\cdot\DeltaS]=0$ with $\QQ^\circ\in\Mset$ forces $e\cdot\DeltaS=0$, which contradicts (C2). For the closure, $W_{(1-s)h}=(1-s)W_h+s>0$ if $W_h\ge0$ and $s\in(0,1]$. (c) If $W_h=W_{h'}$, then $(h-h')\cdot\DeltaS=0$, so $h=h'$ by (C2). (d) Because $\Pz>0$, $\kl{\PP}{\Pz}=\sum_\omega[\PP(\omega)\ln\PP(\omega)-\PP(\omega)\ln\Pz(\omega)]$ is continuous and strictly convex on $\cl\mathcal P$, and it is at most $-\sum_\omega\PP(\omega)\ln\Pz(\omega)\le\ln(1/\min_\omega\Pz(\omega))$. \Halmos
\end{proof}

\begin{lemma}[Facts about KL]\label{lem:ec-rob-kl}
(a) $\kl{\PP}{\QQ}\ge0$, with equality iff $\PP=\QQ$. (b) KL is jointly convex and jointly lower semicontinuous on $\cl\mathcal P\times\cl\mathcal P$. (c) If $\PP\in\mathcal P$, then $\QQ\mapsto\kl{\PP}{\QQ}$ is strictly convex where it is finite. (d) (Jensen bound.) If $\kl{\PP}{\QQ}<\infty$ and $W>0$ on $\Omega$, then $\EE_\PP[\ln W]\le\kl{\PP}{\QQ}+\ln\EE_\QQ[W]$, with equality iff $\PP=W\QQ/\EE_\QQ[W]$.
\end{lemma}

\begin{proof}{Proof.}
(a)--(c) are standard \citep{CsiszarShields2004}. (d) Let $S=\{\PP>0\}$, on which $\QQ>0$. Concavity of the logarithm gives $\EE_\PP[\ln W]-\kl{\PP}{\QQ}=\sum_{S}\PP\ln(W\QQ/\PP)\le\ln\sum_SW\QQ\le\ln\EE_\QQ[W]$. The first inequality is an equality if and only if $W\QQ/\PP$ is constant on $S$, and the second if and only if $\QQ=0$ off $S$; together, if and only if $\PP=W\QQ/\EE_\QQ[W]$. \Halmos
\end{proof}

The next lemma extends log duality, Lemma~\ref{lem:duality}, to models with zero entries, which plausible models may have. For such a model the supremum over $\mathcal H$ may not be attained, because the best portfolio may want zero wealth in an ignored state; part (c) says that its value is still the distance to $\cl\Mset$.

\begin{lemma}[Log duality on the closed simplex]\label{lem:ec-rob-logdual}
Assume (C2)--(C3), and let $u(\PP)=\sup_{h\in\mathcal H}G(h,\PP)$ for $\PP\in\cl\mathcal P$.
(a) (Weak duality.) $G(h,\PP)\le\kl{\PP}{\QQ}$ for all $h\in\mathcal H$, $\PP\in\cl\mathcal P$ and $\QQ\in\cl\Mset$.
(b) (Strictly positive models.) For $\PP\in\mathcal P$, $G(\cdot,\PP)$ has a unique maximizer $\hstar(\PP)\in\mathcal H$, characterized by $\EE_\PP[\DeltaS/\Wstar(\PP)]=0$; $\QL(\PP)=\PP/\Wstar(\PP)\in\Mset$; and $u(\PP)=\kl{\PP}{\QL(\PP)}=\min_{\QQ\in\cl\Mset}\kl{\PP}{\QQ}$, with $\QL(\PP)$ the unique minimizer. The map $\PP\mapsto\hstar(\PP)$ is $C^\infty$ on $\mathcal P$.
(c) (All models.) For every $\PP\in\cl\mathcal P$, $u(\PP)=\min_{\QQ\in\cl\Mset}\kl{\PP}{\QQ}<\infty$.
(d) $u$ is convex and lower semicontinuous on $\cl\mathcal P$.
\end{lemma}

\begin{proof}{Proof.}
(a) Lemma~\ref{lem:ec-rob-kl}(d) gives $G(h,\PP)\le\kl{\PP}{\QQ}+\ln\EE_\QQ[W_h]=\kl{\PP}{\QQ}$, because $\EE_\QQ[\DeltaS]=0$. (b) This is Lemma~\ref{lem:duality} \citep{GollRuschendorf2001}. In short: $G(\cdot,\PP)$ is strictly concave on the bounded set $\mathcal H$ (by $\PP>0$ and Lemma~\ref{lem:ec-rob-sets}(c)) and tends to $-\infty$ at its boundary, so it has a unique maximizer, which solves $\EE_\PP[\DeltaS/\Wstar]=0$. Then $\QL=\PP/\Wstar$ is strictly positive, prices $\DeltaS$ at zero, sums to $\EE_\PP[(\Wstar-\hstar\cdot\DeltaS)/\Wstar]=1$, and has $\kl{\PP}{\QL}=\EE_\PP[\ln\Wstar]$; with (a) and Lemma~\ref{lem:ec-rob-kl}(c), it is the unique minimizer. Smoothness follows from the implicit function theorem, because $\partial_h\EE_\PP[\DeltaS/W_h]=-\EE_\PP[\DeltaS\DeltaS^\top/W_h^2]$ is negative definite by (C2). (c) Let $D(\PP)=\min_{\QQ\in\cl\Mset}\kl{\PP}{\QQ}$, which is attained and finite. By (a) and (b), $u\le D$, with equality on $\mathcal P$. Joint lower semicontinuity of KL and compactness of $\cl\Mset$ make $D$ lower semicontinuous. The function $u$ is convex and finite, as a supremum of affine functions with $0=G(0,\PP)\le u\le D$. For $\PP\in\cl\mathcal P\setminus\mathcal P$, let $\PP_s=(1-s)\PP+s\PP^\circ$ with $\PP^\circ$ uniform; convexity gives $\limsup_{s\downarrow0}u(\PP_s)\le u(\PP)$, so $D(\PP)\le\liminf_{s\downarrow0}D(\PP_s)=\liminf_{s\downarrow0}u(\PP_s)\le u(\PP)\le D(\PP)$. (d) $u$ is a supremum of continuous affine functions. \Halmos
\end{proof}

\begin{lemma}[Minimal-entropy measure and threshold]\label{lem:ec-rob-qe}
Assume (C1)--(C3).
(a) $\kl{\cdot}{\Pz}$ has a unique minimizer $\QE$ on $\cl\Mset$. It is strictly positive, and $\QE=\Pz e^{-h_E\cdot\DeltaS}/\EE_{\Pz}[e^{-h_E\cdot\DeltaS}]$, where $h_E$ is the unique minimizer of $\eta\mapsto\EE_{\Pz}[e^{-\eta\cdot\DeltaS}]$ on $\RR^d$.
(b) $\cbar=\kl{\QE}{\Pz}=\max_{\eta\in\RR^d}\{-\ln\EE_{\Pz}[e^{-\eta\cdot\DeltaS}]\}$, and $\kl{\QQ}{\Pz}=\kl{\QQ}{\QE}+\cbar$ for every $\QQ\in\cl\Mset$.
(c) $\cbar>0$ iff (C4) holds.
(d) $\cl\Mset\cap\Bc\neq\emptyset$ iff $c\ge\cbar$, and $\cl\Mset\cap\Ball_{\cbar}=\{\QE\}$.
\end{lemma}

\begin{proof}{Proof.}
(a)--(b) are the I-projection on a linear family \citep{Csiszar1975} and its link with exponential utility \citep{Frittelli2000}; Section~\ref{sec:incomplete} states them as Lemma~\ref{lem:duality} and Theorem~\ref{thm:twodecomp}(a) for the strictly positive model $\Pz$. The identity extends from $\Mset$ to $\cl\Mset$ by continuity, since $\Pz>0$ and $\QE>0$. For the maximum, Lemma~\ref{lem:ec-rob-kl}(d) with the pair $(\QQ,\Pz)$ and $W=e^{-\eta\cdot\DeltaS}$ gives $-\ln\EE_{\Pz}[e^{-\eta\cdot\DeltaS}]\le\EE_\QQ[\eta\cdot\DeltaS]+\kl{\QQ}{\Pz}=\kl{\QQ}{\Pz}$ for every $\eta$ and $\QQ\in\cl\Mset$, with equality at $(h_E,\QE)$. (c) $\cbar=0$ if and only if $\QE=\Pz$ if and only if $\EE_{\Pz}[\DeltaS]=0$. (d) By (b), $\kl{\QQ}{\Pz}\ge\cbar$ on $\cl\Mset$, with equality only at $\QE$. \Halmos
\end{proof}

The worst-case model comes from the Gibbs (Donsker--Varadhan) variational formula \citep[see][]{DupuisEllis1997}; part (c) below is the KL dual of distributionally robust optimization \citep{HuHong2012,BenTalDenHertogDeWaegenaereEtAl2013}.

\begin{lemma}[Gibbs formula and linear objectives on the ball]\label{lem:ec-rob-gibbs}
Assume (C1). Let $\xi\in\RR^M$, $\theta>0$, and $\PP^\theta=\Pz e^{-\xi/\theta}/\EE_{\Pz}[e^{-\xi/\theta}]$.
(a) For $\PP\in\cl\mathcal P$, $\EE_\PP[\xi]+\theta\,\kl{\PP}{\Pz}=\theta\,\kl{\PP}{\PP^\theta}-\theta\ln\EE_{\Pz}[e^{-\xi/\theta}]$. Hence $\min_{\PP\in\cl\mathcal P}\{\EE_\PP[\xi]+\theta\,\kl{\PP}{\Pz}\}=-\theta\ln\EE_{\Pz}[e^{-\xi/\theta}]$, attained only at $\PP^\theta$.
(b) Let $\xi$ be non-constant, $m=\min_\omega\xi(\omega)$, $\Omega_m=\{\omega:\xi(\omega)=m\}$ and $c_m=-\ln\Pz(\Omega_m)>0$. Then $K(\theta)=\kl{\PP^\theta}{\Pz}$ is continuous and strictly decreasing on $(0,\infty)$, with limits $c_m$ as $\theta\downarrow0$ and $0$ as $\theta\to\infty$.
(c) For all $c\ge0$, $\PP\in\Bc$ and $\theta>0$, $\EE_\PP[\xi]\ge-\theta\ln\EE_{\Pz}[e^{-\xi/\theta}]-\theta c$. If $\xi$ is non-constant and $0<c<c_m$, let $\theta_c$ be the root of $K(\theta)=c$. Then $\min_{\PP\in\Bc}\EE_\PP[\xi]$ is attained only at $\PP^{\theta_c}$, which is strictly positive and has $\kl{\PP^{\theta_c}}{\Pz}=c$, and it equals $\max_{\theta>0}\{-\theta\ln\EE_{\Pz}[e^{-\xi/\theta}]-\theta c\}$, with $\theta_c$ the unique maximizer. If $c\ge c_m$, the minimum is $m$. In every case, $\min_{\PP\in\Bc}\EE_\PP[\xi]=\sup_{\theta>0}\{-\theta\ln\EE_{\Pz}[e^{-\xi/\theta}]-\theta c\}$.
\end{lemma}

\begin{proof}{Proof.}
(a) On $\{\PP>0\}$, $\theta\ln(\PP/\PP^\theta)=\theta\ln(\PP/\Pz)+\xi+\theta\ln\EE_{\Pz}[e^{-\xi/\theta}]$; take expectations under $\PP$ and use Lemma~\ref{lem:ec-rob-kl}(a). (b) With $\beta=1/\theta$ and $\psi(\beta)=\ln\EE_{\Pz}[e^{-\beta\xi}]$, the divergence $\kappa(\beta)=\kl{\PP^{1/\beta}}{\Pz}=\beta\psi'(\beta)-\psi(\beta)$ has $\kappa'(\beta)=\beta\psi''(\beta)=\beta\Var_{\PP^{1/\beta}}(\xi)>0$ and $\kappa(0^+)=0$. As $\beta\to\infty$, $\PP^{1/\beta}\to\Pz(\cdot\mid\Omega_m)$, so $\kappa(\beta)\to c_m$; and $K(\theta)=\kappa(1/\theta)$. (c) The inequality follows from (a), because $\kl{\PP}{\PP^\theta}\ge0$ and $\kl{\PP}{\Pz}\le c$. At $\theta=\theta_c$, both steps are equalities if and only if $\PP=\PP^{\theta_c}$, so $\PP^{\theta_c}$ is the unique minimizer, and the minimum equals $D(\theta_c)$, where $D(\theta)=-\theta\ln\EE_{\Pz}[e^{-\xi/\theta}]-\theta c$ never exceeds the minimum. Since $D'(\theta)=K(\theta)-c$, which by (b) is positive below $\theta_c$ and negative above it, $\theta_c$ is the unique maximizer of $D$. If $c\ge c_m$, then $\EE_\PP[\xi]\ge m$, and $\Pz(\cdot\mid\Omega_m)\in\Bc$ attains $m$. For the last claim, the inequality gives $\sup_\theta D(\theta)\le\min_{\PP\in\Bc}\EE_\PP[\xi]$, with equality in the case just treated. Otherwise, write $D(\theta)=m-\theta\ln\EE_{\Pz}[e^{-(\xi-m)/\theta}]-\theta c$. If $\xi$ is non-constant and $c\ge c_m$, the expectation tends to $\Pz(\Omega_m)>0$ as $\theta\downarrow0$, so $D(\theta)\to m$, the minimum; if $\xi$ is constant, $D(\theta)=m-\theta c\to m$; and if $c=0$, $D(\theta)\to\EE_{\Pz}[\xi]$ as $\theta\to\infty$, the minimum over $\Ball_0=\{\Pz\}$. \Halmos
\end{proof}

\begin{lemma}[The P-step: I-projection onto the ball]\label{lem:ec-rob-pstep}
Assume (C1). Let $\QQ\in\mathcal P$, $c>0$, and let $e_t\propto\Pz^{1-t}\QQ^t$, $t\in[0,1]$, be the exponential line from $\Pz$ to $\QQ$. The problem $\min_{\PP\in\Bc}\kl{\PP}{\QQ}$ has a unique solution: $\QQ$ itself if $\kl{\QQ}{\Pz}\le c$, and otherwise $e_{t^*}$, where $t^*\in(0,1)$ is the unique root of $\kl{e_t}{\Pz}=c$. Moreover, for $\beta\ge0$ and $t=1/(1+\beta)$, and for every $\PP\in\cl\mathcal P$,
\[
\kl{\PP}{\QQ}+\beta\,\kl{\PP}{\Pz}=(1+\beta)\,\kl{\PP}{e_t}-(1+\beta)\ln\textstyle\sum_\omega\Pz(\omega)^{1-t}\QQ(\omega)^t,
\]
so $e_t$ is the unique minimizer of $\kl{\PP}{\QQ}+\beta\,\kl{\PP}{\Pz}$ over $\cl\mathcal P$.
\end{lemma}

\begin{proof}{Proof.}
The identity follows from $(1+\beta)(1-t)=\beta$ and $(1+\beta)t=1$, by taking expectations under $\PP$ of $(1+\beta)\ln(\PP/e_t)$. If $\kl{\QQ}{\Pz}\le c$, then $\PP=\QQ$ attains the value $0$. Otherwise $\ell=\ln(\QQ/\Pz)$ is not constant, $e_t\propto\Pz e^{t\ell}$, and, as in Lemma~\ref{lem:ec-rob-gibbs}(b), $t\mapsto\kl{e_t}{\Pz}$ increases strictly from $0$ to $\kl{\QQ}{\Pz}>c$, so $t^*$ is unique. With $\beta^*=(1-t^*)/t^*$, the identity gives $\kl{\PP}{\QQ}\ge\kl{e_{t^*}}{\QQ}+\beta^*(c-\kl{\PP}{\Pz})\ge\kl{e_{t^*}}{\QQ}$ for $\PP\in\Bc$, with equality if and only if $\PP=e_{t^*}$. \Halmos
\end{proof}

\begin{lemma}[From blockwise to joint optimality]\label{lem:ec-rob-block}
Let $\lambda\ge0$, $\PP^\star\in\mathcal P$ and $\QQ^\star\in\Mset$. If $\PP^\star$ minimizes $L_\lambda(\cdot,\QQ^\star)$ over $\cl\mathcal P$ and $\QQ^\star$ minimizes $L_\lambda(\PP^\star,\cdot)$ over $\cl\Mset$, then $(\PP^\star,\QQ^\star)$ minimizes $L_\lambda$ over $\cl\mathcal P\times\cl\Mset$.
\end{lemma}

\begin{proof}{Proof.}
$L_\lambda$ is convex (Lemma~\ref{lem:ec-rob-kl}(b)) and continuously differentiable near $z^\star=(\PP^\star,\QQ^\star)$, where all entries are positive. For $z$ in the product with $L_\lambda(z)<\infty$, the convex function $j(s)=L_\lambda(z^\star+s(z-z^\star))$ on $[0,1]$ has $j'(0^+)=\nabla_\PP L_\lambda(z^\star)\cdot(\PP-\PP^\star)+\nabla_\QQ L_\lambda(z^\star)\cdot(\QQ-\QQ^\star)\ge0$, because each term is the right derivative of a one-block function that is minimal at $s=0$. So $L_\lambda(z)=j(1)\ge j(0)+j'(0^+)\ge L_\lambda(z^\star)$. \Halmos
\end{proof}

\begin{lemma}[CRRA investors]\label{lem:ec-rob-crra}
Assume (C1)--(C3) and let $\gamma\ge1$.
(a) $h^{(\gamma)}$ exists, is unique, lies in $\mathcal H$, and is characterized by $\EE_{\Pz}[(\Wg)^{-\gamma}\DeltaS]=0$.
(b) $h^{(\gamma)}=\hstar(\Pg)$, and $\Qg=\QL(\Pg)=\Pg/\Wg\in\Mset$; in particular $\EE_{\Pz}[(\Wg)^{1-\gamma}]=\EE_{\Pz}[(\Wg)^{-\gamma}]$.
(c) $\gamma\mapsto h^{(\gamma)}$ is $C^\infty$ on $[1,\infty)$.
(d) Under (C4), $h^{(\gamma)}\neq0$, so $\Wg$ is not constant.
\end{lemma}

\begin{proof}{Proof.}
(a) For $\gamma=1$ this is Lemma~\ref{lem:ec-rob-logdual}(b). For $\gamma>1$, $h\mapsto\EE_{\Pz}[U_\gamma(W_h)]$ is strictly concave on the bounded set $\mathcal H$ and tends to $-\infty$ at its boundary, so its unique maximizer solves the first-order condition. (b) $\EE_{\Pg}[\DeltaS/\Wg]$ is proportional to $\EE_{\Pz}[(\Wg)^{-\gamma}\DeltaS]=0$, the log first-order condition under $\Pg\in\mathcal P$, so Lemma~\ref{lem:ec-rob-logdual}(b) applies; the normalization follows from $\EE_{\Pz}[W^{-\gamma}(W-1)]=h\cdot\EE_{\Pz}[W^{-\gamma}\DeltaS]=0$ with $W=\Wg$ and $h=h^{(\gamma)}$. (c) The map $(h,\gamma)\mapsto\EE_{\Pz}[W_h^{-\gamma}\DeltaS]$ is smooth for $\gamma$ in an open interval around $[1,\infty)$, and its derivative in $h$, $-\gamma\,\EE_{\Pz}[W_h^{-\gamma-1}\DeltaS\DeltaS^\top]$, is negative definite; apply the implicit function theorem. (d) At $h=0$ the first-order condition reads $\EE_{\Pz}[\DeltaS]=0$, which (C4) excludes. \Halmos
\end{proof}

\subsection{Proof of Theorem~\ref{thm:minimax}}\label{sec:ec-rob-minimax}

\begin{repeattheorem}[Theorem~\ref{thm:minimax} (Robust value as a KL distance).]
Assume (C1)--(C3) and $c\ge0$. Then
\[
v(c)=\sup_{h\in\mathcal H}\ \min_{\PP\in\Bc}\EE_\PP[\ln W_h]
=\min_{\PP\in\Bc}\ \sup_{h\in\mathcal H}\EE_\PP[\ln W_h]
=\min_{\PP\in\Bc,\ \QQ\in\cl\Mset}\kl{\PP}{\QQ}.
\]
All minima are attained, and the supremum over $h$ is attained in $\mathcal H$. If $h^\dagger$ attains the supremum and $(\PP^\dagger,\QQ^\dagger)$ attains the last minimum, then $(h^\dagger,\PP^\dagger)$ is a saddle point,
\[
G(h,\PP^\dagger)\le G(h^\dagger,\PP^\dagger)=v(c)\le G(h^\dagger,\PP)\qquad\text{for all }h\in\mathcal H\text{ and }\PP\in\Bc,
\]
and $\QQ^\dagger=\QL(\PP^\dagger)$ whenever $\PP^\dagger$ is strictly positive.
\end{repeattheorem}

\begin{proof}{Proof.}
\textit{Minimax.} The minimax theorem of \citet{Sion1958} applies to $G(h,\PP)$ on $\Bc\times\mathcal H$: $\Bc$ is compact and convex (Lemma~\ref{lem:ec-rob-sets}(d)), $\mathcal H$ is convex, and $G$ is real valued, affine and continuous in $\PP$, and concave and continuous in $h$. Sion's theorem needs compactness of one set only, and it gives $\inf_\PP\sup_hG=\sup_h\inf_\PP G$. On the right, $\inf_\PP G(h,\PP)$ is a minimum because $G(h,\cdot)$ is continuous; on the left, the infimum of $u=\sup_hG(h,\cdot)$ is a minimum because $u$ is lower semicontinuous (Lemma~\ref{lem:ec-rob-logdual}(d)).
\textit{Distance.} By Lemma~\ref{lem:ec-rob-logdual}(c), $\min_{\Bc}u$ equals the minimum of KL over $\Bc\times\cl\Mset$, which exists because KL is jointly lower semicontinuous on this compact set and finite at $(\Pz,\QL(\Pz))$.
\textit{Attainment in $h$.} The worst-case growth $G_c$ is concave, as an infimum of concave functions, and finite on the open set $\mathcal H$, since $G(h,\PP)\ge\min_\omega\ln W_h(\omega)$; so it is continuous. Also $G_c(h)\le G(h,\Pz)\to-\infty$ at the boundary of the bounded set $\mathcal H$. So a sequence $h_n$ with $G_c(h_n)\to\sup G_c\ge G_c(0)=0$ has a limit point $h^\dagger\in\mathcal H$, and $G_c(h^\dagger)=v(c)$.
\textit{Saddle point.} For a nearest pair $(\PP^\dagger,\QQ^\dagger)$, $u(\PP^\dagger)\le\kl{\PP^\dagger}{\QQ^\dagger}=v(c)=\min_{\Bc}u$, so $u(\PP^\dagger)=v(c)$. Then $v(c)=u(\PP^\dagger)\ge G(h^\dagger,\PP^\dagger)\ge G_c(h^\dagger)=v(c)$, so both inequalities are equalities. The first says that $h^\dagger$ maximizes $G(\cdot,\PP^\dagger)$, and the second that $\PP^\dagger$ minimizes $G(h^\dagger,\cdot)$ on $\Bc$. Finally $\kl{\PP^\dagger}{\QQ^\dagger}=u(\PP^\dagger)=\min_{\QQ\in\cl\Mset}\kl{\PP^\dagger}{\QQ}$, so $\QQ^\dagger=\QL(\PP^\dagger)$ when $\PP^\dagger>0$ (Lemma~\ref{lem:ec-rob-logdual}(b)). \Halmos
\end{proof}

\subsection{Proof of Theorem~\ref{thm:robust}}\label{sec:ec-rob-structure}

\begin{repeattheorem}[Theorem~\ref{thm:robust} (Structure of the robust solution).]
Assume (C1)--(C4), so that $\cbar>0$.
\begin{enumerate}
\item[(i)] (Value.) $v$ is convex, nonincreasing and continuous on $[0,\infty)$, with $v(0)=\kl{\Pz}{\QL(\Pz)}=\EE_{\Pz}[\ln\Wstar(\Pz)]$. It is strictly decreasing and strictly convex on $[0,\cbar]$, and $v\equiv0$ on $[\cbar,\infty)$; so $v(c)>0$ iff $c<\cbar$. It is continuously differentiable on $(0,\infty)$, with $v'<0$ on $(0,\cbar)$, $v'=0$ on $[\cbar,\infty)$, and $v'(0^+)=-\infty$.
\item[(ii)] (Nearest pair below the threshold.) For $0\le c<\cbar$ the nearest pair $(\Pc,\Qc)$ is unique. $\Pc$ is strictly positive and $\kl{\Pc}{\Pz}=c$, so the constraint binds. The robust portfolio $h_c$ is unique, $h_c=\hstar(\Pc)\neq0$, and $\Qc=\QL(\Pc)=\Pc/\Wc\in\Mset$, where $\Wc=W_{h_c}$.
\item[(iii)] (Power tilt.) For $0<c<\cbar$ there is a unique $\lambda_c>0$ such that
\[
\frac{\dd\Pc}{\dd\Pz}=\frac{\Wc^{-1/\lambda_c}}{\EE_{\Pz}\bigl[\Wc^{-1/\lambda_c}\bigr]},\qquad
\frac{\dd\Qc}{\dd\Pz}=\frac{\Wc^{-(1+1/\lambda_c)}}{\EE_{\Pz}\bigl[\Wc^{-(1+1/\lambda_c)}\bigr]}.
\]
Equivalently, with $\gamma_c=1+1/\lambda_c$, $\Pc\propto\Pz^{1/\gamma_c}\,\Qc^{1-1/\gamma_c}$: the worst-case model lies on the exponential line from $\Pz$ to $\Qc$. The multiplier is $\lambda_c=-v'(c)$, and $(\Pc,\Qc)$ minimizes $\kl{\PP}{\QQ}+\lambda_c\,\kl{\PP}{\Pz}$ over $\cl\mathcal P\times\cl\Mset$.
\item[(iv)] (Robustness acts as risk aversion in a given market.) For $0\le c<\cbar$, with $\gamma_0=1$, we have $h_c=h^{(\gamma_c)}$, $\Pc=\PP^{(\gamma_c)}$ and $\Qc=\QQ^{(\gamma_c)}$: the robust portfolio is the optimal portfolio under $\Pz$ of a CRRA investor with risk aversion $\gamma_c$, and the robust pricing measure is that investor's marginal-utility pricing measure. Conversely, for every $\gamma\ge1$, $(\Pg,\Qg)$ is the nearest pair for the radius $c(\gamma)$, $h^{(\gamma)}$ is the robust portfolio for that radius, and $\gamma_{c(\gamma)}=\gamma$. The map $\gamma\mapsto c(\gamma)$ is a continuous, strictly increasing bijection from $[1,\infty)$ onto $[0,\cbar)$, with inverse $c\mapsto\gamma_c$. Hence $\lambda_c$ decreases continuously and strictly from $+\infty$ to $0$ as $c$ increases from $0$ to $\cbar$.
\item[(v)] (Endpoints.) At $c=0$ the nearest pair is $(\Pz,\QL(\Pz))$. As $c\uparrow\cbar$: $\Pc\to\QE$, $\Qc\to\QE$, $h_c\to0$ and $h_c/\lambda_c\to h_E$. Equivalently, as $\gamma\to\infty$: $\Pg\to\QE$, $\Qg\to\QE$ and $\gamma h^{(\gamma)}\to h_E$, the optimal position of an investor with exponential utility; all three errors are $O(1/\gamma)$.
\item[(vi)] (No trade at and above the threshold.) For $c\ge\cbar$: $v(c)=0$, the unique robust portfolio is $h_c=0$, and the nearest pairs are the pairs $(\QQ,\QQ)$ with $\QQ\in\cl\Mset\cap\Bc$. This set is nonempty, compact and convex, and it reduces to $\{(\QE,\QE)\}$ at $c=\cbar$.
\end{enumerate}
\end{repeattheorem}

\begin{proof}{Proof.}
\textit{Step 1 (part (i): convexity, monotonicity, continuity, positivity).} For nearest pairs $(\PP^i,\QQ^i)$ at radii $c_i$ and $\alpha\in[0,1]$, the mixtures $\PP^\alpha$ and $\QQ^\alpha$ lie in $\Ball_{\alpha c_1+(1-\alpha)c_2}$ and in $\cl\Mset$, and joint convexity gives $v(\alpha c_1+(1-\alpha)c_2)\le\kl{\PP^\alpha}{\QQ^\alpha}\le\alpha v(c_1)+(1-\alpha)v(c_2)$. So $v$ is convex; it is nonincreasing because $\Bc$ grows with $c$, and continuous on $(0,\infty)$ because it is finite and convex. At $0$: nearest pairs at $c_n\downarrow0$ have limit points $(\Pz,\QQ')$, by lower semicontinuity, so $v(0)\le\kl{\Pz}{\QQ'}\le\liminf_nv(c_n)\le v(0)$. The formula for $v(0)$ is Lemma~\ref{lem:ec-rob-logdual}(b), since $\Ball_0=\{\Pz\}$. Finally, $v(c)=0$ if and only if some $\PP\in\Bc$ equals some $\QQ\in\cl\Mset$ (Lemma~\ref{lem:ec-rob-kl}(a)), that is, if and only if $c\ge\cbar$ (Lemma~\ref{lem:ec-rob-qe}(d)).

\textit{Step 2 (the worst-case model).} For $c=0$, part (ii) is Lemma~\ref{lem:ec-rob-logdual}(b), with $\hstar(\Pz)\neq0$ by (C4). Fix $0<c<\cbar$, a maximizer $h^\dagger$ of $G_c$, and any minimizer $\PP^\dagger$ of $u$ on $\Bc$; with a minimizer $\QQ^\dagger$ of $\kl{\PP^\dagger}{\cdot}$, $(\PP^\dagger,\QQ^\dagger)$ is a nearest pair. By the saddle property, $\PP^\dagger$ minimizes $\EE_\PP[\xi]$ over $\Bc$, where $\xi=\ln W_{h^\dagger}$. Since $G_c(h^\dagger)=v(c)>0=G_c(0)$, $h^\dagger\neq0$ and $\xi$ is not constant. Since $\EE_\QQ[W_{h^\dagger}]=1$ on $\Mset$, $\min_\omega\xi(\omega)\le0<v(c)$, so Lemma~\ref{lem:ec-rob-gibbs}(c) excludes $c\ge c_m$. Hence $\PP^\dagger=\Pz W_{h^\dagger}^{-1/\theta_c}/\EE_{\Pz}[W_{h^\dagger}^{-1/\theta_c}]$ with $\kl{\PP^\dagger}{\Pz}=c$. So every minimizer of $u$ on $\Bc$ is this one strictly positive measure on the sphere $\{\kl{\cdot}{\Pz}=c\}$.

\textit{Step 3 (part (ii)).} Call this minimizer $\Pc$. By the saddle property every maximizer of $G_c$ maximizes $G(\cdot,\Pc)$, whose maximizer $\hstar(\Pc)$ is unique (Lemma~\ref{lem:ec-rob-logdual}(b)); so $h_c=\hstar(\Pc)\neq0$ is the unique robust portfolio. A nearest pair $(\PP,\QQ)$ has $u(\PP)\le\kl{\PP}{\QQ}=v(c)$, so $\PP=\Pc$, and $\QQ$ minimizes $\kl{\Pc}{\cdot}$, so $\QQ=\QL(\Pc)=\Pc/\Wc\in\Mset$.

\textit{Step 4 (part (iii)).} Let $\lambda_c=\theta_c$ from Step 2 with $h^\dagger=h_c$. Then $\dd\Pc/\dd\Pz\propto\Wc^{-1/\lambda_c}$ and $\dd\Qc/\dd\Pz=(\dd\Pc/\dd\Pz)/\Wc\propto\Wc^{-(1+1/\lambda_c)}$. The exponent is unique because $\Wc$ is not constant. With $\lambda=\lambda_c$, $\Pz^{\lambda/(1+\lambda)}\Qc^{1/(1+\lambda)}\propto\Pz\Wc^{-1/\lambda}\propto\Pc$, which is the exponential-line form, since $\lambda/(1+\lambda)=1/\gamma_c$; the position on the line is $t_c=1/(1+\lambda_c)=1-1/\gamma_c$. By Lemma~\ref{lem:ec-rob-pstep} with $\beta=\lambda_c$, $\Pc$ minimizes $L_{\lambda_c}(\cdot,\Qc)$; by Lemma~\ref{lem:ec-rob-logdual}(b), $\Qc$ minimizes $L_{\lambda_c}(\Pc,\cdot)$; and Lemma~\ref{lem:ec-rob-block} gives joint minimality.

\textit{Step 5 ($v'(c)=-\lambda_c$ for $0<c<\cbar$).} For $c'\ge0$ and $(\PP,\QQ)\in\Ball_{c'}\times\cl\Mset$, Step 4 gives $\kl{\PP}{\QQ}\ge L_{\lambda_c}(\Pc,\Qc)-\lambda_cc'=v(c)-\lambda_c(c'-c)$, so $-\lambda_c$ is a subgradient. Conversely, let $-s$ be a subgradient; $s\ge0$ since $v$ is nonincreasing. For every $(\PP,\QQ)$, $\kl{\PP}{\QQ}\ge v(\kl{\PP}{\Pz})\ge v(c)-s(\kl{\PP}{\Pz}-c)$, with equality at $(\Pc,\Qc)$; so $\Pc$ minimizes $L_s(\cdot,\Qc)$. If $s=0$, then $\Pc=\Qc$ and $v(c)=0$, a contradiction. If $s>0$, Lemma~\ref{lem:ec-rob-pstep} gives $\Pc\propto\Pz^{s/(1+s)}\Qc^{1/(1+s)}$, that is, $\Pc\propto\Pz\Wc^{-1/s}$, so $s=\lambda_c$. A convex function with a single subgradient at an interior point is differentiable there \citep{Rockafellar1970}.

\textit{Step 6 (part (iv), first claim).} For $c=0$, $h_0=\hstar(\Pz)=h^{(1)}$. For $0<c<\cbar$, the condition $\EE_{\Pc}[\DeltaS/\Wc]=0$ reads $\EE_{\Pz}[\Wc^{-1/\lambda_c-1}\DeltaS]=\EE_{\Pz}[\Wc^{-\gamma_c}\DeltaS]=0$, the CRRA($\gamma_c$) first-order condition; so $h_c=h^{(\gamma_c)}$ (Lemma~\ref{lem:ec-rob-crra}(a)), $\PP^{(\gamma_c)}\propto\Pz\Wc^{1-\gamma_c}\propto\Pc$ and $\QQ^{(\gamma_c)}\propto\Pz\Wc^{-\gamma_c}\propto\Qc$.

\textit{Step 7 (part (iv), converse).} The case $\gamma=1$ is $c=0$. Let $\gamma>1$ and $\lambda=1/(\gamma-1)$. Given $\Qg$, Lemma~\ref{lem:ec-rob-pstep} says that $L_\lambda(\cdot,\Qg)$ is minimized at $\PP\propto\Pz^{1/\gamma}(\Qg)^{1-1/\gamma}\propto\Pz(\Wg)^{1-\gamma}$, that is, at $\Pg$. Given $\Pg$, $L_\lambda(\Pg,\cdot)$ is minimized at $\QL(\Pg)=\Qg$ (Lemmas~\ref{lem:ec-rob-logdual}(b) and~\ref{lem:ec-rob-crra}(b)). By Lemma~\ref{lem:ec-rob-block}, $(\Pg,\Qg)$ minimizes $L_\lambda$, so for $(\PP,\QQ)\in\Ball_{c(\gamma)}\times\cl\Mset$, $\kl{\PP}{\QQ}\ge\kl{\Pg}{\Qg}+\lambda(c(\gamma)-\kl{\PP}{\Pz})\ge\kl{\Pg}{\Qg}$. Since $\Wg$ is not constant (Lemma~\ref{lem:ec-rob-crra}(d)), $\Pg\neq\Pz$ and $\Pg=\Wg\Qg\neq\Qg$, so $0<c(\gamma)<\cbar$ by Step 1. By part (ii) the nearest pair at $c(\gamma)$ is unique, hence equal to $(\Pg,\Qg)$, $h_{c(\gamma)}=h^{(\gamma)}$, and the uniqueness of the exponent in part (iii) gives $\lambda_{c(\gamma)}=\lambda$, that is, $\gamma_{c(\gamma)}=\gamma$.

\textit{Step 8 (part (iv), the bijection).} The map $\gamma\mapsto c(\gamma)$ is continuous (Lemma~\ref{lem:ec-rob-crra}(c)), injective (Step 7), and onto $[0,\cbar)$, since $c(1)=0$ and $c=c(\gamma_c)$ for $c\in(0,\cbar)$ by Step 6. A continuous bijection between intervals is strictly monotone, and $c(1)=0<c(\gamma)$ for $\gamma>1$, so it is strictly increasing, with a continuous, strictly increasing inverse $c\mapsto\gamma_c$. The limits of $\gamma_c$ and $\lambda_c=1/(\gamma_c-1)$ follow.

\textit{Step 9 (rest of part (i)).} On $(0,\cbar)$, $v'=-\lambda_c$ is negative, continuous and strictly increasing, so $v$ is continuously differentiable, strictly decreasing and strictly convex there, and by continuity on $[0,\cbar]$ (a convex function affine on some $[0,y]$ would be affine on $(0,y)$). At $\cbar$ the left derivative is $\lim_{c\uparrow\cbar}(-\lambda_c)=0$ and the right derivative is $0$, so $v'(\cbar)=0$; and $v'(0^+)=\lim_{c\downarrow0}(-\lambda_c)=-\infty$.

\textit{Step 10 (part (v): limits).} Let $c_n\uparrow\cbar$. Every limit point $(\PP',\QQ')$ of the nearest pairs has $\kl{\PP'}{\Pz}\le\cbar$ and $\kl{\PP'}{\QQ'}\le\liminf_nv(c_n)=0$, so $\PP'=\QQ'\in\cl\Mset\cap\Ball_{\cbar}=\{\QE\}$. Hence $\Pc,\Qc\to\QE$, $\Wc=\Pc/\Qc\to1$, and $h_c\to0$ by (C2); by part (iv), the same limits hold for $\Pg,\Qg$ as $\gamma\to\infty$.

\textit{Step 11 (part (v): the exponential position).} The vector $\eta_\gamma=\gamma h^{(\gamma)}$ minimizes the convex function $F_\gamma(\eta)=\EE_{\Pz}[(1+\eta\cdot\DeltaS/\gamma)^{1-\gamma}]$ over $\gamma\mathcal H$, because $1-\gamma<0$. As $\gamma\to\infty$, $F_\gamma\to F_\infty(\eta)=\EE_{\Pz}[e^{-\eta\cdot\DeltaS}]$ uniformly on compact sets, and $\gamma\mathcal H$ eventually contains every compact set. Since $F_\infty$ is strictly convex with unique minimizer $h_E$, it exceeds $F_\infty(h_E)$ by some $\kappa>0$ on the boundary of any closed ball $K$ around $h_E$; for large $\gamma$, $F_\gamma>F_\gamma(h_E)$ on that boundary, and convexity puts the minimizer $\eta_\gamma$ in $K$. So $\eta_\gamma\to h_E$, and $h_c/\lambda_c=(\gamma_ch_c)/(1+\lambda_c)\to h_E$.

\textit{Step 12 (part (v): the rate).} Put $\iota=1/\gamma$. The first-order condition in $\eta$ is $F(\eta,\iota)=\EE_{\Pz}[(1+\iota\,\eta\cdot\DeltaS)^{-1/\iota}\DeltaS]=0$, with $F(\eta,0)=\EE_{\Pz}[e^{-\eta\cdot\DeltaS}\DeltaS]$. Since $\iota^{-1}\ln(1+\iota x)=\sum_{i\ge1}(-1)^{i+1}\iota^{i-1}x^i/i$ for $|\iota x|<1$, $F$ is real-analytic near $(h_E,0)$, and $\partial_\eta F(h_E,0)=-\EE_{\Pz}[e^{-h_E\cdot\DeltaS}\DeltaS\DeltaS^\top]$ is negative definite. The implicit function theorem gives an analytic solution $\eta(\iota)$ with $\eta(0)=h_E$, which equals $\eta_{1/\iota}$ for small $\iota>0$ by local uniqueness. So $\gamma h^{(\gamma)}=h_E+O(1/\gamma)$. The densities $\dd\Pg/\dd\Pz\propto(1+\iota\,\eta(\iota)\cdot\DeltaS)^{1-1/\iota}$ and $\dd\Qg/\dd\Pz\propto(1+\iota\,\eta(\iota)\cdot\DeltaS)^{-1/\iota}$ are analytic in $\iota$ and proportional to $e^{-h_E\cdot\DeltaS}$ at $\iota=0$, so $\Pg$ and $\Qg$ equal $\QE+O(1/\gamma)$.

\textit{Step 13 (part (vi)).} For $c\ge\cbar$, $\QE\in\Bc$, so $v(c)=0$, and the nearest pairs are the pairs $(\QQ,\QQ)$ with $\QQ\in\cl\Mset\cap\Bc$, a nonempty compact convex set equal to $\{\QE\}$ at $c=\cbar$ (Lemma~\ref{lem:ec-rob-qe}(d)). For $h\in\mathcal H$, $G_c(h)\le\EE_{\QE}[\ln W_h]\le\ln\EE_{\QE}[W_h]=0$, and $G_c(h)=0$ forces equality in Jensen's inequality under $\QE>0$, so $W_h\equiv1$ and $h=0$ by (C2). \Halmos
\end{proof}

\textit{The limit in other settings.} For $q>1$ the $q$-optimal martingale measure minimizes $\EE_{\Pz}[(\dd\QQ/\dd\Pz)^q]$ over the pricing measures, and for $0<q<1$ it maximizes this expectation. By the duality of \citet{GollRuschendorf2001} (Section~\ref{sec:ec-rob-literature}), for $\gamma>1$ the measure $\Qg$ is therefore the $q$-optimal measure with $q=1-1/\gamma$. As $q\to1$, $q$-optimal martingale measures converge to the minimal-entropy measure: from above in discrete time \citep{Grandits1999} and for continuous semimartingales \citep{GranditsRheinlander2002}, and from below, which is the CRRA case $q=1-1/\gamma\uparrow1$, for continuous semimartingales (\citealt{Santacroce2005}; \citealt{ManiaTevzadze2003}, who also give an entropy rate of order $|q-1|$). Theorem~\ref{thm:robust}(v) is the finite one-period version, with rate $O(1/\gamma)$ (Step 12), and \citet{Nutz2012} proves the primal counterpart of $\gamma h^{(\gamma)}\to h_E$ for a continuous price process.

\subsection{Relation of Theorem~\ref{thm:robust} to the Literature}\label{sec:ec-rob-literature}

Parts of Theorem~\ref{thm:robust} are known in other forms. The power tilt is the worst case of KL-constrained robust optimization \citep{HuHong2012,BenTalDenHertogDeWaegenaereEtAl2013} and of KL-ball model risk \citep{GlassermanXu2014,BreuerCsiszar2016}. The CRRA form of the multiplier investor in Remark~\ref{rem:scope} is known \citep{Tallarini2000,Maenhout2004,BarillasHansenSargent2009}; in a Savage setting, multiplier preferences are expected-utility preferences with a more concave utility \citep{Strzalecki2011}. The ball (constraint) form and the multiplier form are both variational preferences \citep{MaccheroniMarinacciRustichini2006}, and they are dual (Proposition~\ref{prop:ec-rob-mult}). Part (iv) is the counterpart for log utility in a general finite market of the mean--variance result of \citet{Taboga2006}.

The robust path follows a known family of measures. By the duality of \citet{GollRuschendorf2001}, for $\gamma>1$ the measure $\Qg$ maximizes $\EE_{\Pz}[(\dd\QQ/\dd\Pz)^{1-1/\gamma}]$ over the pricing measures, and thus minimizes a power (Cressie--Read) divergence from $\Pz$. Here log utility ($\gamma=1$) gives $\QL$, while the limit $\gamma \to \infty$ gives $\QE$. Theorem~\ref{thm:robust}(v) gives this limit in a finite one-period market, with rate $O(1/\gamma)$; the end of Section~\ref{sec:ec-rob-structure} compares it with continuous-time results. When $\Pz$ is an empirical distribution, $\QL$ yields the empirical-likelihood weights and $\QE$ is the canonical valuation measure of \citet{Stutzer1996}, while \citet{HaleyWalker2010} compare the tilts of this family for option pricing. What the ball adds is the map from the doubt radius $c$ to the tilt with exponent $1-1/\gamma_c$.

To our knowledge, there are three consequences of the ball form that are new; the first two require log utility. The first is the exact map $c \mapsto \gamma_c$ in a given market, with $\lambda_c = -v'(c)$. The second is the identification of $\Qc$ as the CRRA($\gamma_c$) marginal-utility measure on a path from $\QL(\Pz)$ to $\QE$. The third is the identification of the threshold with the minimal relative entropy of the market, $\cbar = \kl{\QE}{\Pz}$; that some threshold exists follows from the classical non-participation argument \citep{DowWerlang1992,CaoWangZhang2005}, which Proposition~\ref{prop:utilityfree}(a)--(c) applies to the ball. The ball form creates a no-trade region because the guaranteed growth $G_c$ has a kink at $h=0$. In contrast, no CRRA investor has a no-trade region (Lemma~\ref{lem:ec-rob-crra}(d)), nor does the multiplier form, nor do smooth ambiguity preferences \citep{KlibanoffMarinacciMukerji2005}.

\subsection{Proof of Theorem~\ref{thm:pricing}, No-Trade Prices, and Implied Doubt}\label{sec:ec-rob-pricing}

\begin{repeattheorem}[Theorem~\ref{thm:pricing} (Robust prices of a new claim).]
Assume (C1)--(C4). Let $X$ be a claim and $p\in\RR$ a price.
\begin{enumerate}
\item[(a)] (Below the threshold.) If $c<\cbar$, then $p$ is a no-trade price of $X$ iff $p=p_c(X)=\EE_{\Qc}[X]$.
\item[(b)] (At and above the threshold.) If $c\ge\cbar$, then $p$ is a no-trade price of $X$ iff $p\in I_c(X)$.
\item[(c)] (The interval.) For $c\ge\cbar$, $I_c(X)$ is a nonempty compact interval, and $I_c(X)=\{p:\cbar_X(p)\le c\}=J_{c-\cbar}(X)$, the entropy-budget interval \eqref{eq:entropy-budget} with budget $c-\cbar$. It equals $\{\EE_{\QE}[X]\}$ at $c=\cbar$, it grows with $c$, and $\bigcap_{c>\cbar}I_c(X)=\{\EE_{\QE}[X]\}$; also $p_c(X)\to\EE_{\QE}[X]$ as $c\uparrow\cbar$.
\item[(d)] (Direction of trade.) Let $c<\cbar$, let $p$ lie in the no-arbitrage interval of $X$, and let $m(k)=\sup_{h:(h,k)\in\mathcal H^+}\min_{\PP\in\Bc}\EE_\PP[\ln(W_h+kY)]$ ($\sup\emptyset=-\infty$), the guaranteed growth with $k$ units of $X$ when the traded position is re-optimized; $m(0)=v(c)$. If $p<p_c(X)$, then $m(k)>v(c)$ for all small $k>0$ and $m(k)<v(c)$ for all $k<0$; if $p>p_c(X)$, the roles of $k>0$ and $k<0$ are exchanged.
\end{enumerate}
\end{repeattheorem}

\begin{proof}{Proof.}
Write $Y=X-p$ and $G^+_c(h,k)=\min_{\PP\in\Bc}\EE_\PP[\ln(W_h+kY)]$ on $\mathcal H^+$, so that $v^+(c;p)=\sup G^+_c$ and $G^+_c(h,0)=G_c(h)$.

(a) If $p=p_c(X)$, Lemma~\ref{lem:ec-rob-kl}(d) with $(\Pc,\Qc)$ gives, for $(h,k)\in\mathcal H^+$,
\[
G^+_c(h,k)\le\EE_{\Pc}[\ln(W_h+kY)]\le\kl{\Pc}{\Qc}+\ln\EE_{\Qc}[W_h+kY]=v(c)+\ln\bigl(1+k(p_c(X)-p)\bigr)=v(c),
\]
so $v^+(c;p)=v(c)$. If $p\neq p_c(X)$, let $e=p_c(X)-p$, $f(k)=G^+_c(h_c,k)$ for small $|k|$, and let $\PP_k$ minimize $\EE_\PP[\ln(\Wc+kY)]$ over $\Bc$. The objectives converge uniformly as $k\to0$, and $\Pc$ is the unique minimizer at $k=0$ (Step 2 of the proof of Theorem~\ref{thm:robust}), so $\PP_k\to\Pc$. A Taylor bound, uniform over the finite $\Omega$, gives $\ln(\Wc+kY)\ge\ln\Wc+kY/\Wc-Ck^2$, and $\EE_\PP[\ln\Wc]\ge G_c(h_c)=v(c)$ on $\Bc$, so
\[
f(k)\ge v(c)+k\,\EE_{\PP_k}[Y/\Wc]-Ck^2,\qquad \EE_{\PP_k}[Y/\Wc]\to\EE_{\Pc}[Y/\Wc]=\EE_{\Qc}[Y]=e.
\]
So $f(k)>v(c)$ for every small $k$ with the sign of $e$, and $v^+(c;p)>v(c)$.

(b) Let $c\ge\cbar$, so $v(c)=0$. If $p\in I_c(X)$, some $\QQ'\in\cl\Mset\cap\Bc$ has $\EE_{\QQ'}[X]=p$; since $\QQ'$ is plausible, $G^+_c(h,k)\le\EE_{\QQ'}[\ln(W_h+kY)]\le\ln\EE_{\QQ'}[W_h+kY]=0$. If $p\notin I_c(X)$, Sion's theorem applies to the extended market as in the proof of Theorem~\ref{thm:minimax}; it needs neither boundedness of $\mathcal H^+$ nor absence of arbitrage. So $v^+(c;p)=\min_{\PP\in\Bc}u^+(\PP)$ with the lower semicontinuous $u^+(\PP)=\sup_{\mathcal H^+}\EE_\PP[\ln(W_h+kY)]\in[0,\infty]$. Now $u^+>0$ on $\Bc$: if $\PP\notin\cl\Mset$, then $u^+(\PP)\ge u(\PP)>0$ (Lemma~\ref{lem:ec-rob-logdual}(c)); if $\PP\in\cl\Mset\cap\Bc$, then $\EE_\PP[Y]\neq0$, and $h=0$ with a small $k$ of the sign of $\EE_\PP[Y]$ gives $\EE_\PP[\ln(1+kY)]=k\,\EE_\PP[Y]+O(k^2)>0$. Hence $v^+(c;p)>0$.

(c) $I_c(X)$ is the linear image of the nonempty compact convex set $\cl\Mset\cap\Bc$, hence a nonempty compact interval, and $p\in I_c(X)$ if and only if $\cbar_X(p)\le c$, the minimum in $\cbar_X$ being attained. Lemma~\ref{lem:ec-rob-qe}(b) gives $\kl{\QQ}{\Pz}=\kl{\QQ}{\QE}+\cbar$ on $\cl\Mset$, so $\QQ\in\cl\Mset$ lies in $\Bc$ if and only if $\kl{\QQ}{\QE}\le c-\cbar$; this is $I_c(X)=J_{c-\cbar}(X)$, and it also gives $\cbar_X\ge\cbar$, with equality only at $\EE_{\QE}[X]$. As a partial minimum of a jointly convex function, $\cbar_X$ is convex, so the sets $I_c(X)$ are nested intervals, equal to $\{\EE_{\QE}[X]\}$ at $c=\cbar$. If $p_n\in I_{c_n}(X)$ with $c_n\downarrow\cbar$, measures $\QQ_n$ attaining $\cbar_X(p_n)$ have all limit points in $\cl\Mset\cap\Ball_{\cbar}=\{\QE\}$, so $p_n\to\EE_{\QE}[X]$. Also $p_c(X)\to\EE_{\QE}[X]$ as $c\uparrow\cbar$ by Theorem~\ref{thm:robust}(v).

(d) The function $m$ is concave, because $G^+_c$ is jointly concave on the convex set $\mathcal H^+$, and it is finite wherever it is not $-\infty$: some $\QQ'\in\Mset$ has $\EE_{\QQ'}[Y]=0$, because $p$ lies in the no-arbitrage interval, so $G^+_c(h,k)\le\EE_{\Pc}[\ln(W_h+kY)]\le\kl{\Pc}{\QQ'}$ by Lemma~\ref{lem:ec-rob-kl}(d). Also $m(0)=v(c)$, and the proof of (a) gives $m(k)\ge f(k)>m(0)$ for every small $k$ with the sign of $e$. For $k'$ of the opposite sign with $m(k')>-\infty$, take such a $k$ and $\alpha=k'/(k'-k)\in(0,1)$; concavity gives $m(0)\ge\alpha m(k)+(1-\alpha)m(k')$, so $m(k')<m(0)$. Since $e>0$ when $p<p_c(X)$, this is the claim. \Halmos
\end{proof}

\begin{lemma}[Forms of the no-trade condition]\label{lem:ec-rob-notrade}
Assume (C1)--(C4), let $X\in\RR^M$ and $p\in\RR$, and recall that $h_c=0$ for $c\ge\cbar$.
(a) $v^+(c;p)=v(c)$ iff $(h_c,0)$ attains $v^+(c;p)$.
(b) If $X$ is not spanned and $p$ lies in the no-arbitrage interval of $X$, the extended market satisfies (C1)--(C4), its robust problem has a unique optimal strategy, and $v^+(c;p)=v(c)$ iff this strategy has $k=0$.
(c) Under the conditions of (b), $\cbar_X(p)$ is the threshold of the extended market and equals $\max_{(\eta,k)\in\RR^{d+1}}\{-\ln\EE_{\Pz}[e^{-\eta\cdot\DeltaS-k(X-p)}]\}$, the best certainty-equivalent gain of an investor with exponential utility who can also trade $X$ at $p$. So $\cbar_X(p)-\cbar$ is the gain from access to $X$ at $p$.
\end{lemma}

\begin{proof}{Proof.}
(a) If $v^+(c;p)=v(c)$, then $G^+_c(h_c,0)=G_c(h_c)=v^+(c;p)$. Conversely, if some $(h,0)$ is optimal, then $v^+(c;p)=G_c(h)\le v(c)\le v^+(c;p)$. (b) The extended market has the gains $(\DeltaS,Y)$. Condition (C2) holds because $\mathrm{span}\{\mathbf 1,\DeltaS^1,\dots,\DeltaS^d,Y\}=\mathrm{span}\{\mathbf 1,\DeltaS^1,\dots,\DeltaS^d,X\}$ has dimension $d+2$, (C3) because some $\QQ\in\Mset$ has $\EE_\QQ[X]=p$, and (C1) and (C4) are inherited. Its closed set of martingale measures is $\{\QQ\in\cl\Mset:\EE_\QQ[X]=p\}$, so its threshold is $\cbar_X(p)\ge\cbar$. Theorem~\ref{thm:minimax} gives an optimal strategy, unique by Theorem~\ref{thm:robust}(ii) if $c<\cbar_X(p)$ and by Theorem~\ref{thm:robust}(vi) if $c\ge\cbar_X(p)$, both applied to the extended market. If it has $k=0$, then $v^+(c;p)=v(c)$ by (a); conversely, if $v^+(c;p)=v(c)$, then $(h_c,0)$ is optimal by (a), hence it is the optimal strategy. (c) The formula is Lemma~\ref{lem:ec-rob-qe}(b) for the extended market, and the reading is Lemma~\ref{lem:duality}. \Halmos
\end{proof}

The first-order width of $I_c(X)$ needs no new proof. By Theorem~\ref{thm:pricing}(c), $I_c(X)$ is the entropy-budget interval $J_{c-\cbar}(X)$ of \eqref{eq:entropy-budget} for the model $\Pz$. So Proposition~\ref{prop:ec-width} gives its end points, $\EE_{\QE}[X]\pm\mathrm{sd}_{\QE}(\tilde X)\sqrt{2(c-\cbar)}+O(c-\cbar)$ as $c\downarrow\cbar$, where $\tilde X$ is the unhedgeable part of $X$ defined there.

\begin{corollary}[Implied doubt]\label{cor:ec-rob-implied}
Assume (C1)--(C4), let $X\in\RR^M$ and $p\in\RR$, and let $\mathcal N_X(p)=\{c\in[0,\cbar):p_c(X)=p\}$. If $\mathcal N_X(p)\neq\emptyset$, it has a smallest element and $c^{\mathrm{imp}}_X(p)=\min\mathcal N_X(p)<\cbar$; otherwise $c^{\mathrm{imp}}_X(p)=\cbar_X(p)\ge\cbar$. In both cases $c^{\mathrm{imp}}_X(p)<\infty$ iff $p\in[\min_{\QQ\in\cl\Mset}\EE_\QQ[X],\,\max_{\QQ\in\cl\Mset}\EE_\QQ[X]]$.
\end{corollary}

\begin{proof}{Proof.}
By Theorem~\ref{thm:pricing}(a)--(b), the no-trade radii form $\mathcal N_X(p)\cup[\cbar_X(p),\infty)$, since $\cbar_X(p)\ge\cbar$. The map $c\mapsto p_c(X)=\EE_{\QQ^{(\gamma_c)}}[X]$ is continuous on $[0,\cbar)$ (Theorem~\ref{thm:robust}(iv), Lemma~\ref{lem:ec-rob-crra}(c)), so $\mathcal N_X(p)$ is closed relative to $[0,\cbar)$ and contains its infimum when it is nonempty. If $p$ is in the closed interval, some $\QQ\in\cl\Mset$ prices $X$ at $p$, so $\cbar_X(p)\le\kl{\QQ}{\Pz}<\infty$. Otherwise $\cbar_X(p)=\infty$, and $\mathcal N_X(p)=\emptyset$ because every $p_c(X)$ lies in the no-arbitrage interval. \Halmos
\end{proof}

\begin{remark}[Spanned claims, and why the closure is needed]\label{rem:ec-rob-closure}
If $X=a+b\cdot\DeltaS$, then $p_c(X)=a$ and $I_c(X)=\{a\}$, the unique arbitrage-free price. The interval needs the closed set $\cl\Mset\cap\Bc$. For $M=3$, $\DeltaS=(-0.1,0,0.1)$, $\Pz=(0.2,0.3,0.5)$, $c=\ln5$ (so $\Bc=\cl\mathcal P$) and $X=\mathbf 1_{\{\omega_2\}}$, we have $\cl\Mset=\{(q,1-2q,q):q\in[0,\frac12]\}$: the prices under $\Mset$ fill the open interval $(0,1)$, while $I_c(X)=[0,1]$. At $p=1$ the investor does not trade, because the plausible model $(0,1,0)\in\cl\Mset\cap\Bc$ ignores the two states in which a sale of $X$ gains.
\end{remark}

\subsection{Proof of Proposition~\ref{prop:utilityfree}}\label{sec:ec-rob-utilityfree}

\begin{repeatproposition}[Proposition~\ref{prop:utilityfree} (Threshold and no-trade interval beyond log utility).]
Assume (C1)--(C4). Let $U:(0,\infty)\to\RR$ be strictly increasing, strictly concave, and differentiable at $1$. For $c\ge0$ let $v_U(c)=\sup_{h\in\mathcal H}\min_{\PP\in\Bc}\EE_\PP[U(W_h)]$, and for a claim $X$ and a price $p$ let $v_U^+(c;p)=\sup_{(h,k)\in\mathcal H^+}\min_{\PP\in\Bc}\EE_\PP[U(W_h+k(X-p))]$.
\begin{enumerate}
\item[(a)] If $c\ge\cbar$, then $v_U(c)=U(1)$, and $h=0$ is the unique robust portfolio.
\item[(b)] If $c<\cbar$, then $v_U(c)>U(1)$, so $h=0$ is not a robust portfolio.
\item[(c)] If $c\ge\cbar$, then $v_U^+(c;p)=v_U(c)$ if and only if $p\in I_c(X)$.
\item[(d)] Let $c<\cbar$, and let $U$ be continuously differentiable with $U(x)\to-\infty$ as $x\downarrow0$. Then the robust portfolio $h_c^U$ is unique and nonzero, the worst-case model $\PP_c^U$ against it is unique and charges every state, and $v_U^+(c;p)=v_U(c)$ if and only if $p=\EE_{\PP_c^U}[U'(W_c^U)X]/\EE_{\PP_c^U}[U'(W_c^U)]$, where $W_c^U=W_{h_c^U}$.
\end{enumerate}
\end{repeatproposition}

\begin{proof}{Proof.}
For $h\in\mathcal H$ write $G^U(h)=\min_{\PP\in\Bc}\EE_\PP[U(W_h)]$; the minimum is attained because $\PP\mapsto\EE_\PP[U(W_h)]$ is linear and $\Bc$ is compact. First, $U'(1)>0$: $U'(1)\ge0$ because $U$ is increasing, and if $U'(1)=0$, concavity would give $U(x)\le U(1)+U'(1)(x-1)=U(1)$ for all $x>0$, against strict increase.

\textit{(a)} Let $c\ge\cbar$. By Lemma~\ref{lem:ec-rob-qe}(a),(d), $\QE\in\cl\Mset\cap\Bc$ and $\QE>0$. For $h\in\mathcal H$, Jensen's inequality gives $G^U(h)\le\EE_{\QE}[U(W_h)]\le U(\EE_{\QE}[W_h])=U(1+h\cdot\EE_{\QE}[\DeltaS])=U(1)$. Since $U$ is strictly concave and $\QE$ charges every state, the second inequality is an equality only if $W_h$ is constant on $\Omega$; then $h\cdot\DeltaS$ is a constant with $\QE$-mean zero, so $h\cdot\DeltaS=0$ and $h=0$ by (C2). As $G^U(0)=U(1)$, $v_U(c)=U(1)$, and $h=0$ is the only portfolio that attains it.

\textit{(b)} Let $c<\cbar$. The set $K=\{\EE_\PP[\DeltaS]:\PP\in\Bc\}\subset\RR^d$ is compact and convex, as the image of $\Bc$ (Lemma~\ref{lem:ec-rob-sets}(d)) under a linear map, and $0\notin K$, since otherwise some $\PP\in\Bc$ would lie in $\cl\Mset$, against Lemma~\ref{lem:ec-rob-qe}(d). By strict separation there are a unit vector $e\in\RR^d$ and $m>0$ with $e\cdot\EE_\PP[\DeltaS]\ge m$ for every $\PP\in\Bc$. Let $s=\max_\omega|e\cdot\DeltaS(\omega)|$, which is positive, and write $U(1+x)=U(1)+U'(1)x+r(x)$, where $r(x)/x\to0$ as $x\to0$ because $U$ is differentiable at $1$. For $0<t<1/s$ the portfolio $te$ lies in $\mathcal H$, and for every $\PP\in\Bc$,
\[
\EE_\PP\bigl[U(1+t\,e\cdot\DeltaS)\bigr]\ge U(1)+tU'(1)m-\sup_{|x|\le ts}|r(x)|.
\]
The last term is $o(t)$ and $U'(1)m>0$, so for small $t>0$ the right side exceeds $U(1)$ uniformly in $\PP$. Hence $v_U(c)\ge G^U(te)>U(1)=G^U(0)$.

\textit{(c)} Let $c\ge\cbar$, so $v_U(c)=U(1)$ by (a), and write $Y=X-p$. If $p\in I_c(X)$, some $\QQ'\in\cl\Mset\cap\Bc$ has $\EE_{\QQ'}[X]=p$, and for $(h,k)\in\mathcal H^+$ Jensen's inequality gives $\min_{\PP\in\Bc}\EE_\PP[U(W_h+kY)]\le\EE_{\QQ'}[U(W_h+kY)]\le U(1+h\cdot\EE_{\QQ'}[\DeltaS]+k\,\EE_{\QQ'}[Y])=U(1)$. So $v_U^+(c;p)\le v_U(c)$, and the reverse inequality holds because $(h,0)\in\mathcal H^+$ for $h\in\mathcal H$. If $p\notin I_c(X)$, the compact convex set $\{(\EE_\PP[\DeltaS],\EE_\PP[Y]):\PP\in\Bc\}\subset\RR^{d+1}$ does not contain $0$, because a model $\PP\in\Bc$ with $\EE_\PP[\DeltaS]=0$ and $\EE_\PP[X]=p$ would put $p$ in $I_c(X)$. The argument of (b), with the gains $(\DeltaS,Y)$ in place of $\DeltaS$ and the open set $\mathcal H^+$, which contains $0$, in place of $\mathcal H$, gives a strategy with guaranteed utility above $U(1)$; so $v_U^+(c;p)>v_U(c)$.

\textit{(d)} Let $c<\cbar$, and let $U$ be continuously differentiable with $U(0^+)=-\infty$. Then $U'>0$ on $(0,\infty)$, because $U'$ is nonincreasing and nonnegative, and it cannot vanish on a half-line $[x,\infty)$ without contradicting strict increase.

\emph{Existence.} $G^U$ is concave, as a minimum of concave functions, and finite on the open set $\mathcal H$, so it is continuous there. It tends to $-\infty$ at the boundary of the bounded set $\mathcal H$ (Lemma~\ref{lem:ec-rob-sets}(b)): if $h_n\to h^\circ$ with $W_{h^\circ}(\omega_0)=0$, then $G^U(h_n)\le\EE_{\Pz}[U(W_{h_n})]\to-\infty$, because $\Pz\in\Bc$, $\Pz(\omega_0)>0$, $U(W_{h_n}(\omega_0))\to-\infty$ and the other terms are bounded above. So $G^U$ attains its supremum $v_U(c)$ on $\mathcal H$, and every maximizer is nonzero by (b).

\emph{Uniqueness.} Let $h\neq0$ be a maximizer and $\xi=U(W_h)$. As in (a), $W_h$ is not constant, and $U$ is strictly increasing, so $\xi$ is not constant. Since $\EE_\QQ[W_h]=1$ for $\QQ\in\Mset$, $\min_\omega W_h(\omega)\le1$, so $\min_\omega\xi(\omega)\le U(1)<v_U(c)=G^U(h)$. By Lemma~\ref{lem:ec-rob-gibbs}(c), the case $c\ge c_m$ would give $G^U(h)=\min_\omega\xi(\omega)$; so $c<c_m$, and the minimizer of $\EE_\PP[\xi]$ over $\Bc$ is unique and strictly positive: $\Pz$ if $c=0$, and the Gibbs tilt $\PP^{\theta_c}$ of that lemma if $c>0$. Now suppose that $h_1\neq h_2$ are maximizers. Their midpoint $h_m$ is a maximizer by concavity and is nonzero by (b), so its worst case $\PP^*$ is strictly positive. Since $W_{h_1}\neq W_{h_2}$ (Lemma~\ref{lem:ec-rob-sets}(c)) and $U$ is strictly concave, $U(W_{h_m})\ge\frac12U(W_{h_1})+\frac12U(W_{h_2})$ in every state, with strict inequality in some state, which $\PP^*$ charges. Hence
\[
v_U(c)=G^U(h_m)=\EE_{\PP^*}[U(W_{h_m})]>\tfrac12\EE_{\PP^*}[U(W_{h_1})]+\tfrac12\EE_{\PP^*}[U(W_{h_2})]\ge\tfrac12G^U(h_1)+\tfrac12G^U(h_2)=v_U(c),
\]
a contradiction. So the robust portfolio $h_c^U$ is unique and nonzero, and the worst case $\PP_c^U$ against it is unique and charges every state.

\emph{Saddle point.} The function $F(h,\PP)=\EE_\PP[U(W_h)]$ is concave and continuous in $h$ on the convex set $\mathcal H$, and linear and continuous in $\PP$ on the compact convex set $\Bc$. So the minimax theorem of \citet{Sion1958} gives $\sup_h\min_\PP F=\min_\PP\sup_hF$, where the minimum on the right is attained because $\PP\mapsto\sup_hF(h,\PP)$ is lower semicontinuous. For a minimizer $\hat\PP$, $v_U(c)=\sup_hF(h,\hat\PP)\ge F(h_c^U,\hat\PP)\ge G^U(h_c^U)=v_U(c)$. So $\hat\PP$ minimizes $F(h_c^U,\cdot)$ over $\Bc$, that is, $\hat\PP=\PP_c^U$, and $h_c^U$ maximizes $F(\cdot,\PP_c^U)$ over the open set $\mathcal H$. The first-order condition is $\EE_{\PP_c^U}[U'(W_c^U)\DeltaS]=0$.

\emph{No-trade price.} Let $p^*=\EE_{\PP_c^U}[U'(W_c^U)X]/\EE_{\PP_c^U}[U'(W_c^U)]$ and $Y=X-p$. For $(h,k)\in\mathcal H^+$, concavity gives $U(W_h+kY)\le U(W_c^U)+U'(W_c^U)\bigl((h-h_c^U)\cdot\DeltaS+kY\bigr)$ in every state, so, with the first-order condition,
\[
\min_{\PP\in\Bc}\EE_\PP[U(W_h+kY)]\le\EE_{\PP_c^U}[U(W_h+kY)]\le v_U(c)+k\,\EE_{\PP_c^U}[U'(W_c^U)]\,(p^*-p).
\]
At $p=p^*$ this gives $v_U^+(c;p^*)\le v_U(c)$, hence equality. Let $p\neq p^*$, and for small $|k|$ let $\PP_k$ minimize $\EE_\PP[U(W_c^U+kY)]$ over $\Bc$. The objectives converge uniformly to the one at $k=0$, whose minimizer $\PP_c^U$ is unique, so $\PP_k\to\PP_c^U$. A Taylor bound, uniform over the finite $\Omega$ because $U$ is continuously differentiable near the values of $W_c^U$, gives $U(W_c^U+kY)\ge U(W_c^U)+kU'(W_c^U)Y-o(k)$, and $\EE_{\PP_k}[U(W_c^U)]\ge G^U(h_c^U)=v_U(c)$. So
\[
\min_{\PP\in\Bc}\EE_\PP[U(W_c^U+kY)]\ge v_U(c)+k\,\EE_{\PP_k}[U'(W_c^U)Y]-o(k),
\]
and $\EE_{\PP_k}[U'(W_c^U)Y]\to\EE_{\PP_c^U}[U'(W_c^U)]\,(p^*-p)\neq0$.
A small $k$ with the sign of $p^*-p$ gives a guaranteed utility above $v_U(c)$, so $v_U^+(c;p)>v_U(c)$. \Halmos
\end{proof}

The proof of (a)--(c) uses only that $\Bc$ is compact and convex and meets $\cl\Mset$ exactly when $c\ge\cbar$; the uniqueness of $h=0$ in (a) also uses that $\cl\Mset\cap\Bc$ contains the strictly positive measure $\QE$.

\subsection{Constraint and Multiplier Forms}\label{sec:ec-rob-mult}

For $\gamma>1$ and $W>0$, let $\mathrm{CE}_\gamma(W)=(\EE_{\Pz}[W^{1-\gamma}])^{1/(1-\gamma)}$ be the certainty equivalent of the CRRA($\gamma$) investor, and $g^*(\gamma)=\max_{h\in\mathcal H}\ln\mathrm{CE}_\gamma(W_h)$ their best certainty-equivalent growth rate.

\begin{proposition}[Multiplier preferences and conjugacy]\label{prop:ec-rob-mult}
Assume (C1)--(C3).
(a) For $W>0$ and $\theta>0$, $\min_{\PP\in\cl\mathcal P}\{\EE_\PP[\ln W]+\theta\,\kl{\PP}{\Pz}\}=-\theta\ln\EE_{\Pz}[W^{-1/\theta}]=\ln\mathrm{CE}_{1+1/\theta}(W)$, attained only at $\PP\propto\Pz W^{-1/\theta}$.
(b) For $\theta>0$ and $\gamma=1+1/\theta$, the multiplier value $V(\theta)=g^*(\gamma)$ equals $\min_{\PP\in\cl\mathcal P,\,\QQ\in\cl\Mset}\{\kl{\PP}{\QQ}+\theta\,\kl{\PP}{\Pz}\}$, attained at $h^{(\gamma)}$ and at $(\Pg,\Qg)$.
(c) With $V(0)=0$, $V(\theta)=\min_{c\ge0}\{v(c)+\theta c\}$ for $\theta\ge0$, and $v(c)=\sup_{\theta\ge0}\{V(\theta)-\theta c\}$, a maximum when $c>0$. Under (C4): for $c\in(0,\cbar)$ the maximizing $\theta$ is unique and equals $\lambda_c$; for $c\ge\cbar$ it is $0$; and for $\theta>0$ the minimizing $c$ is unique and equals $c(1+1/\theta)$. Equivalently, with $g^*(\infty)=0$, $v(c)=\max_{1<\gamma\le\infty}\{g^*(\gamma)-c/(\gamma-1)\}$ for $c>0$, with the maximum at $\gamma_c$ when $c<\cbar$.
\end{proposition}

\begin{proof}{Proof.}
(a) is Lemma~\ref{lem:ec-rob-gibbs}(a) with $\xi=\ln W$, since $\ln\mathrm{CE}_\gamma(W)=\frac{1}{1-\gamma}\ln\EE_{\Pz}[W^{1-\gamma}]$ and $1-\gamma=-1/\theta$. (b) By (a), $V(\theta)=\sup_h\min_\PP\{G(h,\PP)+\theta\,\kl{\PP}{\Pz}\}$. Sion's theorem on the compact set $\cl\mathcal P$ (the function is convex and continuous in $\PP$ and concave in $h$) and Lemma~\ref{lem:ec-rob-logdual}(c) turn this into the stated minimum. Step 7 of the proof of Theorem~\ref{thm:robust}, which uses only (C1)--(C3), shows that $(\Pg,\Qg)$ attains it, and $h^{(\gamma)}$ attains the supremum because maximizing $\ln\mathrm{CE}_\gamma(W_h)$ is maximizing $\EE_{\Pz}[U_\gamma(W_h)]$. (c) For each $c$, pairs with $\kl{\PP}{\Pz}\le c$ give at least $v(c)$, and the nearest pair at $c$ gives at most $v(c)+\theta c$; so $V(\theta)=\inf_c\{v(c)+\theta c\}$, attained since $v$ is continuous and $v(c)+\theta c\to\infty$ for $\theta>0$. As $v$ is convex, continuous and finite on $[0,\infty)$, the Fenchel--Moreau theorem \citep{Rockafellar1970} gives $v(c)=\sup_\theta\{V(\theta)-\theta c\}$, where $V(\theta)=-\infty$ for $\theta<0$ because $v$ is bounded. A value $\theta$ attains the supremum if and only if $-\theta$ is a subgradient of $v$ at $c$; by Theorem~\ref{thm:robust}(i) and (iii), the subdifferential is $\{-\lambda_c\}$ on $(0,\cbar)$ and $\{0\}$ on $[\cbar,\infty)$, and it is empty at $0$, where the supremum is approached as $\theta\to\infty$. For $\theta>0$, $c\mapsto v(c)+\theta c$ has derivative $\theta-\lambda_c$ on $(0,\cbar)$ and $\theta$ beyond, so its unique minimizer solves $\lambda_c=\theta$, that is, $c=c(1+1/\theta)$ by Theorem~\ref{thm:robust}(iv). The form in $\gamma$ substitutes $\theta=1/(\gamma-1)$. \Halmos
\end{proof}

In words, the ball investor with radius $c\in(0,\cbar)$ and the penalty investor with $\theta=\lambda_c$ hold the same portfolio $h^{(\gamma_c)}$, while the penalty form has no no-trade region: every penalty investor with $\theta>0$ holds $h^{(1+1/\theta)}\neq0$. Part (a) is the Gibbs formula, whose dynamic versions are \citet{Skiadas2003} and \citet{Maenhout2004}, and Lagrange duality links the two forms \citep{HansenSargentTurmuhambetovaEtAl2006}, also for control problems \citep{PetersenJamesDupuis2000}. In continuous time these criteria take the form of stochastic differential utility, which \citet{LazrakQuenez2003} generalize so that risk aversion can depend on the direction of risk.

\subsection{Small Premia}\label{sec:ec-rob-asym}

\begin{remark}[Small premia: an approximation, not a theorem]\label{rem:ec-rob-smallpremium}
When the risk premium is small, $\gamma_c\approx1/(1-\sqrt{c/\cbar})$ and $v(c)\approx(\sqrt{\cbar}-\sqrt c)^2$ in every market. We give a heuristic reason, not a proof. To second order, $\kl{\PP}{\QQ}\approx\frac12\|\dd\PP/\dd\Pz-\dd\QQ/\dd\Pz\|^2_{L^2(\Pz)}$. Then $\Bc$ is a round ball of radius $\sqrt{2c}$ around $\Pz$, $\cl\Mset$ is an affine set at distance $\sqrt{2\cbar}$ from $\Pz$, and half the squared distance between them is $(\sqrt{\cbar}-\sqrt c)^2$; so $\lambda_c\approx\sqrt{\cbar/c}-1$ and $\gamma_c\approx1/(1-\sqrt{c/\cbar})$. To the same order $v(0)\approx\cbar$, so $v(c)\approx(\sqrt{\cbar}-\sqrt c)^2\approx v(0)(1-\sqrt{c/\cbar})^2$, and the second form is exact at both ends of the path. In the Black--Scholes model KL is exactly quadratic in the drift, $v(0)=\cbar$, and the formulas are exact (Proposition~\ref{prop:bsdoubt}). Section~\ref{sec:ec-numerics-robust} reports their accuracy in the empirical market, and Section~\ref{sec:ec-numerics-e2} in a synthetic market with a traded call.
\end{remark}

\subsection{Black--Scholes With Drift Doubt}\label{sec:ec-rob-bs}

\textit{Relation to earlier results.} In Proposition~\ref{prop:bsdoubt}, the investor acts as if the Sharpe ratio were $\SR-a$, with $a=\sqrt{2c/T}$. For $a<\SR$, the robust strategy is the one under the $\kappa$-ignorance of \citet{ChenEpstein2002} with $\kappa=a$, where drift distortions satisfy $|\phi_t|\le a$; for $a \ge \SR$, the zero position is the classical non-participation result \citep{DowWerlang1992,GarlappiUppalWang2007}. The weight $w^*(1-z/\tau)$ in Proposition~\ref{prop:bsdoubt}(d) is the single-asset multi-prior rule of \citet{GarlappiUppalWang2007} \citep[see also][]{Maenhout2004,BiaginiPinar2017,KallbladOblojZariphopoulou2018}. \citet{TrojaniVanini2002} derive closed forms for a robust investor in Merton's model with a bound on the rate of relative entropy; for log utility and the rate $c/T$ they give the weight $w_c$ of Proposition~\ref{prop:bsdoubt}(a). The closed forms of $v(c)$ and $c\mapsto\gamma_c$ for the KL ball, and their reading in years of data, are our additions (Section~\ref{sec:rob-bs}). Outside this model, these closed forms serve as approximations (Remark~\ref{rem:ec-rob-smallpremium}), and Section~\ref{sec:ec-numerics-robust} measures their accuracy in the empirical market.

We work in the setting of Section~\ref{sec:rob-bs}, with these technical conditions. The filtration is the augmented filtration of $Z$. A strategy $w$ is bounded and progressively measurable, and the wealth solves $\dd W_t=W_t[(r+w_t(\mu-r))\,\dd t+w_t\sigma\,\dd Z_t]$, $W_0=1$. An alternative model is $\PP^\phi$ with $\dd\PP^\phi/\dd\Pz=\exp(\int_0^T\phi_t\,\dd Z_t-\frac12\int_0^T\phi_t^2\,\dd t)$ for a bounded progressively measurable $\phi$; by Girsanov's theorem, the stock drift under $\PP^\phi$ is $\mu+\sigma\phi_t$.

\begin{lemma}[Drift changes: entropy, growth, and the constant worst case]\label{lem:ec-rob-bsconst}
Let $w$ and $\phi$ be bounded and progressively measurable, let $c\ge0$, and let $a=\sqrt{2c/T}$.
(a) $\PP^\phi$ is a probability measure equivalent to $\Pz$, $Z^\phi_t=Z_t-\int_0^t\phi_s\,\dd s$ is a $\PP^\phi$-Brownian motion, $\kl{\PP^\phi}{\Pz}=\frac12\EE_{\PP^\phi}[\int_0^T\phi_t^2\,\dd t]$, and
\[
J(w,\phi):=\EE_{\PP^\phi}\Bigl[\ln\frac{W_T}{B_T}\Bigr]=\EE_{\PP^\phi}\Bigl[\int_0^T\Bigl(\sigma w_t(\SR+\phi_t)-\tfrac12\sigma^2w_t^2\Bigr)\dd t\Bigr].
\]
(b) For every $\PP^\phi\in\mathcal A_c$, $\EE_{\PP^\phi}[\int_0^T\phi_t\,\dd t]\ge-aT$, with equality iff $\phi\equiv-a$. Hence, for a constant strategy $w\equiv y/\sigma$ with $y\ge0$, $J(y/\sigma,\phi)\ge T[y(\SR-a)-\frac12y^2]$, with equality, when $y>0$, iff $\phi\equiv-a$.
(c) For a constant distortion $\phi\equiv-b$ with $\frac12b^2T\le c$, $J(w,-b)\le\frac12T(\SR-b)^2$ for every strategy $w$, with equality iff $\sigma w_t=\SR-b$, $\dd t\otimes\dd\Pz$-a.e.
\end{lemma}

\begin{proof}{Proof.}
(a) Since $\phi$ is bounded, Novikov's criterion and Girsanov's theorem apply \citep[see][]{Bjork2019}. Then $\ln(\dd\PP^\phi/\dd\Pz)=\int_0^T\phi_t\,\dd Z^\phi_t+\frac12\int_0^T\phi_t^2\,\dd t$, and the stochastic integral has mean zero under $\PP^\phi$. By It\^{o}'s formula, $\dd\ln W_t=(r+\sigma w_t(\SR+\phi_t)-\frac12\sigma^2w_t^2)\,\dd t+\sigma w_t\,\dd Z^\phi_t$, and $\ln B_T=rT$; take expectations. (b) The Cauchy--Schwarz inequality on $[0,T]\times\Omega$ with $\dd t\otimes\PP^\phi$ gives $\EE_{\PP^\phi}[\int_0^T\phi_t\,\dd t]\ge-(T\,\EE_{\PP^\phi}[\int_0^T\phi_t^2\,\dd t])^{1/2}\ge-(2cT)^{1/2}=-aT$. The first inequality is an equality if and only if $\phi$ is almost everywhere a nonpositive constant $-b$, and the second if and only if $b=a$. The bound on $J$ follows from (a), since $J(y/\sigma,\phi)=T(y\,\SR-\frac12y^2)+y\,\EE_{\PP^\phi}[\int_0^T\phi_t\,\dd t]$. (c) Pointwise, $y(\SR-b)-\frac12y^2\le\frac12(\SR-b)^2$, with equality if and only if $y=\SR-b$; apply this with $y=\sigma w_t$ and use (a) and the equivalence of $\PP^{-b}$ and $\Pz$. \Halmos
\end{proof}

\begin{repeatproposition}[Proposition~\ref{prop:bsdoubt} (Black--Scholes drift doubt).]
Let $a=\sqrt{2c/T}$, the doubt radius in Sharpe-ratio units.
\begin{enumerate}
\item[(a)] If $c<\cbar$, then $v(c)=(\sqrt{\cbar}-\sqrt c)^2$. The robust strategy is unique ($\dd t\otimes\dd\Pz$-a.e.) and constant,
\[
w_c=\frac{\SR-a}{\sigma}=\frac{\mu-r}{\gamma_c\sigma^2},\qquad \gamma_c=\frac{1}{1-\sqrt{c/\cbar}},
\]
which is the Merton fraction of a CRRA investor with risk aversion $\gamma_c$. The worst-case model lowers the drift by the constant $\sigma a$ (that is, $\phi\equiv-a$), and it is the unique worst case against $w_c$.
\item[(b)] If $c\ge\cbar$, then $v(c)=0$, $w\equiv0$ is the unique robust strategy, and $\QQ$ (that is, $\phi\equiv-\SR$) is a worst case.
\item[(c)] (Agreement with Theorem~\ref{thm:robust}.) For $0<c<\cbar$, $\lambda_c=-v'(c)=\sqrt{\cbar/c}-1$, so $\gamma_c=1+1/\lambda_c$, and $c(\gamma)=\cbar(1-1/\gamma)^2$; $v(0)=\cbar=\kl{\Pz}{\QQ}$, and $\QQ$ is the robust pricing measure for every $c$.
\item[(d)] (Calibration to the length of the data.) Suppose the investor estimates $\SR$ from $n$ years of data with $\sigma$ known, so that the estimate has standard error $1/\sqrt n$, and that they doubt every constant drift shift in the confidence interval $|\phi|\le z/\sqrt n$. The matching radius is $c_n=z^2T/(2n)$. With the $t$-statistic $\tau=\SR\sqrt n$ of the mean excess return, $c_n/\cbar=(z/\tau)^2$. If $\tau>z$, then $\gamma_{c_n}=\tau/(\tau-z)$ and $w_{c_n}=w^*(1-z/\tau)$, where $w^*=(\mu-r)/\sigma^2$ is the log-optimal fraction. If $\tau\le z$, that is, if $n\le(z/\SR)^2$, the robust investor holds no stock. The robust strategy is the same whether the doubt set is $\mathcal A_{c_n}$ or only the constant drifts with $|\phi|\le z/\sqrt n$.
\end{enumerate}
\end{repeatproposition}

\begin{proof}{Proof.}
(a) Let $c<\cbar$, so $a<\SR$; put $y=\SR-a>0$ and $w_c=y/\sigma$. Lemma~\ref{lem:ec-rob-bsconst}(b)--(c), with $b=a$ allowed because $\frac12a^2T=c$, gives
\[
\tfrac12T(\SR-a)^2\le\inf_{\mathcal A_c}J(w_c,\cdot)\le v(c)\le\inf_{\PP^\phi\in\mathcal A_c}\sup_wJ(w,\phi)\le\sup_wJ(w,-a)\le\tfrac12T(\SR-a)^2.
\]
So $(w_c,-a)$ is a saddle point and $v(c)=\frac12T(\SR-a)^2=(\sqrt{\cbar}-\sqrt c)^2$, since $\cbar=\SR^2T/2$ and $c=a^2T/2$. Any robust $w$ has $v(c)\le J(w,-a)\le v(c)$, so $\sigma w_t=\SR-a$ almost everywhere by Lemma~\ref{lem:ec-rob-bsconst}(c), and the worst case against $w_c$ is unique by Lemma~\ref{lem:ec-rob-bsconst}(b). Finally $\SR-a=\SR(1-\sqrt{c/\cbar})=\SR/\gamma_c$, and $(\mu-r)/(\gamma_c\sigma^2)$ is the optimal constant fraction of a CRRA($\gamma_c$) investor in this model \citep[see][]{Bjork2019}.
(b) If $c\ge\cbar$, then $\QQ=\PP^{-\SR}\in\mathcal A_c$, $J(w,-\SR)=-\frac12\sigma^2\EE_\QQ[\int_0^Tw_t^2\,\dd t]\le0$ with equality if and only if $w=0$ almost everywhere, and $J(0,\phi)=0$ for every $\phi$. So $v(c)=0$, every robust $w$ has $0\le J(w,-\SR)\le0$, hence $w=0$ almost everywhere, and $\QQ$ is a worst case.
(c) Differentiating $(\sqrt{\cbar}-\sqrt c)^2$ gives $\lambda_c=\sqrt{\cbar/c}-1$ and $1+1/\lambda_c=\sqrt{\cbar}/(\sqrt{\cbar}-\sqrt c)=\gamma_c$. Let $\Wc=W_T/B_T$ under $w_c$. Under $\Pz$, $\ln\Wc=yZ_T+(y\,\SR-\frac12y^2)T$ with $y=\SR-a$, so $\dd\PP^{-a}/\dd\Pz\propto e^{-aZ_T}\propto\Wc^{-a/y}$ with $a/y=\sqrt c/(\sqrt{\cbar}-\sqrt c)=1/\lambda_c$, and $\dd\QQ/\dd\Pz\propto e^{-\SR\,Z_T}\propto\Wc^{-\SR/y}$ with $\SR/y=\gamma_c=1+1/\lambda_c$: the worst case and the pricing measure are the power tilts of Theorem~\ref{thm:robust}(iii). The CRRA($\gamma$) investor holds the constant fraction $\SR/(\gamma\sigma)$, so $\Pg\propto\Pz(\Wg)^{1-\gamma}\propto\Pz e^{-\SR(1-1/\gamma)Z_T}$, the drift change $\phi\equiv-\SR(1-1/\gamma)$, with entropy $c(\gamma)=\frac12\SR^2(1-1/\gamma)^2T=\cbar(1-1/\gamma)^2$. Also $v(0)=\frac12\SR^2T=\kl{\Pz}{\QQ}$, and $\QQ$ is the only pricing measure of the complete market.
(d) The constant shifts with $|\phi|\le z/\sqrt n$ are those with $\frac12\phi^2T\le c_n=z^2T/(2n)$, and $c_n/\cbar=z^2/(n\,\SR^2)=(z/\tau)^2$. If $\tau>z$, part (a) gives $\gamma_{c_n}=1/(1-z/\tau)=\tau/(\tau-z)$ and $w_{c_n}=\frac{\mu-r}{\gamma_{c_n}\sigma^2}=w^*(1-z/\tau)$; if $\tau\le z$, that is, $n\le(z/\SR)^2$, then $c_n\ge\cbar$ and part (b) applies. The smaller set of constant shifts contains the worst cases used above, $\phi\equiv-z/\sqrt n$ and, when $\tau\le z$, $\phi\equiv-\SR$, and the lower bounds of Lemma~\ref{lem:ec-rob-bsconst}(b) hold on it a fortiori; so the same saddle points give the same value and the same unique robust strategy. \Halmos
\end{proof}

Volatility cannot change under an equivalent model, because the quadratic variation of $\ln S$ on $[0,T]$ is $\sigma^2T$ under every model equivalent to $\Pz$, so drift changes are the natural doubt here. The doubt set $\mathcal A_c$ is a ball on path space, not rectangular in the sense of \citet{EpsteinSchneider2003}, but the robust strategy is dynamically consistent because the worst case is constant: at a date $t$ on the worst-case path the remaining budget is $c-\frac12a^2t=\frac12a^2(T-t)$, and re-solving on $[t,T]$ gives the same $a$ and the same constant strategy.

The radius $c_n$ of part (d) also has a detection-error reading, the usual calibration of the penalty in the multiplier form \citep{AndersonHansenSargent2003,BarillasHansenSargent2009}. With $n$ years of data, the log-likelihood ratio of the worst case $\phi\equiv-z/\sqrt n$ against $\Pz$ is normal with mean $z^2/2$ and variance $z^2$ under the worst case, and with mean $-z^2/2$ and the same variance under $\Pz$. Choosing the model with the higher likelihood is therefore wrong, under either model, with the probability that a standard normal variable falls below $-z/2$; this is about 0.16 at $z=1.96$ and 0.31 at $z=1$.

\subsection{Computing the Robust Solution}\label{sec:ec-rob-algorithms}

The structure results give two ways to compute the robust solution. The first, alternating projections, moves between the two sets. The Q-step replaces a model $\PP$ by its reverse I-projection $\QL(\PP)$ on $\cl\Mset$, which needs the log-optimal portfolio under $\PP$. The P-step replaces a pricing measure $\QQ$ by its I-projection $\argmin_{\PP\in\Bc}\kl{\PP}{\QQ}$ on the ball: $\QQ$ itself if $\kl{\QQ}{\Pz}\le c$, and otherwise the point $e_t\propto\Pz^{1-t}\QQ^t$ of the exponential line with $\kl{e_t}{\Pz}=c$ (Lemma~\ref{lem:ec-rob-pstep}). The second method searches over risk aversion.

\begin{proposition}[Two ways to compute the robust solution]\label{prop:algorithms}
Assume (C1)--(C4) and $0<c<\cbar$.
\begin{enumerate}
\item[(a)] (Alternating projections.) Start from the reference model $\Pz$ and, for $j\ge0$, let $\QQ_j=\QL(\PP_j)$ and $\PP_{j+1}=\argmin_{\PP\in\Bc}\kl{\PP}{\QQ_j}$. Then $\kl{\PP_j}{\QQ_j}$ decreases to $v(c)$, with $\kl{\PP_j}{\QQ_j}-v(c)\le\kl{\Pc}{\QL(\Pz)}/j$ for $j\ge1$, and $(\PP_j,\QQ_j)\to(\Pc,\Qc)$. Near $\Pc$ the convergence is linear: the error in $\ln\PP_j$ shrinks by an asymptotic factor of at most $1-1/\gamma_c$ per iteration. If $M\ge d+3$, this bound cannot be improved, because the linearized iteration multiplies the errors in a subspace of dimension at least $M-d-2$ by exactly $1-1/\gamma_c$.
\item[(b)] (Search over risk aversion.) The pair $(h_c,\lambda_c)$ is the unique maximizer of the jointly concave function $\Phi_c(h,\lambda)=-\lambda\ln\EE_{\Pz}[W_h^{-1/\lambda}]-\lambda c$ over $\mathcal H\times(0,\infty)$, and $v(c)=\Phi_c(h_c,\lambda_c)$. For fixed $\lambda=1/(\gamma-1)$, the maximizer of $\Phi_c(\cdot,\lambda)$ is the CRRA portfolio $h^{(\gamma)}$, and $\gamma_c$ is the unique root of $c(\gamma)=c$.
\end{enumerate}
\end{proposition}

In words, near the solution each iteration of alternating projections removes, when $M\ge d+3$, only the fraction $1/\gamma_c$ of the error in the slowest direction, so an accuracy $\mathrm{tol}$ of the iterates takes about $\gamma_c\ln(1/\mathrm{tol})$ iterations. Part (a) is the alternating minimization of \citet{CsiszarTusnady1984}, the em-algorithm of information geometry \citep{Amari1995}, with the log-optimal step of \citet{Cover1984}; the local rate is our addition. Part (b) is the KL dual of distributionally robust optimization \citep{HuHong2012,BenTalDenHertogDeWaegenaereEtAl2013}, read as risk aversion. The factor $1-1/\gamma_c$ is the position of the worst-case model on the exponential line from $\Pz$ to $\Qc$ (Theorem~\ref{thm:robust}(iii)), and it tends to $1$ as $c\uparrow\cbar$ (Theorem~\ref{thm:robust}(iv)); so, when $M\ge d+3$, the method slows down without bound. The joint Newton method (Section~\ref{sec:ec-alg-jointnewton}) maximizes $\Phi_c$ directly, with a safeguard, and the $\gamma$-search (Section~\ref{sec:ec-alg-gamma}) solves one CRRA problem per step; neither is a new algorithm. Section~\ref{sec:ec-numerics-e4} compares the three methods, details what each timing includes, and recommends the joint Newton method as the fastest in the median run even when every input is timed.

\begin{proof}{Proof.}
(a) \textit{Global convergence.} This is the alternating minimization argument of \citet{CsiszarTusnady1984}; we verify its two properties. Let $a_j=\kl{\PP_j}{\QQ_{j-1}}$ and $b_j=\kl{\PP_j}{\QQ_j}$. All iterates are strictly positive (Lemmas~\ref{lem:ec-rob-pstep} and~\ref{lem:ec-rob-logdual}(b)), and since each step minimizes KL in one argument, $b_{j-1}\ge a_j\ge b_j\ge v(c)$. Three-point property: for $\PP\in\Bc$, the right derivative at $s=0$ of $s\mapsto\kl{(1-s)\PP_j+s\PP}{\QQ_{j-1}}$, which is $\EE_\PP[\ln(\PP_j/\QQ_{j-1})]-a_j$, is nonnegative (Lemma~\ref{lem:ec-rob-pstep}), so $\kl{\PP}{\QQ_{j-1}}\ge\kl{\PP}{\PP_j}+a_j$. Four-point property: for $\QQ\in\cl\Mset$ with $\kl{\PP}{\QQ}<\infty$, $\PP_j/\QQ_j=W_j$ (the growth-optimal wealth under $\PP_j$) and $\EE_\QQ[W_j]=1$, so Jensen's inequality gives $\kl{\PP}{\QQ_j}-\kl{\PP}{\QQ}-\kl{\PP}{\PP_j}=\EE_\PP[\ln(\QQ\PP_j/(\QQ_j\PP))]\le\ln\EE_\QQ[W_j]=0$. Together they give the five-point inequality $a_j-\kl{\PP}{\QQ}\le\kl{\PP}{\QQ_{j-1}}-\kl{\PP}{\QQ_j}$. Summing over $j\le J$ with $(\PP,\QQ)=(\Pc,\Qc)$, and using that $a_j$ is nonincreasing, gives $b_J-v(c)\le a_J-v(c)\le\kl{\Pc}{\QL(\Pz)}/J$; the same inequality shows that $\kl{\Pc}{\QQ_j}$ is nonincreasing. Limit points of $(\PP_j,\QQ_j)$ are nearest pairs, and the nearest pair is unique (Theorem~\ref{thm:robust}(ii)).

\textit{Local rate: the linearized map.} Write models near $\Pc$ as $\PP\propto\Pz e^{-\zeta}$, with $\zeta$ defined up to a constant and perturbations normalized to mean zero under $\Pc$. The iteration map $\Gamma(\PP)=\argmin_{\PP'\in\Bc}\kl{\PP'}{\QL(\PP)}$ is smooth near $\Pc$: $\QL$ is smooth (Lemma~\ref{lem:ec-rob-logdual}(b)), and the P-step position $t$ depends smoothly on $\QQ$ near $\Qc$, because $\kl{\Qc}{\Pz}>c$ and $\partial_t\kl{e_t}{\Pz}=t\Var_{e_t}(\ln(\QQ/\Pz))>0$. The P-step reads $\zeta_{j+1}=t_j(\zeta_j+\ln W_j)$ up to a constant, with fixed point $\zeta_c=t_c(\zeta_c+\ln\Wc)$, $t_c=1/(1+\lambda_c)=1-1/\gamma_c$, that is, $\zeta_c=\ln\Wc/\lambda_c$. Put $\zeta_j=\zeta_c+\varepsilon_j$. Linearizing $\EE_{\PP_j}[\DeltaS/W_j]=0$ gives the portfolio change $\delta h=-\Sigma_c^{-1}\EE_{\Pc}[\varepsilon_j\DeltaS/\Wc]$ with $\Sigma_c=\EE_{\Pc}[\DeltaS\DeltaS^\top/\Wc^2]$, so the change of $\ln W_j$ is $-\Pi_S\varepsilon_j$, where $\Pi_S$ is the $L^2(\Pc)$-orthogonal projection onto $\mathrm{span}\{\DeltaS^i/\Wc\}_{i\le d}$ (these functions have mean zero under $\Pc$). Hence $\varepsilon_{j+1}=t_c(I-\Pi_S)\varepsilon_j+\beta_j\ln\Wc$ to first order, where $\beta_j$ is proportional to the change of $t_j$. The constraint $\kl{\PP_{j+1}}{\Pz}=c$ linearizes to $\mathrm{Cov}_{\Pc}(\varepsilon_{j+1},\ln\Wc)=0$, because the derivative of $\zeta\mapsto\kl{\PP}{\Pz}$ in a direction $\varepsilon$ is $\mathrm{Cov}_\PP(\zeta,\varepsilon)$ and $\zeta_c\propto\ln\Wc$. So $\beta_j$ removes the component along $\ell_c=\ln\Wc-\EE_{\Pc}[\ln\Wc]$, and on the tangent space $\mathcal V=\{\varepsilon:\EE_{\Pc}[\varepsilon]=0,\ \mathrm{Cov}_{\Pc}(\varepsilon,\ln\Wc)=0\}$ of the sphere the derivative of $\Gamma$ at $\Pc$ is $L=t_c(I-\Pi_\ell)(I-\Pi_S)$, where $\Pi_\ell$ is the orthogonal projection onto $\mathrm{span}\{\ell_c\}$.

\textit{Local rate: the factor.} Both projections have norm one, so the spectral radius $\rho$ of $L$ is at most $t_c$. For every $\vartheta>0$ some norm on $\mathcal V$ gives $L$ an operator norm at most $\rho+\vartheta$, so $\Gamma$ is a contraction with factor $\rho+2\vartheta<1$ near $\Pc$; the global part brings the iterates there, and the asymptotic factor is $\rho\le1-1/\gamma_c$. If $M\ge d+3$, the mean-zero functions orthogonal to $\DeltaS^1/\Wc,\dots,\DeltaS^d/\Wc$ and to $\ell_c$ form a subspace of $\mathcal V$ of dimension at least $(M-1)-d-1=M-d-2\ge1$, on which $L\varepsilon=t_c\varepsilon$.

(b) \textit{Joint concavity.} The map $(\xi,\lambda)\mapsto\lambda\ln\EE_{\Pz}[e^{-\xi/\lambda}]$ is the perspective of the convex function $\xi\mapsto\ln\EE_{\Pz}[e^{-\xi}]$, hence jointly convex \citep{BoydVandenberghe2004}. So $\Psi(\xi,\lambda)=-\lambda\ln\EE_{\Pz}[e^{-\xi/\lambda}]$ is jointly concave and nondecreasing in each coordinate of $\xi$, and composing it with the concave map $h\mapsto\ln W_h$ keeps joint concavity; so $\Phi_c(h,\lambda)=\Psi(\ln W_h,\lambda)-\lambda c$ is jointly concave. \textit{Maximizer.} By Lemma~\ref{lem:ec-rob-gibbs}(c), $\Phi_c(h,\lambda)\le G_c(h)\le v(c)$. At $h=h_c$, Step 2 of the proof of Theorem~\ref{thm:robust} shows that $\ln\Wc$ is not constant and that $c<c_m$, so Lemma~\ref{lem:ec-rob-gibbs}(c) gives $\max_\lambda\Phi_c(h_c,\lambda)=G_c(h_c)=v(c)$, attained only at $\lambda=\theta_c=\lambda_c$. If $\Phi_c(h,\lambda)=v(c)$, then $G_c(h)=v(c)$, so $h=h_c$ (Theorem~\ref{thm:robust}(ii)) and $\lambda=\lambda_c$. \textit{CRRA step and root.} For $\lambda=1/(\gamma-1)$, $-\lambda\ln\EE_{\Pz}[W_h^{-1/\lambda}]=\frac{1}{1-\gamma}\ln\EE_{\Pz}[W_h^{1-\gamma}]$ is an increasing function of $\EE_{\Pz}[U_\gamma(W_h)]$, so its maximizer is $h^{(\gamma)}$ and its maximum is $g^*(\gamma)$, as in Proposition~\ref{prop:ec-rob-mult}. By Theorem~\ref{thm:robust}(iv), $c(\gamma)=c$ has the unique root $\gamma_c$. \Halmos
\end{proof}

\begin{remark}[At and above the threshold]\label{rem:ec-rob-start}
For $c\ge\cbar$ the argument of part (a), with any nearest pair $(\QQ^\circ,\QQ^\circ)$, $\QQ^\circ\in\cl\Mset\cap\Bc$, in the five-point inequality, still gives $\kl{\PP_j}{\QQ_j}\to0$. The iterates also converge. By compactness of $\Bc\times\cl\Mset$, a subsequence $(\PP_{j_i},\QQ_{j_i})$ converges to a pair $(\PP^\circ,\QQ^\circ)$; lower semicontinuity of KL and $\kl{\PP_j}{\QQ_j}\to0$ give $\PP^\circ=\QQ^\circ\in\cl\Mset\cap\Bc$, a nearest pair (Theorem~\ref{thm:robust}(vi)). The five-point inequality with $(\QQ^\circ,\QQ^\circ)$ makes $\kl{\QQ^\circ}{\QQ_j}$ nonincreasing in $j$. Along the subsequence it tends to $0$, because $\QQ_{j_i}\to\QQ^\circ$ and $\kl{\QQ^\circ}{\cdot}$ is continuous at $\QQ^\circ$ (its terms run over the support of $\QQ^\circ$, where $\QQ^\circ>0$, and every $\QQ_j$ is strictly positive). Hence $\kl{\QQ^\circ}{\QQ_j}\to0$ for the whole sequence, and Pinsker's inequality \citep{Csiszar1967} gives $\QQ_j\to\QQ^\circ$ and, with $\kl{\PP_j}{\QQ_j}\to0$, $\PP_j\to\QQ^\circ$. The nearest pair is not unique (Theorem~\ref{thm:robust}(vi)), so the limit depends on the start, and the meaningful output above the threshold is the interval $I_c(X)$. Below the threshold, the search of part (b) is a root search in $s=1-1/\gamma\in[0,1]$, with $c(\gamma)\to\cbar$ as $s\to1$ (Theorem~\ref{thm:robust}(v)), so its number of steps for a given accuracy in $s$ does not grow as $c\uparrow\cbar$, apart from the bracketing step, which grows like $\log_{10}\gamma_c$.
\end{remark}

\subsection{The Chi-Square Radius Rule}\label{sec:ec-rob-chi2}

\begin{proposition}[Chi-square rule for the radius]\label{prop:ec-rob-chi2}
Let $\PP\in\mathcal P$ be the true model, $\hat\PP_N$ the empirical distribution of $N$ independent draws from $\PP$, and $\chi^2_{M-1,1-\alpha}$ the $(1-\alpha)$-quantile of the $\chi^2_{M-1}$ distribution. Then $2N\,\kl{\hat\PP_N}{\PP}$ and $2N\,\kl{\PP}{\hat\PP_N}$ converge in distribution to $\chi^2_{M-1}$; the second is defined on $\{\hat\PP_N>0\}$, whose probability tends to one. Hence, with $\Pz=\hat\PP_N$ and $c_{\mathrm{full}}=\chi^2_{M-1,1-\alpha}/(2N)$, the probability that $\PP\in\Ball_{c_{\mathrm{full}}}$ tends to $1-\alpha$.
\end{proposition}

\begin{proof}{Proof.}
A second-order Taylor expansion of $x\ln(x/y)$ at $x=y$ shows that both divergences equal $\frac12\sum_\omega(\hat\PP_N(\omega)-\PP(\omega))^2/\PP(\omega)+O(|\hat\PP_N-\PP|^3)$. By the multinomial central limit theorem, $|\hat\PP_N-\PP|=O_p(N^{-1/2})$ and Pearson's statistic $N\sum_\omega(\hat\PP_N(\omega)-\PP(\omega))^2/\PP(\omega)$ converges in distribution to $\chi^2_{M-1}$. So both scaled divergences converge to $\chi^2_{M-1}$, and $\{\PP\in\Ball_{c_{\mathrm{full}}}\}=\{2N\,\kl{\PP}{\hat\PP_N}\le\chi^2_{M-1,1-\alpha}\}$ has probability tending to $1-\alpha$. \Halmos
\end{proof}

This is a first-order rule, and \citet{BenTalDenHertogDeWaegenaereEtAl2013} use the same calibration for divergence balls. The classical statistic has the empirical measure in the first slot of KL, while our ball has it in the second; the proposition shows that both give the same first-order radius. The radius grows like $(M-1)/(2N)$, so finer state grids need more data. For doubt about $\mathrm{df}$ parameters of a model fitted by maximum likelihood, $\chi^2_{\mathrm{df}}$ replaces $\chi^2_{M-1}$; Proposition~\ref{prop:bsdoubt}(d) is the case $\mathrm{df}=1$, with quantile $z^2$.

\begin{remark}[Cells built from the same data]\label{rem:ec-rob-cells}
Proposition~\ref{prop:ec-rob-chi2} assumes fixed cells. If the $M$ cells are sample quantiles with about $N/M$ observations each, $\Pz$ is (nearly) uniform by construction, and the true probabilities of the realized cells are spacings of uniform order statistics, which are Dirichlet distributed with parameters about $N/M$. Their scaled fluctuations have the same Gaussian limit as multinomial frequencies, so $2N\,\kl{\PP}{\Pz}$ is again approximately $\chi^2_{M-1}$. This is a first-order heuristic, not a proof. The ball then covers the cell probabilities given the cell values; it covers neither the error of evaluating payoffs at cell values nor the dependence of returns over time.
\end{remark}

\textit{Several risky assets and the value bound.} The investor with radius $c_{\mathrm{one}}$ trades exactly when $c_{\mathrm{one}}<\cbar$, that is, when $2N\cbar>\chi^2_{1,1-\alpha}$. With $d>1$ risky assets, $2N\cbar$ is approximately $\chi^2_d$ under zero premia, so this investor trades under zero premia with a probability close to the chance that a $\chi^2_d$ variable exceeds $\chi^2_{1,1-\alpha}$, which is 0.15 for $d=2$ and $\alpha=0.05$; the test radius $\chi^2_{d,1-\alpha}/(2N)$ of Section~\ref{sec:rob-radius} avoids this. The value bound of \citet{DuchiGlynnNamkoong2021} concerns only one number, the optimal growth, so it keeps $\chi^2_1$. At zero premia, the true optimal growth is zero and $v(c_{\mathrm{one}})\ge0$, so the bound holds only when $v(c_{\mathrm{one}})=0$. With $d=1$, this happens with probability close to $1-\alpha$, not the nominal level $1-\alpha/2$ of Section~\ref{sec:rob-radius}. With $d=2$ and $\alpha=0.05$, this probability is about 0.85, and it falls further as $d$ grows.

\section{Algorithms}\label{sec:ec-algorithms}

This section states the algorithms behind Sections~\ref{sec:robust} and~\ref{sec:numerics}; Section~\ref{sec:ec-rob-algorithms} proves their main properties (Proposition~\ref{prop:algorithms}). The market is the one-period finite market of Section~\ref{sec:robust}, and Assumption~\ref{ass:C} holds, so $\Pz$ charges each of the $M$ states and $\cbar>0$. We write $\cl\mathcal P$ for the closed simplex of probability vectors on the states and $\mathcal P$ for the strictly positive ones, $\mathcal H:=\{h\in\RR^d:W_h>0\text{ in every state}\}$ for the solvent portfolios, and $G_c(h):=\min_{\PP\in\Bc}\EE_{\PP}[\ln W_h]$ for the worst-case growth of $h$, so that $v(c)=\sup_{h\in\mathcal H}G_c(h)$ (Theorem~\ref{thm:minimax}). For $\gamma>1$, $\mathrm{CE}_\gamma(W):=(\EE_{\Pz}[W^{1-\gamma}])^{1/(1-\gamma)}$ is the certainty equivalent of a CRRA investor with relative risk aversion $\gamma$ who trusts $\Pz$; their optimal portfolio is $h^{(\gamma)}$, and $\Wg$, $\Pg\propto\Pz(\Wg)^{1-\gamma}$, $\Qg\propto\Pz(\Wg)^{-\gamma}$ and $c(\gamma):=\kl{\Pg}{\Pz}$ are as in Theorem~\ref{thm:robust}.

Three methods compute $v(c)$ and the robust objects for $0<c<\cbar$: alternating projections (A1; Proposition~\ref{prop:algorithms}(a)), the $\gamma$-search (A2; Algorithm~\ref{alg:gammasearch}) and the joint Newton method (JN; Algorithm~\ref{alg:jointnewton}). All three first compute $\cbar$, because the robust solution changes its nature there (Theorem~\ref{thm:robust}); Section~\ref{sec:ec-alg-interval} computes the end points of the no-trade interval for $c>\cbar$.

\subsection{Building Blocks}\label{sec:ec-alg-blocks}

Three convex problems appear inside the algorithms. The first gives the threshold. By Section~\ref{sec:inc-measures}, $\QE\propto\Pz e^{-h_E\cdot\DeltaS}$, where $h_E$ minimizes the strictly convex function $\eta\mapsto\ln\EE_{\Pz}[e^{-\eta\cdot\DeltaS}]$, and $\cbar=-\ln\EE_{\Pz}[e^{-h_E\cdot\DeltaS}]$; we compute $h_E$ by a damped Newton method. The other two are the steps of the alternating method. The Q-step replaces a model $\PP\in\mathcal P$ by its reverse I-projection on $\cl\Mset$, the log-utility measure $\QL(\PP)=\PP/\Wstar(\PP)$, so it needs only the growth-optimal portfolio under $\PP$ (Lemma~\ref{lem:ec-rob-logdual}(b)). The P-step replaces a pricing measure $\QQ\in\mathcal P$ by its I-projection on the ball (Lemma~\ref{lem:ec-rob-pstep}): $\QQ$ itself if $\kl{\QQ}{\Pz}\le c$, and otherwise the point $e^{\QQ}_{t^*}$ of the exponential line $e^{\QQ}_t\propto\Pz^{1-t}\QQ^t$ with $\kl{e^{\QQ}_{t^*}}{\Pz}=c$. Along this line, $\kappa(t):=\kl{e^{\QQ}_t}{\Pz}=t\psi'(t)-\psi(t)$ with $\psi(t):=\ln\EE_{\Pz}[e^{t\ln(\QQ/\Pz)}]$, so $\kappa'(t)=t\Var_{e^{\QQ}_t}(\ln(\QQ/\Pz))>0$, and a safeguarded Newton method finds $t^*$. In words, the Q-step replaces a model by its log-utility pricing measure, and the P-step moves from $\Pz$ toward a pricing measure along the exponential line until it reaches the boundary of the ball.

The certificates below rest on one inequality, the finite form of the Donsker--Varadhan variational formula \citep[Proposition 1.4.2]{DupuisEllis1997}. For every $f\in\RR^M$ and $\PP\in\cl\mathcal P$,
\begin{equation}\label{eq:ec-gibbs}
\EE_{\PP}[f]\le\kl{\PP}{\Pz}+\ln\EE_{\Pz}[e^f],\qquad\text{with equality if and only if }\PP=\Pz e^f/\EE_{\Pz}[e^f];
\end{equation}
this is Lemma~\ref{lem:ec-rob-kl}(d) with $\QQ=\Pz$ and $W=e^f$. In words, the mean of a payoff under any model is at most its entropic certainty equivalent under $\Pz$ plus the model's distance from $\Pz$. So every multiplier gives a guaranteed bound.

\subsection{Alternating Projections}\label{sec:ec-alg-altproj}

Alternating projections minimize $\kl{\PP}{\QQ}$ over $\Bc\times\cl\Mset$ one argument at a time, as in \citet{CsiszarTusnady1984}. The method is the em algorithm of information geometry \citep{Amari1995}, and its Q-step is the log-optimal portfolio problem, for which \citet{Cover1984} gives an alternating-minimization algorithm. It first computes $h_E$ and $\cbar$ by the damped Newton method of Section~\ref{sec:ec-alg-blocks}, and it returns $v(c)=0$ and $h_c=0$ if $c\ge\cbar$. For $0<c<\cbar$, the first iterate is the reference model $\Pz$. For $j\ge0$, the Q-step computes the log-optimal portfolio $h_j$ under $\PP_j$ by Newton's method started at $h_{j-1}$ (with $h_{-1}=0$), and it sets $\QQ_j=\PP_j/(1+h_j\cdot\DeltaS)$ and $b_j=\kl{\PP_j}{\QQ_j}=\EE_{\PP_j}[\ln(1+h_j\cdot\DeltaS)]$. The P-step sets $\PP_{j+1}=e^{\QQ_j}_{t_j}$, where $t_j\in(0,1)$ solves $\kl{e^{\QQ_j}_t}{\Pz}=c$ and is found by a safeguarded Newton method started at $t_{j-1}$ (with $t_{-1}=1/2$). Each $\QQ_j$ is a pricing measure, so $\kl{\QQ_j}{\Pz}\ge\cbar>c$, and this root exists by Lemma~\ref{lem:ec-rob-pstep}. The method stops when $|b_{j-1}-b_j|\le\mathrm{tol}\cdot b_j$ (with $b_{-1}=+\infty$) and returns $v(c)\approx b_j$, $(\Pc,\Qc)\approx(\PP_j,\QQ_j)$, $h_c\approx h_j$ and $\gamma_c\approx1/(1-t_{j-1})$.

Proposition~\ref{prop:algorithms}(a), proved in Section~\ref{sec:ec-rob-algorithms}, gives the convergence of alternating projections for $0<c<\cbar$: the values $\kl{\PP_j}{\QQ_j}$ decrease to $v(c)$, the iterates converge to $(\Pc,\Qc)$, and near $\Pc$ the error shrinks by an asymptotic factor of at most $1-1/\gamma_c$ per iteration, attained if $M\ge d+3$. When it is attained, an accuracy $\mathrm{tol}$ of the iterates takes about $\gamma_c\ln(1/\mathrm{tol})$ iterations, so the method slows down without bound as $c\uparrow\cbar$; Remark~\ref{rem:ec-rob-start} covers $c\ge\cbar$.

\subsection{The Dual in $(h,\lambda)$ and the $\gamma$-Search}\label{sec:ec-alg-gamma}

Theorem~\ref{thm:robust} turns the robust problem into a CRRA problem, and convex duality explains why. For $h\in\mathcal H$ and $\lambda>0$, the function of Proposition~\ref{prop:algorithms}(b) is
\begin{equation}\label{eq:ec-dual}
\Phi_c(h,\lambda)=-\lambda\ln\EE_{\Pz}\big[W_h^{-1/\lambda}\big]-\lambda c=\ln\mathrm{CE}_{1+1/\lambda}(W_h)-\lambda c ,
\end{equation}
where the second form uses $1/(1-\gamma)=-\lambda$ for $\gamma=1+1/\lambda$. This is the dual of KL-constrained distributionally robust optimization \citep{HuHong2012,BenTalDenHertogDeWaegenaereEtAl2013}, applied to the log growth $\ln W_h$. For $\PP\in\Bc$, \eqref{eq:ec-gibbs} with $f=-\ln W_h/\lambda$ gives $\EE_{\PP}[\ln W_h]\ge\Phi_c(h,\lambda)$, so $\Phi_c(h,\lambda)\le G_c(h)\le v(c)$ for every pair $(h,\lambda)$. Proposition~\ref{prop:algorithms}(b) adds that $\Phi_c$ is jointly concave, that $(h_c,\lambda_c)$ is its unique maximizer with maximum $v(c)$, and that for a fixed multiplier the maximizer is the CRRA portfolio $h^{(1+1/\lambda)}$. So the robust problem is a concave program in $d+1$ variables, the portfolio and the multiplier of the ball, and every pair $(h,\lambda)$ comes with the guaranteed growth rate $\Phi_c(h,\lambda)$.

Two methods use this program. The joint Newton method (Algorithm~\ref{alg:jointnewton}, Section~\ref{sec:ec-alg-jointnewton}) takes Newton steps in all $d+1$ variables at once. The $\gamma$-search (Algorithm~\ref{alg:gammasearch}) nests it: an outer search over the multiplier, written as $t=1-1/\gamma=1/(1+\lambda)\in[0,1)$, and an inner CRRA solve; it is the KL dual with the multiplier read as risk aversion, not a new algorithm.

\begin{algorithm}[t]
\caption{$\gamma$-search (A2) for $v(c)$}\label{alg:gammasearch}
\begin{algorithmic}[1]\small
\Require $\Pz\in\mathcal P$, $\DeltaS$, radius $c>0$, root tolerance $\mathrm{tol}$
\Ensure $v(c)$, $(\Pc,\Qc)$, $h_c$, $\gamma_c$, $\lambda_c$, and a dual lower bound on $v(c)$
\State compute $h_E$ and $\cbar$ (Section~\ref{sec:ec-alg-blocks}); \textbf{if} $c\ge\cbar$ \textbf{return} $v(c)=0$, $h_c=0$
\Function{Radius}{$t$} \Comment{$t=1-1/\gamma\in[0,1)$; for $\gamma=1$, $h^{(1)}$ is the log-optimal portfolio and \Call{Radius}{0} $=0$}
  \State $\gamma\gets1/(1-t)$
  \State $h^{(\gamma)}\gets\argmin_{h\in\mathcal H}\ln\EE_{\Pz}[(1+h\cdot\DeltaS)^{1-\gamma}]$ \Comment{CRRA$(\gamma)$; warm start with $\gamma h$ fixed}
  \State $\Wg\gets1+h^{(\gamma)}\cdot\DeltaS$; \ $\Pg\propto\Pz(\Wg)^{1-\gamma}$
  \State \Return $c(\gamma)=\kl{\Pg}{\Pz}$
\EndFunction
\State $t_{\mathrm{lo}}\gets0$; \ $k\gets1$
\While{\Call{Radius}{$1-10^{-k}$} $\le c$}
  \State $t_{\mathrm{lo}}\gets1-10^{-k}$; \ $k\gets k+1$
\EndWhile
\State $t_c\gets$ root of $\sqrt{\Call{Radius}{t}}-\sqrt c=0$ on $[t_{\mathrm{lo}},1-10^{-k}]$ \Comment{bracketing root finder}
\State $\gamma_c\gets1/(1-t_c)$; \ $\lambda_c\gets1/(\gamma_c-1)$; \ $h_c\gets h^{(\gamma_c)}$; \ $\Pc\gets\PP^{(\gamma_c)}$; \ $\Qc\gets\Pc/\Wc$
\State \Return $v(c)=\EE_{\Pc}[\ln\Wc]$ and the certificate $G_c(h_c)\le v(c)$ (below)
\end{algorithmic}
\end{algorithm}

\begin{proposition}[Correctness and cost of the $\gamma$-search]\label{prop:ec-gammasearch}
Let $0<c<\cbar$.
\begin{enumerate}
\item[(a)] The map $t\mapsto c(1/(1-t))$ is continuous and strictly increasing on $[0,1)$, is $0$ at $t=0$, and tends to $\cbar$ as $t\uparrow1$. The equation $c(1/(1-t))=c$ has the unique root $t_c=1-1/\gamma_c$, and Algorithm~\ref{alg:gammasearch} returns the $\gamma_c$, $\lambda_c$, $h_c$, $(\Pc,\Qc)$ and $v(c)$ of Theorem~\ref{thm:robust}.
\item[(b)] Each evaluation of $c(\gamma)$ solves one strictly convex minimization over $\mathcal H\subset\RR^d$; one Newton step costs $O(Md^2+d^3)$ operations.
\item[(c)] The bracketing loop stops after $\lfloor\log_{10}\gamma_c\rfloor+1$ evaluations, and its final bracket contains $t_c$. If the root is then located by bisection, $k$ steps locate $t_c$ within $2^{-k}$ times the bracket length, for every $c$.
\end{enumerate}
\end{proposition}

In words, the $\gamma$-search is a one-dimensional root search over risk aversion, and each step solves a CRRA problem of the size of the market. Apart from a bracketing step that grows like $\log_{10}\gamma_c$, the number of steps for a fixed accuracy in $t=1-1/\gamma$ does not depend on how close $c$ is to $\cbar$; a fixed relative accuracy in $\gamma_c$ needs about $\log_2\gamma_c$ more bisection steps, because $\dd\gamma/\gamma=\gamma\,\dd t$. Part (c) is stated for bisection, as a reference; our code uses Brent's method, which keeps a bracket at every step. In practice a robust portfolio costs about a dozen CRRA portfolios (Section~\ref{sec:ec-numerics-e4}).

\begin{proof}{Proof.}
(a) By Theorem~\ref{thm:robust}, $\gamma\mapsto c(\gamma)$ is a continuous, strictly increasing bijection from $[1,\infty)$ onto $[0,\cbar)$ with inverse $c\mapsto\gamma_c$; compose it with the increasing bijection $t\mapsto1/(1-t)$ from $[0,1)$ onto $[1,\infty)$. The same theorem gives $h_c=h^{(\gamma_c)}$ and $\Pc=\PP^{(\gamma_c)}$. The CRRA first-order condition $\EE_{\Pz}[(\Wg)^{-\gamma}\DeltaS]=0$ says that $h^{(\gamma)}$ is log-optimal under $\Pg$, so Lemma~\ref{lem:ec-rob-crra}(b) gives $\QL(\Pg)=\Pg/\Wg\propto\Pz(\Wg)^{-\gamma}=\Qg$, which is $\Qc$ at $\gamma_c$; and $v(c)=\kl{\Pc}{\Qc}=\EE_{\Pc}[\ln\Wc]$. The square root keeps the root and improves the scaling: in the Black--Scholes market, $\gamma_c=1/(1-\sqrt{c/\cbar})$ (Proposition~\ref{prop:bsdoubt}) gives $c(\gamma)=\cbar t^2$, so $\sqrt{c(\gamma)}$ is linear in $t$.
(b) Since $1-\gamma<0$, the CRRA problem minimizes $F_\gamma(h):=\ln\sum_j\Pz(\omega_j)\exp\{(1-\gamma)\ln W_h(\omega_j)\}$, which is convex because its exponents are convex in $h$ and log-sum-exp is convex and increasing, strictly convex by non-redundancy, and tends to $+\infty$ at the boundary of the bounded set $\mathcal H$. Gradient and Hessian are sums over $M$ states, which costs $O(Md^2)$, and the Newton system costs $O(d^3)$.
(c) By strict monotonicity, $c(10^k)>c$ if and only if $10^k>\gamma_c$, that is, if and only if $k>\log_{10}\gamma_c$; at the end of the loop the radius is at most $c$ at $t_{\mathrm{lo}}$ and above $c$ at $1-10^{-k}$, so the bracket contains $t_c$. Each bisection step halves the bracket. \Halmos
\end{proof}

The value of Algorithm~\ref{alg:gammasearch} is bounded from both sides. Inequality~\eqref{eq:ec-gibbs} gives $\Phi_c(h,\lambda)\le v(c)$ for every $(h,\lambda)$, and every feasible pair gives $v(c)\le\kl{\PP}{\QQ}$ (Theorem~\ref{thm:minimax}). The certificate below computes the best lower bound for the output portfolio, $G_c(h_c)$, and the output pair is feasible up to the root tolerance. So, in exact arithmetic and up to that tolerance, $G_c(h_c)\le v(c)\le\EE_{\Pc}[\ln\Wc]$, and the relative gap $(\EE_{\Pc}[\ln\Wc]-G_c(h_c))/\EE_{\Pc}[\ln\Wc]$ bounds the relative error. In the 250 runs of Section~\ref{sec:ec-numerics-e4} this gap, computed in double precision, is at most $7.6\times10^{-10}$ in absolute value, and $|c(\gamma_c)-c|$ is at most $8.2\times10^{-15}$.

The certificate computes the worst-case growth $G_c(h)=\min_{\PP\in\Bc}\EE_{\PP}[\ln W_h]$ of a portfolio $h\in\mathcal H$ with $\ln W_h$ not constant, and a worst-case model. Let $A$ be the set of states in which $W_h$ is smallest. If $c\ge-\ln\Pz(A)$, the ball contains $\Pz(\cdot\mid A)$, which is then a worst case, and $G_c(h)=\min_\omega\ln W_h(\omega)$. Otherwise, for $\lambda>0$ let $\PP^h_\lambda\propto\Pz W_h^{-1/\lambda}$ and $\kappa_h(\lambda):=\kl{\PP^h_\lambda}{\Pz}$. By Lemma~\ref{lem:ec-rob-gibbs}(b), $\kappa_h$ falls from $-\ln\Pz(A)$ as $\lambda\downarrow0$ to $0$ as $\lambda\to\infty$, so $\kappa_h(\lambda)=c$ has a root $\lambda_h$. Halving $\lambda$ from 1 until $\kappa_h(\lambda)\ge c$ brackets the root, a safeguarded Newton method with bisection finds it, and $G_c(h)=\Phi_c(h,\lambda_h)$ with worst-case model $\PP^h_{\lambda_h}$ (Lemma~\ref{lem:ec-rob-gibbs}(c)). This is the finite-state form of the KL-ball worst case of \citet{HuHong2012}. The halving stops at the floor $\lambda_{\min}$ of Section~\ref{sec:ec-alg-impl}, which can be reached only when $c$ is within rounding error of $-\ln\Pz(A)$; the value $\Phi_c(h,\lambda_{\min})$ returned there is still a valid lower bound, because every $\Phi_c(h,\lambda)$ is one.

\subsection{End Points of the No-Trade Interval}\label{sec:ec-alg-interval}

Above the threshold a claim $X\in\RR^M$ has the no-trade interval $I_c(X)=\{\EE_{\QQ}[X]:\QQ\in\cl\Mset\cap\Bc\}$ (Theorem~\ref{thm:pricing}), and its end points come from tilting the minimal-entropy problem toward the claim. For $\beta\in\RR$, let $\QQ^X_\beta$ maximize $\beta\,\EE_{\QQ}[X]-\kl{\QQ}{\Pz}$ over $\QQ\in\cl\Mset$. Since $\kl{\QQ}{\Pz e^{\beta X}/\EE_{\Pz}[e^{\beta X}]}=\kl{\QQ}{\Pz}-\beta\,\EE_{\QQ}[X]+\ln\EE_{\Pz}[e^{\beta X}]$, $\QQ^X_\beta$ is the minimal-entropy martingale measure of the tilted model $\Pz e^{\beta X}/\EE_{\Pz}[e^{\beta X}]$, and $\QQ^X_0=\QE$.

\begin{proposition}[End points of the no-trade interval]\label{prop:ec-interval}
Let $X\in\RR^M$.
\begin{enumerate}
\item[(a)] For each $\beta$, $\QQ^X_\beta$ exists, is unique, and equals $\Pz e^{\beta X-\theta_\beta\cdot\DeltaS}/\EE_{\Pz}[e^{\beta X-\theta_\beta\cdot\DeltaS}]\in\Mset$, where $\theta_\beta$ is the unique minimizer of the strictly convex function $\theta\mapsto\EE_{\Pz}[e^{\beta X-\theta\cdot\DeltaS}]$.
\item[(b)] $\kappa_X(\beta):=\kl{\QQ^X_\beta}{\Pz}$ is continuous, nonincreasing on $(-\infty,0]$ and nondecreasing on $[0,\infty)$, with $\kappa_X(0)=\cbar$, and $\beta\mapsto\EE_{\QQ^X_\beta}[X]$ is nondecreasing.
\item[(c)] (Certificate.) For all $\beta\neq0$, $\theta\in\RR^d$ and $\QQ\in\cl\Mset\cap\Bc$,
\begin{equation}\label{eq:ec-interval-cert}
\beta\,\EE_{\QQ}[X]\le c+\ln\EE_{\Pz}\big[e^{\beta X-\theta\cdot\DeltaS}\big],
\end{equation}
so $U_X(\beta,\theta):=(c+\ln\EE_{\Pz}[e^{\beta X-\theta\cdot\DeltaS}])/\beta$ bounds $\max I_c(X)$ from above when $\beta>0$ and $\min I_c(X)$ from below when $\beta<0$.
\item[(d)] Let $c>\cbar$. If $\kappa_X(\beta_c)=c$ for some $\beta_c>0$, then $\max I_c(X)=\EE_{\QQ^X_{\beta_c}}[X]=U_X(\beta_c,\theta_{\beta_c})$. If $\kappa_X(\beta)<c$ for all $\beta>0$, then $\max I_c(X)=\max_{\QQ\in\cl\Mset}\EE_{\QQ}[X]$, the upper no-arbitrage bound. The same holds for $\min I_c(X)$ with $\beta<0$.
\end{enumerate}
\end{proposition}

In words, each end point is the price under the minimal-entropy measure of the reference model tilted toward or away from the claim. The tilt is chosen so that this measure sits on the boundary of the ball, and~\eqref{eq:ec-interval-cert} certifies the end point from the other side. Economically these are entropic good-deal bounds \citep{CochraneSaaRequejo2000,Cerny2003}: since $\kl{\QQ}{\Pz}=\kl{\QQ}{\QE}+\cbar$ on $\cl\Mset$ (Theorem~\ref{thm:twodecomp}(a)), they are the end points of the entropy-budget interval $J_{c-\cbar}(X)$ of~\eqref{eq:entropy-budget}.

\begin{proof}{Proof.}
(a) The tilted model is strictly positive, and no arbitrage and non-redundancy are properties of $\DeltaS$ alone, so Lemma~\ref{lem:duality} gives the Gibbs form; uniqueness follows from the strict convexity of $\kl{\cdot}{\Pz}$.
(b) $\phi(\beta):=\max_{\QQ\in\cl\Mset}\{\beta\,\EE_{\QQ}[X]-\kl{\QQ}{\Pz}\}$ is convex, as a maximum of affine functions. The objective is jointly continuous ($\Pz>0$), $\cl\Mset$ is compact and the maximizer is unique, so $\beta\mapsto\QQ^X_\beta$ and $\kappa_X$ are continuous. Comparing each maximizer with the other gives
\[
(\beta'-\beta)\,\EE_{\QQ^X_\beta}[X]\le\phi(\beta')-\phi(\beta)\le(\beta'-\beta)\,\EE_{\QQ^X_{\beta'}}[X],
\]
so $\EE_{\QQ^X_\beta}[X]$ is nondecreasing and, by continuity, equals $\phi'(\beta)$. Since $\kappa_X=\beta\phi'-\phi$, for $0\le\beta_1<\beta_2$ the right inequality gives $\kappa_X(\beta_2)-\kappa_X(\beta_1)\ge\beta_1(\phi'(\beta_2)-\phi'(\beta_1))\ge0$, and for $\beta_1<\beta_2\le0$ the left inequality gives $\kappa_X(\beta_1)-\kappa_X(\beta_2)\ge\beta_2(\phi'(\beta_1)-\phi'(\beta_2))\ge0$. Finally $\kappa_X(0)=\kl{\QE}{\Pz}=\cbar$.
(c) Apply~\eqref{eq:ec-gibbs} with $f=\beta X-\theta\cdot\DeltaS$, and use $\EE_{\QQ}[\DeltaS]=0$ and $\kl{\QQ}{\Pz}\le c$.
(d) If $\kappa_X(\beta_c)=c$, then $\QQ^X_{\beta_c}$ is feasible, and~\eqref{eq:ec-interval-cert} at $(\beta_c,\theta_{\beta_c})$ holds with equality there, because this measure is proportional to $\Pz e^f$ and lies on the sphere; so $\EE_{\QQ^X_{\beta_c}}[X]=U_X(\beta_c,\theta_{\beta_c})\ge\max I_c(X)\ge\EE_{\QQ^X_{\beta_c}}[X]$. If $\kappa_X(\beta)<c$ for all $\beta>0$, every $\QQ^X_\beta$ with $\beta>0$ is feasible. If $\QQ^\dagger$ maximizes $\EE_{\QQ}[X]$ over $\cl\Mset$, optimality of $\QQ^X_\beta$ and $\kl{\QQ^\dagger}{\Pz}\le\ln(1/\min_j\Pz(\omega_j))$ give $\EE_{\QQ^X_\beta}[X]\ge\EE_{\QQ^\dagger}[X]-\ln(1/\min_j\Pz(\omega_j))/\beta$, which tends to the no-arbitrage bound as $\beta\to\infty$. For $\beta<0$, replace $X$ by $-X$. \Halmos
\end{proof}

The computation applies Proposition~\ref{prop:ec-interval}(d) to the upper end; the lower end is the upper end for $-X$, with the sign changed. Starting from $\beta=1/(\max_jX(\omega_j)-\min_jX(\omega_j))$, it doubles $\beta$ until $\kappa_X(\beta)\ge c$ or $\beta\ge\beta_{\max}$, and each evaluation of $\kappa_X$ computes $\theta_\beta$ by a damped Newton method. If $\kappa_X$ is still below $c$ at the cap, the code treats the ball constraint as inactive and returns the upper no-arbitrage bound $\max_{\QQ\in\cl\Mset}\EE_{\QQ}[X]$, a linear program. Otherwise Brent's method finds the root $\beta_c$ of $\kappa_X(\beta)=c$ on the last bracket, and the end point is $\EE_{\QQ^X_{\beta_c}}[X]$, with the certificate $U_X(\beta_c,\theta_{\beta_c})$. At $c=\cbar$ both end points equal $\EE_{\QE}[X]$, and for $c<\cbar$ the robust price is $\EE_{\Qc}[X]$ from the $\gamma$-search or the joint Newton method. In the synthetic market of Section~\ref{sec:ec-numerics-e2} the gap to the certificates is at most $5.4\times10^{-16}$.

\subsection{The Joint Newton Method}\label{sec:ec-alg-jointnewton}

Algorithm~\ref{alg:jointnewton} maximizes the concave function $\Phi_c$ of~\eqref{eq:ec-dual} by Newton's method in all $d+1$ variables. Write $\PP^h_\lambda\propto\Pz W_h^{-1/\lambda}$ for the tilted model of the certificate (Section~\ref{sec:ec-alg-gamma}), $a=\DeltaS/W_h$ and $\ell=\ln W_h$. Then
\[
\nabla_h\Phi_c=\EE_{\PP^h_\lambda}[a],\qquad \partial_\lambda\Phi_c=\kl{\PP^h_\lambda}{\Pz}-c,
\]
and the Hessian has the blocks $-\EE_{\PP^h_\lambda}[aa^\top]-\mathrm{Cov}_{\PP^h_\lambda}(a)/\lambda$, $\mathrm{Cov}_{\PP^h_\lambda}(a,\ell)/\lambda^2$ and $-\Var_{\PP^h_\lambda}(\ell)/\lambda^3$. For a direction $(u,s)$ the quadratic form equals $-\EE_{\PP^h_\lambda}[(a\cdot u)^2]-\Var_{\PP^h_\lambda}(a\cdot u-s\ell/\lambda)/\lambda$, which shows the concavity directly; it is negative whenever $h\neq0$ and $(u,s)\neq0$, by non-redundancy. Each step costs $O(Md^2+d^3)$ operations, as one step of the CRRA solve of Algorithm~\ref{alg:gammasearch}. The method is Newton's method applied to the KL dual, not a new algorithm; \citet{BartlDrapeauTangpi2020} study the computation of robust optimized certainty equivalents and option prices under divergence penalties.

The warm start comes from the Black--Scholes shapes of Proposition~\ref{prop:bsdoubt}, $h_0=(1-\sqrt{c/\cbar})\hstar$ and $\lambda_0=\sqrt{\cbar/c}-1$. Damped Newton from $(h_0,\lambda_0)$ stays in the superlevel set $\{\Phi_c\ge\Phi_c(h_0,\lambda_0)\}$, which is compact in $\mathcal H\times(0,\infty)$ if $\Phi_c(h_0,\lambda_0)>0$: $\Phi_c(h,\lambda)\le\min_\omega\ln W_h(\omega)-\lambda\ln\min_\omega\Pz(\omega)-\lambda c$ with $\min_\omega\ln W_h(\omega)\le0$, because $W_h$ has mean one under every pricing measure, and, by Jensen's inequality, $\Phi_c(h,\lambda)\le\EE_{\Pz}[\ln W_h]-\lambda c$ with $\EE_{\Pz}[\ln W_h]$ bounded above on $\mathcal H$, so $\Phi_c\to-\infty$ as $\lambda\to\infty$ or as $W_h$ approaches zero in some state. On this set $h\neq0$, because $\Phi_c(0,\lambda)=-\lambda c<0$, so the Hessian is negative definite there, and the standard analysis of damped Newton methods for strongly concave functions applies \citep[Section 9.5]{BoydVandenberghe2004}: the iterates converge to the unique maximizer $(h_c,\lambda_c)$. If instead $\Phi_c(h_0,\lambda_0)\le0$, the iterates may drift toward $\lambda=0$, where $\Phi_c$ tends to $\min_\omega\ln W_h(\omega)$. Line 2 of Algorithm~\ref{alg:jointnewton} then uses the near-threshold start $(\lambda_0h_E,\lambda_0)$, from $h_c/\lambda_c\to h_E$ (Theorem~\ref{thm:robust}(v)); in every run of Section~\ref{sec:ec-numerics-e4} one of the two starts had $\Phi_c>0$, and the code keeps a last fallback (the exact multiplier $\lambda_h$ of the certificate, with $h$ halved until $G_c(h)>0$) that was never needed.

\begin{algorithm}[t]
\caption{Joint Newton method (JN) for $v(c)$}\label{alg:jointnewton}
\begin{algorithmic}[1]\small
\Require $\Pz\in\mathcal P$, $\DeltaS$, radius $0<c<\cbar$, $\hstar$, $h_E$, tolerance $\mathrm{tol}$, cap $N_{\max}$
\Ensure $v(c)\in[G,U]$ with $(U-G)/U\le\mathrm{tol}$, and $(h_c,\lambda_c)$ up to the tolerance
\State $(h,\lambda)\gets\big((1-\sqrt{c/\cbar})\hstar,\ \sqrt{\cbar/c}-1\big)$ \Comment{Black--Scholes warm start}
\State \textbf{if} $\Phi_c(h,\lambda)\le0$ \textbf{then} $h\gets\lambda h_E$ \Comment{safeguard: near-threshold start}
\For{$j=0,1,\dots,N_{\max}$}
  \State $g\gets\nabla\Phi_c(h,\lambda)$; \ $H\gets\nabla^2\Phi_c(h,\lambda)$; \ $\Delta\gets-H^{-1}g$ \Comment{Newton step in $\RR^{d+1}$}
  \If{$g\cdot\Delta\le\mathrm{tol}\,|\Phi_c(h,\lambda)|$}
    \State $G\gets G_c(h)$ and its worst case $\PP^h$ (certificate of Section~\ref{sec:ec-alg-gamma}); \ $U\gets\max_{h'\in\mathcal H}\EE_{\PP^h}[\ln W_{h'}]$
    \State \textbf{if} $(U-G)/U\le\mathrm{tol}$ \textbf{return} $v(c)\in[G,U]$, $h_c\approx h$, $\lambda_c\approx\lambda$
  \EndIf
  \State $t\gets$ largest step $\le1$ with $W_{h+t\Delta_h}>0$ (fraction 0.99 of the boundary) and $\lambda+t\Delta_\lambda\ge\lambda/10$
  \State halve $t$ until $\Phi_c$ rises by at least $0.1\,t\,g\cdot\Delta$; \ $(h,\lambda)\gets(h,\lambda)+t\Delta$
\EndFor
\end{algorithmic}
\end{algorithm}

The certificate of the joint Newton method is two-sided. The lower bound is $G_c(h)\le v(c)$. The upper bound uses the worst-case model $\PP^h$ of the current portfolio, which lies in $\Bc$: by Theorem~\ref{thm:minimax}, $v(c)=\min_{\PP\in\Bc}\sup_{h'}\EE_{\PP}[\ln W_{h'}]\le\sup_{h'}\EE_{\PP^h}[\ln W_{h'}]$, a log-optimal portfolio problem under $\PP^h$. Both bounds tend to $v(c)$ at a quadratic rate in the error of $h$: $G_c$ is smooth and maximal at $h_c$, and $\PP\mapsto\sup_{h'}\EE_\PP[\ln W_{h'}]$ is smooth and minimal over the sphere $\{\kl{\PP}{\Pz}=c\}$ at $\Pc$, which is where $\PP^h$ lies. Section~\ref{sec:ec-numerics-e4} reports the times, the steps and the gaps (Table~\ref{tab:algorithms}).

\subsection{Implementation and the Generic Baseline}\label{sec:ec-alg-impl}

All inner problems are solved by damped Newton methods. Each step is cut so that $W_h$ stays positive in every state and then halved until an Armijo condition holds, and sums of exponentials are computed in log space. The P-step and the certificate use a safeguarded Newton method with bisection fallback; the floor of the certificate is $\lambda_{\min}=10^{-12}$, and the cap of the interval computation is $\beta_{\max}=10^8$. Alternating projections stop when the relative change of $\kl{\PP_j}{\QQ_j}$ is at most $10^{-10}$, after 100,000 iterations, or after 300 seconds; no run in Section~\ref{sec:ec-numerics-e4} reached a cap. The $\gamma$-search and the interval computation use Brent's method with tolerance $10^{-15}$. Algorithm~\ref{alg:jointnewton} uses $\mathrm{tol}=10^{-10}$ and $N_{\max}=100$. The generic baseline minimizes $\kl{\PP}{\QQ}$ over $(\PP,\QQ)\in\RR^{2M}$ subject to $\sum_j\PP(\omega_j)=\sum_j\QQ(\omega_j)=1$, $\EE_{\QQ}[\DeltaS]=0$ and $\kl{\PP}{\Pz}\le c$ with sequential least-squares quadratic programming (SLSQP, in SciPy), exact gradients and the start $(\Pz,\QL(\Pz))$, for $M\le200$ only.

\section{Additional Numerical Results and Code}\label{sec:ec-numerics}\label{sec:numerics-synthetic}

This section reports two synthetic studies, the robustness checks of the empirical results of Section~\ref{sec:numerics}, and the code and data. The synthetic studies, an incomplete market in which a call also trades and a comparison of the algorithms of Section~\ref{sec:ec-algorithms}, use illustrative parameters, taken neither from data nor from a paper.

\subsection{A Synthetic Incomplete Market}\label{sec:ec-numerics-e2}

The first study builds a fat-tailed one-month market in discounted units (bond rate zero, $S_0=1$, $T=1/12$). There are $M=41$ states with monthly log excess returns $y_k=-0.36+0.015k$, $k=0,\dots,40$; state probabilities are bin probabilities, and the two end bins absorb the tails. The reference model $\Pz$ mixes a normal law with mean $m_1=0.010813$ and standard deviation 0.04 (weight 0.9) and a normal law with mean $-0.06$ and standard deviation 0.08 (weight 0.1); $m_1$ solves $\EE_{\Pz}[e^y]=1.005$. A dealer measure $\QQ^{\mathrm{mkt}}\propto\Pz\exp(-0.102307\,y+20\min(y,0)^2)$ makes the stock a martingale and adds weight to losses. Three assets trade: the bond, the stock with gain $e^y-1$, and an at-the-money call at the dealer price $\EE_{\QQ^{\mathrm{mkt}}}[(e^y-1)_+]=0.020462$. So $d=2$, and the market is incomplete. The non-traded claims are a 10\% out-of-the-money put with payoff $(0.9-e^y)_+$ and a digital put that pays one if $e^y<0.9$. Under $\Pz$ the annualized Sharpe ratio is 0.349 and the skewness $-0.632$.

The log-optimal portfolio is $\hstar=(3.0921,-3.6348)$ in the stock and the call, and $\Wstar$ ranges over $[0.1396,1.074]$. The robust value at zero doubt is $v(0)=\kl{\Pz}{\QL}=0.00709158$ per month, and the threshold is $\cbar=\kl{\QE}{\Pz}=0.00816209$, so $v(0)/\cbar=0.8688$; in the Black--Scholes model this ratio is one. The exponential investor who trusts $\Pz$ holds $h_E=(3.872,-4.636)$. Both decompositions of Theorem~\ref{thm:twodecomp} hold for the dealer measure: $\kl{\QQ^{\mathrm{mkt}}}{\Pz}=0.010702=0.00253992+0.00816209$, with residual $-5.274\times10^{-16}$, and $\kl{\Pz}{\QQ^{\mathrm{mkt}}}=0.00778225=0.00709158+6.90674\times10^{-4}$, with residual $7.709\times10^{-17}$. The natural sum $\kl{\Pz}{\QL}+\kl{\QL}{\QQ^{\mathrm{mkt}}}$ exceeds $\kl{\Pz}{\QQ^{\mathrm{mkt}}}$ by $3.485\times10^{-4}$. Table~\ref{tab:ec-e2-measures} compares the pricing measures.

\begin{table}[htbp]
\TABLE{Pricing measures in the synthetic incomplete market.\label{tab:ec-e2-measures}}
{\footnotesize\begin{tabular}{@{}lrrrr@{}}
\toprule
 & $\QL$ & $\QE$ & $\QQ^{\mathrm{mkt}}$ (dealer) & $\Pz$ \\
\midrule
10\% out-of-the-money put (bp) & 24.0 & 22.2 & 28.1 & 12.4 \\
Digital put (\%) & 4.44 & 4.33 & 4.72 & 2.65 \\
Model-free volatility $\sqrt{2\kl{\QQ}{\QS}/T}$ & 19.55\% & 19.36\% & 19.97\% & -- \\
Defect $\defect(\Pz,\QQ,\QS)$ & 0.00532422 & 0.00529318 & 0.00539243 & -- \\
\bottomrule
\end{tabular}}
{Monthly horizon, discounted units. The column $\Pz$ gives expected payoffs without risk adjustment. No-arbitrage bounds: $[0.0,136.9]$ bp for the put and $[0.00,18.10]\%$ for the digital put.}
\end{table}

Table~\ref{tab:ec-e2-path} reports the robust path. The robust value falls from $v(0)$ to zero at $c=\cbar$, and $\gamma_c$ explodes there. The Black--Scholes shapes of Proposition~\ref{prop:bsdoubt} are close in this fat-tailed market: $v(c)$ stays within 2.1\% of $v(0)(1-\sqrt{c/\cbar})^2$, and $\gamma_c$ exceeds $1/(1-\sqrt{c/\cbar})$ by 4.8\%, 9.6\%, 12.2\% and 12.7\% at $c/\cbar=0.1$, 0.5, 0.9 and 0.99. The robust prices move from the $\QL$ prices to the $\QE$ prices and then open into no-trade intervals that widen like $\sqrt{c-\cbar}$. As in the data, the put price falls with doubt below $\cbar$, from 24.0 to 22.2 basis points: the growth-optimal portfolio holds a large position, and $\Wstar=0.1396$ in the worst state, so $\QL=\Pz/\Wstar$ weights crash states more than $\QE$ does. The dealer quotes enter the no-trade interval at $c/\cbar=1.282$ for the put and $1.055$ for the digital put, below $\kl{\QQ^{\mathrm{mkt}}}{\Pz}/\cbar=1.311$: one quote needs less doubt than the whole dealer measure.

\begin{table}[htbp]
\TABLE{The robust path in the synthetic incomplete market.\label{tab:ec-e2-path}}
{\footnotesize\begin{tabular}{@{}rrrrlll@{}}
\toprule
$c/\cbar$ & $v(c)$ & $\gamma_c$ & $\lambda_c$ & $h_c$ (stock, call) & Put (bp) & Digital put (\%) \\
\midrule
0 & 0.0070916 & 1 & $\infty$ & $(3.092,-3.635)$ & 24.0 & 4.44 \\
0.1 & 0.0033429 & 1.532 & 1.8796 & $(2.203,-2.614)$ & 23.3 & 4.40 \\
0.5 & $6.1814\times10^{-4}$ & 3.7418 & 0.36473 & $(0.9813,-1.172)$ & 22.6 & 4.36 \\
0.9 & $1.9052\times10^{-5}$ & 21.865 & 0.04793 & $(0.1755,-0.2101)$ & 22.3 & 4.33 \\
0.99 & $1.8191\times10^{-7}$ & 224.79 & 0.0044685 & $(0.0172,-0.0206)$ & 22.2 & 4.33 \\
1 & 0 & $\infty$ & 0 & $(0,0)$ & [22.2, 22.2] & [4.33, 4.33] \\
1.1 & 0 & $\infty$ & 0 & $(0,0)$ & [18.9, 25.7] & [3.81, 4.85] \\
1.5 & 0 & $\infty$ & 0 & $(0,0)$ & [15.0, 30.1] & [3.18, 5.51] \\
\bottomrule
\end{tabular}}
{Below $\cbar$ the price is the unique robust price $\EE_{\Qc}[X]$; from $\cbar$ on, the no-trade interval of Theorem~\ref{thm:pricing}. For $c\ge\cbar$ we write $\gamma_c=\infty$ and $\lambda_c=0$, since $h_c=0$. Computed by the $\gamma$-search (Algorithm~\ref{alg:gammasearch}) and the interval computation of Section~\ref{sec:ec-alg-interval}.}
\end{table}

The computations pass every check: the multiplier $\lambda_c$ from the $\gamma$-search agrees with $-v'(c)$ from central differences to a relative error of $1.8\times10^{-7}$, and alternating projections and SLSQP reproduce the $\gamma$-search value at $c/\cbar\in\{0.25,0.5,0.75,0.9\}$ to relative errors of at most $6.5\times10^{-12}$ and $9.0\times10^{-12}$ in absolute value. The interval end points agree with SLSQP to $3.7\times10^{-9}$ and with their dual certificates to $5.4\times10^{-16}$, the radius constraint holds to $1.0\times10^{-15}$, and the minimax certificate has a relative gap of at most $9.9\times10^{-10}$ on the whole path.

\subsection{Algorithm Performance}\label{sec:ec-numerics-e4}

The second study compares alternating projections (A1), the $\gamma$-search (A2, Algorithm~\ref{alg:gammasearch}) and the joint Newton method (JN, Algorithm~\ref{alg:jointnewton}) with the generic SLSQP baseline of Section~\ref{sec:ec-alg-impl}. The instances have $M\in\{20,100,1000,10{,}000,100{,}000\}$ scenarios and $d\in\{1,5\}$ assets. The gains are $\DeltaS=e^y-1$, where $y$ is multivariate Student-$t$ with 4 degrees of freedom and correlation 0.5. The monthly volatilities lie between 4\% and 6\%, and the monthly Sharpe ratios between 0.08 and 0.12. The reference model is uniform over the scenarios, a sample-average approximation. With five seeds and $c/\cbar\in\{0.25,0.5,0.75,0.9,0.99\}$ there are 250 runs. A1, A2 and JN ran in one session, so their times are comparable; times are wall-clock seconds of a single process. SLSQP ran in an earlier session on the same instances, only for $M\le200$, that is, for 100 runs. The gap of A1 or SLSQP is $(\text{objective}-v_{\mathrm{ref}})/v_{\mathrm{ref}}$, where $v_{\mathrm{ref}}$ is the A2 value, which the certificate of Section~\ref{sec:ec-alg-gamma} bounds from below; the rerun reproduces every A2 value and certificate of the earlier session exactly.

\begin{table}[htbp]
\TABLE{Algorithm performance: medians over five seeds at $c/\cbar=0.5$ and $0.99$.\label{tab:algorithms}}
{\footnotesize\begin{tabular}{@{}rr rrrr rrrr rrrr@{}}
\toprule
 & & \multicolumn{4}{c}{Alternating projections (A1)} & \multicolumn{4}{c}{$\gamma$-search (A2)} & \multicolumn{4}{c}{Joint Newton (JN)} \\
\cmidrule(lr){3-6}\cmidrule(lr){7-10}\cmidrule(l){11-14}
 & & \multicolumn{2}{c}{Iterations} & \multicolumn{2}{c}{Time (s)} & \multicolumn{2}{c}{CRRA solves} & \multicolumn{2}{c}{Time (s)} & \multicolumn{2}{c}{Newton steps} & \multicolumn{2}{c}{Time (s)} \\
\cmidrule(lr){3-4}\cmidrule(lr){5-6}\cmidrule(lr){7-8}\cmidrule(lr){9-10}\cmidrule(lr){11-12}\cmidrule(l){13-14}
$d$ & $M$ & $c/\cbar = 0.5$ & 0.99 & 0.5 & 0.99 & 0.5 & 0.99 & 0.5 & 0.99 & 0.5 & 0.99 & 0.5 & 0.99 \\
\midrule
1 & 20 & 28 & 1689 & 0.0033 & 0.150 & 11 & 10 & 0.0013 & 0.0012 & 4 & 4 & 0.0004 & 0.0004 \\
1 & 100 & 29 & 1796 & 0.0036 & 0.184 & 14 & 15 & 0.0014 & 0.0015 & 3 & 3 & 0.0004 & 0.0004 \\
1 & 1\,000 & 28 & 1753 & 0.0073 & 0.444 & 15 & 13 & 0.0028 & 0.0026 & 3 & 3 & 0.0007 & 0.0009 \\
1 & 10\,000 & 31 & 1845 & 0.030 & 1.59 & 14 & 10 & 0.014 & 0.013 & 3 & 4 & 0.0029 & 0.0038 \\
1 & 100\,000 & 32 & 1826 & 0.268 & 13.0 & 9 & 10 & 0.102 & 0.105 & 4 & 3 & 0.031 & 0.031 \\
\addlinespace
5 & 20 & 29 & 1710 & 0.0031 & 0.133 & 9 & 8 & 0.0013 & 0.0015 & 4 & 4 & 0.0004 & 0.0005 \\
5 & 100 & 30 & 1820 & 0.0039 & 0.180 & 11 & 11 & 0.0015 & 0.0015 & 4 & 5 & 0.0005 & 0.0006 \\
5 & 1\,000 & 32 & 1885 & 0.012 & 0.561 & 12 & 12 & 0.0039 & 0.0040 & 4 & 4 & 0.0012 & 0.0013 \\
5 & 10\,000 & 32 & 1861 & 0.056 & 2.54 & 12 & 12 & 0.022 & 0.022 & 4 & 3 & 0.0057 & 0.0055 \\
5 & 100\,000 & 34 & 1848 & 0.515 & 20.4 & 11 & 10 & 0.200 & 0.197 & 6 & 3 & 0.063 & 0.044 \\
\bottomrule
\end{tabular}}
{Medians over five seeds, one session for A1, A2 and JN. A2: CRRA solves, including the bracketing step; its time includes one log-optimal solve and no certificate. JN: Newton systems solved in $(h,\lambda)$, each followed by a line search; its time includes its two-sided certificate but not its inputs $\hstar$, $\cbar$ and $h_E$. Section~\ref{sec:ec-numerics-e4} gives the time ratios when both methods do the same work. SLSQP (earlier session, $M\le200$): median time 0.011, 2.02, 0.010 and 1.90 seconds at $c/\cbar=0.5$ for $(d,M)=(1,20)$, $(1,100)$, $(5,20)$ and $(5,100)$.}
\end{table}

Table~\ref{tab:algorithms} gives the results. JN stops with its two-sided certificate in all 250 runs, after a median of 4 Newton steps (range 2 to 7). Its warm start needed the safeguard of line 2 of Algorithm~\ref{alg:jointnewton} in 41 runs, all near the threshold; without the safeguard the method stalls toward $\lambda=0$ in 6 runs, all at $c/\cbar=0.99$. The bounds of both methods are computed in double precision; in exact arithmetic $G\le v(c)\le U$. For both methods, the two bounds agree to $10^{-10}$ in relative terms in all 250 runs, except where $v(c)$ is so small that rounding limits the relative accuracy (one A2 run, with $v(c)=2.5\times10^{-8}$). The computed gaps $U-G$ are at the level of rounding, and they are negative in 93 JN runs and in 131 A2 runs. The bounds of the two methods lie at most $1.6\times10^{-10}$ apart in relative terms, except for $d=1$, $M=100$, seed 2 and $c/\cbar=0.99$, where $\cbar=4.11\times10^{-5}$ and $v(c)$ is about $10^{-9}$: there the JN bounds lie $3.7\times10^{-9}$ above the A2 bounds, so the bounds of at least one method miss the exact value. The computed intervals $[G,U]$ of the two methods are disjoint in 138 of the 250 runs, with a median distance between them of $7.3\times10^{-13}$ relative to $v(c)$, so each pair of bounds holds only up to rounding.

What a time includes matters. In Table~\ref{tab:algorithms} the time of A2 includes one log-optimal solve and no certificate, and the time of JN includes its certificate but not its inputs $\hstar$, $\cbar$ and $h_E$. Timed so, JN takes a median of 0.30 times the time of A2 (range 0.15 to 0.55), and A1 a median of 18 times the time of JN. A second session timed both methods with the same work, each piece as the minimum of three repetitions. With a two-sided certificate in both methods and the shared inputs left out, the median ratio of JN to A2 is 0.29. With every input counted, so that JN includes the minimal-entropy and log-optimal solves and A2 includes its log-optimal solve, the computation of $\cbar$ and its certificate, the median ratio is 0.44 (interquartile range 0.36 to 0.48, range 0.07 to 0.94); in the empirical market of Section~\ref{sec:numerics-put} it is also 0.44. We therefore recommend JN, with A2 as a fallback that needs no safeguard; A1 serves to show the local rate of Proposition~\ref{prop:algorithms}(a).

The number of A1 iterations hardly depends on $M$ or $d$ and follows $\gamma_c$, as Proposition~\ref{prop:algorithms}(a) predicts: a median of 9.32 iterations per unit of $\gamma_c$ (range 5.63 to 10.5), and no run reached a cap. The stopping rule of Section~\ref{sec:ec-alg-altproj} uses the objective, whose error is of second order in the error of the iterates. The reason is that the iterates $\PP_j$, $j\ge1$, lie on the sphere $\{\kl{\PP}{\Pz}=c\}$, over which $\PP\mapsto\min_{\QQ\in\cl\Mset}\kl{\PP}{\QQ}$ is smooth and minimal at $\Pc$ (as for the certificate of Section~\ref{sec:ec-alg-jointnewton}). So the error of the objective shrinks by about $(1-1/\gamma_c)^2$ per iteration, and the count is of order $(\gamma_c/2)\ln(1/\mathrm{tol})\approx11.5\,\gamma_c$ at $\mathrm{tol}=10^{-10}$. A2 needs a median of 12 CRRA solves (range 7 to 28); the six runs with 20 or more solves spend the extra solves on bisection steps of Brent's method at the floating-point limit of the root tolerance $10^{-15}$. The largest relative gap between the A2 value and its lower bound $G_c(h_c)$ is $7.6\times10^{-10}$, and the largest radius error $|c(\gamma_c)-c|$ is $8.2\times10^{-15}$.

The generic baseline is fragile: SLSQP ends without success in 12 of its 100 runs, with the ball constraint violated by up to $5.1\times10^{-9}$, and in one run it reports success at a wrong value, with gap 0.241 ($d=5$, $M=20$, seed 2, $c/\cbar=0.99$). Its other successful runs agree with A2 to $3.3\times10^{-7}$, and representing $(\PP,\QQ)$ as vectors limits its relative accuracy to about $10^{-16}/v(c)$, while A2 and JN represent $\Pc$ through $h_c$. Interior-point solvers for the exponential cone were not tested, so this fragility refers to SLSQP.

\subsection{Robustness of the Empirical Results}\label{sec:ec-numerics-robust}

\textit{Binning.} The cell markets replace the $N=1203$ monthly states by $M$ cells of the empirical distribution of $\ln R$: cells with nearly equal counts, whose boundaries are sample quantiles, or 41 equal-width bins, of which 9 are empty and dropped. Each state value is the mean of $R$ in its cell, and $\Pz$ gives each cell its sample frequency. Binning compresses the tails: with 41 cells of 29 or 30 months, the lowest state is 0.849, while the worst month is 0.713, so the 10\% put and the digital pay only in the lowest cell, and with 11 cells no state lies below 0.90 and both pay nothing. With 41 or more cells the $\QL$ and $\QE$ prices of the 10\% put and of the digital differ from the unbinned ones by up to 15\%, and the no-arbitrage bounds shrink sharply. The full-distribution radius $c_{\mathrm{full}}$ grows with $M$, from $0.75\,\cbar$ at $M=11$ to $62\,\cbar$ unbinned, while $c_{\mathrm{one}}$ stays between $0.158\,\cbar$ and $0.186\,\cbar$. So in these data the regime at $c_{\mathrm{full}}$ switches with $M$: with 11 cells $c_{\mathrm{full}}<\cbar$, so each claim has one robust price at $c_{\mathrm{full}}$, while with 32 or more states $c_{\mathrm{full}}>\cbar$, so a claim that the market does not span has a no-trade interval. The number of cells must therefore be fixed before the data are seen.

\begin{table}[htbp]
\TABLE{The empirical market of Section~\ref{sec:numerics-put} and its binned versions.\label{tab:ec-binning}}
{\footnotesize\begin{tabular}{@{}lrrrrr@{}}
\toprule
 & Unbinned & 11 cells & 41 cells & 81 cells & 41 equal-width \\
\midrule
States $M$ & 1203 & 11 & 41 & 81 & 32 \\
$\hstar$ & 2.25 & 2.78 & 2.46 & 2.37 & 2.26 \\
$v(0)$ and $\cbar$ (\% per month) & 0.821, 0.857 & 0.986, 1.009 & 0.884, 0.914 & 0.855, 0.888 & 0.826, 0.864 \\
Lowest and highest state $R$ & 0.713, 1.388 & 0.905, 1.095 & 0.849, 1.152 & 0.814, 1.196 & 0.713, 1.379 \\
$c_{\mathrm{one}}/\cbar$ and $c_{\mathrm{full}}/\cbar$ & 0.186, 62.2 & 0.158, 0.754 & 0.175, 2.54 & 0.180, 4.77 & 0.185, 2.17 \\
\addlinespace
Put, $K=0.90$: $\QL$, $\QE$ (\%) & 0.231, 0.203 & 0, 0 & 0.197, 0.185 & 0.212, 0.195 & 0.231, 0.203 \\
Put, $K=0.95$: $\QL$, $\QE$ (\%) & 0.632, 0.594 & 0.560, 0.547 & 0.603, 0.579 & 0.618, 0.589 & 0.630, 0.591 \\
Put, $K=1.00$: $\QL$, $\QE$ (\%) & 1.976, 1.939 & 1.947, 1.927 & 1.967, 1.940 & 1.972, 1.941 & 1.975, 1.937 \\
Digital, $R<0.90$: $\QL$, $\QE$ (\%) & 4.102, 3.874 & 0, 0 & 3.841, 3.613 & 3.815, 3.585 & 3.683, 3.453 \\
Call, $K=1.05$: $\QL$, $\QE$ (\%) & 0.342, 0.319 & 0.349, 0.340 & 0.348, 0.333 & 0.346, 0.328 & 0.340, 0.318 \\
Upper no-arbitrage bound, 5\% put (\%) & 13.6 & 2.27 & 5.08 & 6.99 & 13.5 \\
\bottomrule
\end{tabular}}
{Monthly relative returns of the CRSP market over the bill, 1926:07--2026:09 ($N=1203$). Cells: nearly equal counts (109 or 110, 29 or 30, and 14 or 15 months), or 41 equal-width bins on $\ln R$ with 9 empty bins dropped. $c_{\mathrm{one}}=\chi^2_{1,0.95}/(2N)$, $c_{\mathrm{full}}=\chi^2_{M-1,0.95}/(2N)$; at $c_{\mathrm{full}}$ in the unbinned market the interval of the 5\% put is $[0,3.14]$ (\%), down to the no-arbitrage bound.}
\end{table}

\textit{Comparison with Black--Scholes.} In the unbinned market, for $0\le c\le0.99\,\cbar$, $\gamma_c$ and $v(c)$ deviate from their Black--Scholes shapes $1/(1-\sqrt{c/\cbar})$ and $v(0)(1-\sqrt{c/\cbar})^2$ (Proposition~\ref{prop:bsdoubt}) by at most 2.2\% and 2.4\%. For the 5\% put, the Black--Scholes price at the realized volatility lies below the $\QE$ price, with implied doubt $1.46\,\cbar$, and the price at the RMS VIX lies above the $\QL$ price, with implied doubt $1.04\,\cbar$. For the 10\% put, the Black--Scholes prices at the realized volatility and at the RMS VIX (Table~\ref{tab:emp-put}) both lie below even its mean payoff under $\Pz$. Their implied doubt, $2.69\,\cbar$ and $1.67\,\cbar$, measures the difference between lognormal and empirical tails, not doubt in the KL sense. Section~\ref{sec:numerics-put} compares the width of the no-trade interval of the 5\% put at $c=1.5\,\cbar$ with the difference of the Black--Scholes prices at the realized volatility and at the RMS VIX (ratio 1.7). With the mean VIX in place of the RMS VIX, the ratio is 3.7.

\textit{Inference and serial dependence.} Table~\ref{tab:ec-emp-robust} reports the standard errors of Table~\ref{tab:emp-triangle} with other bandwidths and block lengths, the calibration of Table~\ref{tab:emp-doubt} with HAC standard errors, and the within-month products of daily log returns of Section~\ref{sec:numerics-triangle} by subsample. For the $\QQ$-side, whose first-order autocorrelation is 0.76, longer bandwidths raise the standard error from 0.23 to 0.29 points; for $\hat\varepsilon$ they matter little. With a Newey--West standard error, $\tau$ falls from 4.54 to 4.25 (12 lags), and $\gamma_{c_n}$ rises from 1.76 to 1.86. Then $\tau^2=18.0$, still far above $\chi^2_{1,0.95}=3.84$, so the rejection of a zero premium in Section~\ref{sec:numerics-put} does not rest on independent months. The within-month products lie below the deterministic-drift benchmark in every decade except the 1990s, significantly in the 2000s and 2010s, and the effect survives when the highest-variance months are dropped. With returns over $k$ days, the variance ratio rises gradually, so the reversal is not a feature of daily closes alone. Across the 2003 change of the VIX method, $\hat\varepsilon$ is stable while half the variance premium falls; the horizon convention is immaterial.

\begin{table}[htbp]
\TABLE{Inference and serial dependence: robustness of Tables~\ref{tab:emp-triangle} and~\ref{tab:emp-doubt}.\label{tab:ec-emp-robust}}
{\footnotesize\begin{tabular}{@{}lrrrrr@{}}
\toprule
\multicolumn{6}{@{}l}{A. Standard errors of Table~\ref{tab:emp-triangle} (\% per year): Newey--West lags; Andrews bandwidth; block bootstrap}\\
 & 12 & 24 & 36 & Andrews (lags) & Block (length) \\
$\QQ$-side & 0.23 & 0.28 & 0.29 & 0.26 (20) & 0.30 (31) \\
P-term & 2.61 & 2.60 & 2.49 & 2.52 (1) & 2.53 (2) \\
Defect $\hat\defect$ & 2.54 & 2.51 & 2.38 & 2.52 (0) & 2.52 (1) \\
Entropy premium $\hat\varepsilon$ & 0.12 & 0.13 & 0.14 & 0.11 (2) & 0.11 (2) \\
\addlinespace
\multicolumn{6}{@{}l}{B. Table~\ref{tab:emp-doubt} at $z=1.96$ with HAC standard errors of the mean excess return}\\
 & $\tau$, i.i.d. & $\tau$, NW 12 & $\tau$, Andrews & $\gamma_{c_n}$: i.i.d., NW 12, Andrews & $\gamma_{c_n}$ over $\tau\pm1.96$ \\
1926:07--2026:09 & 4.54 & 4.25 & 4.39 & 1.76, 1.86, 1.81 & 1.43 to 4.15 \\
1990:01--2026:09 & 3.57 & 3.47 & 3.57 & 2.22, 2.29, 2.22 & 1.55 to $\infty$ \\
\addlinespace
\multicolumn{6}{@{}l}{C. Within-month products of daily log returns minus the deterministic-drift benchmark}\\
 & Months & Mean (\% per year) & NW $t$ & & \\
Full sample & 440 & $-0.533$ & $-2.90$ & & \\
1990s, 2000s & 119, 120 & $0.026$, $-1.174$ & 0.31, $-2.39$ & & \\
2010s, 2020s & 120, 81 & $-0.359$, $-0.758$ & $-2.52$, $-1.63$ & & \\
Without top 1, 5 RV months & 439, 435 & $-0.451$, $-0.334$ & $-2.66$, $-3.52$ & & \\
Without top 10, 20 RV months & 430, 420 & $-0.249$, $-0.237$ & $-3.61$, $-3.90$ & & \\
\addlinespace
\multicolumn{6}{@{}l}{D. Returns over $k$ days in realized variance}\\
 & $k=1$ & $k=2$ & $k=5$ & $k=10$ & \\
$\widehat{\Var}(\ln R)/\overline{\mathrm{RV}_k}$ & 0.708 & 0.757 & 0.803 & 0.899 & \\
Half variance premium (\% per year) & 0.53 & 0.64 & 0.73 & 0.88 & \\
\addlinespace
\multicolumn{6}{@{}l}{E. Split at the 2003 change of the VIX method, and the horizon convention}\\
 & Months & $\hat\varepsilon$ (\%) & Half variance premium (\%) & $\hat\defect$ (\%) & \\
VIX months to 2003:08 & 164 & 1.08 & 0.89 & 8.22 & \\
VIX months from 2003:09 & 276 & 0.99 & 0.32 & 11.28 & \\
Actual days between closes & 440 & 1.020 & & 10.137 & \\
\bottomrule
\end{tabular}}
{Panel A: Newey--West \citep{NeweyWest1987} with 12, 24 and 36 lags; the automatic bandwidth of \citet{Andrews1991} for the Bartlett kernel with an AR(1) plug-in, with the number of lags it implies; moving-block bootstrap \citep{Kunsch1989} with the automatic block length of \citet{PolitisWhite2004}, 10,000 draws; delta-method rows use their linearized series. Panel B: $\tau=\SR\sqrt n$ with the i.i.d. or a HAC standard error; at $\tau-1.96\le1.96$ the robust investor holds no stock ($\gamma_{c_n}=\infty$). Panel C: products $\sum_{i<j}r_ir_j$ of daily log returns $r_i$ within a month, minus their mean under a deterministic drift (Section~\ref{sec:ec-numerics-vrp}), with the drift and the variance proxies estimated in each subsample. Panel D: $\mathrm{RV}_k$ sums the squared log returns over consecutive blocks of $k$ trading days within the month. Panel E: VIX months up to August 2003 against September 2003 on; last row: $T_{t+1}$ from the actual days between the VIX close of month $t$ and the last close of month $t+1$ (28 to 33 days, mean 30.4), against 1.021 and 10.139 with calendar months.}
\end{table}

\textit{Expanding windows.} We apply the calibration of Section~\ref{sec:numerics-doubt} to windows from July 1926 to December $Y$, for $Y=1950,\dots,2025$. At $z=1.96$ the robust investor holds no stock for $Y=1950$ and 1951, and their effective risk aversion exceeds 20 in the next two windows. As data accumulate, $\gamma_{c_n}$ falls, though not monotonically, and reaches 1.77 for data up to December 2025; the robust weight, 1.39 at the end, rises toward the log-optimal weight, 2.45, but stays well below it. At $z=1.645$ and $z=1$ the robust investor holds stock in every window.

\begin{figure}[htbp]
\FIGURE{\includegraphics[width=0.74\textwidth]{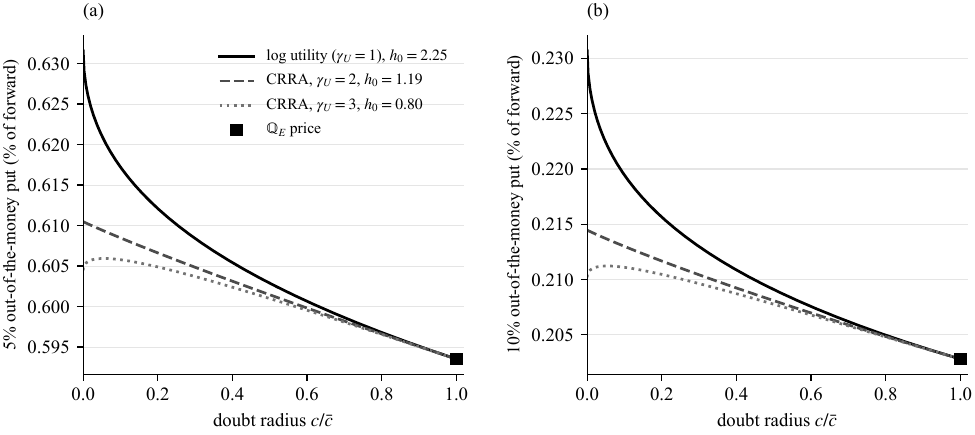}}
{Robust put prices for log and CRRA base utilities, empirical market.\label{fig:ec-crra-path}}
{The market of Section~\ref{sec:numerics-put}. (a) Put with $K=0.95$; (b) put with $K=0.90$. Robust price below $\cbar$ for log utility ($\gamma_U=1$) and for CRRA utilities with relative risk aversion $\gamma_U=2$ and 3, which hold $h_0$ times wealth in the stock at $c=0$; all paths end at the $\QE$ price (square).}
\end{figure}

\textit{A CRRA base utility.} Figure~\ref{fig:ec-crra-path} repeats the robust path of Section~\ref{sec:numerics-put} for an investor with CRRA utility $U(W)=W^{1-\gamma_U}/(1-\gamma_U)$, $\gamma_U\in\{2,3\}$, and the same KL ball. By Proposition~\ref{prop:utilityfree}(d) the robust portfolio $h^U_c$ is unique, and the unique no-trade price below $\cbar$ is $\EE_{\PP^U_c}[U'(W^U_c)X]/\EE_{\PP^U_c}[U'(W^U_c)]$. The worst-case model has the Gibbs form $\PP^U_c\propto\Pz e^{-U(W^U_c)/\theta_c}$ (Lemma~\ref{lem:ec-rob-gibbs}(c), as in the proof of Proposition~\ref{prop:utilityfree}(d) in Section~\ref{sec:ec-rob-utilityfree}). We compute $(h^U_c,\theta_c)$ from the jointly concave dual $\max_{h,\theta>0}\{-\theta\ln\EE_{\Pz}[e^{-U(W_h)/\theta}]-\theta c\}$ of the same lemma: for fixed $\theta$ the maximization over $h$ is concave, and the optimal $\theta$ solves $\kl{\PP_\theta}{\Pz}=c$ for the worst case $\PP_\theta$ at the optimal $h$. At $c/\cbar=0.25$ and 0.75, a direct search over $h$ of the worst-case utility $\min_{\PP\in\Bc}\EE_\PP[U(W_h)]$ agrees with the dual value to a relative $10^{-12}$. At $c=0$ the investor holds 1.19 ($\gamma_U=2$) and 0.80 ($\gamma_U=3$) times their wealth in the stock, against 2.25 with log utility. As $c$ rises to $\cbar$, their position falls to zero, and the robust price of the 5\% put falls to the $\QE$ price, 0.594\%: monotonically from 0.611\% for $\gamma_U=2$, and from 0.604\% for $\gamma_U=3$ after a rise of about 0.14 basis points that peaks near $c=0.06\,\cbar$; the 10\% put behaves in the same way. So the fall below $\cbar$ is not specific to log utility in this market, but its size shrinks with the leverage of the base investor: 3.8, 1.7 and 1.1 basis points for $\gamma_U=1$, 2 and 3. This is a numerical result in one market, not a theorem. The out-of-the-money call with $K=1.05$ also falls with doubt, from 0.342\% under $\QL$ to 0.319\% under $\QE$ with log utility, so in this market both tails lose weight below $\cbar$.

\textit{The VIX reading: bootstrap.} The intervals of Section~\ref{sec:numerics-vix} and Table~\ref{tab:emp-bridge} come from a circular block bootstrap with 12-month blocks and 2,000 draws. With the 1990--2026 reference model, each draw redraws blocks of the 440 pairs $(\VIX_t,R_{t+1})$, so the reference model and the quote move together. With the century model, each draw redraws blocks of the 1203 months, and the quote is the mean of $\frac12\VIX_t^2T_{t+1}$ over the drawn months that are panel return months. Each draw recomputes $\Pz$, $\cbar$, the range of robust prices (over $1\le\gamma\le10^4$ and the limit $\QE$) and $\cbar_X$ at the drawn quote. The 95\% intervals of $\EE_{\QL}[X]$ and $\EE_{\QE}[X]$ are 0.088\% to 0.138\% and 0.082\% to 0.132\% per month for 1990--2026, and 0.124\% to 0.198\% and 0.114\% to 0.187\% for 1926--2026; in the century draws the quote lies above the range in 1,614 of the 2,000 draws, inside it in 213 and below it in 173. Three checks with 1,000 draws each agree on the side of the quote, though independent blocks widen the 1990--2026 gap interval to as much as 13\% to 126\%: independent 12-month blocks for $\Pz$ and for the quotes, 31-month blocks for the quotes (the block length of \citet{PolitisWhite2004} for the $\QQ$-side), and joint 31-month blocks. In these checks the quote lies above the range in at least 99.3\% of the draws for 1990--2026 and in 75.5\% to 79.4\% for 1926--2026. The quote also lies above the range of robust prices with 41 cells of nearly equal counts, where its implied doubt is $2.20\,\cbar$ for 1926--2026 and $7.65\,\cbar$ for 1990--2026.

\textit{The variance of the reference model.} The variance of monthly $\ln R$ is 0.708 times the mean daily realized variance $\overline{\mathrm{RV}}$ in 1990--2026 (Section~\ref{sec:numerics-triangle}) and 1.10 times $\overline{\mathrm{RV}}$ in 1926--2026. The variance-matched reference model keeps the mean of $\ln R$ and rescales its deviations so that the variance of $\ln R$ under $\Pz$ equals $\overline{\mathrm{RV}}$ of the same months. Its bootstrap redraws each month with its realized variance and recomputes the rescaling in each draw. With 1990--2026 the gap falls from 60\% to 17\% (95\% interval 0.2\% to 36\%, also with joint 31-month blocks), and the quote lies above the range in 1,951 of the 2,000 draws, inside it in 36 and below it in 13. The share above, 97.55\%, lies within its Monte Carlo standard error, 0.35 percentage points, of the 97.5\% that sets the lower end of the 95\% interval. So the lower end, 0.2\%, is zero within Monte Carlo error, and this gap is at the edge of significance. For 1990--2026, $\cbar_X(\bar p)/\cbar$, with $\cbar$ the threshold of the variance-matched model, is 1.59 (interval 1.03 to 4.91), and the quote-specific part is 0.59 (0.03 to 3.91). With 1926--2026 the rescaling lowers the variance, and the gap rises from 13\% to 23\% (95\% interval $-3\%$ to 56\%), with $\cbar_X(\bar p)=1.62\,\cbar$ in the same units. In price units the gaps are 0.81\% and 0.25\% per year with the empirical reference models and 0.31\% and 0.40\% with the variance-matched ones. So the size of the gap and of the quote-specific part depends mainly on the variance of $\Pz$, while with the 1990--2026 months the quote stays above the range in most draws.

\textit{The VIX month by month.} Section~\ref{sec:numerics-vix} also reads the quote $p_t=\frac12\VIX_t^2T_{t+1}$ month by month, against a reference model for month $t+1$ built in real time from the months $s\le t$ from October 1926 on. The variance forecast for month $s$, $\hat\sigma^2_s$, is the mean realized variance of the three months before $s$. With $\hat m_t$ the mean of $\ln R_s$ over $s\le t$, the shocks $\xi_s=(\ln R_s-\hat m_t)/\hat\sigma_s$ are standardized to mean zero and unit sample variance, $\tilde\xi_s$, and the model has the states $\ln R=\hat m_t+\varpi_t\hat\sigma_{t+1}\tilde\xi_s$, $s\le t$, with equal probabilities. It forecasts the variance $\varpi_t^2\hat\sigma^2_{t+1}$, with the scale $\varpi_t$ fitted to the last 120 months $s\le t$: $\varpi_t^2=\sum_s(\ln R_s-\hat m_t)^2/\sum_s\hat\sigma^2_s$. For each panel month we compute $\cbar_t$, the range of robust prices below $\cbar_t$ and the signed gap of the quote: $p_t/\max-1$ above the range, 0 inside it and $p_t/\min-1$ below it. Over the 440 panel months, the mean of $(\ln R_{t+1}-\hat m_t)^2$ is 0.91 times the mean forecast. The quote lies above the range in 369 months, inside it in 29 and below it in 42, with a median signed gap of 38\%. With the RiskMetrics forecast instead, an exponentially weighted average of daily squared log returns (decay 0.94), scaled by the number of trading days of the month, the ratio is 0.92 and the counts are 407, 2 and 31. With the scale fitted to the last 60 months, they are 0.91 and 376, 26 and 38. An in-sample check fits the scale to the panel months themselves, so that the ratio is one, and takes $\hat m$ and the standardized shocks from the whole sample; its counts are 366, 42 and 32.

The window of the scale was chosen by this calibration check, which uses returns only, not the quotes. With the scale fitted to all months $s\le t$, the forecast is too high after 1990, when monthly variance is smaller than the mean daily realized variance (Section~\ref{sec:numerics-triangle}). The ratio is then 0.58, and the quote lies above the range in 222 months, inside it in 67 and below it in 151, with a median signed gap of 0.3\%. With $\varpi_t=1$ in real time, the forecast is the mean daily realized variance of the three previous months. The ratio is then 0.71, the counts are 289, 62 and 89, and the median signed gap is 19\%. The scale $\varpi_t$ fitted to the last 120 months has medians 1.08, 0.90, 0.75 and 0.85 in the 1990s, 2000s, 2010s and 2020s, above one in the 1990s and below one from 2000 on. So this scale carries the within-month reversal of daily returns into the forecast, while the daily-variance forecast leaves it out. Shocks that are neither standardized nor scaled ($\varpi_t=1$, with $\hat m$ and the shocks from the whole sample) give a ratio of 0.44 and put the quote below the range in 267 months. So the side of the monthly quote depends on the calibration of the variance forecast; with a nearly unbiased forecast the quote lies above the range in most months, as in the unconditional reading with the 1990--2026 sample.

\begin{figure}[htbp]
\FIGURE{\includegraphics[width=0.72\textwidth]{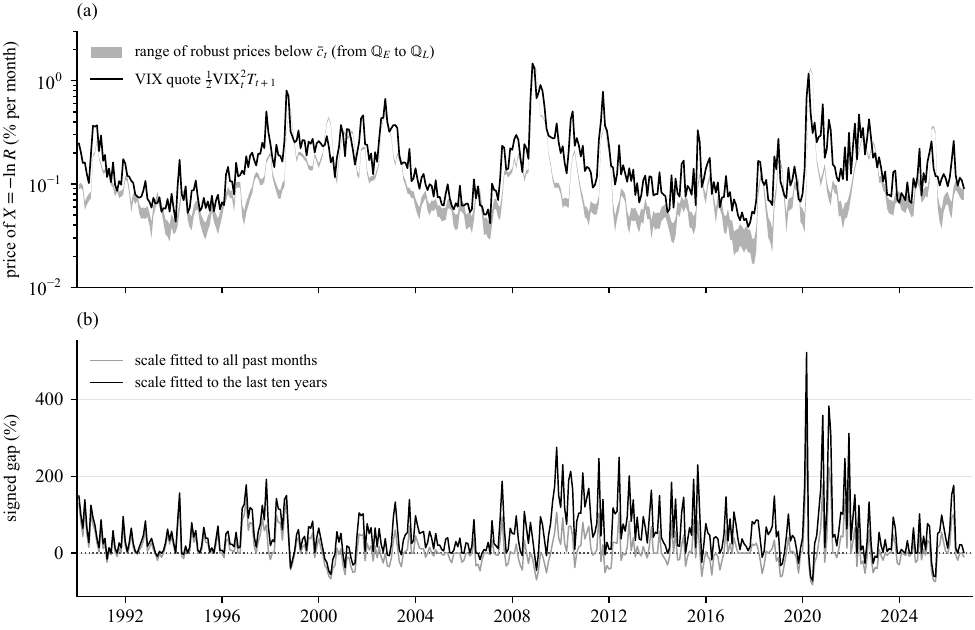}}
{The VIX as a quote of the short log contract, month by month, 1990--2026.\label{fig:ec-vix-monthly}}
{(a) The quote $\frac12\VIX_t^2T_{t+1}$ and the range of robust prices of $X=-\ln R$ below $\cbar_t$, from $\EE_{\QE}[X]$ to $\EE_{\QL}[X]$, in the real-time reference model of month $t+1$ with the scale fitted to the last ten years (\% per month, log scale). (b) The signed gap of the quote, $p_t/\max-1$ above the range, 0 inside it and $p_t/\min-1$ below it, in \%, with the scale fitted to the last ten years (black) and to all past months (gray).}
\end{figure}

\textit{Implied doubt of put quotes.} Implied doubt (Definition~\ref{def:implieddoubt}) converts a price into a radius, also for the puts of Section~\ref{sec:numerics-put}. In general, if the robust prices move monotonically from $\EE_{\QL}[X]$ to $\EE_{\QE}[X]\neq\EE_{\QL}[X]$, implied doubt is $0$ at $p=\EE_{\QL}[X]$, but close to $\cbar_X(\EE_{\QL}[X])>\cbar$ for quotes just outside this range on the side of $\EE_{\QL}[X]$ (Section~\ref{sec:rob-pricing}). Below $\cbar$, the robust 5\% put price falls monotonically, so only quotes in $(0.594\%,0.632\%]$ have implied doubt below $\cbar$. This mapping jumps at the $\QL$ price: a quote 1 basis point below the $\QL$ price has an implied doubt of $0.045\,\cbar$, while a quote just 0.01 basis points above the $\QL$ price has an implied doubt of $1.03\,\cbar$.

\textit{Coverage of the radius rules.} A bootstrap checks the full-distribution rule and the value bound of Section~\ref{sec:rob-radius} when returns depend on each other over time. It treats the empirical distribution of the 1203 months as the true model and redraws 1203 months 10,000 times, either independently or in circular blocks of 12 or 36 consecutive months; circular blocks wrap around the end of the sample, so every month has the same weight, and keep the dependence within each block \citep{PolitisWhite2004}. In each draw the reference model is the empirical distribution of the redrawn months. The robust value $v(c_{\mathrm{one}})$ of the log investor lies at or below the optimal growth under the true model, $v(0)=0.00821$, in 97.1\% of the independent draws, close to the asymptotic level of 97.5\% for this one-sided bound, and in 95.4\% and 94.2\% of the block draws. The full-distribution ball is built on 41 cells, the equal-count cells of the sample, held fixed across draws as cells fixed before the data would be. It contains the true cell probabilities in 92.9\% of the independent draws, against an asymptotic level of 95\%, but only in 84.9\% and 74.9\% of the block draws; with 11 cells the three shares are 94.9\%, 86.3\% and 76.4\%. So volatility clustering lowers the coverage of the full-distribution rule much more than that of the value bound, and, as the binning shows, the regime at $c_{\mathrm{full}}$ switches with the number of cells. With one state per month, no draw contains every month, so no KL ball around a redrawn reference model contains the true model. Moving blocks \citep{Kunsch1989}, which give the months near the ends of the sample less weight, change these shares by at most 3.1 percentage points.

\subsection{The Entropy Premium and Half the Variance Premium}\label{sec:ec-numerics-vrp}

This subsection splits the full-sample difference of 0.49\% per year between $\hat\varepsilon$ and half the variance risk premium (Section~\ref{sec:numerics-triangle}), which equals $\frac12\overline{\mathrm{RV}}-\hat\Lent_{\PP}$ exactly, into two parts. The higher-cumulant part, $\frac12\widehat{\Var}(\ln R)-\hat\Lent_{\PP}$, is 0.012\% per year. The serial-dependence part, $\frac12(\overline{\mathrm{RV}}-\widehat{\Var}(\ln R))$, is 0.477\%, because the variance of monthly log returns is 0.708 times the mean daily realized variance, a variance ratio in the sense of \citet{LoMacKinlay1988}, here below one. The monthly returns come from the monthly French file and the realized variances from the daily file; in the panel months the monthly market return equals the return compounded from the daily file up to 0.133\% (mean absolute difference 0.042\%). Up to this difference, the monthly log market return is the sum of the daily log returns $r_i$ of the month, so its square equals the realized variance plus twice the sum of the products $r_ir_j$ over pairs of days; negative products therefore make the monthly variance smaller than the mean realized variance. Under a deterministic drift of $\dd S/S$, with any stochastic volatility and any jumps, the martingale parts of daily log returns are uncorrelated within the month, but each day's log drift contains the random convexity term $-\sigma_t^2/2$ minus the jump compensator. After a price fall the variance rises (the leverage effect), so the within-month covariances $\Gamma=\sum_{i<j}\mathrm{Cov}(r_i,r_j)$ are positive, 0.0587\% per year with daily squared returns as variance proxies. The serial part implied by a deterministic drift is therefore the drift term 0.0023\% minus $\Gamma$, that is $-0.0565\%$ per year. The same model gives the mean of the within-month products of daily log returns as their drift term plus $\Gamma$, $0.104\%$ per year. In the full sample these products are instead negative on average, $-0.429\%$ per year (Newey--West $t$-statistic $-2.90$ for the difference), so the deterministic drift is rejected. Daily returns partly reverse within the month, which is consistent with a drift that depends on earlier daily returns of the same month. The 10 highest-variance months carry a share of 0.60 of these products, but the mean stays negative without them, and the $t$-statistic stays below $-2.6$ when the 1, 5, 10 or 20 highest-variance months are dropped, so the reversal is not only a crisis effect (Table~\ref{tab:ec-emp-robust}).

Three caveats apply to the measurement of Section~\ref{sec:numerics-triangle} besides the index mismatch. First, the Cboe formula approximates the forward price of the short log contract with finitely many strikes and two expirations, and the link assumes a deterministic dividend yield (Remarks~\ref{rem:ec-dividends} and~\ref{rem:ec-cboe}). Second, $\hat\pi=\ln\overline{R}$ uses the unconditional mean return, so by Jensen's inequality, $\ln\EE[\EE_{\PP_t}R_{t+1}]\ge\EE[\ln\EE_{\PP_t}R_{t+1}]$, it estimates an upper bound on the average conditional log equity premium. Since $\hat\defect$ estimates the average defect itself, $\hat\varepsilon=\hat\defect-\hat\pi$ estimates a lower bound on the average conditional entropy premium. Third, $\mathrm{RV}$ uses daily close-to-close CRSP returns, not intraday S\&P~500 data.

\subsection{Code and Data}\label{sec:ec-code}

\emph{Code.} All computations use Python 3.14.7 with NumPy 2.5.4, SciPy 1.18.1 and Matplotlib 3.11.2, on an arm64 machine running macOS. Our code contains the geometry, solvers and algorithms, continuous-time closed forms, the synthetic studies, the empirical pipeline and the unit tests. Twenty-two unit tests check exact identities of Sections~\ref{sec:setting}--\ref{sec:robust} in finite and closed-form models, with tolerances between $10^{-15}$ and $10^{-3}$. A further 18 tests check solvers and bootstrap code of Sections~\ref{sec:ec-algorithms} and~\ref{sec:ec-numerics-robust}; all 40 pass. One command rebuilds the synthetic study of Section~\ref{sec:ec-numerics-e2}. Another rebuilds the empirical results of Tables~\ref{tab:emp-triangle} and~\ref{tab:emp-doubt} from the raw files in about two minutes, except the HAC column and the Andrews values of Table~\ref{tab:emp-doubt}, and the Politis--White values of Table~\ref{tab:emp-triangle}. The script of Table~\ref{tab:ec-emp-robust} rebuilds these values, and other scripts rebuild the rest. Every computed number in the paper is read from the saved outputs of these scripts. A separate script, with its own file parsers, Newey--West and bootstrap code, and solvers, independently replicates Tables~\ref{tab:emp-triangle} and~\ref{tab:emp-doubt} (without these values), the expanding-window calibration and the 41-cell market of Table~\ref{tab:ec-binning} (without the call prices). It runs 1,439 checks, none of which fails.

\emph{Data.} The empirical results use two public sources: the daily VIX history published by Cboe \citep{CboeVIXHistory2026}, and the monthly and daily Fama--French three-factor files of the Kenneth R.~French Data Library \citep{FrenchDataLibrary2026}, built from the September 2026 vintage of the CRSP database. Cboe adopted the present VIX method in September 2003 and computed the earlier values retroactively \citep{CarrWu2006}. It has changed details since, such as the selection of strikes \citep{Cboe2026}, so the VIX sample spans several methodology vintages. In the French files the one-month Treasury bill return comes from Ibbotson Associates until May 2024 and from the ICE BofA 1-Month Treasury Bill Index afterwards. Both sources were downloaded on October 29, 2026. In two months of the panel of Section~\ref{sec:numerics-data}, January 1997 and December 1999, the last VIX date is one trading day before the last CRSP trading day, and $\VIX_t$ is then that earlier close. We do not redistribute the raw files, which carry their providers' terms. We recorded the SHA-256 hashes of the files we used, and the separate script checks them at every run. The French files are revised from time to time, so a new download can change the numbers slightly.

\end{document}